\documentclass[usenatbib]{mnras}

\usepackage[T1]{fontenc}
\usepackage{ae,aecompl}
\usepackage{graphicx}
\usepackage{amsmath,amssymb}
\usepackage{color,ulem}

\definecolor{webgreen}{rgb}{0,.5,0}
\definecolor{webbrown}{rgb}{.6,0,0}

\defcitealias{cav09}{C09}
\defcitealias{pan14}{P14}

\hypersetup{%
   colorlinks=true,%
   breaklinks=true,%
   plainpages=false, bookmarksnumbered, bookmarksopen=true,
   bookmarksopenlevel=1,%
   urlcolor=webbrown, linkcolor=blue, citecolor=webgreen,
   }

\begin{document}
\title [Decoding the centres of clusters using MCMC] 
{Decoding X-ray observations from centres of galaxy clusters using MCMC}
\author[Kiran Lakhchaura, Tarun Deep Saini and Prateek Sharma]
       {Kiran Lakhchaura$^{1}$\thanks{E-mail : kiru111184@gmail.com}, 
        Tarun Deep Saini$^{1}$\thanks{E-mail : tarun@physics.iisc.ernet.in} and Prateek Sharma$^{1}$\thanks{E-mail : prateek@physics.iisc.ernet.in}\\
$^1$   Department of Physics,
Indian Institute of Science, Bangalore 560012, India \\
} 
\maketitle

\begin{abstract} 
Traditionally the thermodynamic profiles (gas density, temperature, etc.) of galaxy clusters are obtained by assuming spherical symmetry and
modeling projected X-ray spectra in each annulus. The outer annuli contribute to the inner ones and their contribution needs to be subtracted to obtain
the temperature and density of spherical shells. The usual deprojection methods lead to propagation of errors from outside to in and typically do not
model the covariance of parameters in different radial shells.
In this paper we describe a method based on a free-form model of clusters with cluster parameters (density, temperature) given in spherical shells, 
which we {\it jointly} forward fit to the X-ray data by constructing a Bayesian posterior probability distribution that we sample using the MCMC technique. 
By systematically marginalising over the nuisance outer shells, we estimate the inner entropy profiles of clusters and fit them to various models for a sample 
of Chandra X-ray observations of 17 clusters. We show that the entropy profiles in almost all of our clusters are best described as cored power laws. 
A small subsample is found to be either consistent with a power law, or alternatively their cores are not fully resolved (smaller than, or about few kpc). 
We find marginal evidence for bimodality in the central values of entropy 
(and cooling time) corresponding to cool-core and non cool-core clusters. The minimum value of the ratio of the cooling time and the free-fall time 
(min[$t_{\rm cool}/t_{\rm ff}$]; correlation is much weaker with core entropy) is anti-correlated with $H\alpha$ and radio luminosity. $H\alpha$ emitting cold 
gas is absent in our clusters with min$(t_{\rm cool}/t_{\rm ff})\gtrsim 10$. Our lowest core entropies are systematically and substantially lower than 
the values quoted by the ACCEPT sample.
\end{abstract}


\section{Introduction} 

Observations of large scale distribution of galaxies (e.g., \citealt{col01}) and numerical simulations of gravitational structure formation (e.g., \citealt{dav85}) show that the 
mass in the Universe is distributed in
the form of a cosmic foam, with pancakes, filaments, voids, and groups and clusters. Galaxy clusters, which form at the intersections of massive cosmological 
filaments, are the most massive, spherical, virialized objects in the Universe. Galaxy clusters, containing up to 1000s of mostly red galaxies, lie at the exponential
end of the halo mass distribution, and hence their abundance is a sensitive function of cosmological parameters such as the amplitude of perturbations at 
recombination and the cosmological matter fraction (e.g., \citealt{eke96}; see \citealt{all11} for a recent review).

Galaxy cluster observations also shed light on galaxy formation, in particular the role of the extended hot halo gas, which is left behind as a consequence of galaxy
formation (\citealt{whi78}). Unlike galaxy clusters in which the hot gas density is high, the hot gas halo is very difficult to observe in lower mass halos. However, it is
present, and is expected to be spread out beyond the viral radius, accounting for most of the galactic `missing' baryons (e.g., \citealt{and10,sha12b}). The diffuse 
hot halo is also a reservoir from which the cold gas needed for ongoing star formation is accreted, and into which metals due to supernovae are deposited (see \citealt{put12} for a review).

The galaxy cluster core, where the cooling time of the hot gas is shorter than the Hubble time, is most interesting from the perspective of galaxy formation.
Observations show that cluster cores exist in at least two states: cool-cores (CC) and non-cool-cores (NCC). Cool cores have short ($\lesssim$ 1 Gyr) cooling 
times (e.g., \citealt{cro08,cav09,hud10}) and non cool-cores have long cooling times ($\gtrsim $ 1 Gyr ). Some authors also talk about intermediate or weak cool-core clusters (e.g., \citealt{san08,ros10}). There is some merit in 
studying the entropy of hot gas because (at least in simple 1-D models; this breaks down in 2/3-D as mixing, condensation and dropout have a non-trivial impact 
on hot gas entropy) it only changes at the virial shock and in the central cooling/heating regions (e.g., \citealt{toz01,voi03}).  The core entropy is also a measure
of preheating of the intergalactic medium (IGM) at the epoch when the central gas was accreted into dark matter halos (e.g., \citealt{kai91}).

There is disagreement in the literature on the nature of entropy profiles in cluster cores. The extreme views being that all clusters have constant entropy cores 
(\citealt{cav09}; $K\equiv T_{\rm keV}/n_e^{2/3} = K_0$ at smallest radii; $T_{\rm keV}$ is the temperature in keV and $n_e$ is the electron number density) 
and all centres have power law entropy profiles (we call them cusps, $K \propto r^{\alpha1}$ at 
smallest radii; \citealt{pan14}). Of course, latter is a more general form and includes former as a special case ($\alpha_1=0$). The slope of the entropy/density profile
in the core constrains the models of active galactic nucleus (AGN) feedback in cluster cores. For example, a decreasing entropy towards the centre implies that 
the Bondi accretion rate onto the black hole is higher; similar conditions may also lead to (the more plausible) cold feedback (\citealt{piz05}), triggered when the ratio
of the cooling time and the free-fall time ($t_{\rm cool}/t_{\rm ff}$; $t_{\rm cool}\equiv 1.5p/[n_en_i \Lambda]$, where $p$ is gas pressure, $n_e/n_i$ is electron/ion number density and $\Lambda$ is the temperature-dependent cooling function; $t_{\rm ff} \equiv (2r/g)^{1/2}$ is the local free-fall time) becomes $\lesssim 10$ (\citealt{sha12a,gas13}). 
Also, there are predictions from numerical simulations for the hot gas 
entropy profile; e.g., \citet{pra15} suggest that the entropy profile for lower core entropy is steeper at $\sim 10$ kpc scales (top-left panel of 
their Fig. 11; middle panel of our Fig. \ref{fig:flat_core_prof_full_sampl} shows some evidence for this). 
Some papers have argued that $t_{\rm cool}/t_{\rm ff}$ (\citealt{mcc12,sha12a,voi15}) is the physical parameter that 
delineates hot gas halos in which cold gas condensation, star-formation and 
AGN activity occur from those in which they are absent. Such theoretical predictions can be tested by appropriately analyzing the X-ray data from cluster 
cores and by comparing with cold gas and radio observations.

Since the clusters are seen in projection, to reconstruct the three dimensional thermodynamic profiles of a cluster requires deprojection of data. To keep the 
reconstruction bias free, a free-form model of the cluster (that allows $n_e$ and $T$ to vary arbitrarily with radius) is preferred in comparison to analytical fits. 
The cluster can be divided into $N$ spherical shells that are seen in projection as $N$ annuli. The number density and temperature in the shells are 
the $2N$ parameters of the free-form fit to the cluster (we also try some models in which the shell elemental abundance is kept as a free parameter; see 
Appendix \ref{S:optmz_jmcmc}). The inverse method \citep[e.g.,][]{russ08} employs the elegant idea that since the last radial annulus draws 
photons only from the last shell, the last shell parameters can be obtained directly. Then subtracting the contribution of the last shell from the next inner shell, 
one can sequentially obtain the parameters of each shell. The number of parameters fitted at any step is just two (three if elemental abundance is 
allowed to vary in shells), so this method is very efficient.  
Although this method works well for most cases, 
one downside of this method is that the errors in the parameters are not calculated jointly for all shells. For example, outer shell spectra
affect the density and temperature of inner shells but not vice versa. 
Oscillations in deprojected temperature 
sometimes seen due to poor quality X-ray spectra may be manifestations of biased temperatures estimated in outer shells affecting 
the inner shells.

To avoid this, in this paper, we {\it jointly fit the parameters of all the shells to the cluster data through forward fitting}. This increases the number of parameters considerably. We construct the Bayesian posterior probability distribution function for the $2N$ parameters of a cluster. To sample the posterior probability distribution, we use Markov Chain Monte Carlo (MCMC) method and extract the most likely density, temperature, entropy, etc. profiles in cluster cores and their (co)variances. We call our method jMCMC (`j' stands for the joint fitting of all shell parameters). Since the central thermodynamic profiles for clusters are the most interesting for studying cooling and star formation in clusters, we marginalize over model parameters at outer radii to obtain much more precise estimates in the core ($R \sim$ 10 kpc). 

Like all such models, we make the assumption that the X-ray emitting gas is spherically symmetric. This is manifestly a simplification as most cool-core clusters (which are generally more relaxed) show X-ray cavities and radio bubbles (see \citealt{mcn07} for a review). Quantifying systematic bias introduced due to the assumption of spherical symmetry is left for future.

We present our method and its tests in section \ref{S:method}; some tests to optimize our method on the test cluster are shown in Appendix \ref{S:optmz_jmcmc}. Section \ref{S:samp_res} presents our sample and results from from fitting flat core entropy profiles.
Single and double power law profiles are presented in Appendix \ref{S:sgl_powlaw_fit} \& \ref{S:dbl_powlaw_fit}. Readers not interested in technical details may
directly proceed to section \ref{S:discussion} in which we discuss astrophysical implications of our results. We conclude in section \ref{S:conclusions}.

\section{Method}
\label{S:method}

The X-ray data comprise photon counts at a position on the sky as a function of frequency. We divide the projected counts into $N$ annuli and $M$ spectral channels. The number of spectral channels are chosen to ensure that each channel gets a minimum of 25 counts to ensure Gaussian statistics for the photon noise. We define the radius $R_i$ of an annulus as the distance from the centre of the cluster to the centre of the annulus. The three dimensional spherical shells are assigned the same radii as the annuli. The annulus/shell number increases from the centre outwards. The observed counts in the $i^{\rm th}$ annulus is given by
\begin{equation}
{\bf C}_{iJ}= {\bf D}_{iJ} - {\bf B}_{iJ}\,.
\end{equation}
In our notation the lower case Roman letters denote the annuli and the upper case Roman letters denote the spectral bins. The quantities ${\bf D}_{iJ}$ and ${\bf B}_{iJ}$ are the counts detected in the spectrum in the $J^{\rm th}$ spectral bin of the $i^{\rm th}$ annulus of the source, and the corresponding weighted spectrum from the blank sky observation, respectively. 

The photon count (${\bf C}_{iJ}$) in each annulus is a sum of contributions from different spherical shells of radius larger than that of the 
particular annulus. Since the outermost annulus receives photons only from the last shell, its photon count can be taken to be the projected photon count of the 
outermost shell. The standard deprojection technique relies on removing the spectral 
contributions of the outer shells from the given annulus with the appropriate volume factors to obtain the photon count of the shell. 
The procedure is then applied iteratively to the inner annuli to obtain the count rates corresponding to individual shells. The individual 
shells can then be fitted to obtain the gas number density $n_i$ and temperature $T_i$
in the $i^{\rm th}$ shell. 

Since in this method the errors in the photon-count in the outer shells systematically propagate inward, the resulting individual shell counts 
are actually correlated with each other. However, it is difficult to take this correlation into account, since every $i^{\rm th}$ shell spectrum is fitted 
individually for $n_i$ and $T_i$, and not collectively as ${\bf n}, {\bf T}$, where ${\bf n}$ and ${\bf T}$ are the $N$-dimensional 
density and temperature vectors. The usual practice is to use Monte Carlo methods to estimate the range of 
$n$ and $T$ in each shell allowed by the data. 

Since we are primarily interested in the central regions of galaxy clusters, it is useful to be able to systematically isolate the inner 
shells by statistically marginalising over the outer shells to obtain the best possible estimate for the inner shells (the choice of inner shells is described 
in section \ref{S:mcmc_method}). For this we first need 
to construct the likelihood function for data based on the free-form model ${\bf n}, {\bf T}$. Then using Bayesian statistics, we can construct 
the posterior probability for the model parameters.  

To construct the probability distribution function for the unknown parameters of the shell $n_i, T_i$, we begin with the expression for the 
{\it model} of the counts for each annulus through 
\begin{equation}
\begin{split}
{\bf M}_{iJ} = \sum_{k=i}^{N} \sum_{E} {\dot{\bf C}^{\rm th}}(&{n_k},T_k,E) 
R_k(E,J) A_{k}(E) V_{\rm proj}(i,k)\; \Delta t\,,
\end{split}
\end{equation}
where $\dot{{\bf C}}^{\rm th}(n_{k},T_{k},E)$ is the model photon count flux for the $k^{\rm th}$ shell at energy $E$ per unit time per unit source volume 
for gas density $n_{k}$ and temperature $T_{k}$. $\dot{{\bf C}}^{\rm th}$ is calculated using the {\it wabs} photoelectric absorption model \citep{mor83} 
and the Astrophysical Plasma Emission Code \citep[{\it apec};][]{smi01} in the X-ray spectral analysis package \citep[XSPEC; see][]{arn96}. 
We have used Xspec Version 12.8.2 and Atomdb version 2.0.2 for all our analysis. The solar elemental abundance tables 
used for the apec model are from \cite{and89} and those for the wabs model are from \cite{and82}.
The redshift (z), elemental abundance ($Z$) and 
neutral hydrogen column density along the source direction ($N_{H}$), are assumed to be constant (in section \ref{S:optmz_jmcmc} we show that fixing the
elemental abundance across shells does not affect our density and temperature determination). 
$R_k(E,J)$ is the probability of a photon with energy $E$ to be detected in the detector channel $J$, and $A_k(E)$ is the effective area of the 
detector at energy $E$ for the $k^{\rm th}$ annulus. $ V_{\rm proj}(i,k)$ is the projected volume of the $k^{\rm th}$ shell intercepted by the $i^{\rm th}$ annulus 
$(k \geq i)$, and $\Delta t$ is the exposure time of the observation.    
\begin{equation}
V_{\rm proj}(i,k)=V(i,k+1) -V(i+1,k+1)-V(i,k)+V(i+1,k),
\label{eq:vol}
\end{equation}
where $V(i,k)=(4\pi/3)(R_{ko}^2 - R_{ii}^2)^{3/2}$, $R_{ii}$ and $R_{ko}$ being the inner and outer radii of the $i^{\rm th}$ and $k^{\rm th}$ shells, 
respectively \citep[see][]{kri83}. 

In terms of the model parameters ${\bf n}, {\bf T}$, the $\chi^2$  is given by
\begin{equation}
\large{ { \large \chi^2 ({\bf n}, {\bf T})} = \large\sum_{i=1}^{N} \large\sum_{J} \left ( \frac{{\bf C}_{iJ} - {\bf M}_{iJ}} {\sigma_{iJ}} \right)^{2}}\\
\end{equation}
where $\sigma_{iJ} = \sqrt{{\bf D}_{iJ}  + {\bf B}_{iJ} }$ is the Poisson error in the counts in the $J^{\rm th}$ channel of the $i^{\rm th}$ annulus 
in the source observation. In this expression the summation over $J$ matches the observed spectrum for the $i^{\rm th}$ annulus, and 
the summation over $i$ gives the combined match for the full data across all radial bins. The Bayesian posterior probability distribution 
function (PDF) of the model is then given by 
\begin{equation}
P( {\bf n}, {\bf T} \, | \, {\bf D}) \propto \exp \left( -\frac{\chi^2 ({\bf n}, {\bf T}) }{2}  \right) P( {\bf n}, {\bf T} )\,,
\label{eq:pdf}
\end{equation}
where $P({\bf n}, {\bf T})$ is the prior PDF for the model parameters. Since we {\it jointly} fit all shell parameters and use MCMC to sample 
the PDF, we call our approach joint-MCMC or jMCMC.

\subsection{MCMC method}
\label{S:mcmc_method}

We are specifically interested in the behaviour of entropy close to the centre of galaxy clusters, therefore we divide the radial bins roughly into central 
and outer parts, and marginalise over the parameters of the outer part. The observed entropy profiles show departure from single power law (straight line in a
log-log plot) both at small and large radii. We retain all the annuli in the interior of a cluster until the annulus beyond which they depart from a power 
law and treat them as the inner region, which we fit to various entropy models (eqs. \ref{eq:flat_core_model}-\ref{eq:dbl_model}).

Given the large number of radial bins, it is 
hard to use eq.~\ref{eq:pdf} directly, since marginalisation through direct integration of parameters of the outer part of the cluster is computationally expensive. 
A popular method for marginalising is the Markov-Chain Monte Carlo (MCMC) technique using the Metropolis-Hastings algorithm \citep{met53, hast70}. This method 
employs a random walk through the parameter space while maintaining the condition of detailed balance \citep[see, e.g.,][for details]{press03}. After 
some burn-in steps that are discarded, the random walk starts sampling the underlying PDF, so the parameter values from the random walk 
behave as if they were drawn from the PDF. The random walk produces a chain of points in the parameter space 
${\bf n_{\tau}}, {\bf T_{\tau}}$, where $\tau$ denotes the order in the chain. These samples (chains), being drawn from the target PDF (eq.~\ref{eq:pdf}), 
can be used to compute various integrals over the PDF.
From the chain in the parameter space, we estimate the parameters
$n_i, T_i$ for the $i^{\rm th}$ shell by computing the sample means $(\bar{n}_i, \bar{T}_i)$. For example
$$\bar{n}_i = \frac{1}{N_s} \sum_{\tau=1}^{N_s} n_{i\tau}\,, $$
where $N_s$ is the number of samples. Similarly, (co)variances between the shell parameters can also be calculated.

\subsubsection{Implementing MCMC}
\label{S:MCMC}
The calculation of model ${\bf M}_{iJ}$ using XSPEC makes the MCMC runs slow. Fortunately, for most clusters $n$ and $T$ fall into a well-known narrow 
range. Our prior probability function  $P({\bf n}, {\bf T})$ (eq. \ref{eq:pdf}) is chosen to be uniform in the box defined by 
$0.0001 < n_i < 1$ and $0.1 < T_i < 15$, where the 
units of $n$  and $T$ are cm$^{-3}$ and keV, respectively. The number density $n$ is divided into $500$ logarithmic intervals, and $T$ is divided into 
$150$ linear intervals. The grid size for ${\bf n}, {\bf T}$  is much smaller than the typical error bars on the parameter values. We pre-calculate 
the template of ${\dot{\bf C}^{\rm th}}$ on this grid. 

We use Metropolis-Hastings sampling along one parameter at a time for generating chains, so the random walk is constructed by updating 
individual parameters sequentially. 
The proposal density along any single parameter being updated is chosen to be Gaussian
\begin{equation}
q(x) = \frac{1}{\sqrt{2\pi} \sigma} \exp \left[-\frac{(x_\tau-x_{\tau-1})^2}{2\sigma^2} \right]\,.
\end{equation}
Here $x$ denotes any single parameter of the full set of parameters $n_i, T_i$;  $x_\tau$ is the updated value and $x_{\tau-1}$ is the current value. At any 
given step, we update the value of a single parameter,  keeping all the rest fixed, by generating a random number from the proposal distribution given above. 
The proposed point is retained in the chain with the probability 
\begin{equation}
p = \min \left [ 1, \frac{P(s_{k,\tau} ; {\bf n}_{\tau-1}, {\bf T}_{\tau-1} \, | \,{\bf D})}{P(s_{k,\tau-1} ; {\bf n}_{\tau-1}, {\bf T}_{\tau-1} \, | \,{\bf D})}  \right ]\,.
\end{equation}
In this expression we have chosen $s_k$ as the parameter being updated and ${\bf n}_{\tau-1}$ and ${\bf T}_{\tau-1}$ are all the other parameters not equal to $s_k$;
$P$ is the posterior PDF (eq. \ref{eq:pdf}). A comma 
has been added in the subscript for greater clarity. If the point is accepted, then we replace $s_{k,\tau}$ with the nearest grid value in our template, since our 
template ${\dot{\bf C}^{\rm th}}$ are calculated on a grid. The procedure is then repeated for all the other parameters sequentially to construct the chain.

Since our parameter space is compact, if the proposal point falls out of the box then we use periodic boundary condition to bring it back into the box. 
A moment's reflection shows that this does not affect the symmetry of the proposal density $q(x|x^*) = q(x^*|x)$, which is required to keep the condition for detailed balance. Although different $\sigma$s were chosen
for the number density and temperature, their values were kept the same across the radial bins. Their values were chosen to ensure an acceptance rate of roughly 20--30 per cent, and typical $\sigma$ values are larger than the grid size in $n$ and $T$.

\subsubsection{Central entropy profile}
\label{S:cent_entr_prof}
Any physical quantity that depends explicitly on thermodynamic variables can be estimated in a manner similar to the one described above to 
estimate $n_i, T_i$ by using the original chain in the parameter space to construct a subsidiary chain through
\begin{equation}
f_\tau = f(n_{i{\tau}}, T_{i{\tau}})\,,
\end{equation}
where the function can depend on the full range of parameters or on only a few of them. As an example, the entropy in each radial bin can be estimated by constructing the following subsidiary chain ($n_e$ is electron density simply related to the total number density once elemental abundance is known)
\begin{equation}
K_{i{\tau}} = T_{i{\tau}} n_{e, i{\tau}}^{-2/3}.
\end{equation}   

To isolate the behaviour of entropy near the centre of a cluster, we focus our attention on a few central annuli (for how these are chosen, 
see the first paragraph of section \ref{S:mcmc_method}), say $i=1, N_c$, where $N_c < N$. 
We then estimate the 
expectation values and covariance matrix for these entropies $\bar{K}_i$ and ${\rm cov(i,j) \equiv \rm cov}(K_i, K_j)$. In principle we can now plot $\bar{K}_i$ with 
error bars to display the range of entropy profiles allowed by the data. However, this is not optimal, and we find it useful to condense the large information available 
in the covariance matrix by fitting the entropy to an analytic profile. We find that the clusters analyzed by us display two kinds of behaviour: a) central cores and 
b) central cusps. Therefore, we choose the following models for the central entropy profile: (i) a flat core profile after \citealt{cav09}
\begin{equation}
K(R) = K_0 + K_{100} \left(\frac{R}{100~\rm{kpc}} \right)^\alpha;
\label{eq:flat_core_model}
\end{equation}
(ii) a single power law profile
\begin{equation}
K(R) = K_1 \left(\frac{R}{100~\rm{kpc}} \right)^{\alpha_1};
\label{eq:spl_model}
\end{equation}
and (iii) a double power law profile
\begin{equation}
K(R) = K_1 \left(\frac{R}{100~\rm{kpc}} \right)^{\alpha_1} + K_2 \left(\frac{R}{100~\rm{kpc}} \right)^{\alpha_2} .
\label{eq:dbl_model}
\end{equation}

At this point we treat $\bar{K}_i$ as the observed values of entropy. To fit the analytic expressions to these values, we construct the Fisher matrix from the inverse 
of the covariance matrix calculated above, $ F_{ij} = {\rm cov}(i,j)^{-1}$. The $\chi^2$ for the proposed model is then given by
\begin{equation}
\label{eq:fisher}
\chi^2 = \sum^{N_c}_i \sum^{N_c}_j (K(R_i) - \bar{K}_i) F_{ij}(K(R_j) - \bar{K}_j)
\end{equation}
Using this and an expression equivalent to eq.~\ref{eq:pdf}, a Bayesian posterior PDF for the fit parameters (e.g., $K_0$, $K_{100}$ and $\alpha$ for the flat core profile) is obtained 
through a second MCMC analysis. Using the full covariance 
matrix takes into account the radial variation of entropy in individual chains, something that is not captured by using only the marginalised errors on the individual 
shell entropies.

\subsection{Benchmarking the jMCMC method}
\label{S:sim_data_and_test_cluster}

We tested our jMCMC analysis on simulated as well as actual data. Simulated projected spectra were generated using XSPEC. We also used the \textit{Chandra} 
observations of the cool-core cluster A2597 to further test the usefulness of our method for actual data. The procedure used for generating the simulated data, 
and the Chandra observations of A2597 are described in the following sections.

\subsubsection{Simulated Spectra}
\label{S:sim_spectra}

\setcounter{figure}{0}
\begin{figure*}
\begin{center}
\includegraphics[width=0.45\linewidth]{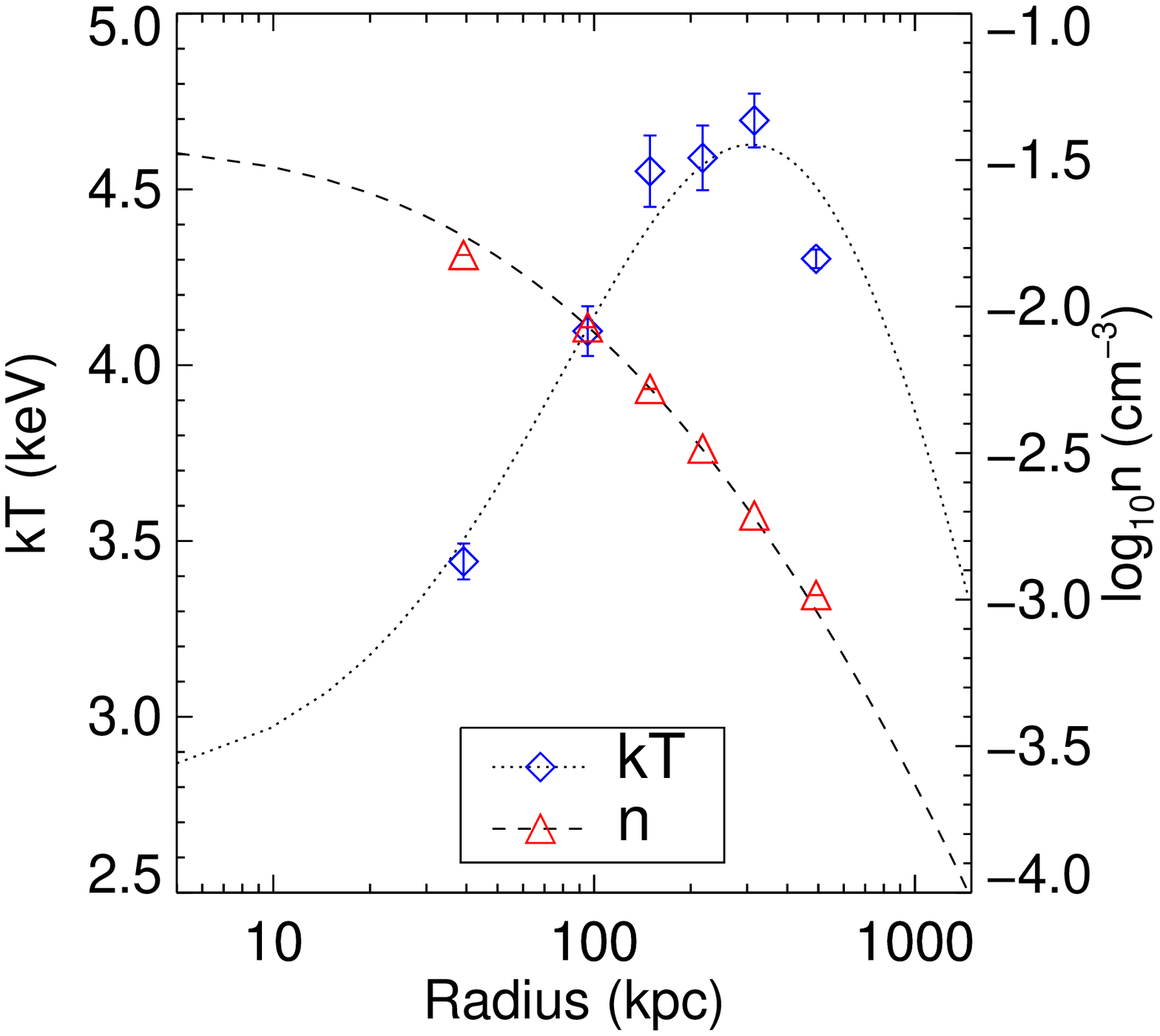}
\includegraphics[width=0.45\linewidth]{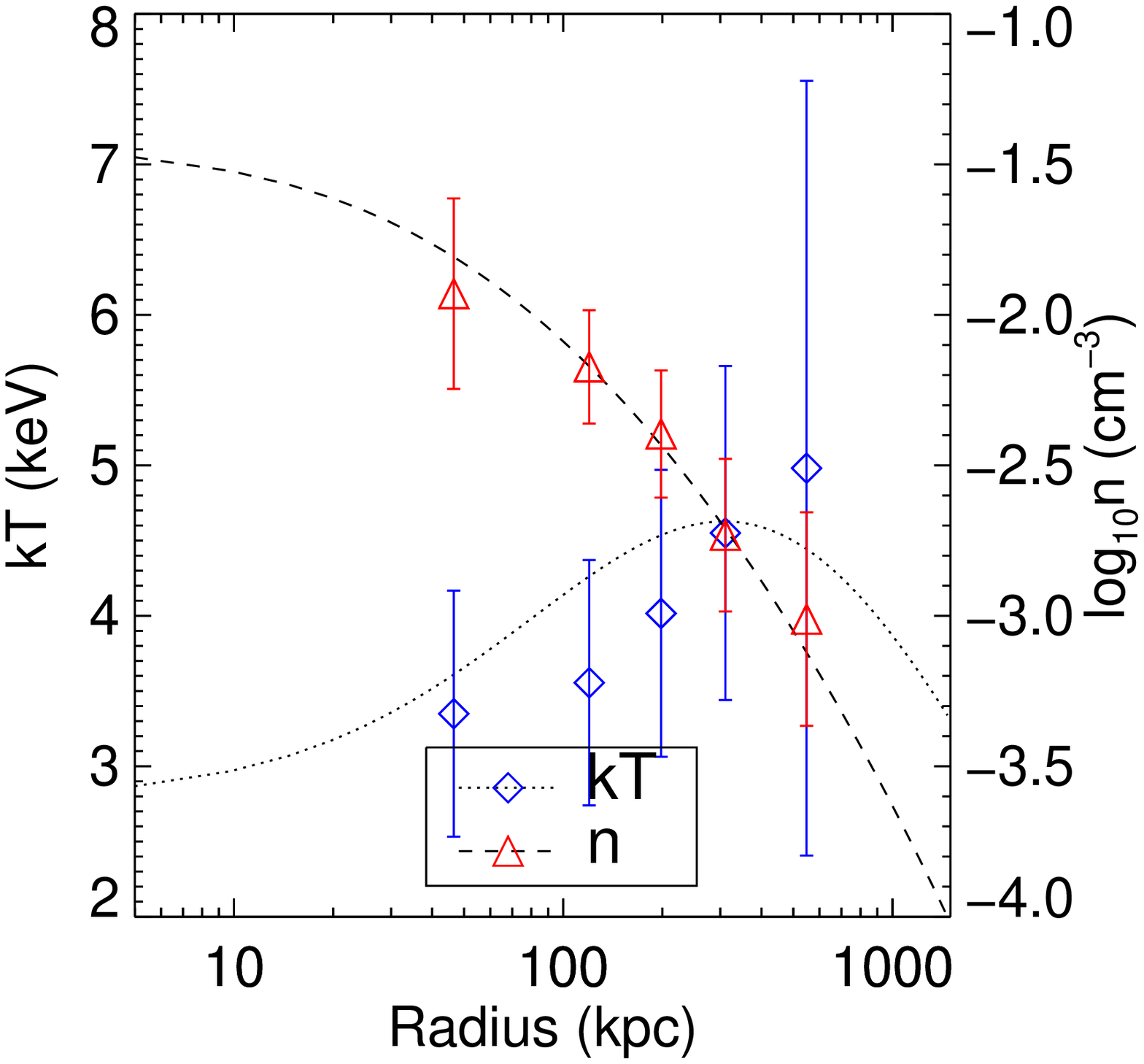}
\caption{Density (red triangles) and temperature (blue diamonds) profiles obtained from the jMCMC analysis of the 100 ks (left) and 20 ks (right) 
simulated cluster spectra described in section \ref{S:sim_spectra}. For comparison the input density and temperature profiles used for 
generating the simulated spectra are also shown.}
\label{fig:100ks_n_T_profs}
\end{center}
\end{figure*}

Simulated spectra for spherically symmetric shells of a cluster were generated using the \textit{apec} model and the 
\textit{fakeit} command in XSPEC. Profiles of intracluster gas density and temperature  (shown with lines in Fig. \ref{fig:100ks_n_T_profs}) 
were generated assuming an NFW dark-matter potential (with concentration parameter $c_{200}=3.3$, $M_{200}=5.24 \times 10^{14} M_\odot$; \citealt{nav96}) 
and a flat-core entropy profile (eq. \ref{eq:flat_core_model}; 
$K_0=37.9$ keV cm$^2$; $K_{100}=117.9$ keV cm$^2$ and $\alpha=1.11$). These density and temperature profiles 
were then used as inputs for simulating spectra from spherical shells, assuming a constant value of density and temperature  in each shell. 
The elemental abundance was frozen to a constant value of one-third solar for all the shells, and a redshift (z) value of 0.05 was assumed for the cluster. 
The instrumental response files (Redistribution Matrix Files and Ancillary Response Files) required by the \textit{fakeit} command were 
produced using the CIAO task \textit{specextract} 
for an actual \textit{Chandra} observation with \textit{weight$=$no}. 
Spectra were simulated for an exposure time of 100 ks. 
The gas number density, $n(R)$, was used for obtaining the normalization of the APEC model ($\eta$), and the two are related through
\begin{equation}
 \eta = \frac{10^{-14} \int{n_e n_p dV} }{4 \pi {D_{A}}^{2} (1+{\rm z})^2},
\end{equation}
where $D_A$ is the angular diameter distance, $n_e$ and $n_p$ are the electron and proton number densities, and for one-third solar abundance 
$n_e=0.53n$ and $n_p=n_e/1.2$. Assuming a 
constant gas density in each shell, the integral in the above equation reduces to ${n_e}^2 V/1.2$; where $V$ is the volume of the shell.

 As the observed emission from an astronomical object is always seen in  projection along the line of sight, we produced projected spectra for all the 
annuli. For the $i^{\rm th}$ shell the projected spectrum was obtained by combining the simulated spectra from all the outer $k \geq i$ shells,  
weighted according to their projected volumes intercepted by the $i^{\rm th}$ shell (eq. \ref{eq:vol}). The projected spectra 
were combined in radial bins in order to contain 
a minimum of 10000 counts in each annulus. For each spectrum, channels were combined together in spectral bins so that each bin contains a minimum of 
25 counts, so that the assumption of a $\chi^2$-distribution with Gaussian errors is valid. We also added a background to all spectral bins 
($=20\%$ of the source spectra with 
Poisson fluctuations) to the individual annuli spectra. The backgrounds (without the fluctuations) were then subtracted from the annuli spectra and the 
errors associated with background subtraction were propagated in the jMCMC analysis.

\subsubsection{Test Cluster}
\label{S:test_cluster}

\begin{table}
 \caption{Cluster sample.}
\label{tab:clus_samp}
\vskip 0.5cm
\centering
{\scriptsize
\begin{tabular}{c c c c c}
\hline
Cluster & z&$\alpha$(J2000), $\delta$(J2000)&Obsn.&Exp.\\
Name&&&ID&Time (ks)\\
\hline
\hline
  A85&0.0557&00 41 37.8, -09 20 33&15173&43.08\\
 A133&0.0603&01 02 39.0, -21 57 15&2203&35.91\\
 A478&0.0881&04 13 20.7, +10 28 35&1669&42.94\\
A1650&0.0823&12 58 36.76, -01 43 34.2&5823&40.13\\
A1795&0.0625&13 49 00.5, +26 35 07 &493&19.88\\
A2029&0.0775&15 10 58.7, +05 45 42&4977&78.91\\
A2142&0.0899&15 58 20.00, +27 14 00.3&15186&91.07\\
A2204& 0.151&16 32 45.7, +05 34 43&7940&78.16\\
A2244&0.0996&17 02 34.01, +34 04 41.1&4179&57.72\\
A2597$^\dag$&0.0824&23 25 19.70, -12 07 27.7&7329&60.90\\
A3112&0.0759&03 17 52.4, -44 14 35&13135&42.80\\
Hydra-A&0.0549&09 18 05.65, -12 05 44&576&19.52\\
 A754&0.0535&09 08 50.1, -09 38 12&577&44.77\\
A2256&0.0581&17 03 43.5, +78 43 03&16129&44.49\\ 
A3158&0.0583&03 42 39.6, -53 37 50&3712&31.35\\
A3667&0.0552&20 12 33.68, -56 50 26.3&5751&130.60\\
ZWCL1215&0.0767&12 17 40.6, +03 39 45&4184&12.22\\
\hline
$^\dag$ Test cluster
\end{tabular}}
\end{table}  

We tested our jMCMC method for actual X-ray data using \textit{Chandra} observations of the test cluster A2597. 
The rationale behind selecting A2597 as the test cluster is that it is a bright cool-core cluster with a deep (60.9 ks) 
\textit{Chandra} observation. 
A log of the \textit{Chandra} Observation of A2597 is given in Table \ref{tab:clus_samp}. We also performed various tests on A2597 to optimize the jMCMC 
algorithm, which are discussed in  Appendix \ref{S:optmz_jmcmc}.

The 
data for this observation were obtained from the High Energy Astrophysics Science Archive Research centre (HEASARC). The CIAO version 4.7 and CALDB 
version 4.6.7 were used for analyzing the data. The data were reprocessed using the standard \textit{chandra\_repro} tool to produce the level 2 
reduced event files (evt2) from the level 1 event files (evt1). Background event files matching with the source observation were produced using 
Maxim Markevitch's blank-sky background database\footnote{see, \url{http://cxc.cfa.harvard.edu/contrib/maxim/acisbg}}. The source observation was cleaned 
of any flare contamination using the \textit{lc\_clean} script so as to match the 
blanksky background maps. Point sources were removed based on a visual inspection of the images. 
Spectra, background spectra and corresponding response files were 
generated for a number of circular annuli centred on the cluster's X-ray peak using the CIAO task \textit{specextract}. The width of the annuli 
was chosen such that each annulus had at least 10000 counts in the 0.7-7.0 keV energy range\footnote{For some of the clusters analyzed in this paper,
this condition was slightly relaxed in order to have at least one annulus within $R\leq$10 kpc while for some of the bright clusters the 
condition was more constrained and the minimum counts were increased in order to have not more than 25 annuli in the cluster.}.

\textit{Background Handling:} The background in the blanksky and source observations is different due to both spatial and temporal variations of its 
different components. The major component of the background is due to the events produced by charged particles that dominates the $>$2keV 
energy range. The difference in the blanksky and source background due to the temporal variation of this component was corrected for by scaling 
the blanksky observations by the ratio of the source and blanksky count rates in the 9.5-12.0 keV energy range, where the Chandra effective area 
is close to zero. Another important background component is the Galactic diffuse soft X-ray background (SXB) that dominates below 1 keV. We 
checked for the SXB fluxes for all our clusters using ROSAT R45 (0.47-1.21 keV energy band) count rates just outside the clusters. In the 
faintest outermost annuli of two of our clusters (A2204 and A1650) the R45 SXB fluxes were found to be a significant fraction ($>$15\%) of the 
total 0.47-1.21 keV fluxes. For these clusters, SXB was modeled using the procedure described in \citet{vik05}. For this, we extracted source and 
blanksky spectra from an off-axis region. The blanksky spectra were rescaled to match particle background in the
source spectra. The residual (source minus rescaled blanksky) spectra from the off-axis regions were modeled using unabsorbed solar abundance 
mekal models. Then for both the clusters, the mekal models were subtracted from the individual annuli spectra (after properly scaling the mekal 
normalizations for the annuli areas). The spectral channels for all annuli spectra were combined in bins to have minimum 25 counts in each bin.

\subsubsection{Cluster Sample}
\label{S:clus_sampl}
  We carried out the jMCMC analysis on the \textit{Chandra} X-ray observations of a sample of clusters selected 
  from the statistically complete low-redshift sample of \cite{san10}. Our sample includes 17 clusters and spans 
  a redshift range of 0.0535 to 0.151. The sample and the details of the observations used, are  given in Table 
  \ref{tab:clus_samp}. The X-ray data was analyzed as described in the previous section, and the results are 
  discussed in section \ref{S:samp_res}. Results for some of the clusters from the low-z sample of \cite{san10} are not 
  shown in this paper mostly due to poor count statistics in their X-ray observations leading to too few annuli in the 
  entropy profiles.

\subsection{Results for simulated spectra and A2597}
\label{S:results}
jMCMC analysis was carried out for the simulated and actual data as described in sections \ref{S:mcmc_method}  \& \ref{S:sim_data_and_test_cluster}. 
\label{S:sim_spec_res}
Fig.~\ref{fig:100ks_n_T_profs} (left) shows the density and temperature 
profiles obtained from the jMCMC analysis of the 100 ks simulated spectra. For comparison the actual input profiles are also shown. 
The jMCMC profiles are found to be consistent with the input profiles, and are unbiased. We also tested our method 
for simulated spectra produced using a lower exposure time of 20 ks. Due to fewer counts available, the minimum counts per spectrum for this 
analysis had to be reduced to 2500. The resulting temperature and density profiles and their comparison 
with the input profiles, are shown in Fig.~\ref{fig:100ks_n_T_profs} (right). Although the results are consistent with the input profiles,  
a lower exposure time leads to larger errors. This is expected as there are fewer counts per spectral bin. The density, temperature 
and entropy profiles obtained for the test cluster are shown in black color in Fig.~\ref{fig:test_clus_different_method}.

\subsection{Comparison with other methods} 
\label{S:method_comp}

\setcounter{figure}{1}
\begin{figure*}
 \includegraphics[width=0.33\linewidth]{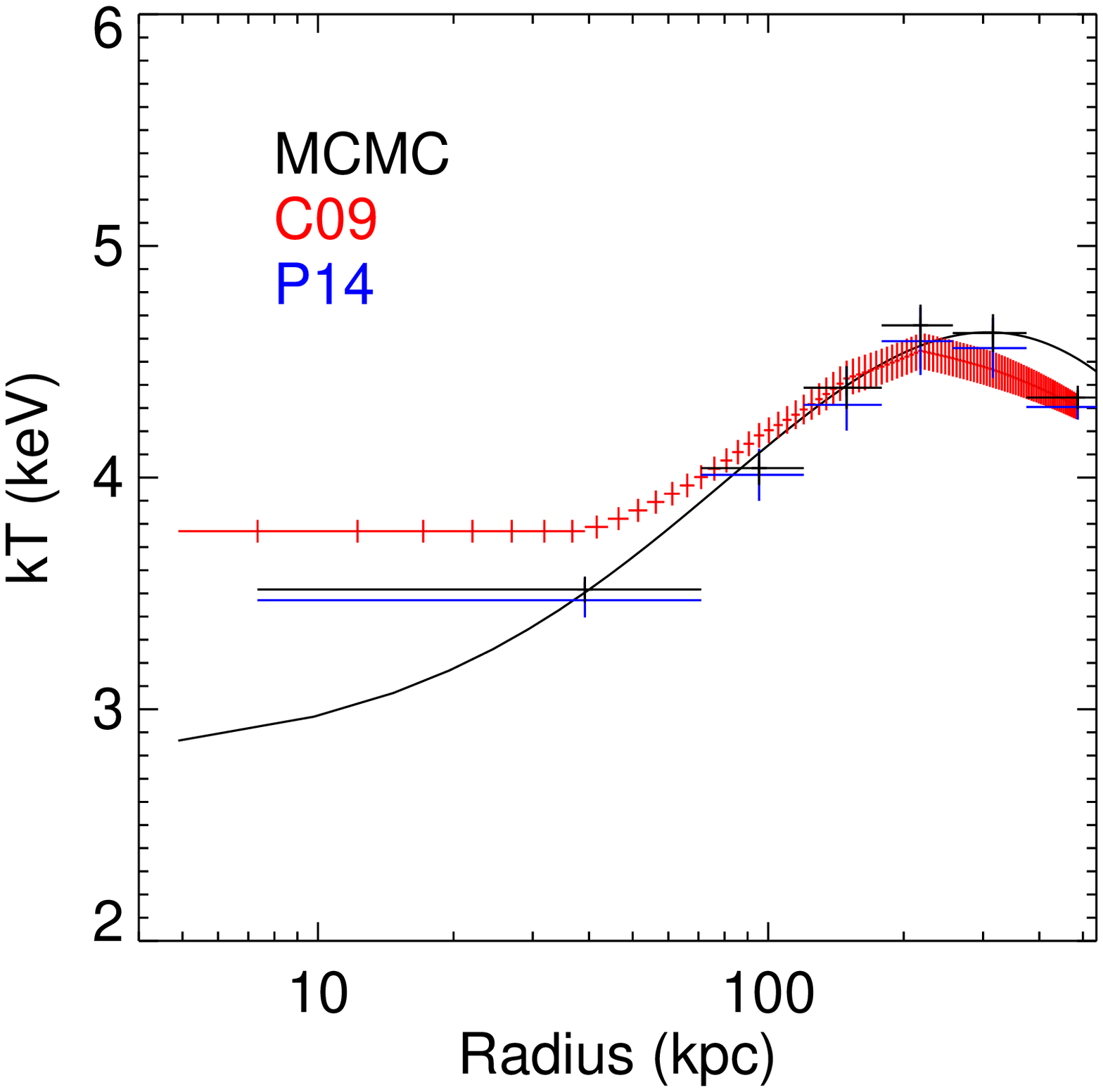}
 \includegraphics[width=0.33\linewidth]{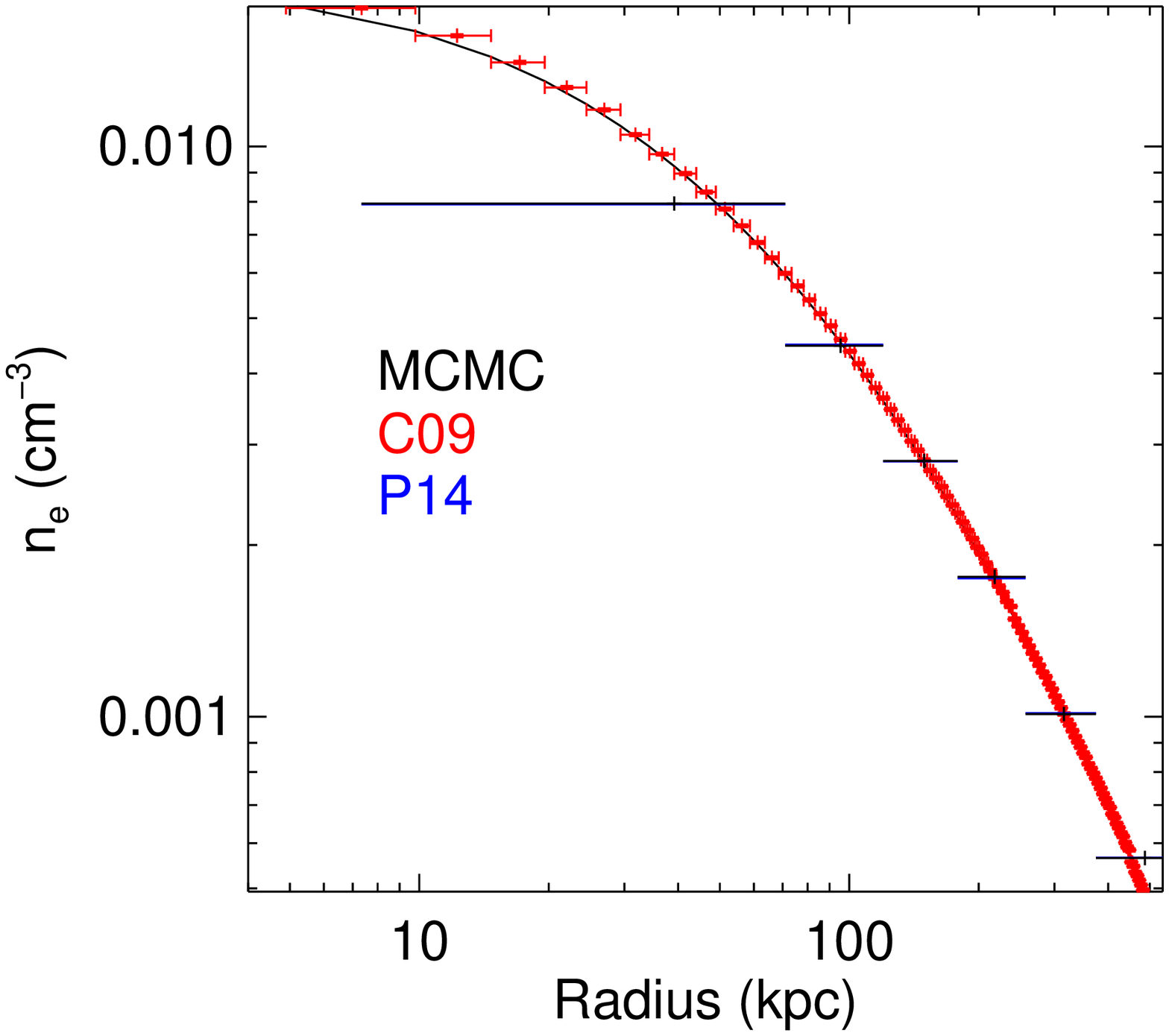}
 \includegraphics[width=0.33\linewidth]{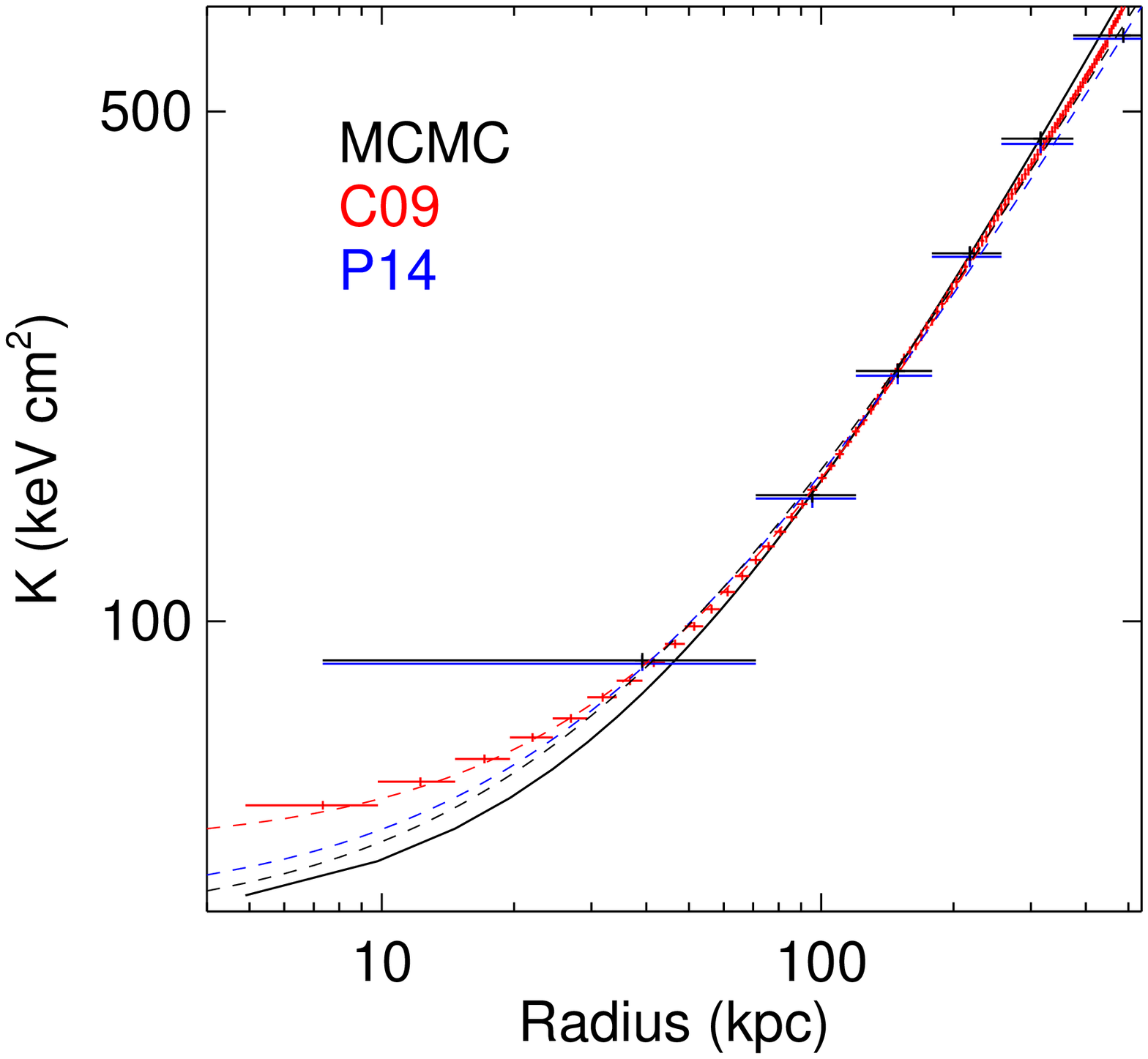} 
\caption{Temperature, electron number density and entropy profiles obtained using different analysis methods 
for the 100 ks simulated spectra. Results from jMCMC, \citetalias{cav09}'s and \citetalias{pan14}'s methods are shown 
using black, red and blue colors. The input electron density, temperature and entropy profiles used for generating 
the simulated spectra are also shown using black solid lines. The dashed black, red and blue lines in the third panel 
show the flat-core model (eq.~\ref{eq:flat_core_model}) fits to the jMCMC, \citetalias{cav09} and \citetalias{pan14} entropy profiles, respectively.}
\label{fig:100ks_sim_different_method}
\end{figure*}

There are two recent analyses of radial entropy profiles which differ from each other: \cite{cav09} (henceforth, \citetalias{cav09}) and \cite{pan14} 
(henceforth, \citetalias{pan14}). In \citetalias{cav09}, temperature profiles are directly obtained from the analysis of the 0.7-7.0 keV {\it projected} spectra. 
High resolution surface-brightness profiles and spectra in the 0.7-2.0 keV range are then used for obtaining high-resolution profiles of 
deprojected electron densities. The projected temperature profiles are interpolated to match the resolution of the density 
profiles. The density and temperature profiles are then combined to obtain the entropy profiles. The main advantage of \citetalias{cav09}'s approach is that 
it gives an excellent resolution in the cluster core. The problem with their analysis is that the use of projected spectra leads to an 
overestimation of the temperature (and hence, entropy) in the central region.\footnote{We thank the anonymous referee for pointing out that the 
earlier Chandra observations also overestimated the temperature in 2-7 keV range; e.g., see section III A.1 in \url{http://cxc.harvard.edu/caldb/downloads/Release_notes/CALDB_v4.1.1.html}. } An earlier paper, \citet{don06},
compared the entropy profiles using projected and deprojected temperatures and found them to be similar (this is not quite correct; see left panels of Figs. \ref{fig:100ks_sim_different_method} \& \ref{fig:test_clus_different_method}).
\cite{pan14} deproject the spectra using the \textit{DSDEPROJ} routine described in \cite{russ08}. Profiles of 
electron density and temperature were obtained from the spectral analysis of the deprojected spectra, and were combined to obtain the entropy profiles. 

\setcounter{figure}{2}
\begin{figure*}
 \includegraphics[width=0.32\linewidth]{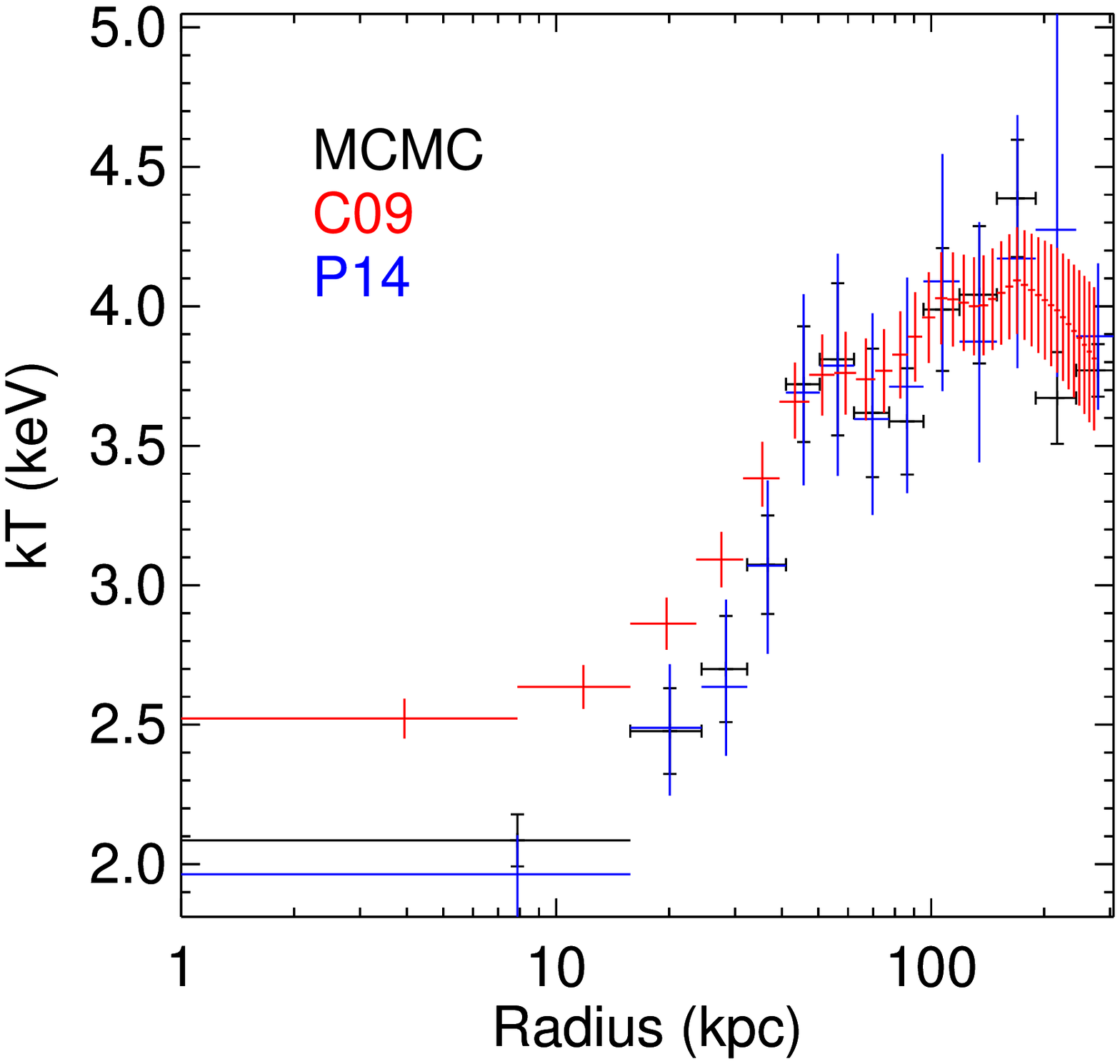}
 \includegraphics[width=0.32\linewidth]{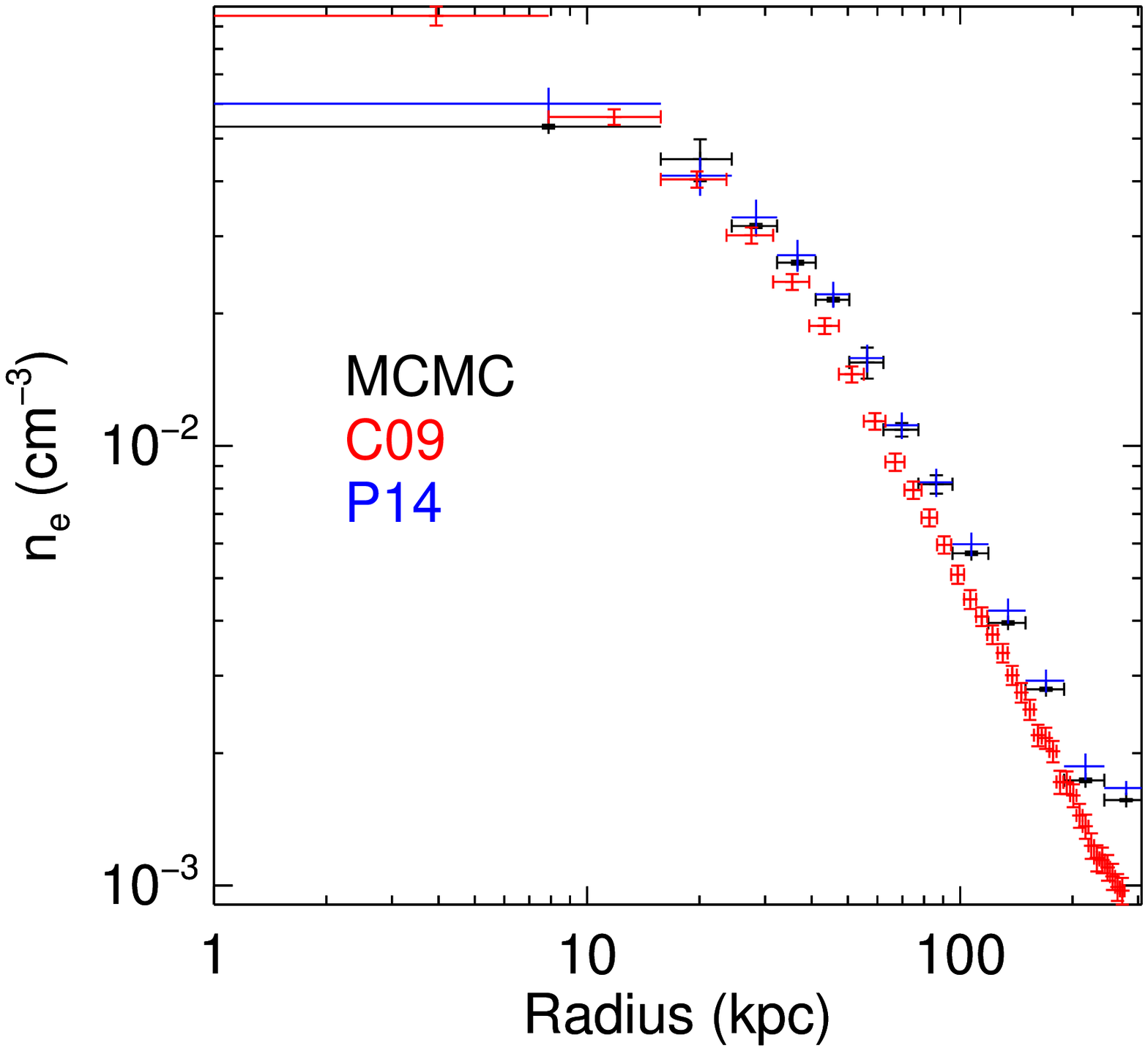}
 \includegraphics[width=0.32\linewidth]{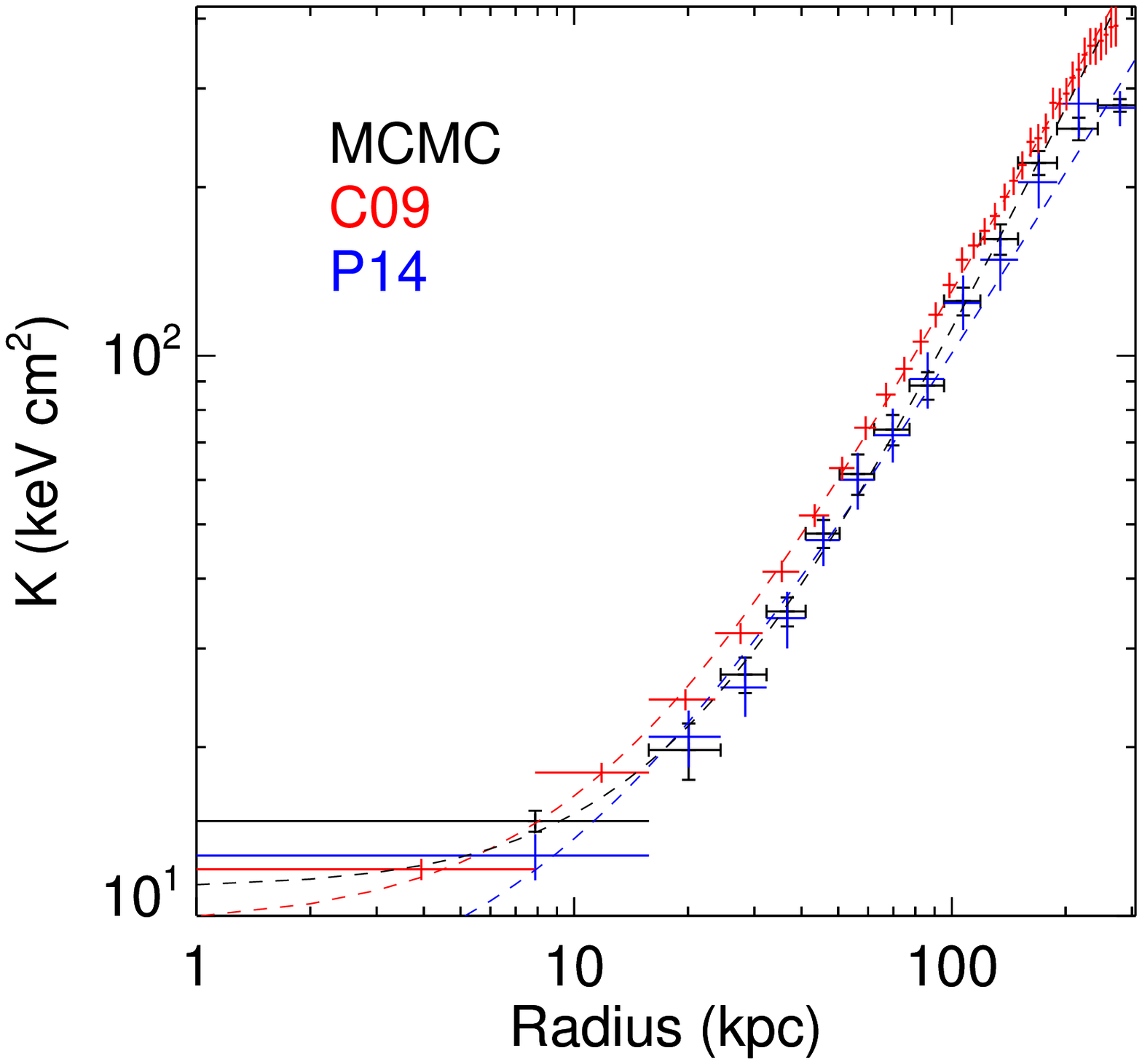}
\caption{Temperature, electron number density and entropy profiles obtained using different analysis methods 
for the test cluster. Results from jMCMC analysis, \citetalias{cav09}'s method and \citetalias{pan14}'s method are shown 
using black, red and blue colors. The dashed black, red and blue lines in the third panel show the flat-core 
model (eq.~\ref{eq:flat_core_model}) fits to the jMCMC, \citetalias{cav09} and \citetalias{pan14} entropy profiles, respectively.}
\label{fig:test_clus_different_method}
\end{figure*}

The temperature, electron density and entropy profiles obtained using 
jMCMC, \citetalias{cav09}'s and \citetalias{pan14}'s analysis methods for the simulated spectra and the test cluster 
are shown in Figs. \ref{fig:100ks_sim_different_method} and \ref{fig:test_clus_different_method}, respectively. 
\citetalias{cav09}'s method is found to overestimate the temperature in the inner annuli which is expected since the method makes 
use of projected spectra and hence inner annuli are contaminated by the hotter shells lying outside. 
Similarly, the temperature at larger radii is underestimated. Results from jMCMC analysis 
and \citetalias{pan14}'s method are found to be in good agreement with each other. For the test cluster, both these methods 
seem to overestimate the density in the outermost annulus. However, this is a common artifact of deprojection analyses 
due to the excess emission from shells outside the outermost annulus contributing to its emission.\\

\subsection{Entropy profile  with/without correlations}
\label{S:entr_fit_witho_correl}

\setcounter{figure}{3}
\begin{figure*}
 \includegraphics[width=0.32\linewidth]{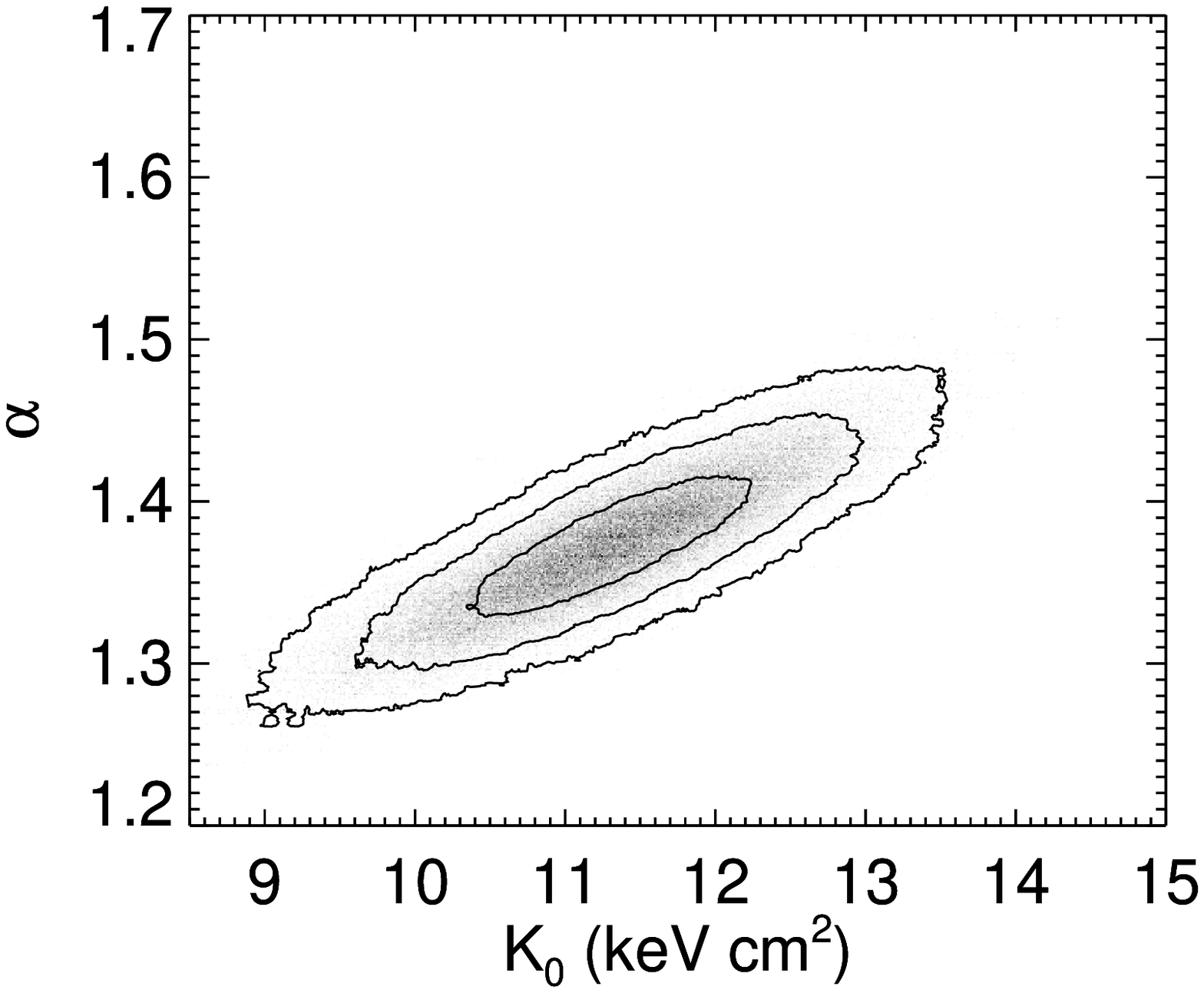}
 \includegraphics[width=0.32\linewidth]{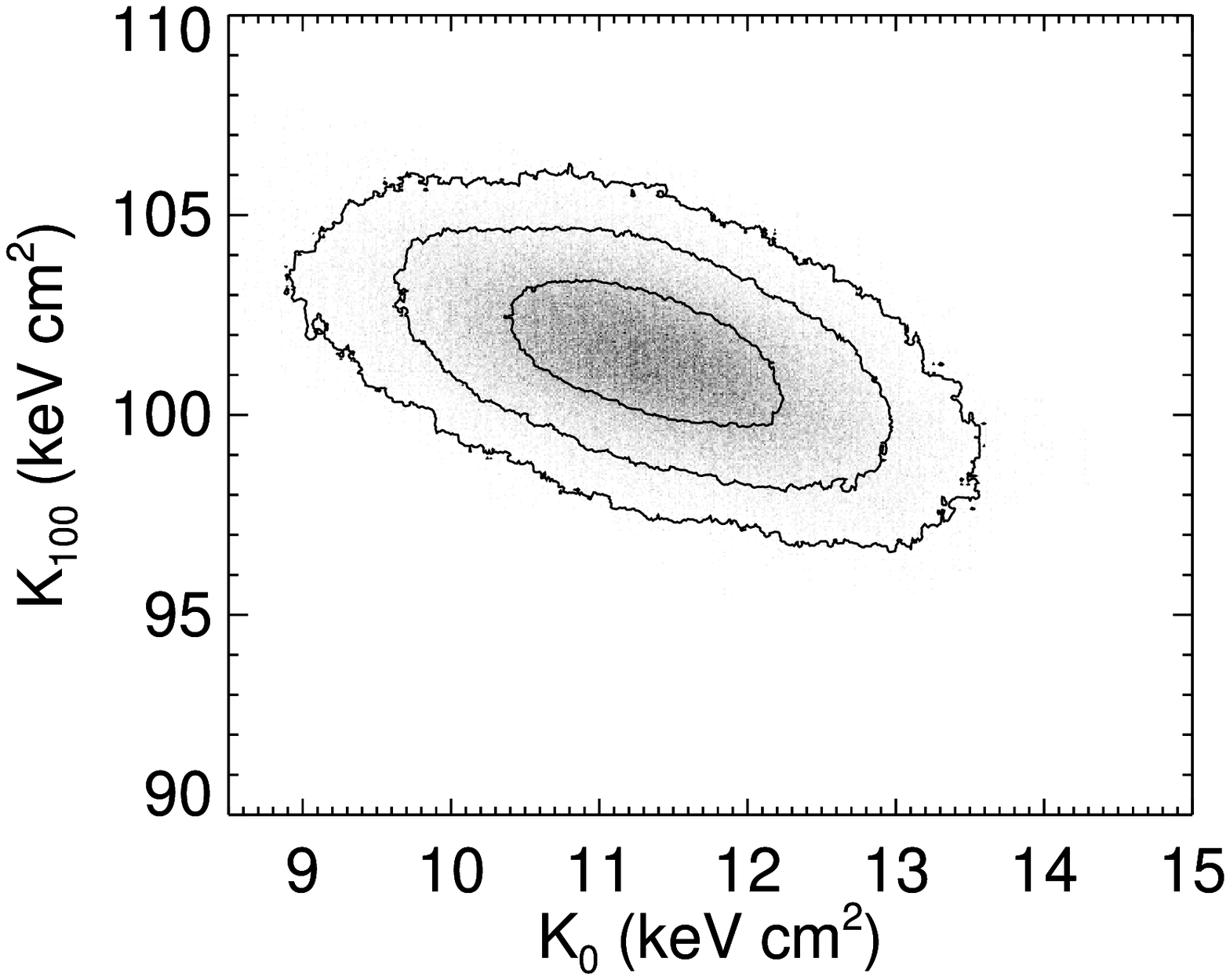}
 \includegraphics[width=0.32\linewidth]{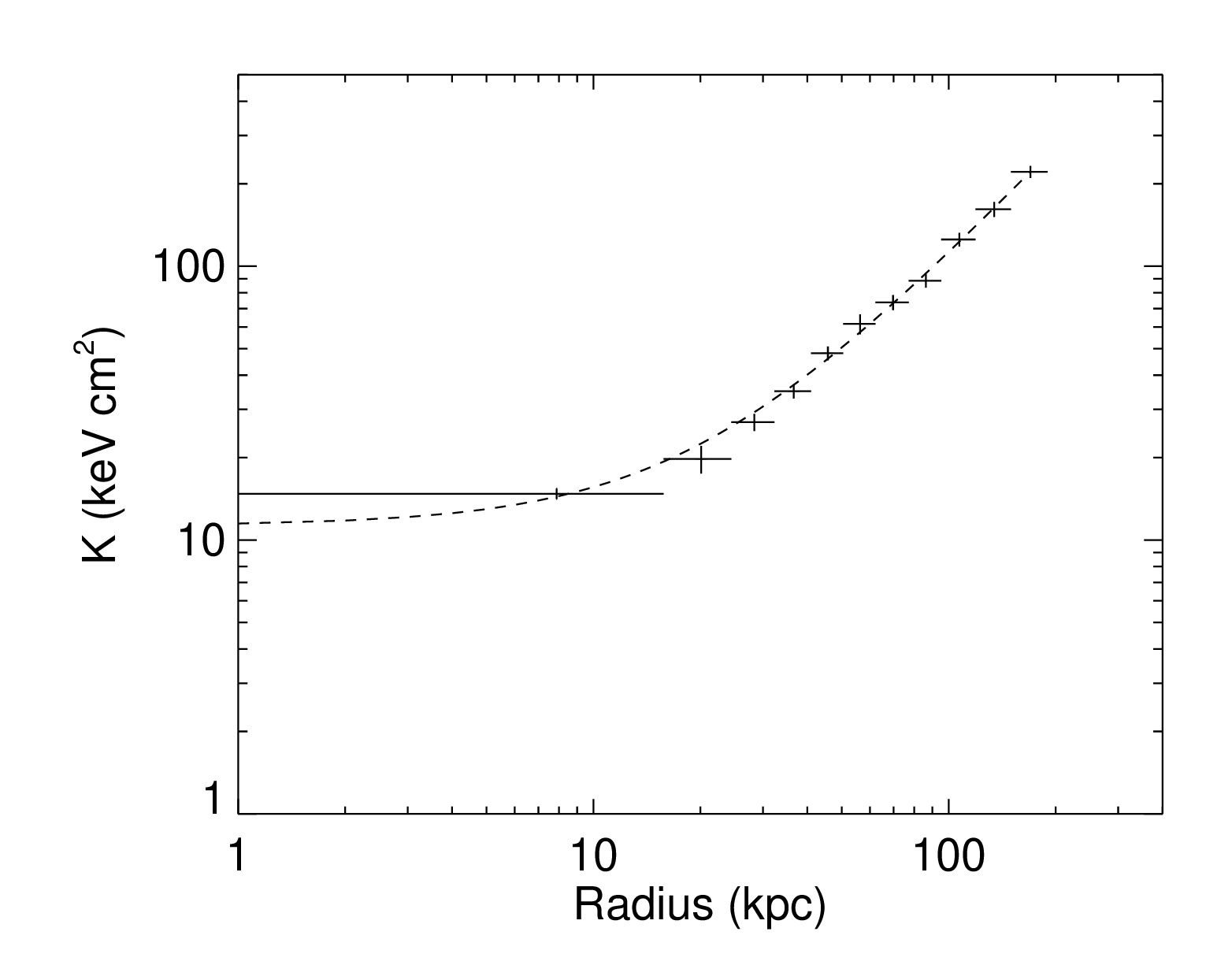}\\
 \includegraphics[width=0.32\linewidth]{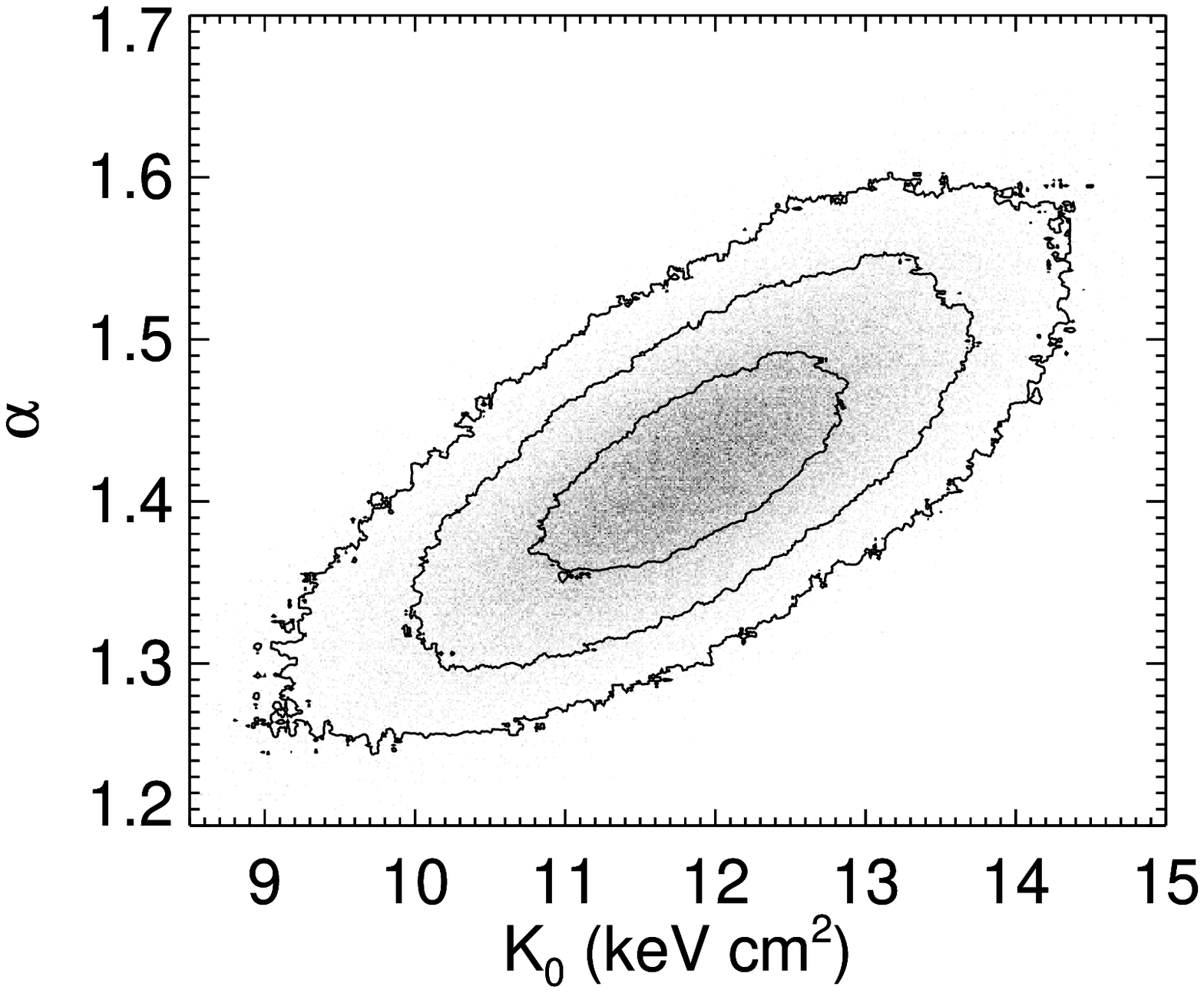}
 \includegraphics[width=0.32\linewidth]{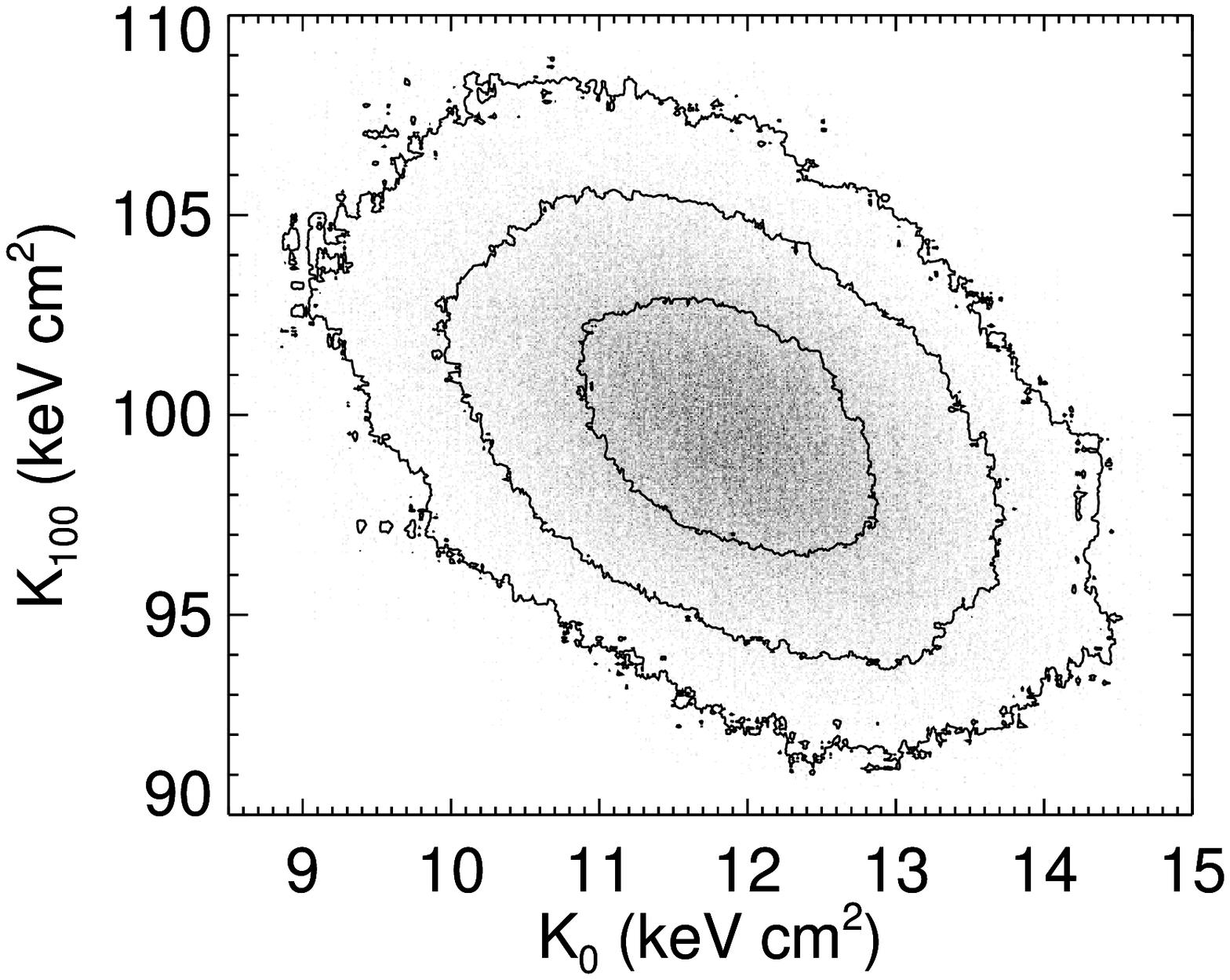}
 \includegraphics[width=0.32\linewidth]{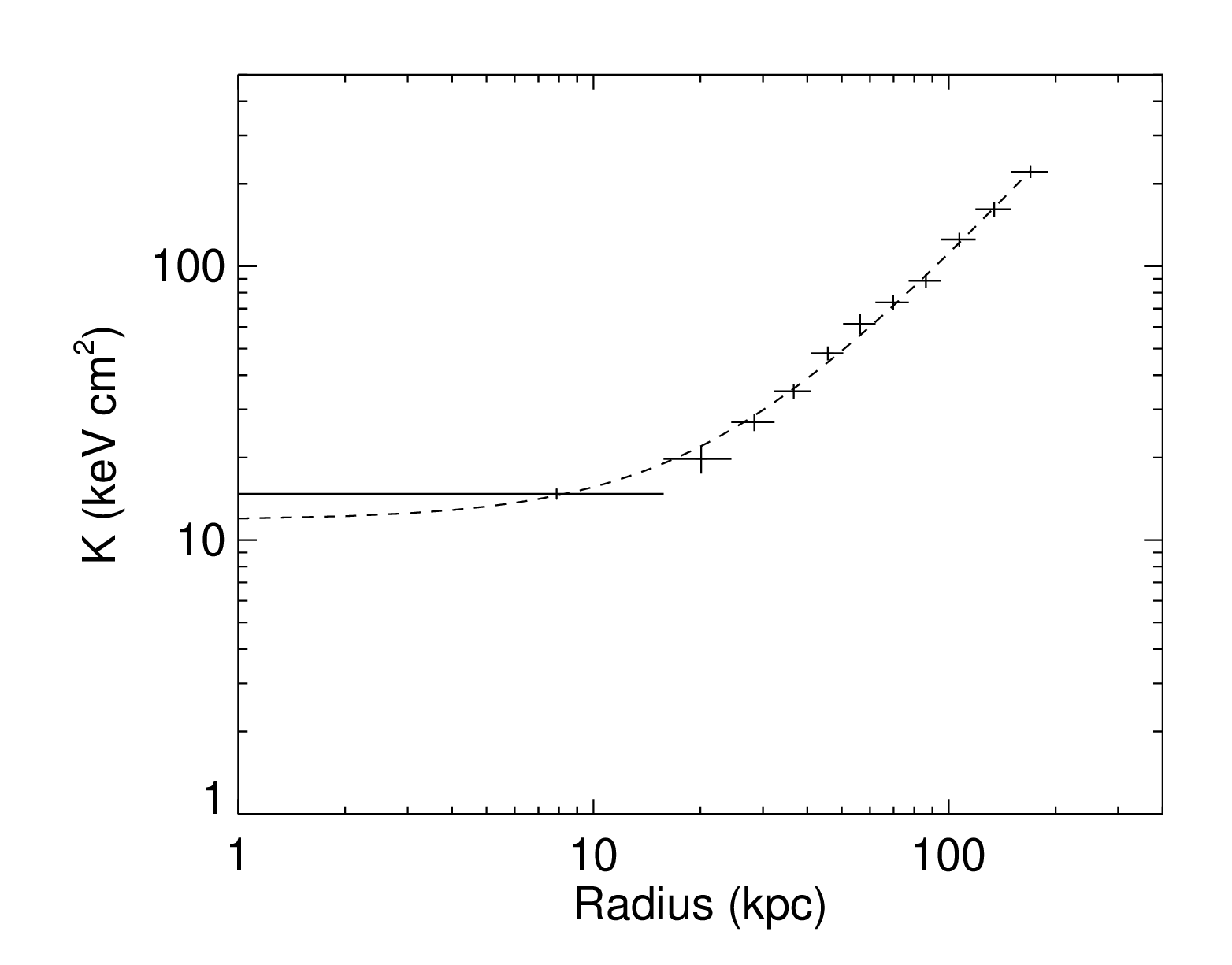}\\
 \includegraphics[width=0.32\linewidth]{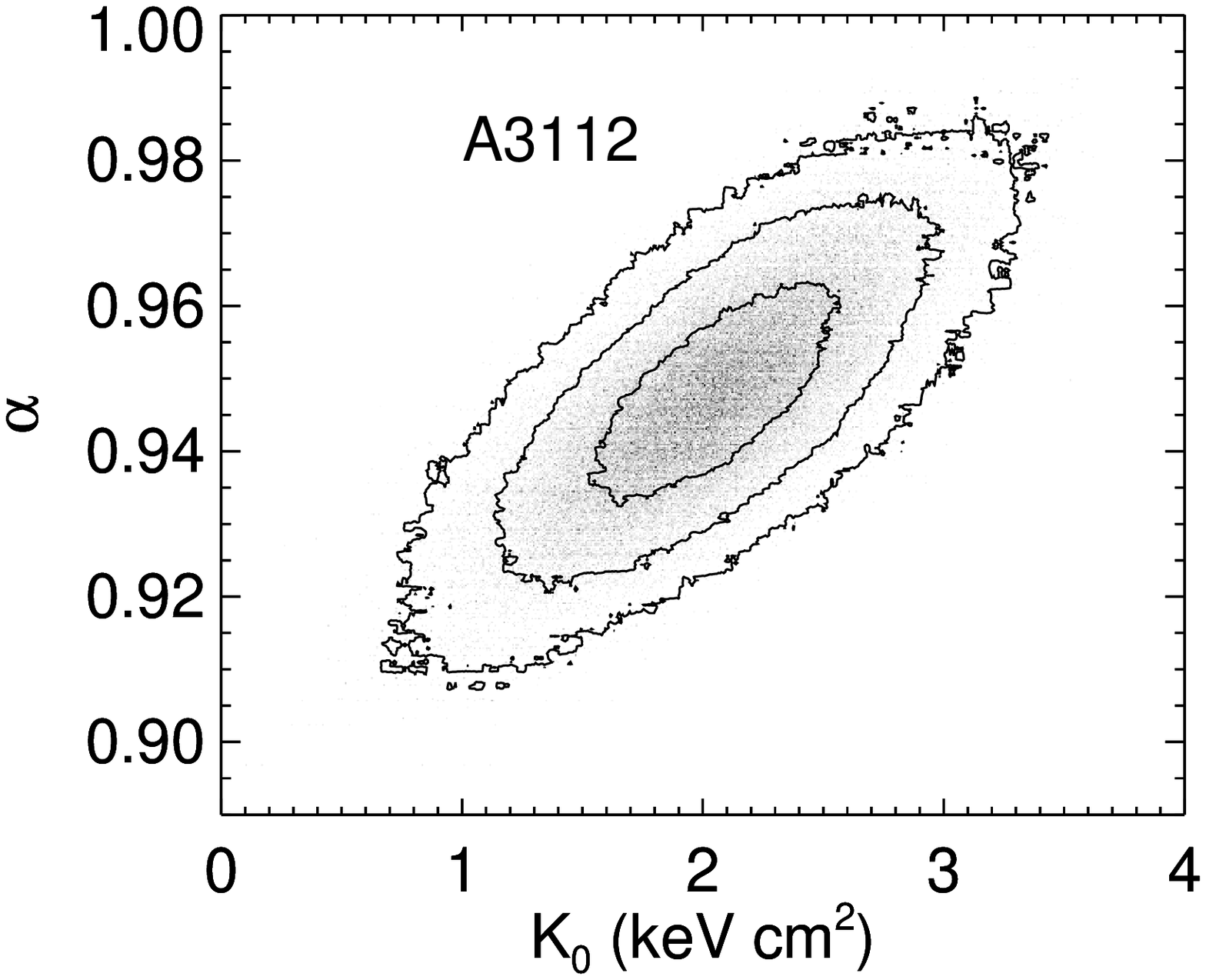}
 \includegraphics[width=0.32\linewidth]{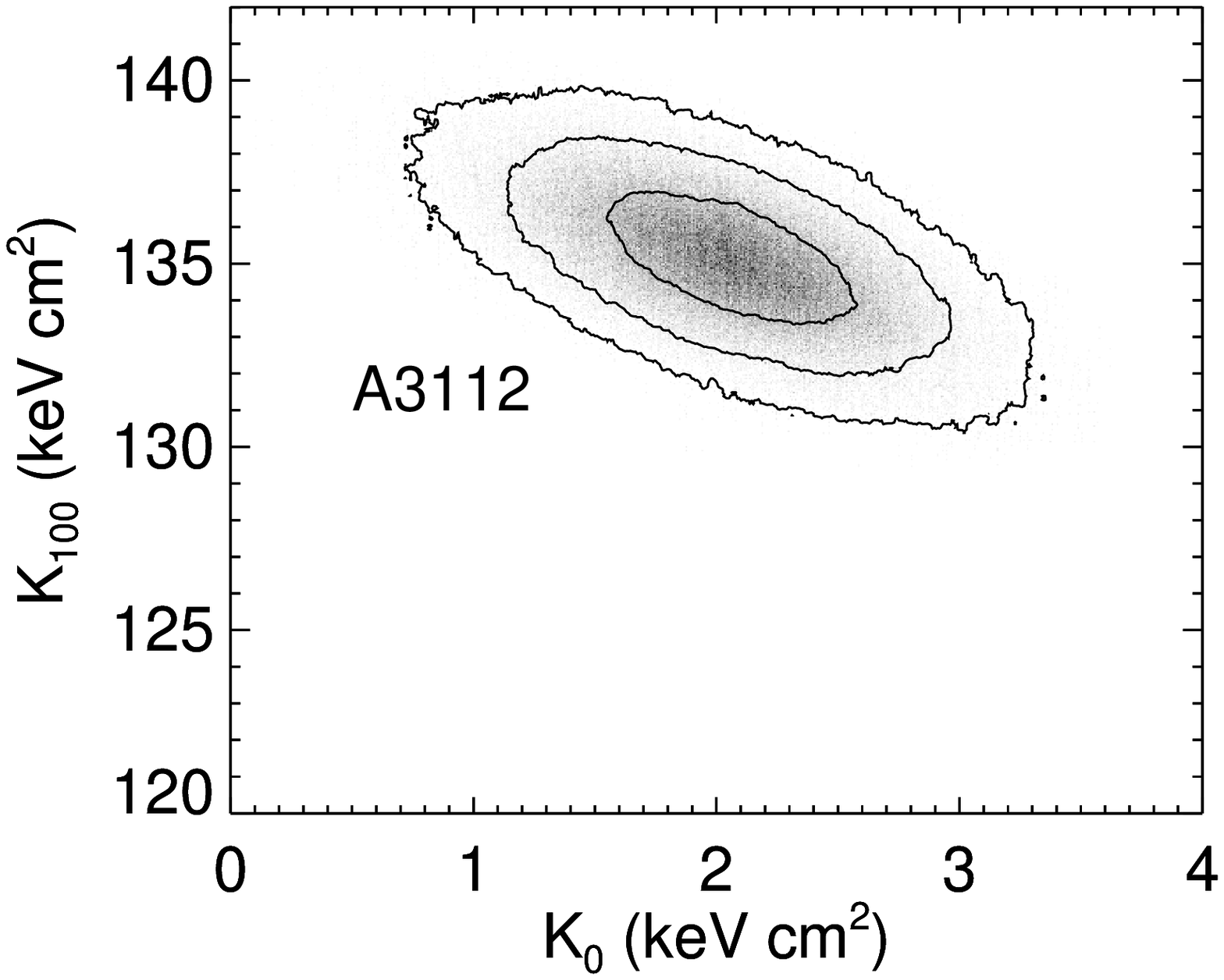}
 \includegraphics[width=0.32\linewidth]{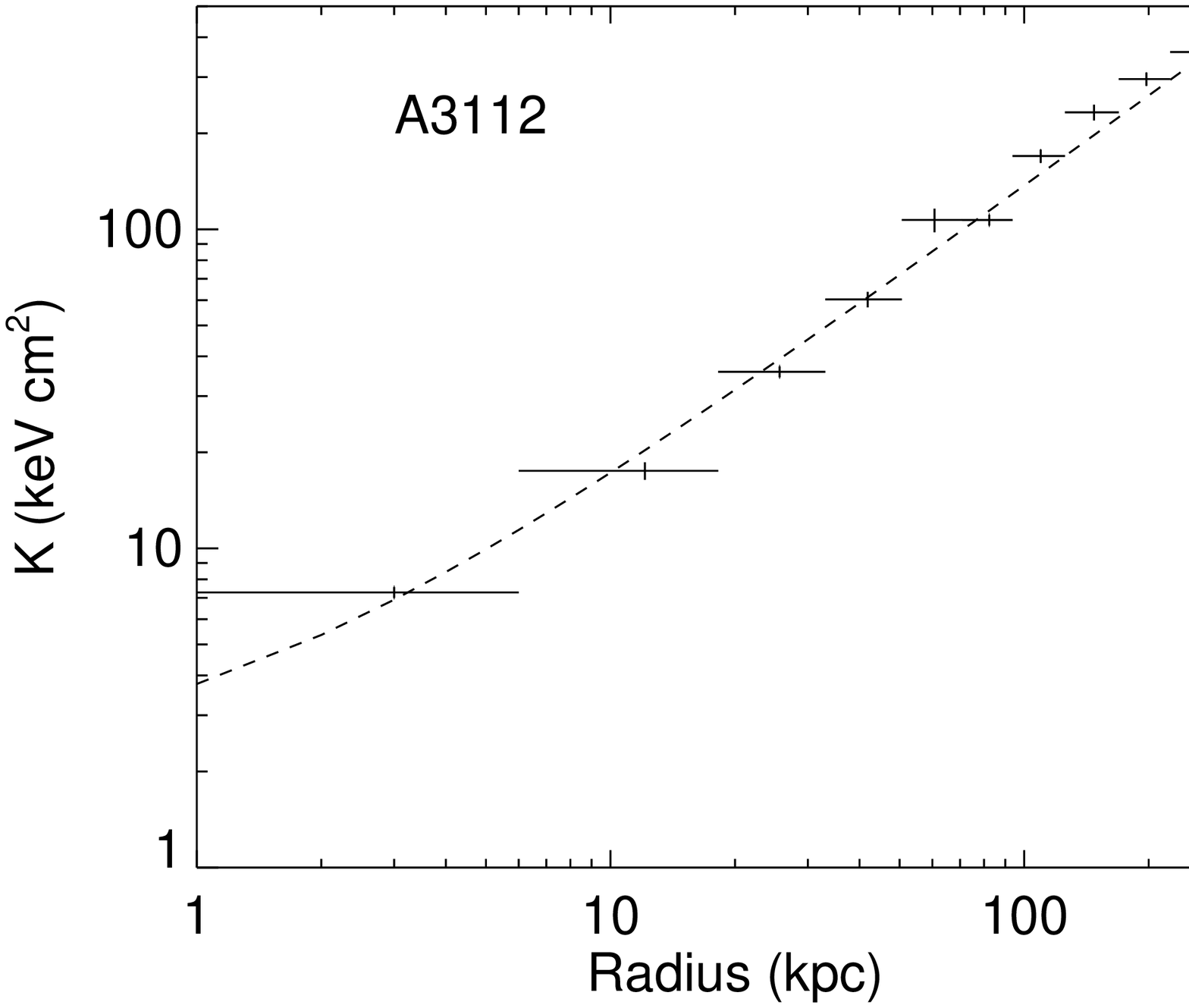}\\
 \includegraphics[width=0.32\linewidth]{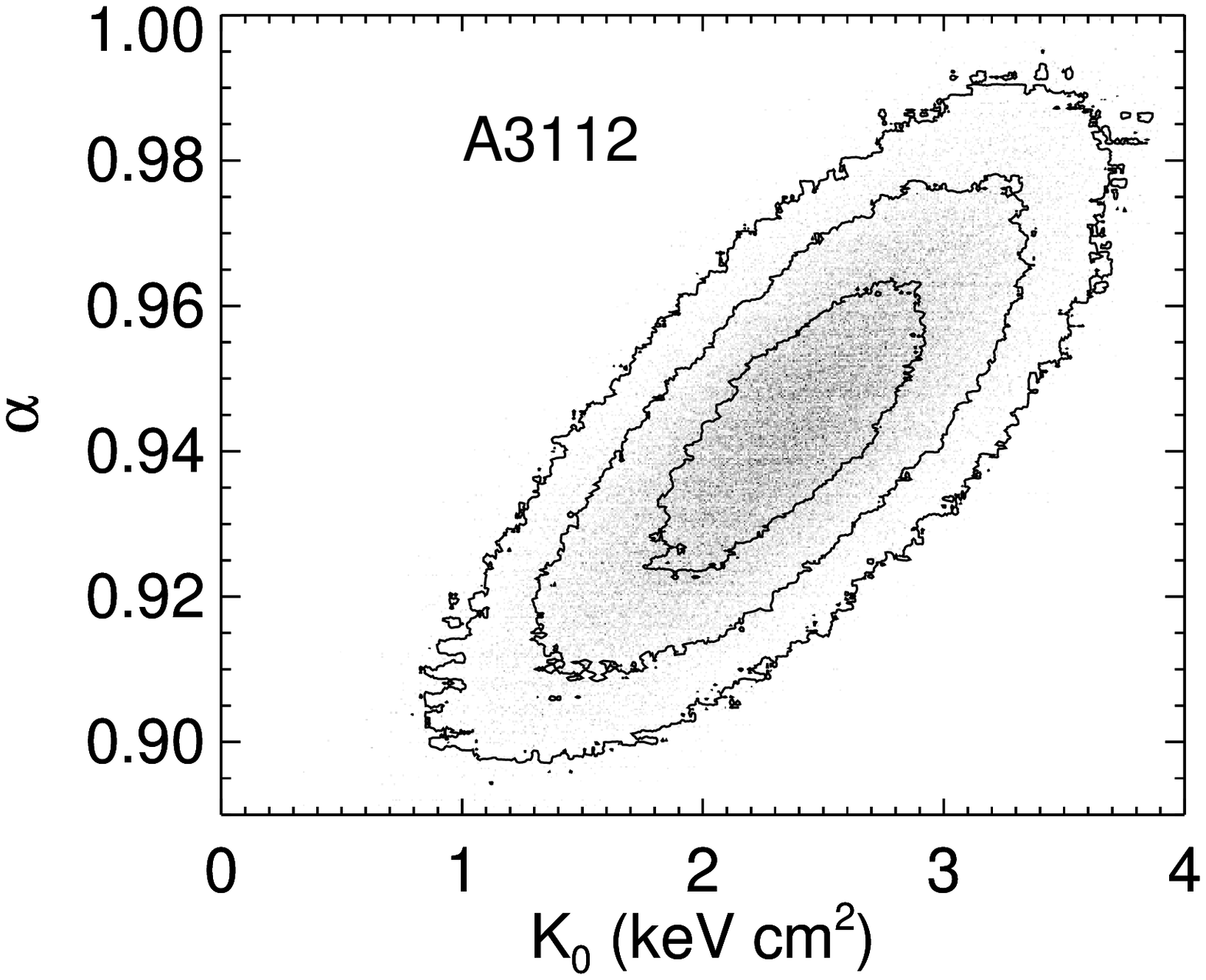}
 \includegraphics[width=0.32\linewidth]{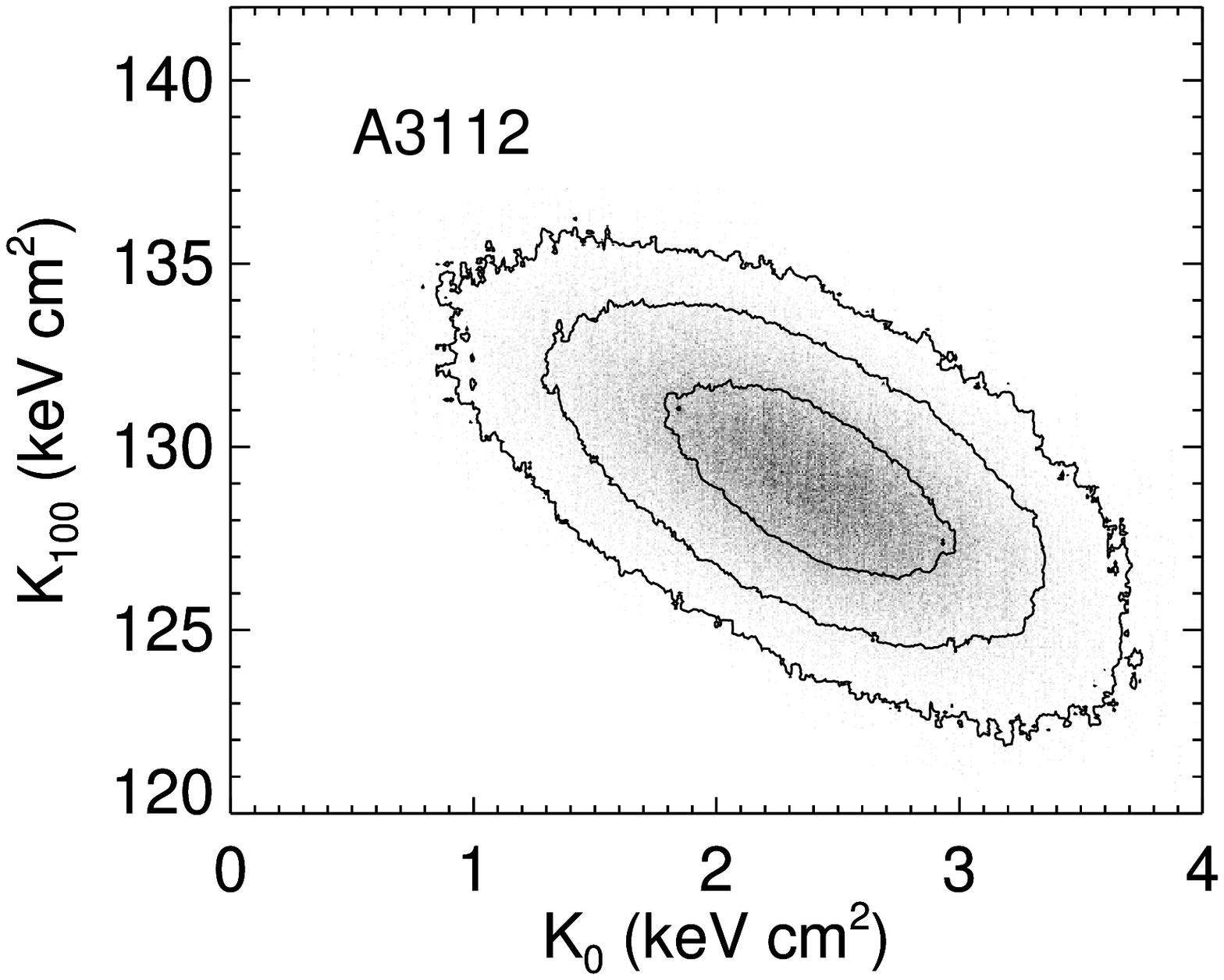}
 \includegraphics[width=0.32\linewidth]{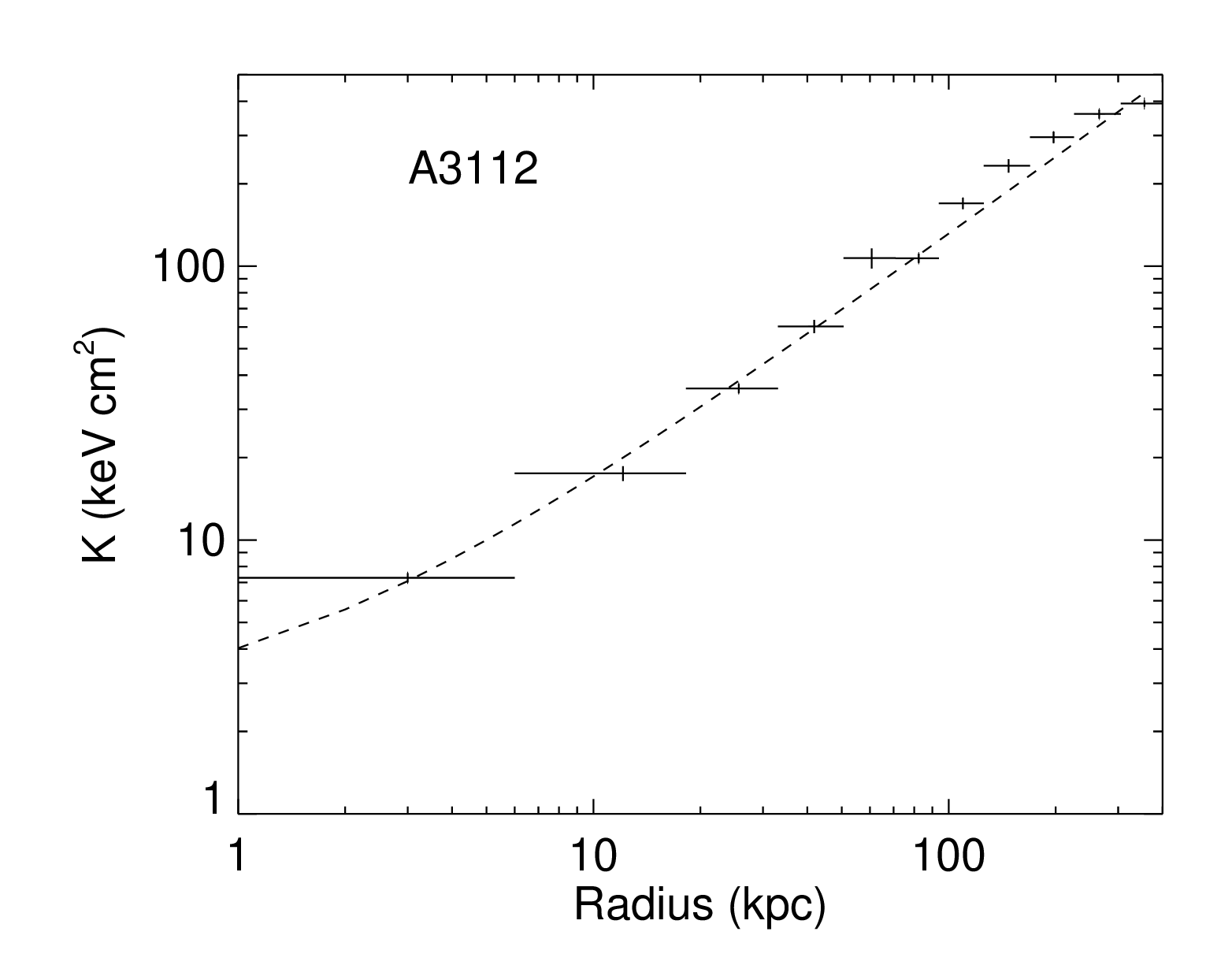}\\ 
\caption{The 2D marginal probability distributions of $\alpha-K_0$ and $K_{100}-K_0$ obtained from 
the flat-core entropy profile fitting,
 with (first row) and without (second row) correlation between the shell entropies ($T_{\rm keV}/n_e^{2/3}$) 
for the test cluster (A2597). Entropy profile of A3112 with and without correlation are also given in the third and fourth rows, respectively. 
The contours mark the 50\%, 90\% and 99\% inclusion levels based on the density of points starting from the innermost contour outwards. 
Greyscale denotes the PDF density.}
\label{fig:entr_fit_witho_correl}
\end{figure*}

One downside of the standard deprojection methods is that they do not take into account the correlations between the 
deprojected parameters obtained for the different shells. To demonstrate this effect we fitted the entropy profile of the test cluster 
obtained from the jMCMC analysis using a flat-core model (eq. \ref{eq:flat_core_model}), 
as described in section \ref{S:cent_entr_prof}. The expectation values of entropy for each shell $\bar{K_i}$ and the entropy covariance 
matrix ${\rm cov}(K_i, K_j)$ were calculated, which were then used in another MCMC analysis to obtain chains of $K_0$, $K_{100}$ and $\alpha$. 
To see the effect of no correlation between the entropies of different shells, we also 
carried out the same exercise with the non-diagonal terms of the entropy covariance matrix set to zero. 

The $K_0$-$\alpha$ 
and $K_0$-$K_{100}$ probability distributions and the resulting flat-core fits obtained from the two methods (with and without cross-covariance) 
are shown in the first two rows 
of Fig. \ref{fig:entr_fit_witho_correl}. The expectation values of $K_0$, $K_{100}$ and $\alpha$ obtained using the full entropy 
covariance matrix are 11.31$\pm$0.78 keV cm$^2$, 101.47$\pm$1.51 keV cm$^2$ and 1.37$\pm$0.04, respectively, and those 
obtained using only the diagonal terms are 11.84$\pm$0.86 keV cm$^2$, 99.71$\pm$2.68 keV cm$^2$ and 1.42$\pm$0.06, respectively. 
Ignoring correlation between the entropies of different shells in the test cluster, therefore, does not seem to have any 
significant effect on the fitted parameters. The $K_0$-$\alpha$ and $K_0$-$K_{100}$ probability distributions obtained from the two methods, 
however, do show some difference. For clusters with large covariances between the shell entropies, the effect of including 
correlations can be significant, as is seen for the cluster A3112, taken from the cluster sample analyzed in this paper. 
The $K_0$, $K_{100}$ and $\alpha$ values obtained for A3112, with and without considering correlation between shell entropies, 
are 2.04$\pm$0.42 keV cm$^2$, 135.14$\pm$1.47 keV cm$^2$ and 0.95$\pm$0.01, and 
2.35$\pm$0.46 keV cm$^2$, 129.17$\pm$2.18 keV cm$^2$ and 0.94$\pm$0.02, respectively. 
The value of $K_{100}$ for A3112, obtained from the two methods are, therefore, found to be significantly different. 
The $K_0$-$\alpha$ and $K_0$-$K_{100}$ probability distributions for A3112 and the resulting flat-core fits obtained from the two methods (with 
and without cross-covariance) are shown in the last two rows of Fig. \ref{fig:entr_fit_witho_correl}.

\section{Results from the cluster sample}
\label{S:samp_res}

The cluster sample described in section \ref{S:clus_sampl} was analyzed using the jMCMC method. The ${\bf n_e}, {\bf T}$ chains resulting 
from the jMCMC analysis were combined 
to produce subsidiary entropy chains using ${\bf K} = {\bf T}{\bf n_e}^{-2/3}$. For each cluster, the elemental abundance in each annulus 
was fixed at the average value obtained from the analysis of the full (across all annuli) cluster spectrum. The resulting entropy profiles were then fitted 
using flat-core and power law models (eqs. \ref{eq:flat_core_model}-\ref{eq:dbl_model}), as described below. 

\subsection{Flat-Core model}
\label{S:flat_core_fit}

\setcounter{figure}{4}
\begin{figure*}
 \centering
  \includegraphics[width=0.32\textwidth]{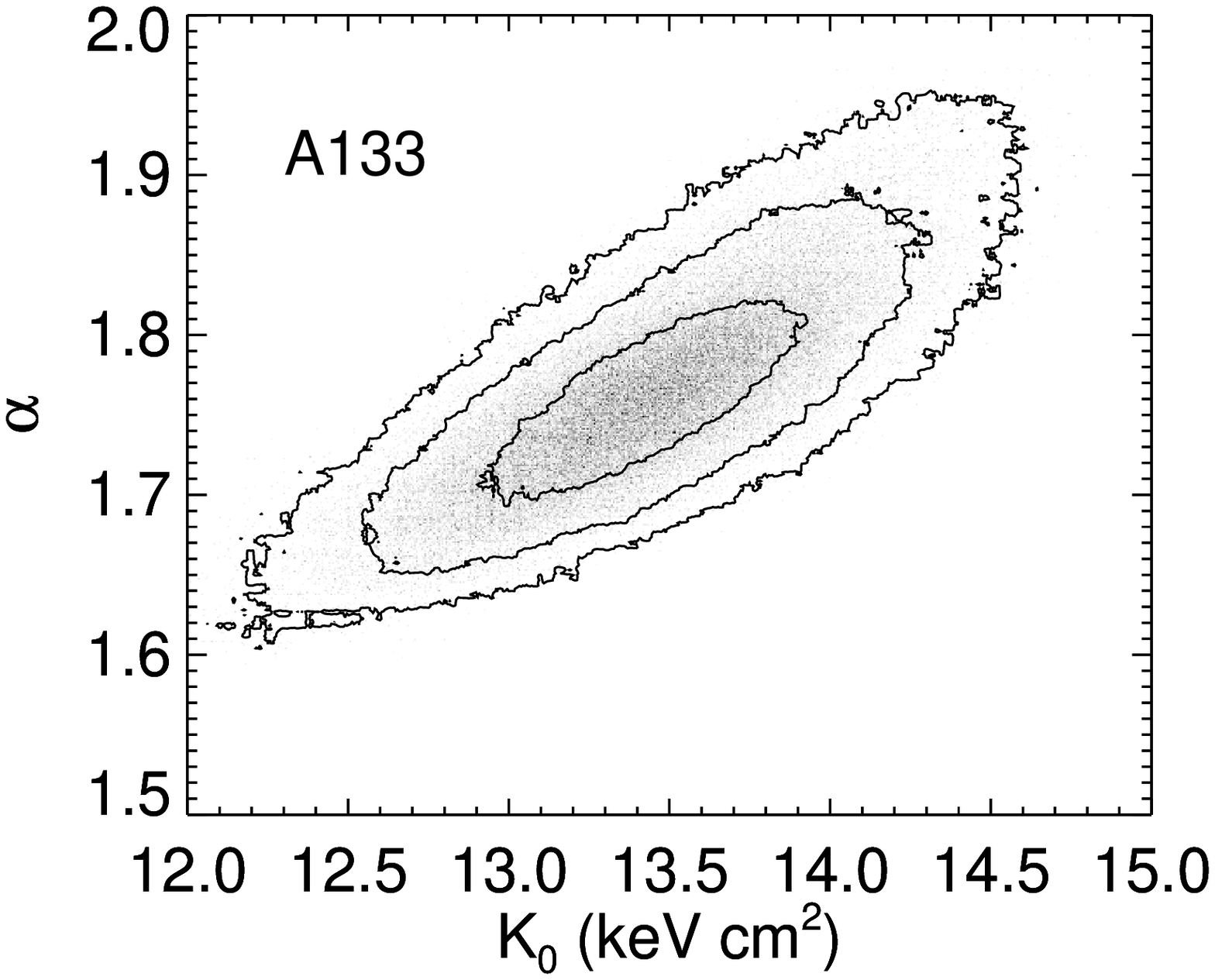}
  \includegraphics[width=0.32\textwidth]{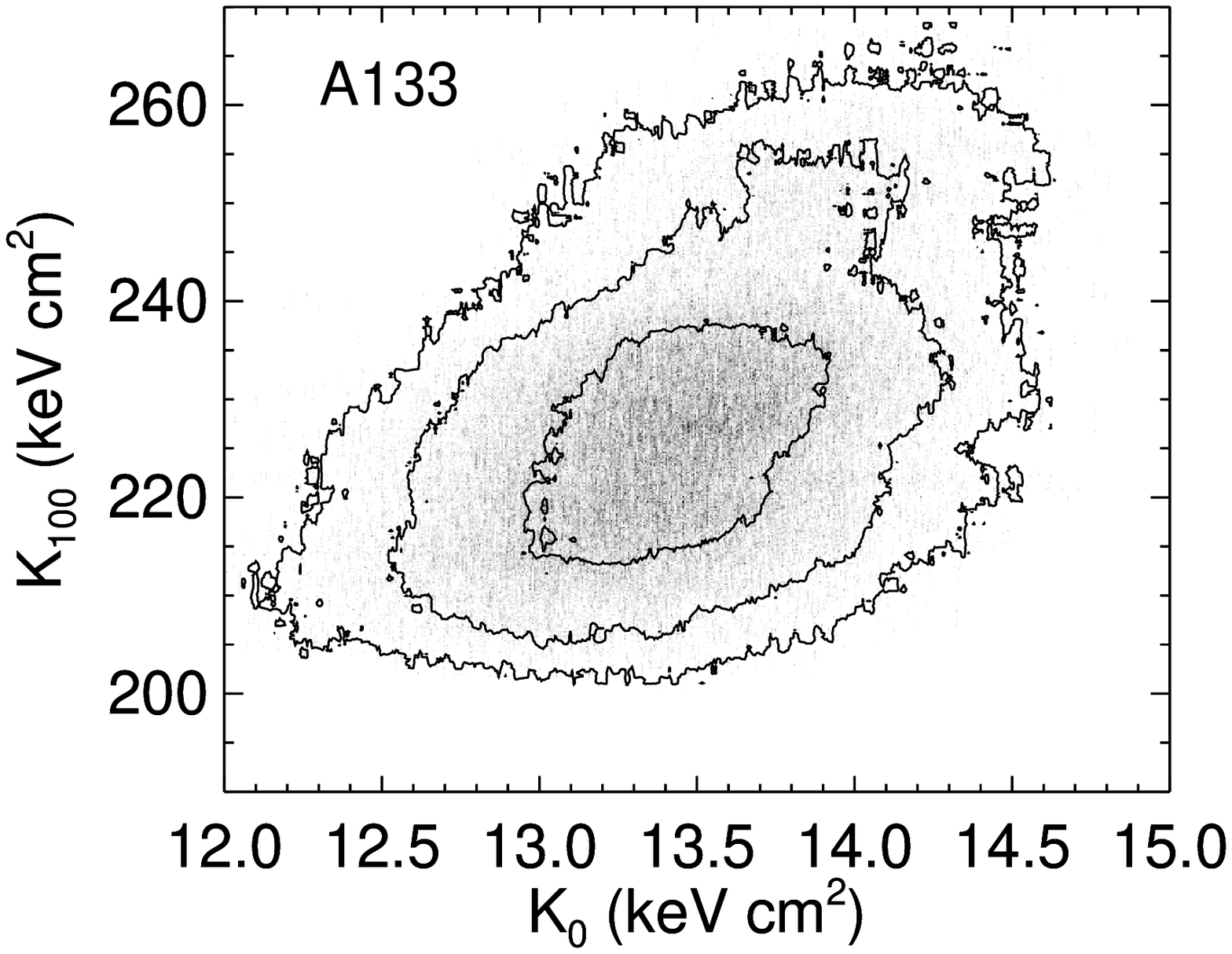}
  \includegraphics[width=0.32\textwidth]{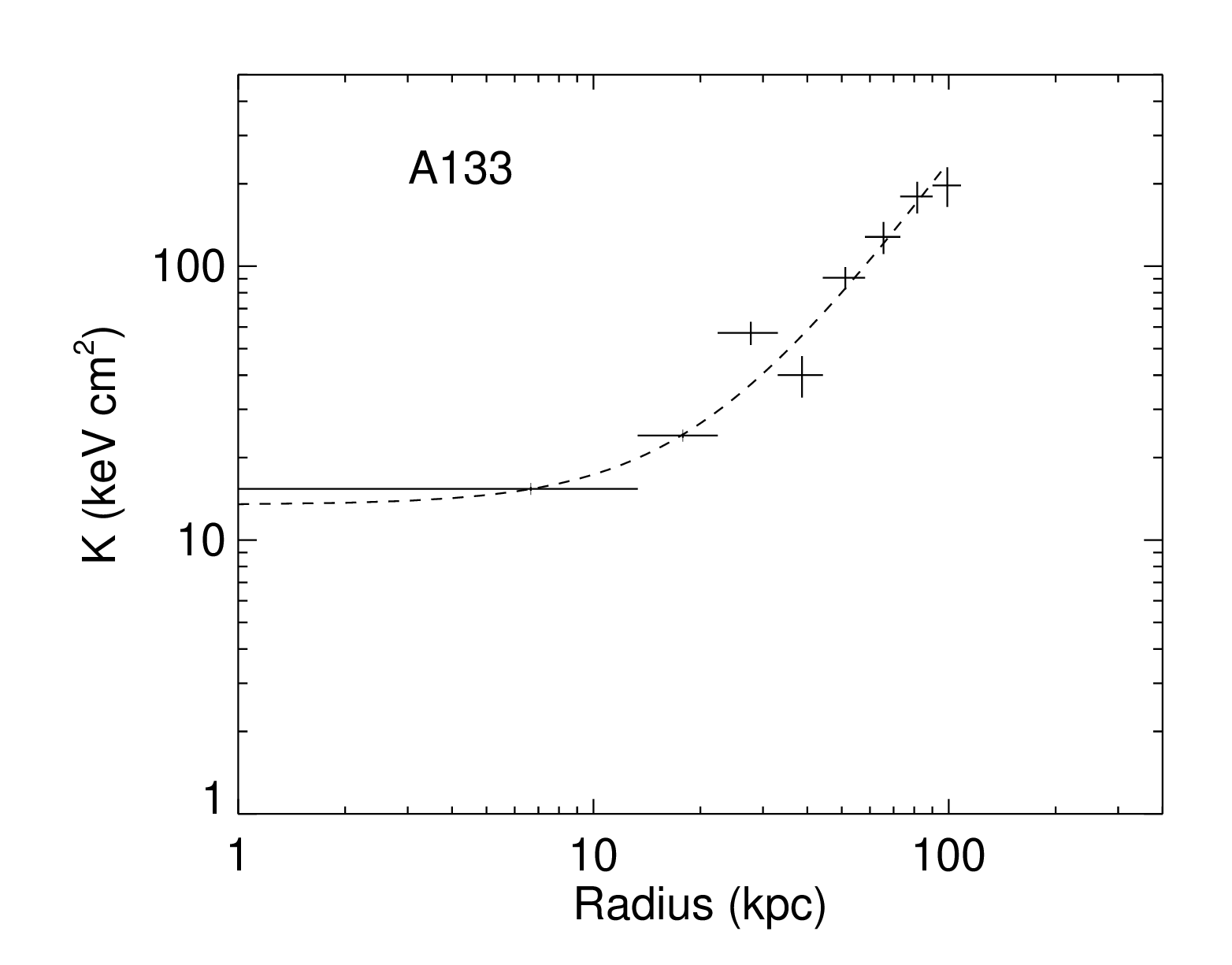}\\
  \includegraphics[width=0.32\textwidth]{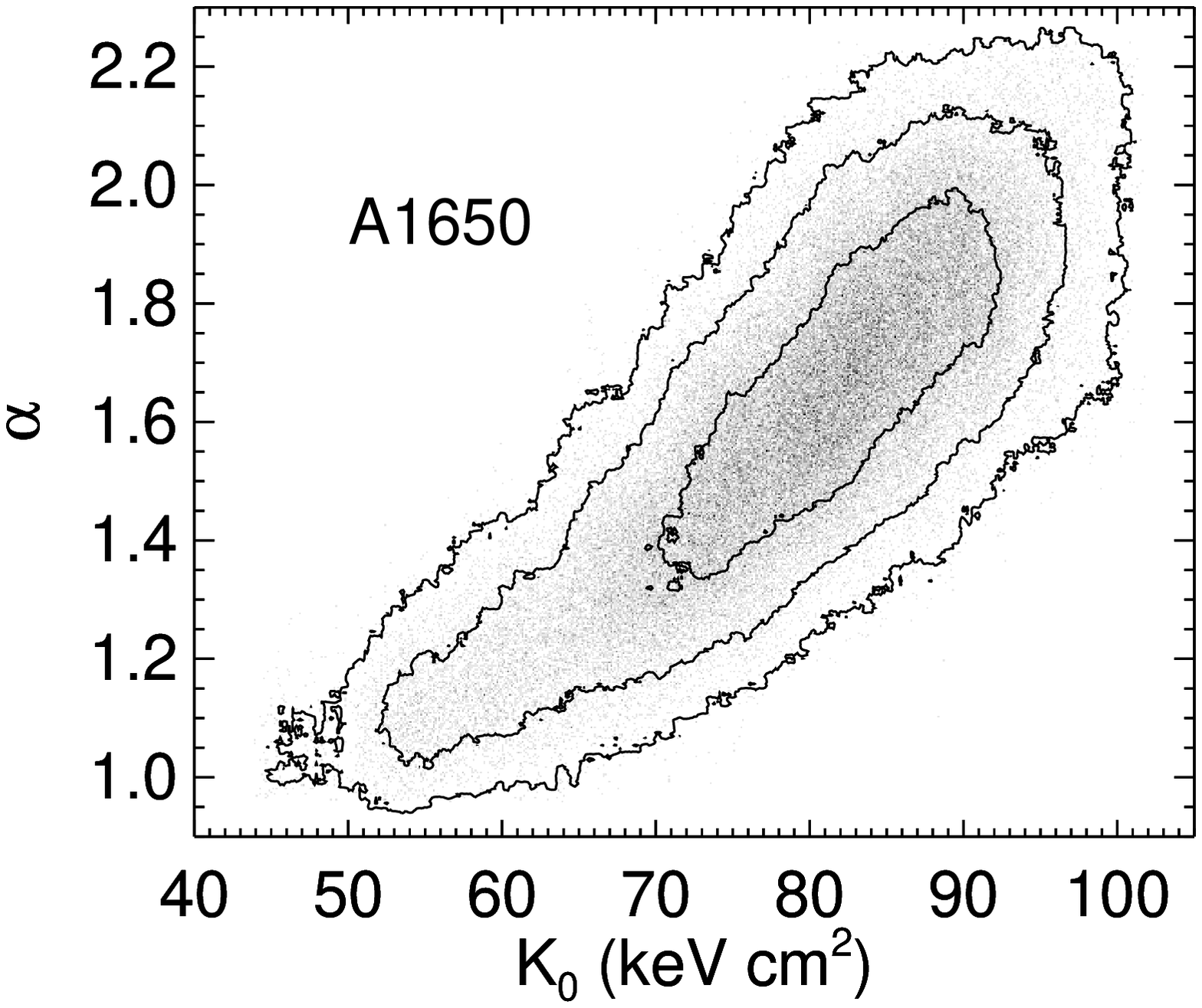}
  \includegraphics[width=0.32\textwidth]{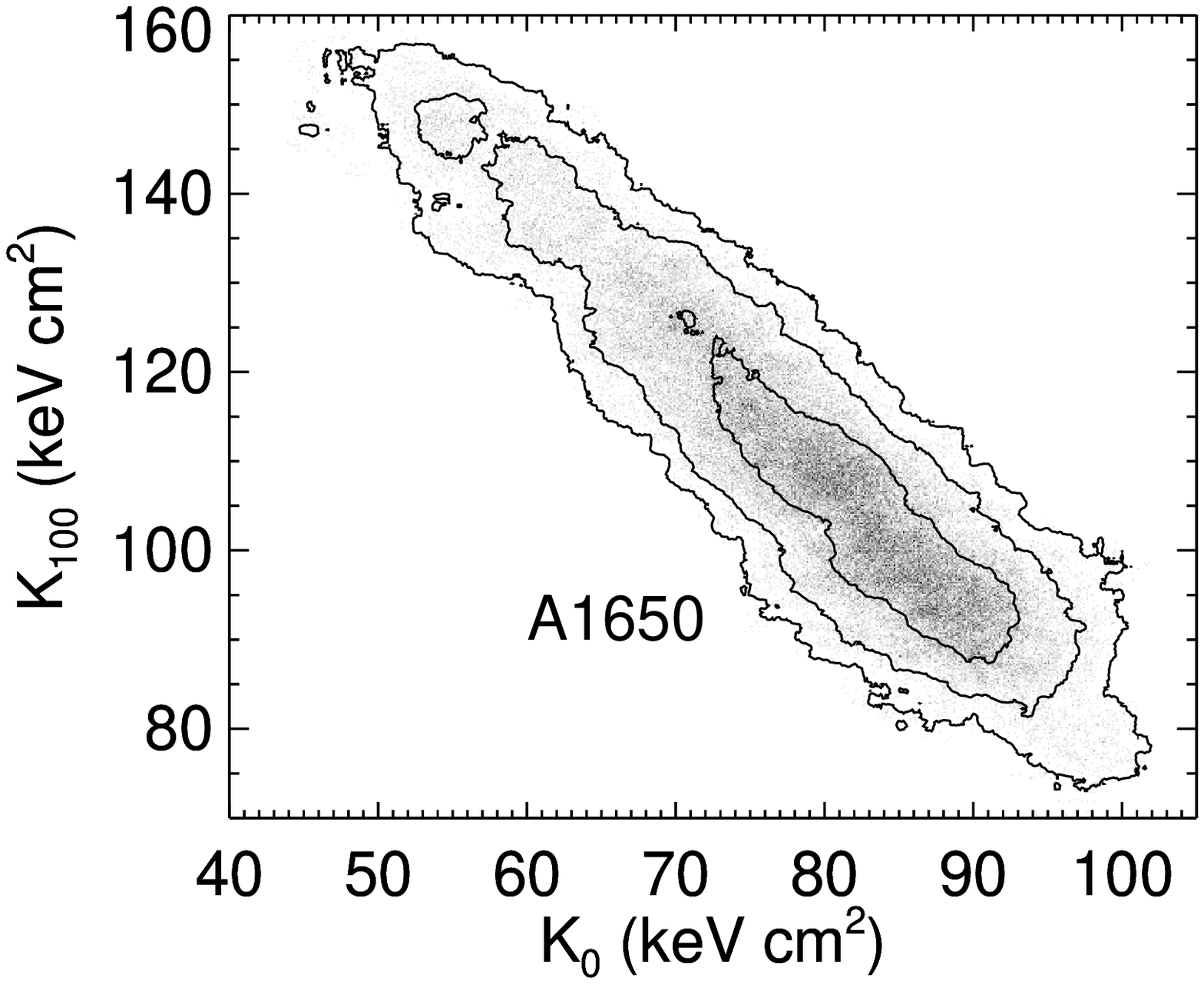}
  \includegraphics[width=0.32\textwidth]{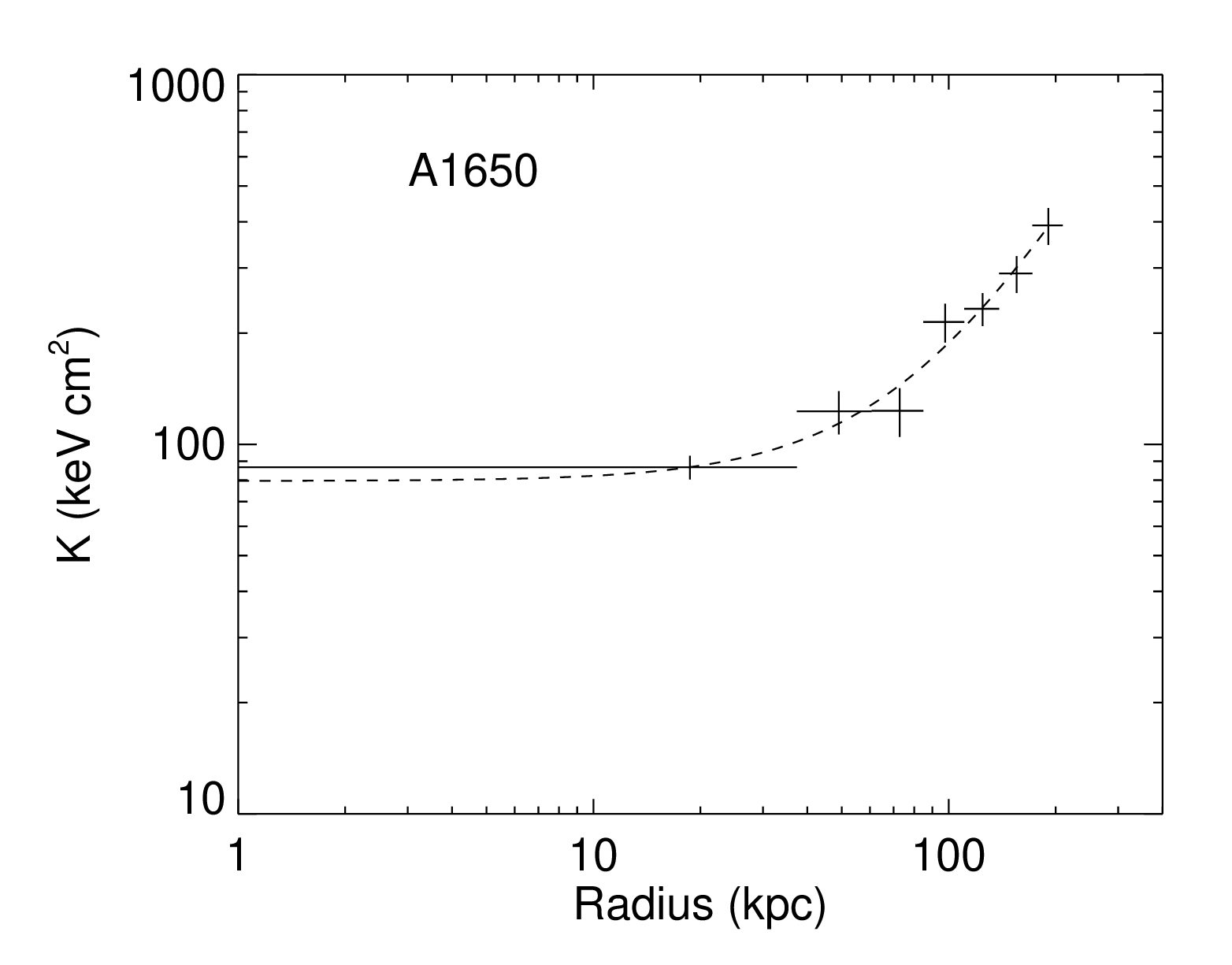}\\   
  \includegraphics[width=0.32\textwidth]{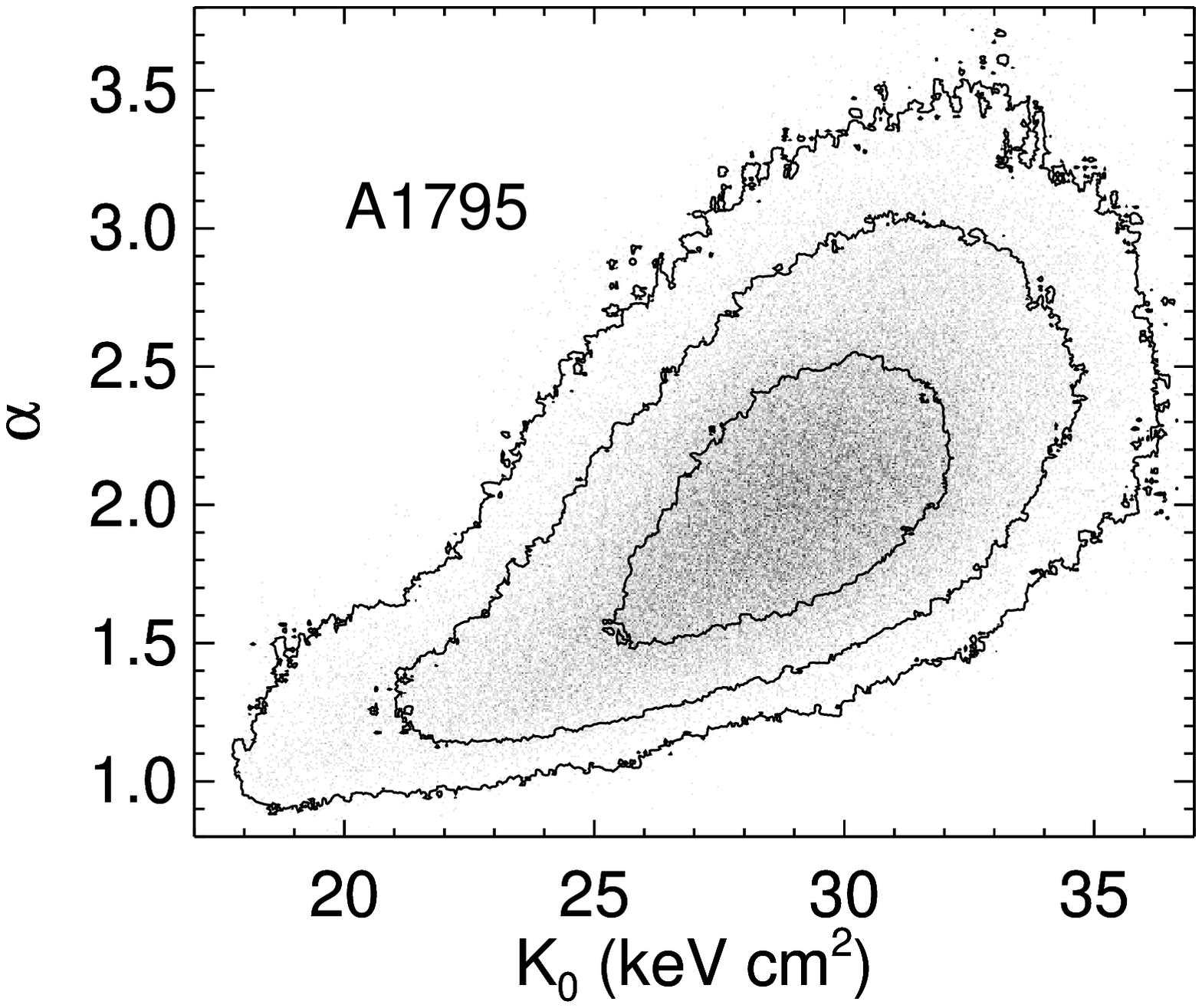}
  \includegraphics[width=0.32\textwidth]{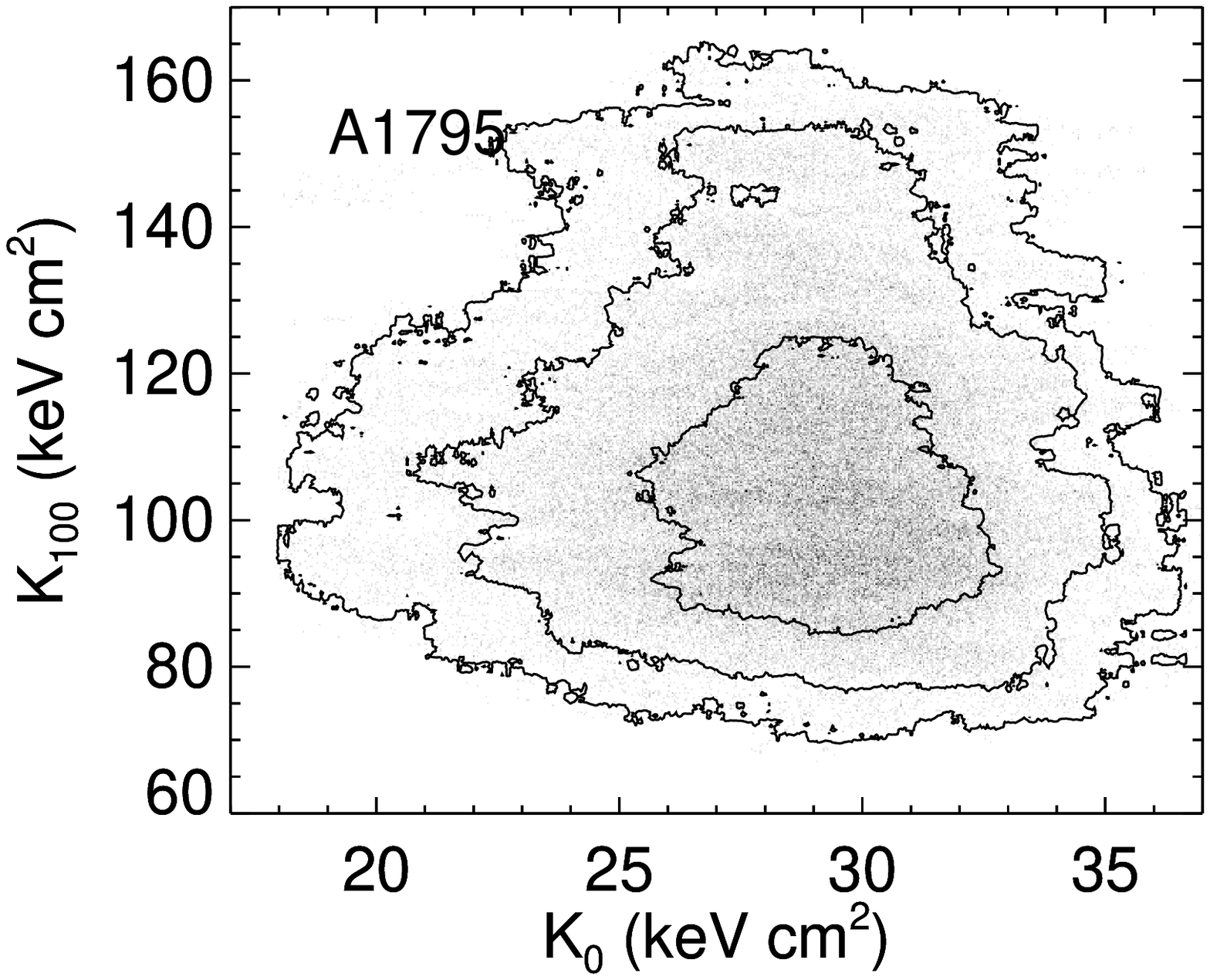}
  \includegraphics[width=0.32\textwidth]{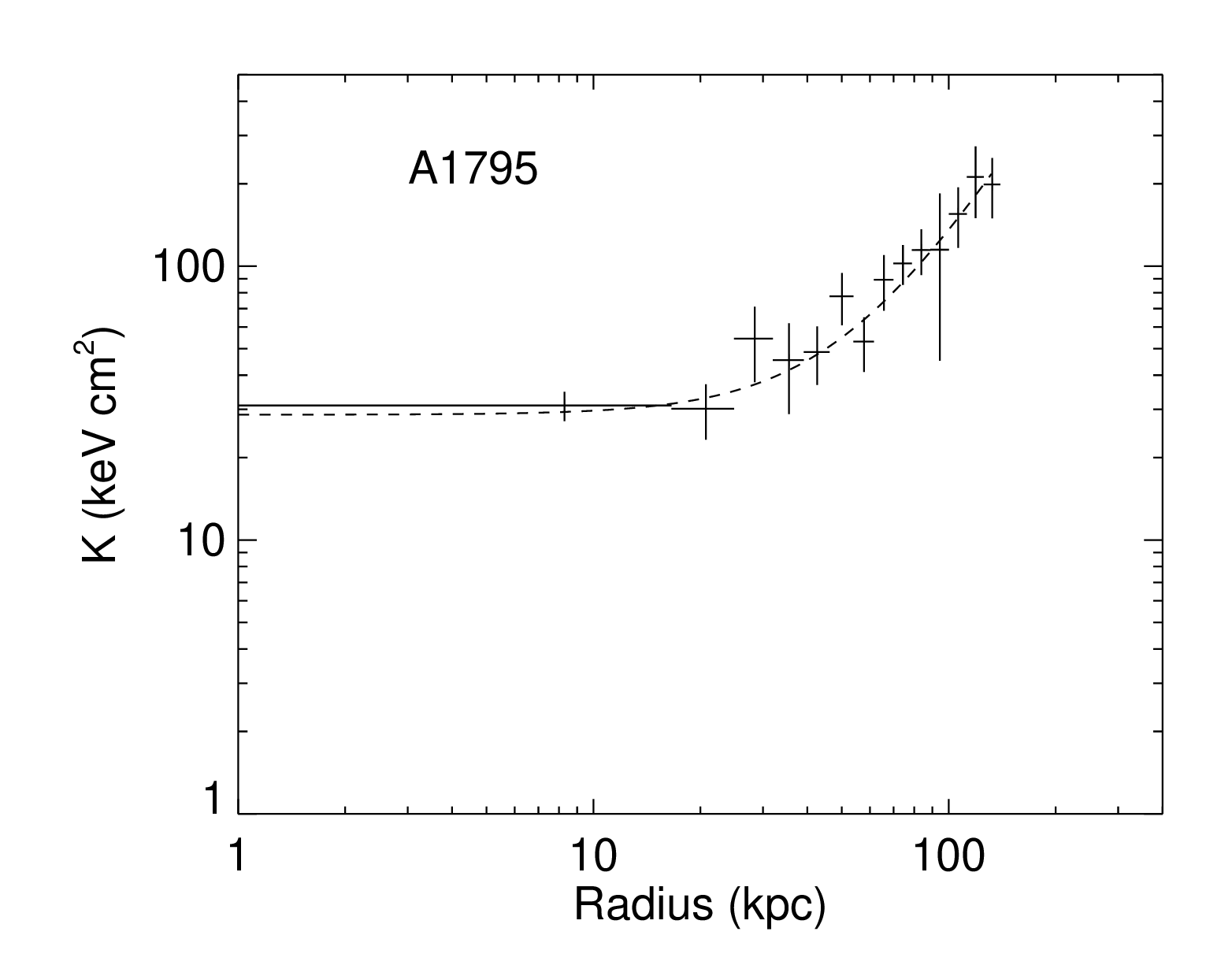}\\  
  \includegraphics[width=0.32\textwidth]{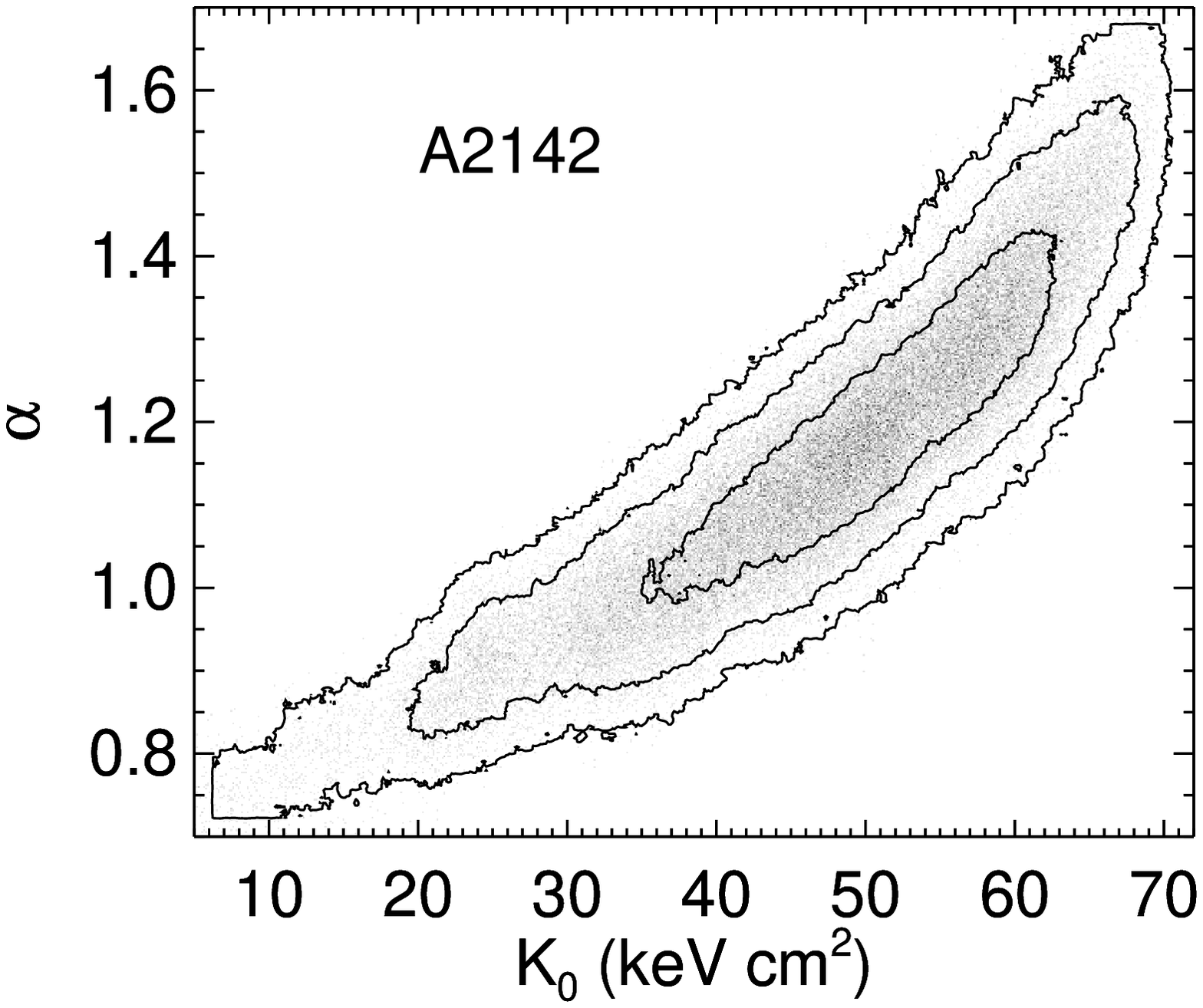}
  \includegraphics[width=0.32\textwidth]{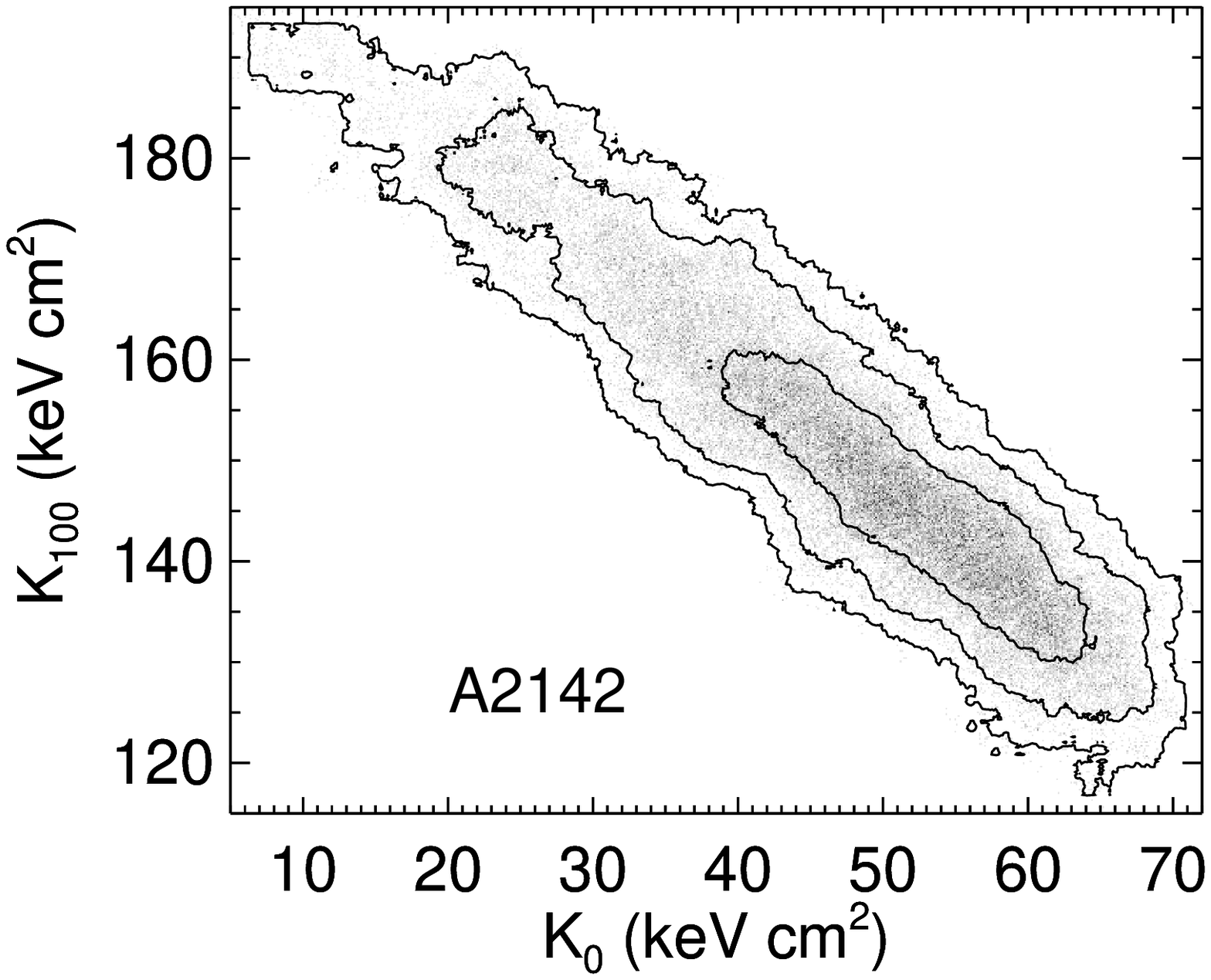}
  \includegraphics[width=0.32\textwidth]{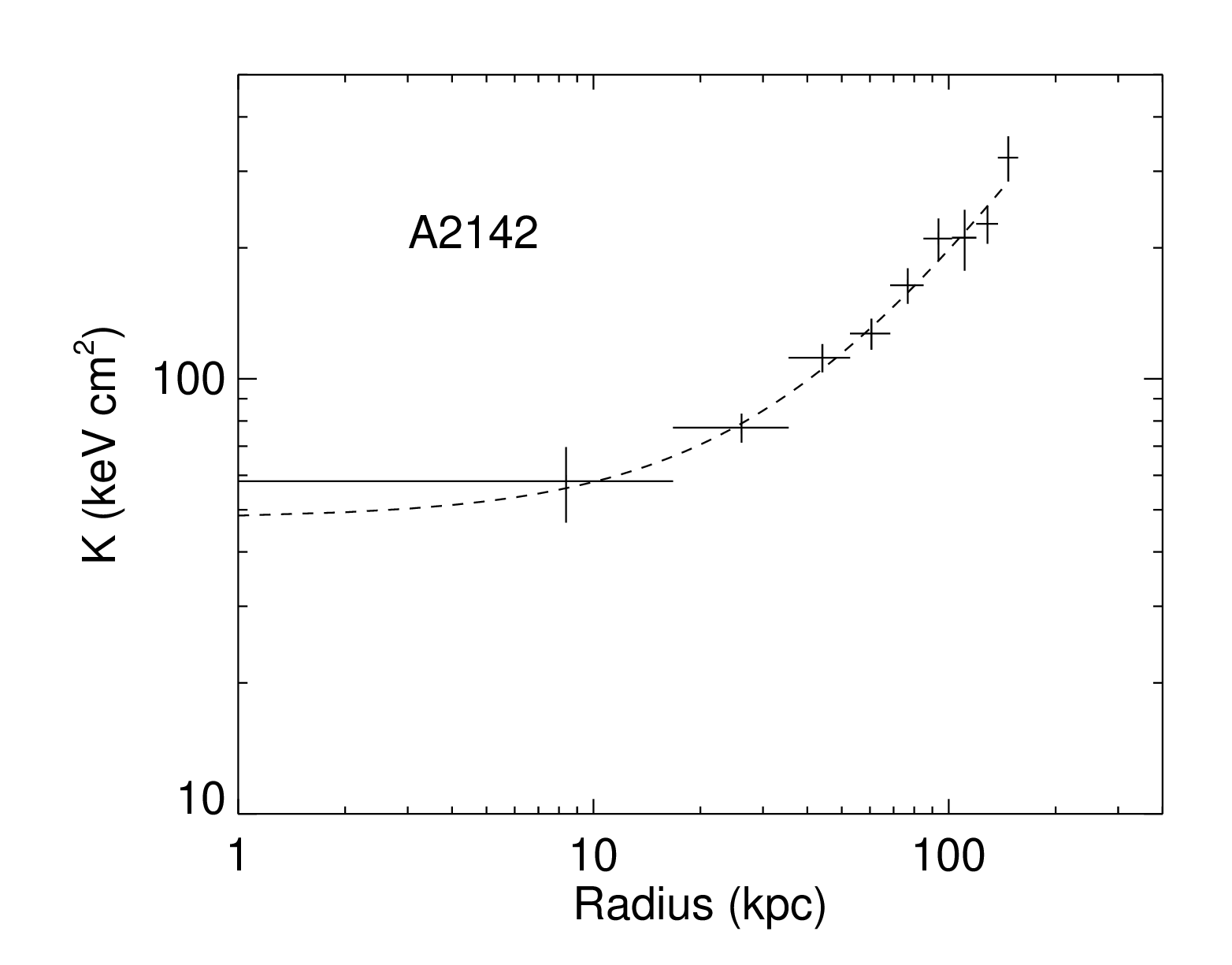}\\
  \includegraphics[width=0.32\textwidth]{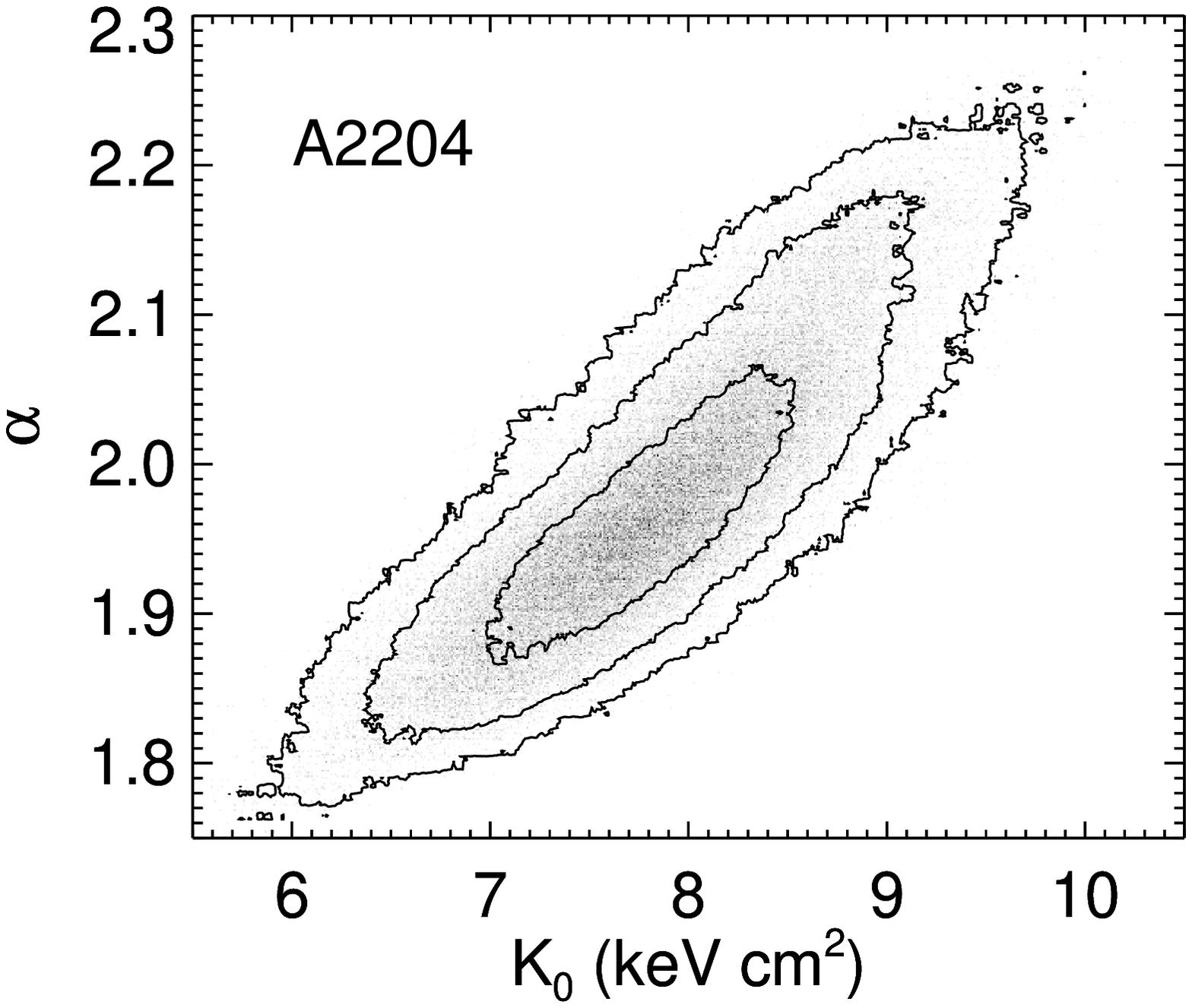}
  \includegraphics[width=0.32\textwidth]{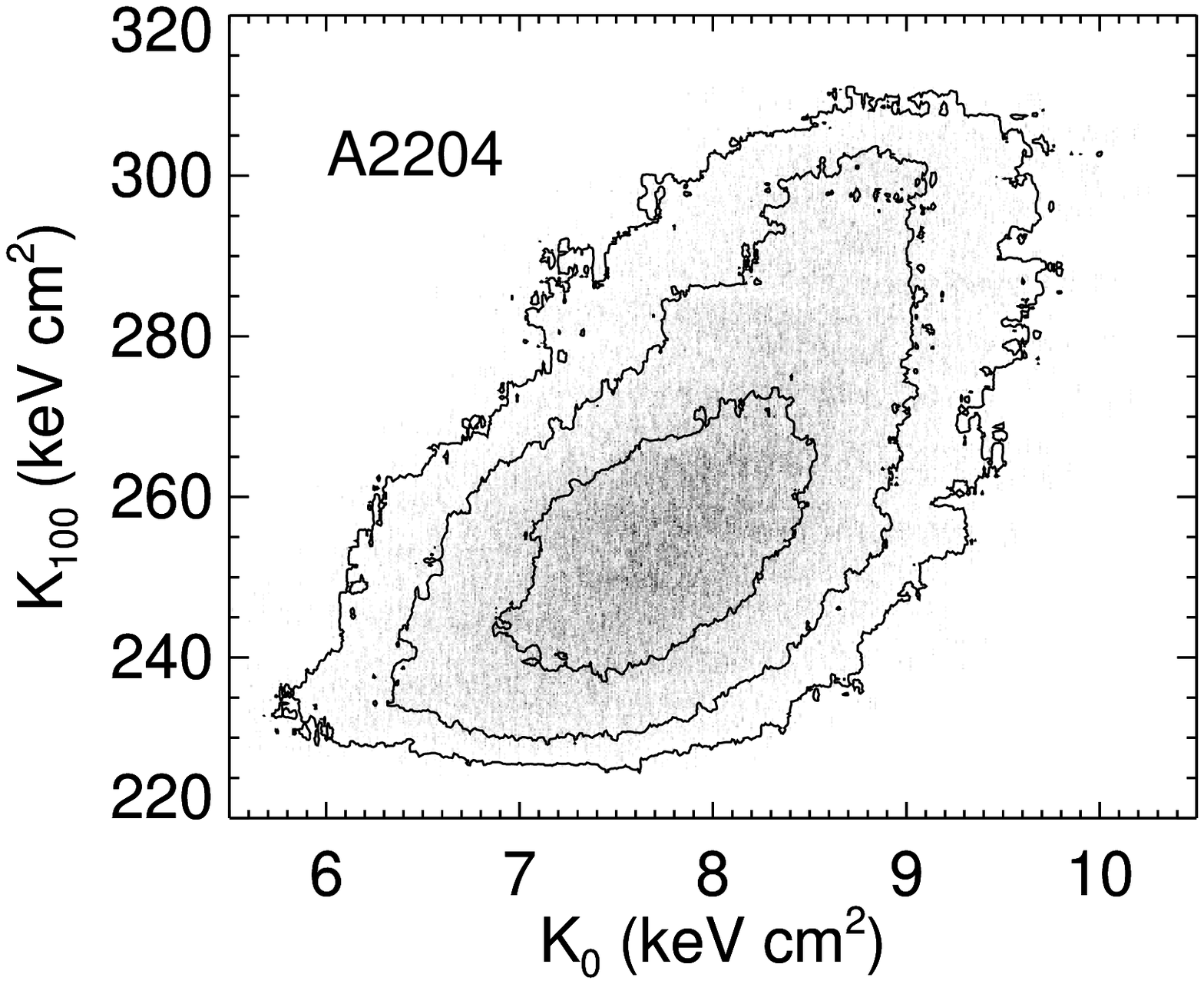}
  \includegraphics[width=0.32\textwidth]{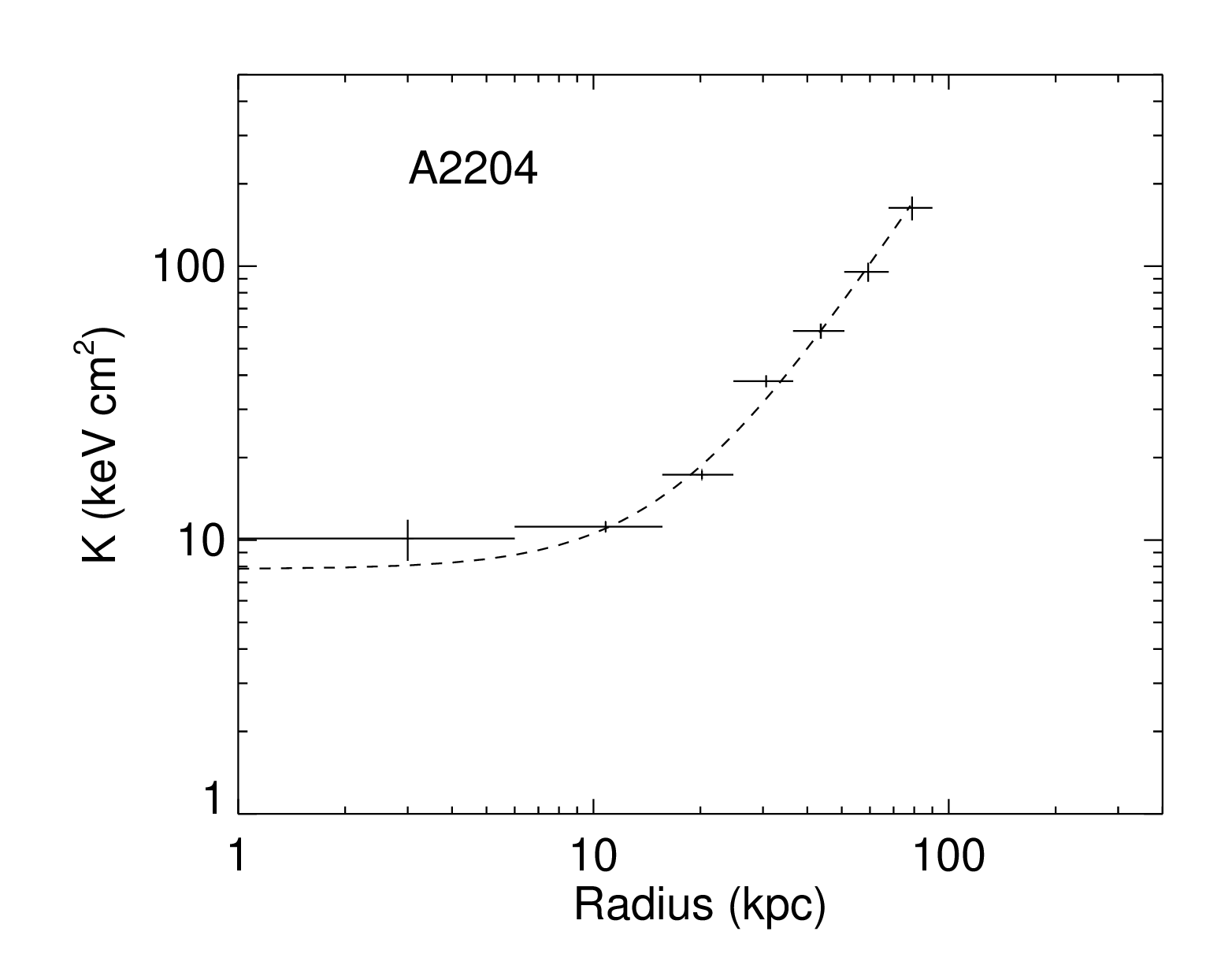}\\  
   \caption{The $K_0$-$\alpha$ and $K_0$-$K_{100}$ marginalized probability distributions obtained from the flat-core entropy model (eq.~\ref{eq:flat_core_model}) fitting for 
   FC (flat-core) sample. 
   The contours mark the 50\%, 90\% and 99\% inclusion levels based on the density of points starting from the innermost contour outwards.
   Greyscale denotes the PDF density.}
  \label{fig:flat_core_entr_fit_samp1}
\end{figure*}

\setcounter{figure}{4}
\begin{figure*}
 \centering
  \includegraphics[width=0.32\textwidth]{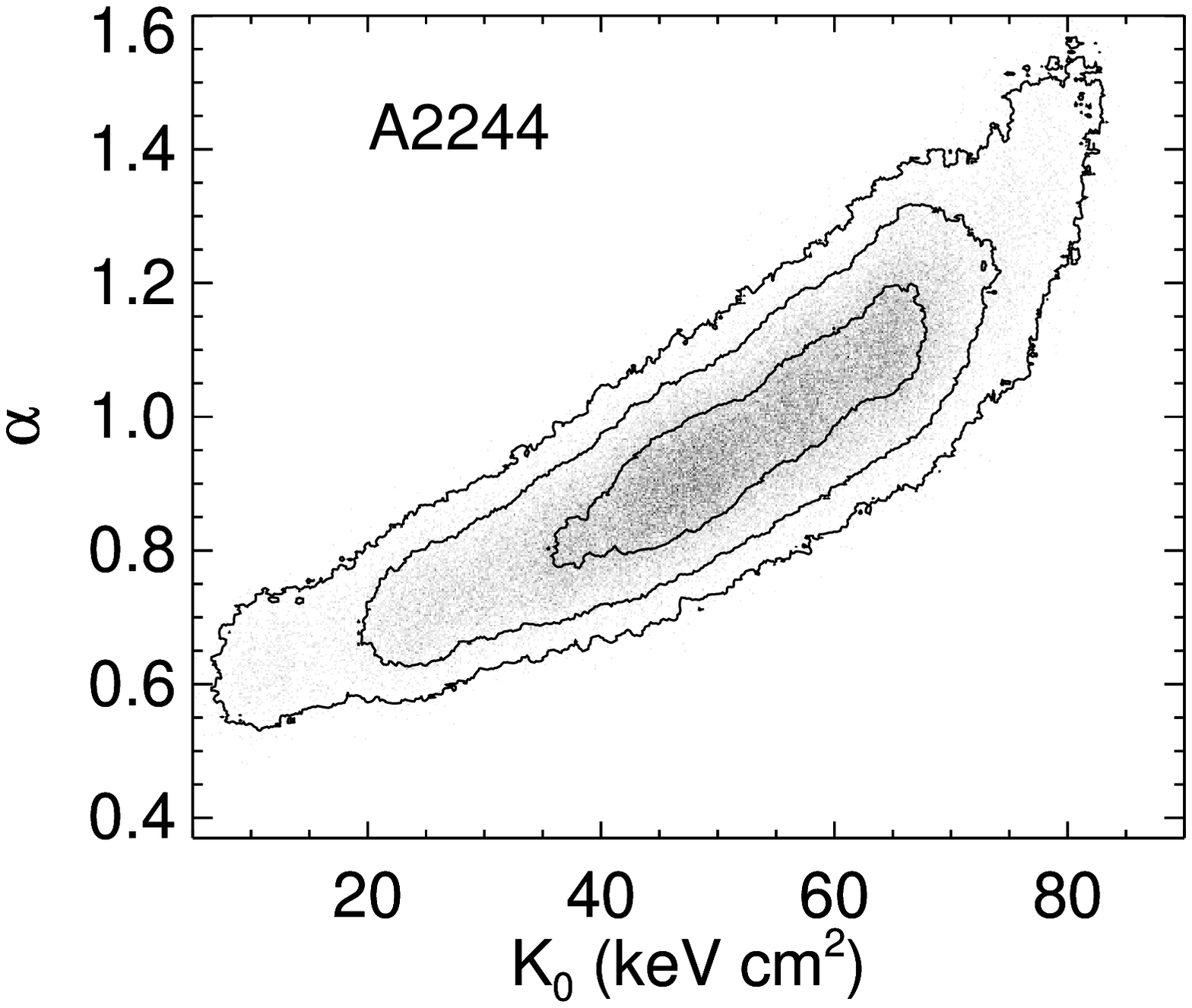}
  \includegraphics[width=0.32\textwidth]{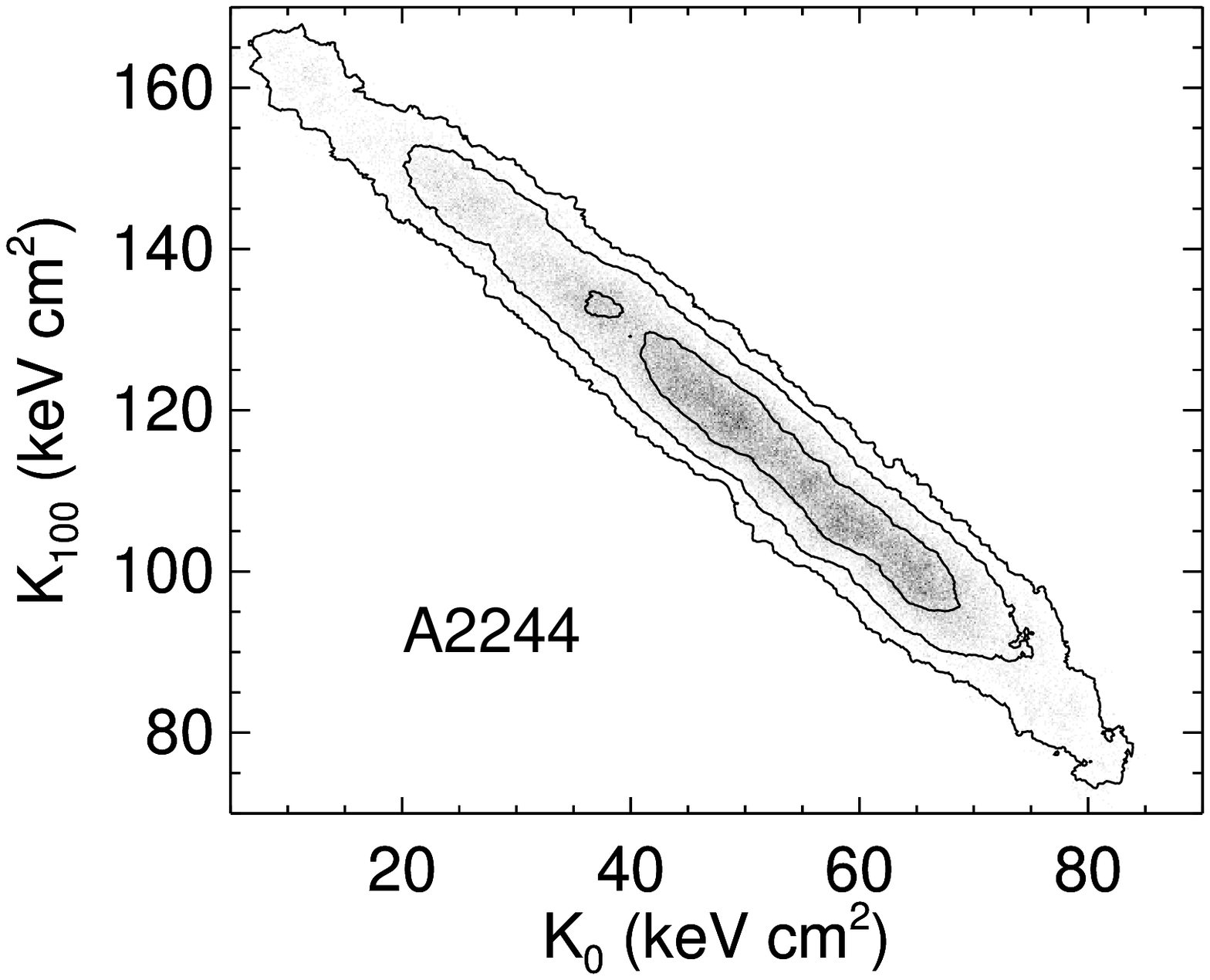}
  \includegraphics[width=0.32\textwidth]{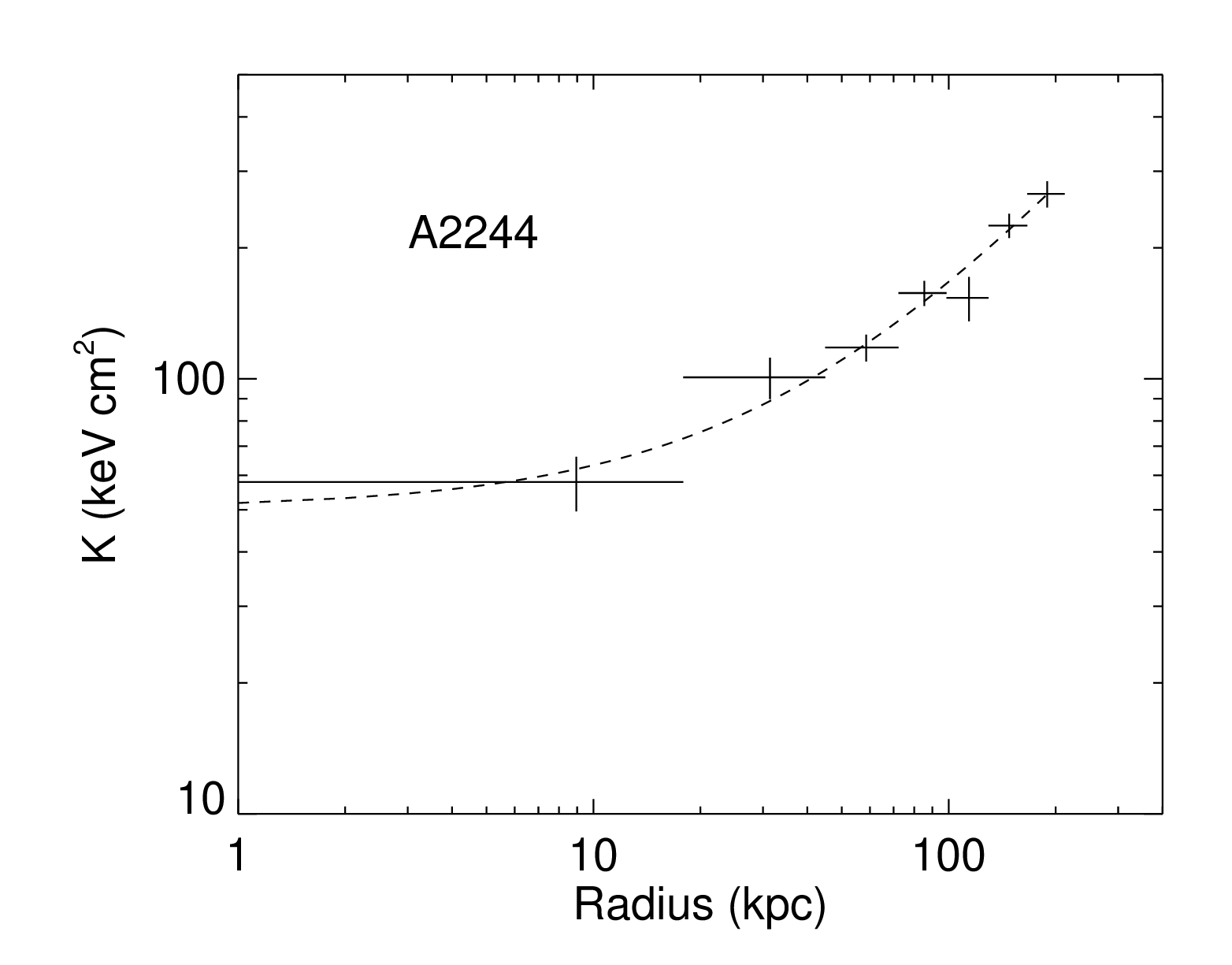}\\ 
  \includegraphics[width=0.32\textwidth]{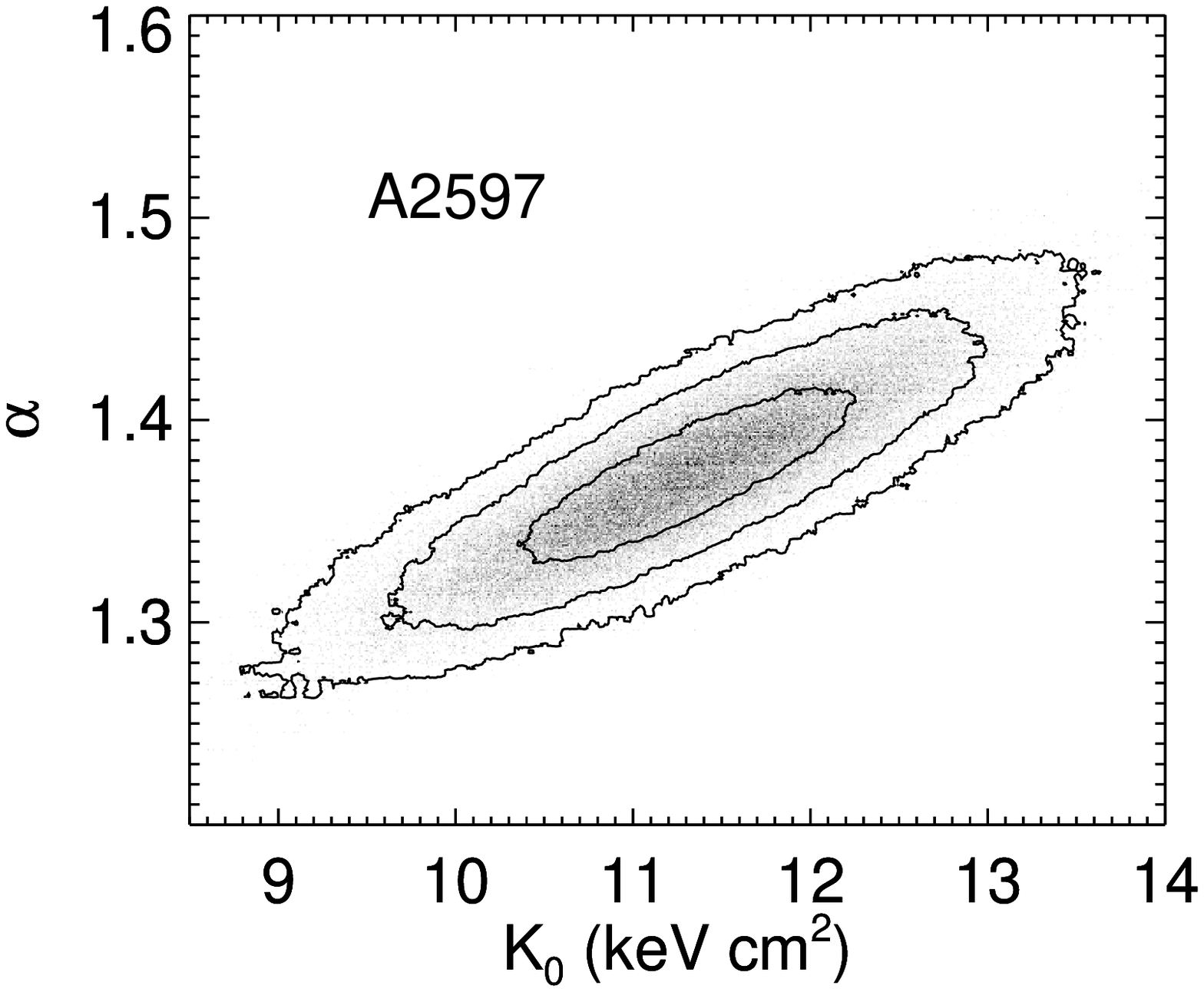}
  \includegraphics[width=0.32\textwidth]{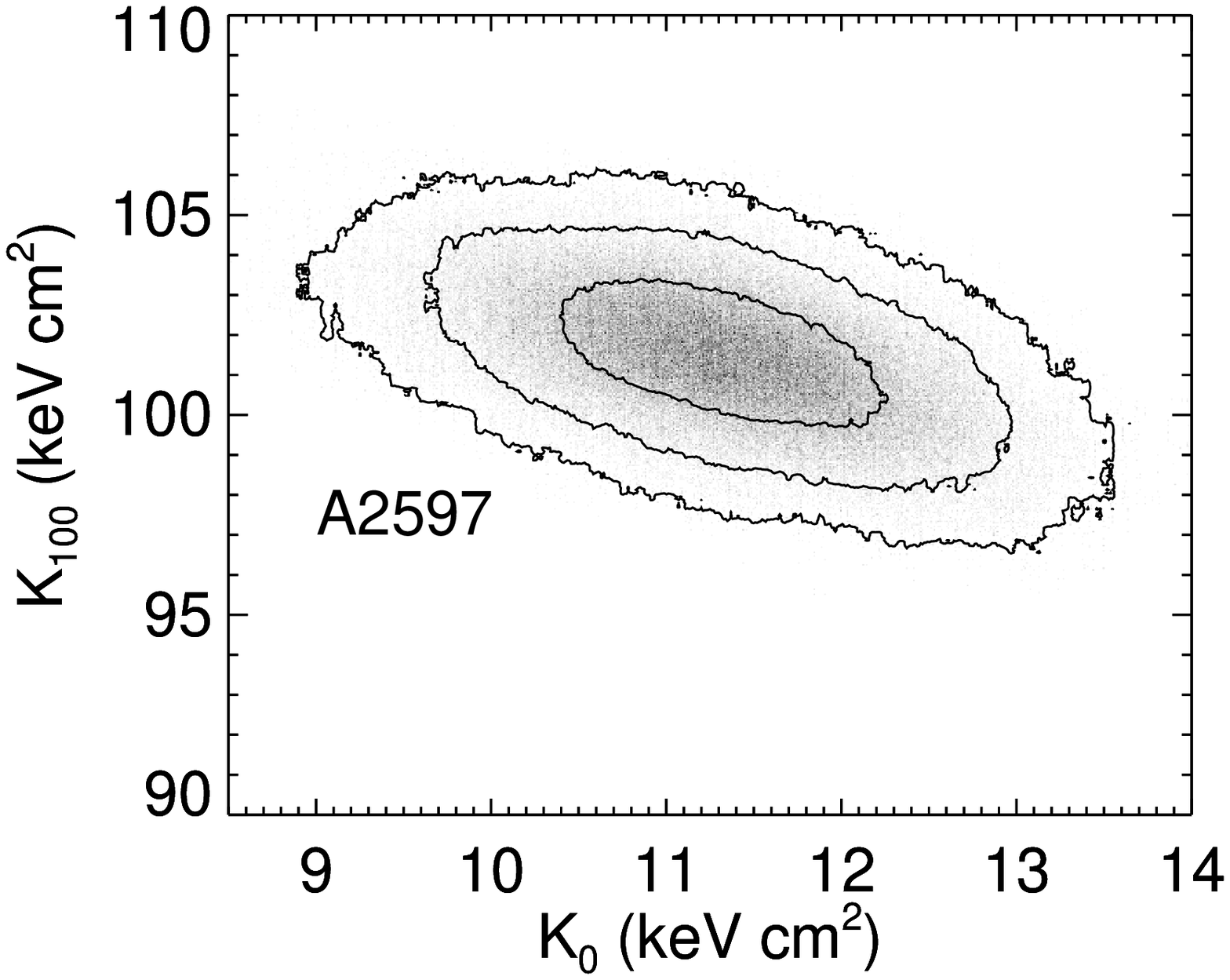}
  \includegraphics[width=0.32\textwidth]{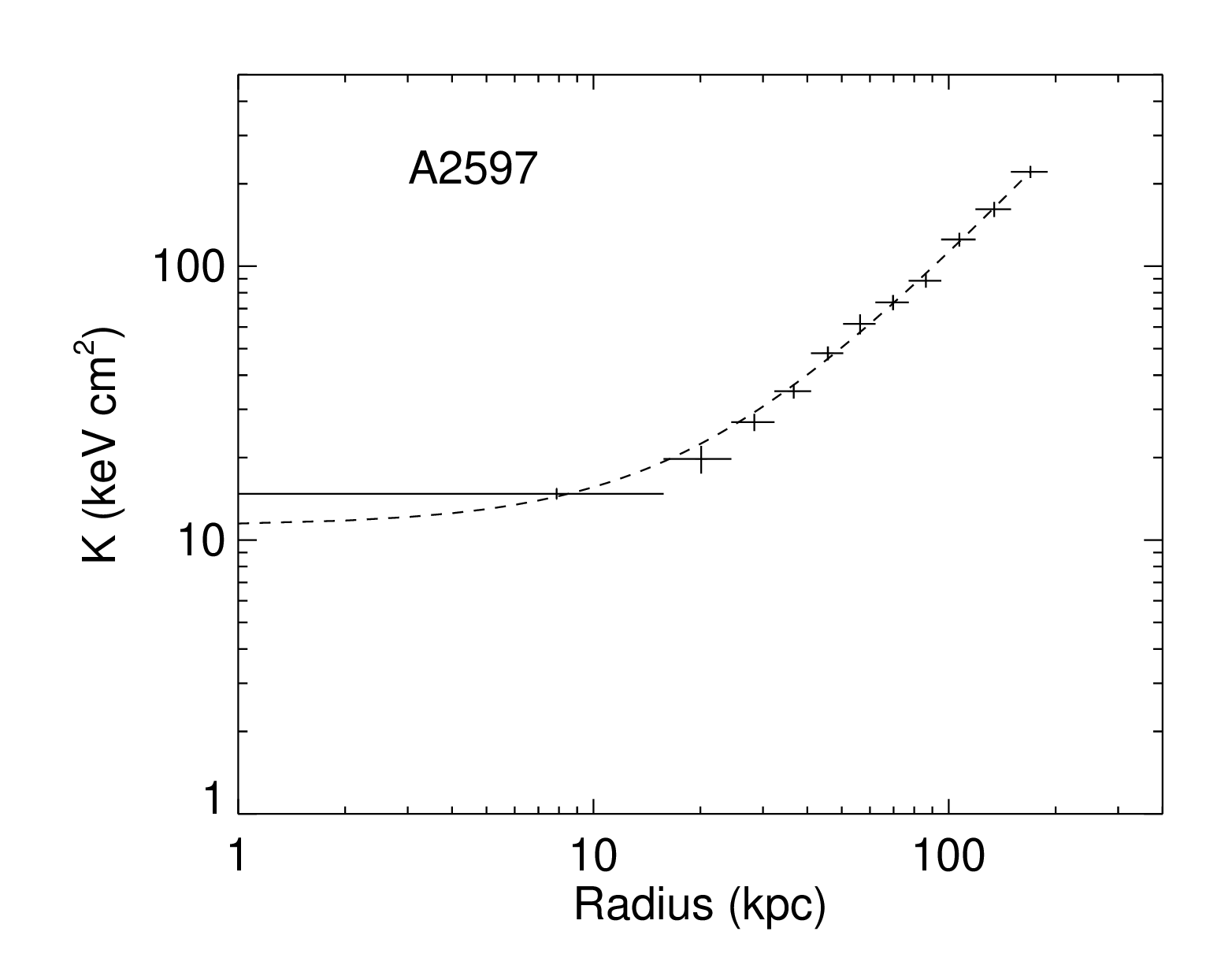}\\
  \includegraphics[width=0.32\textwidth]{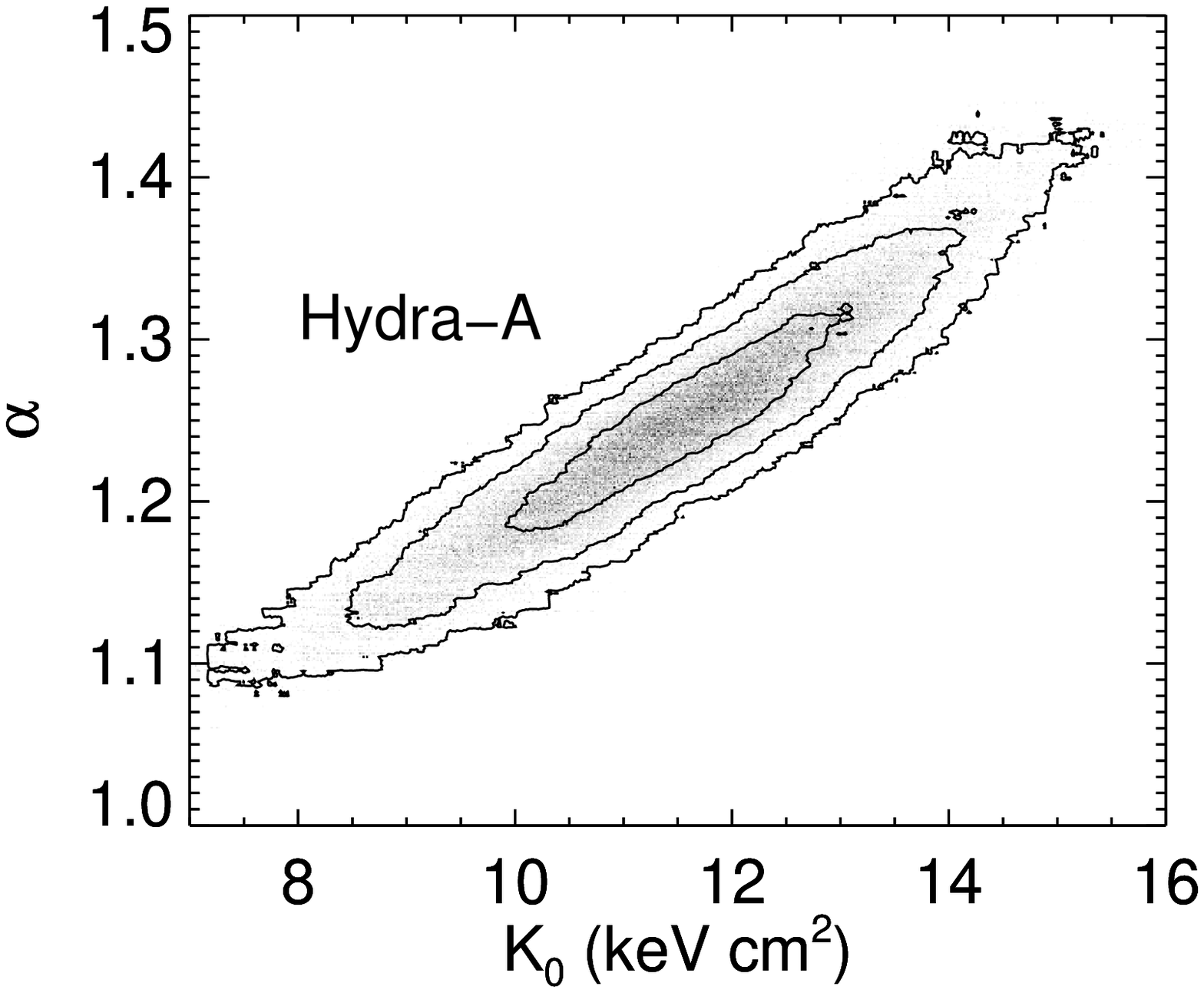}
  \includegraphics[width=0.32\textwidth]{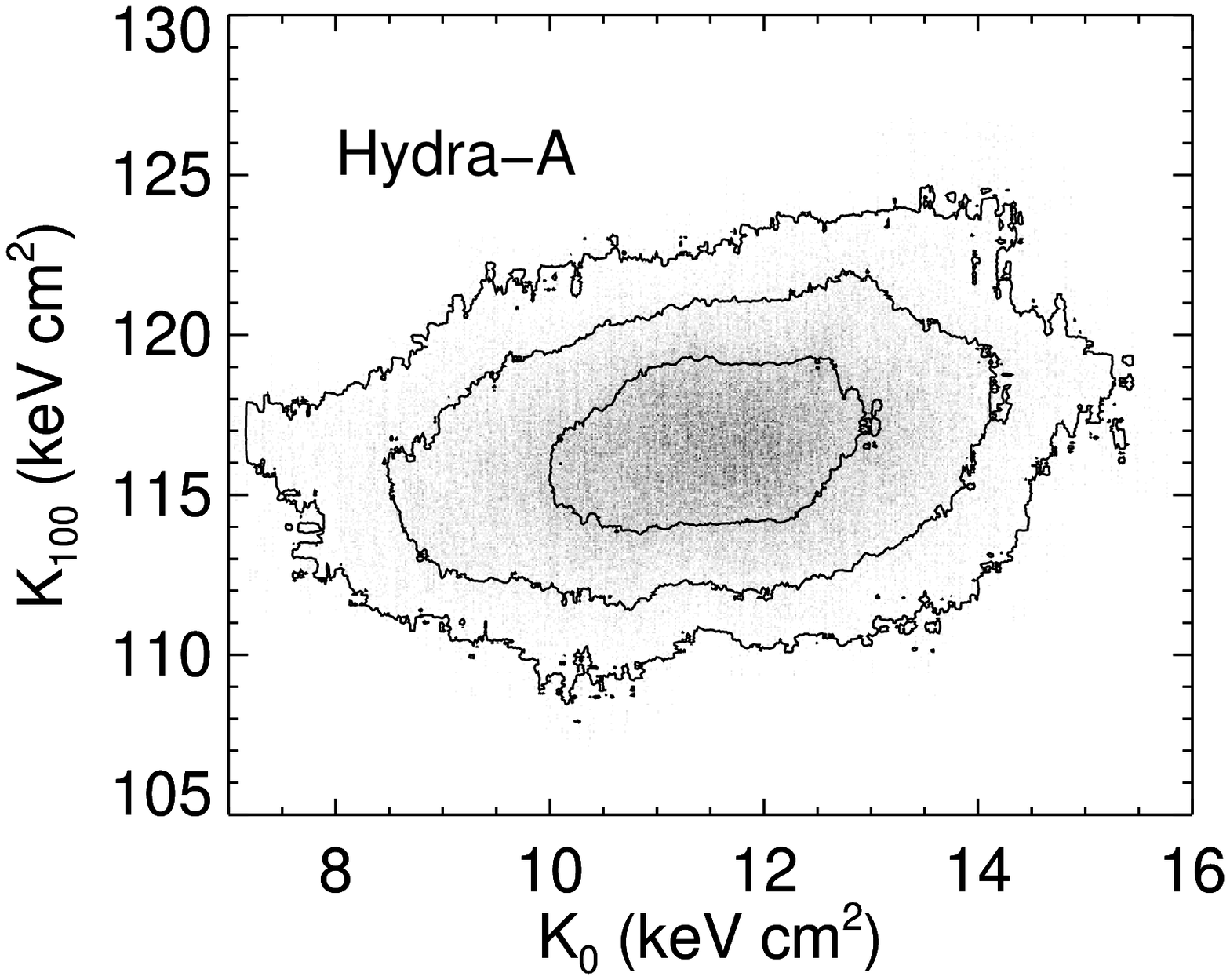}
  \includegraphics[width=0.32\textwidth]{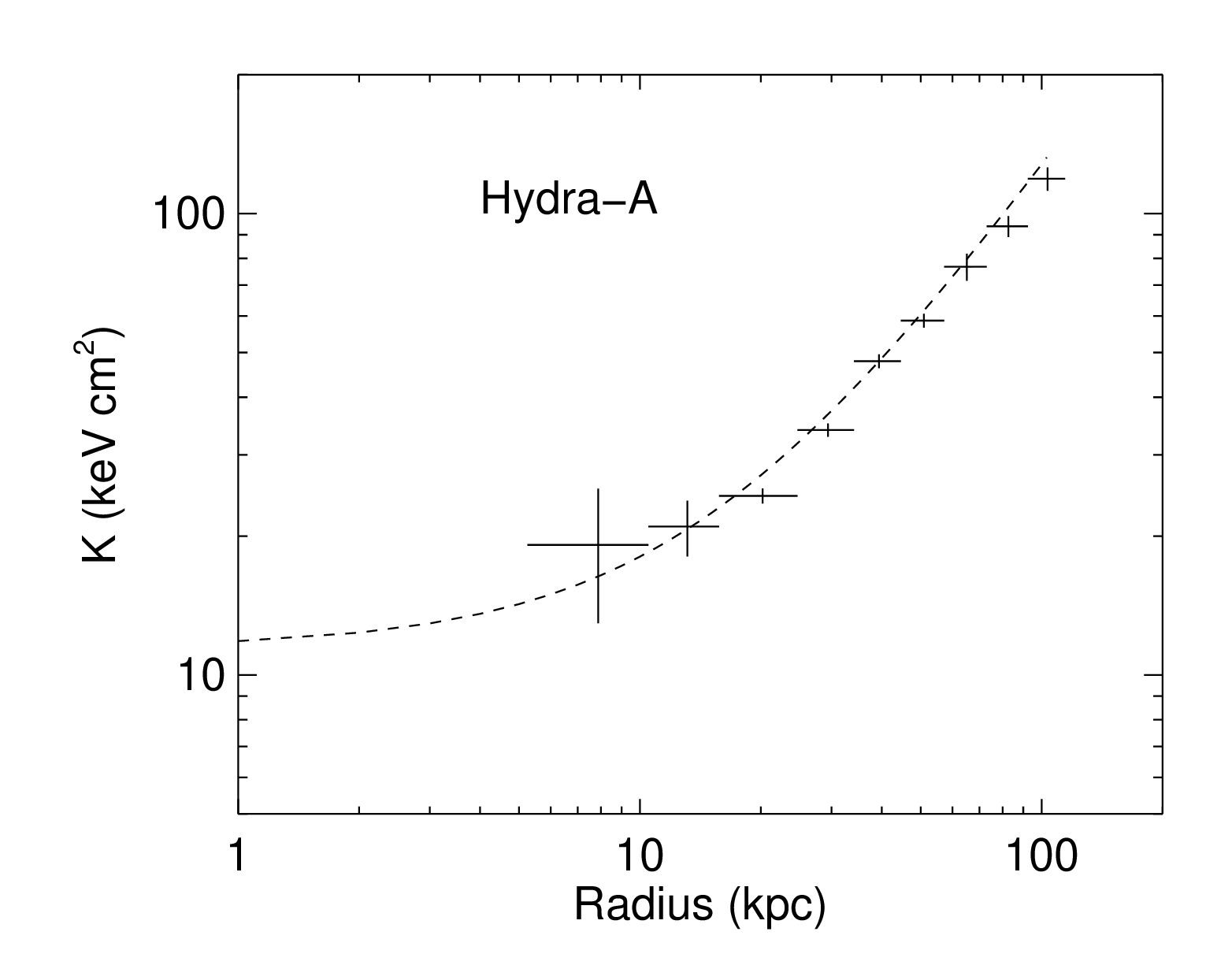}\\
  \includegraphics[width=0.32\textwidth]{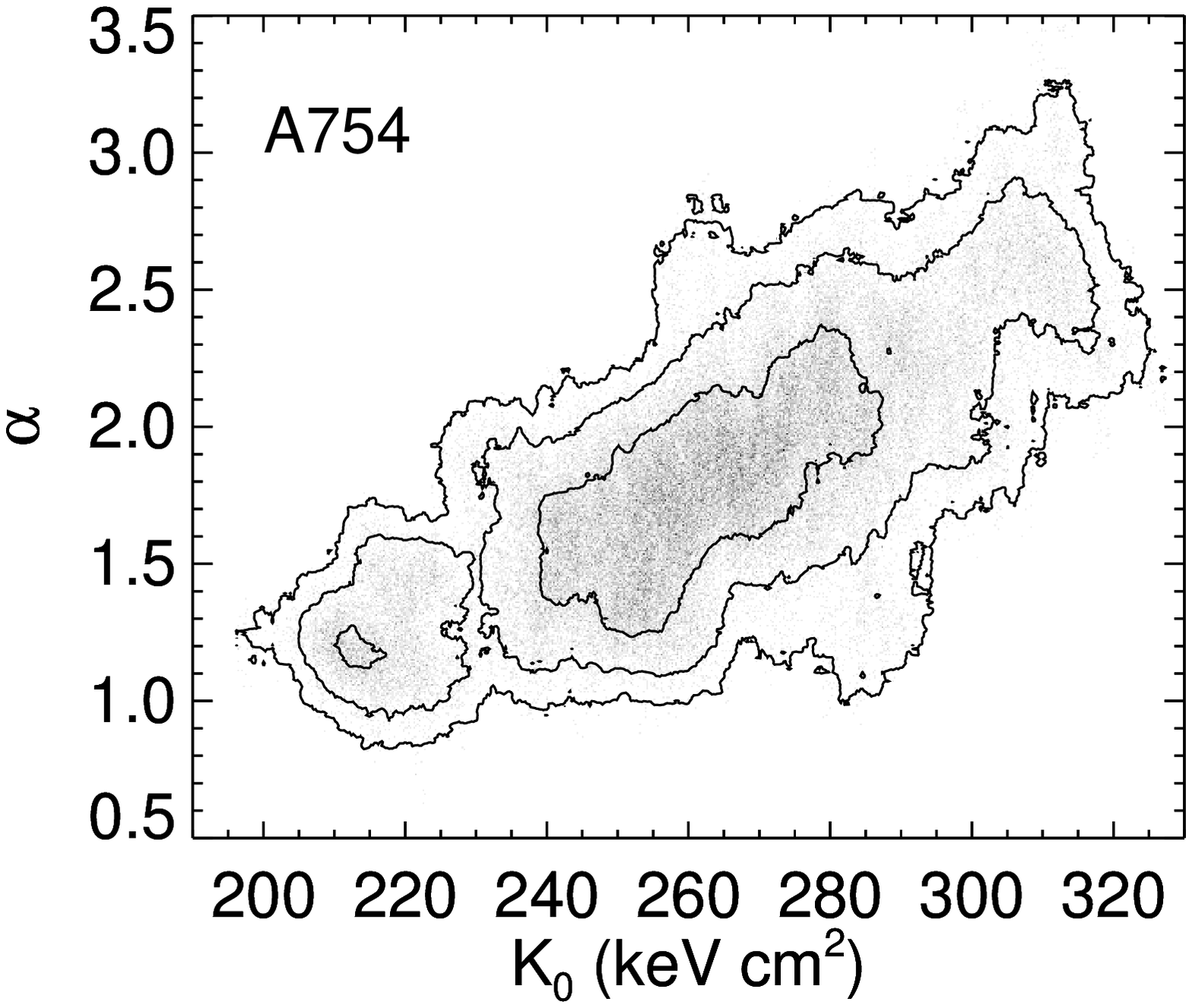}
  \includegraphics[width=0.32\textwidth]{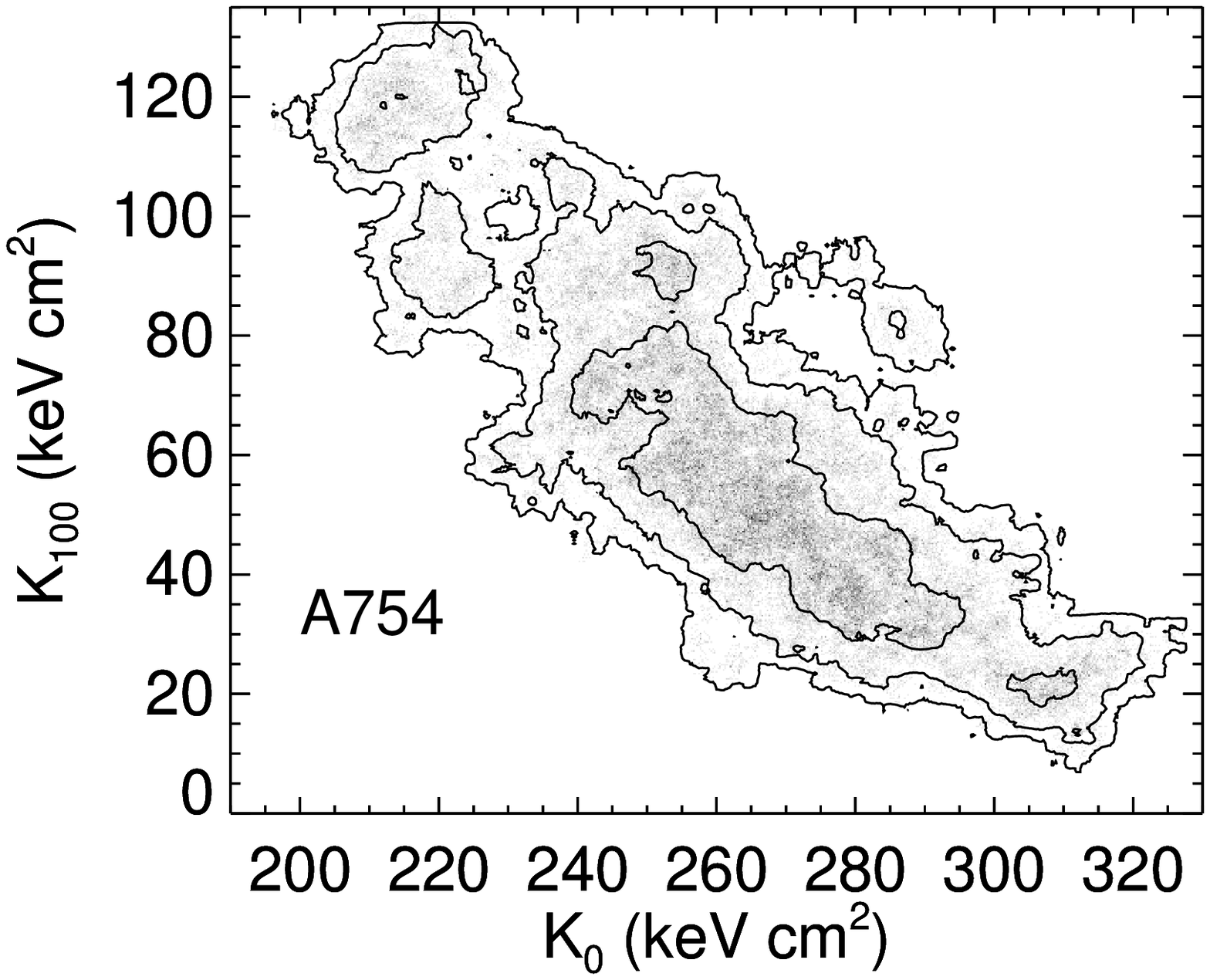}
  \includegraphics[width=0.32\textwidth]{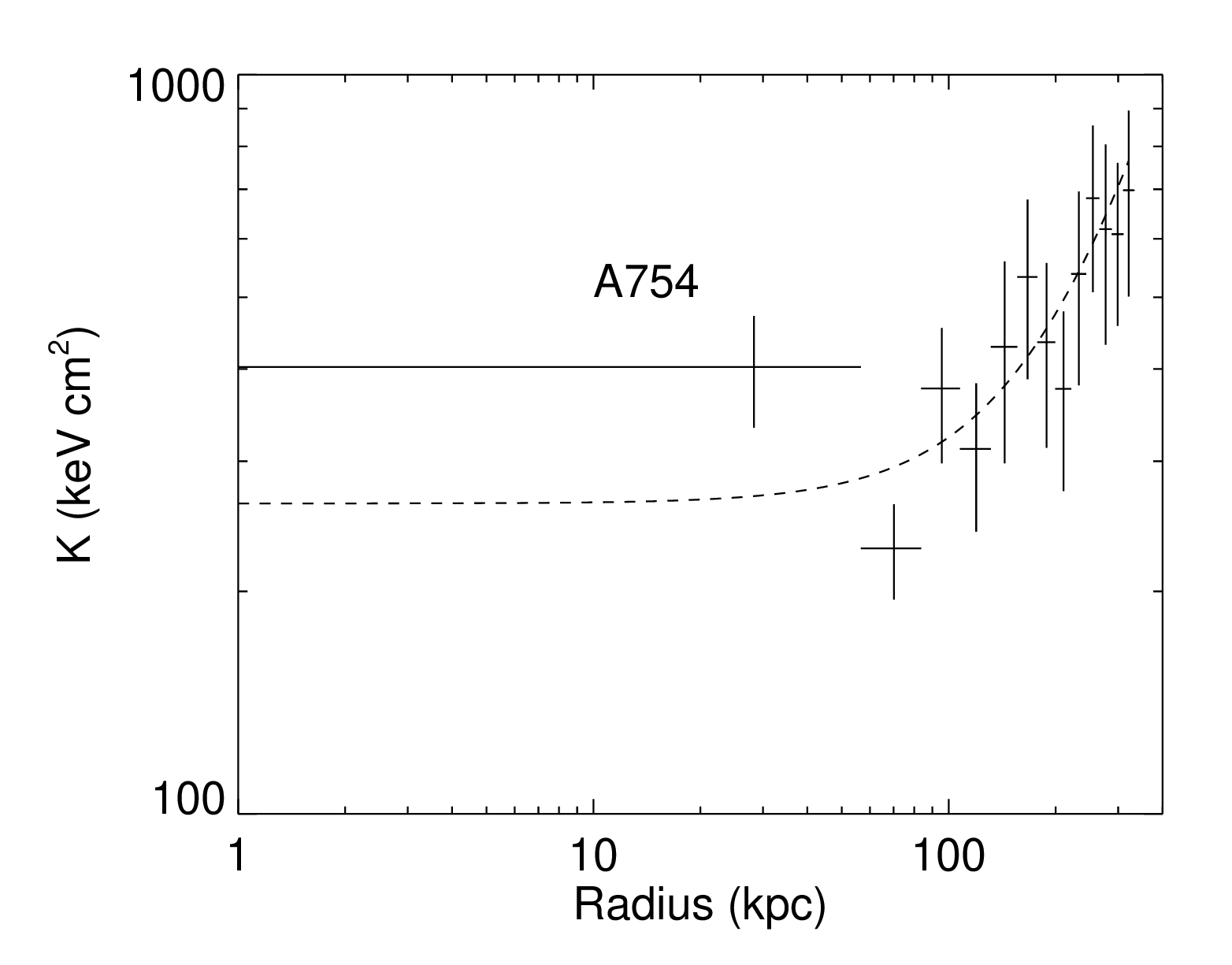}\\
  \includegraphics[width=0.32\textwidth]{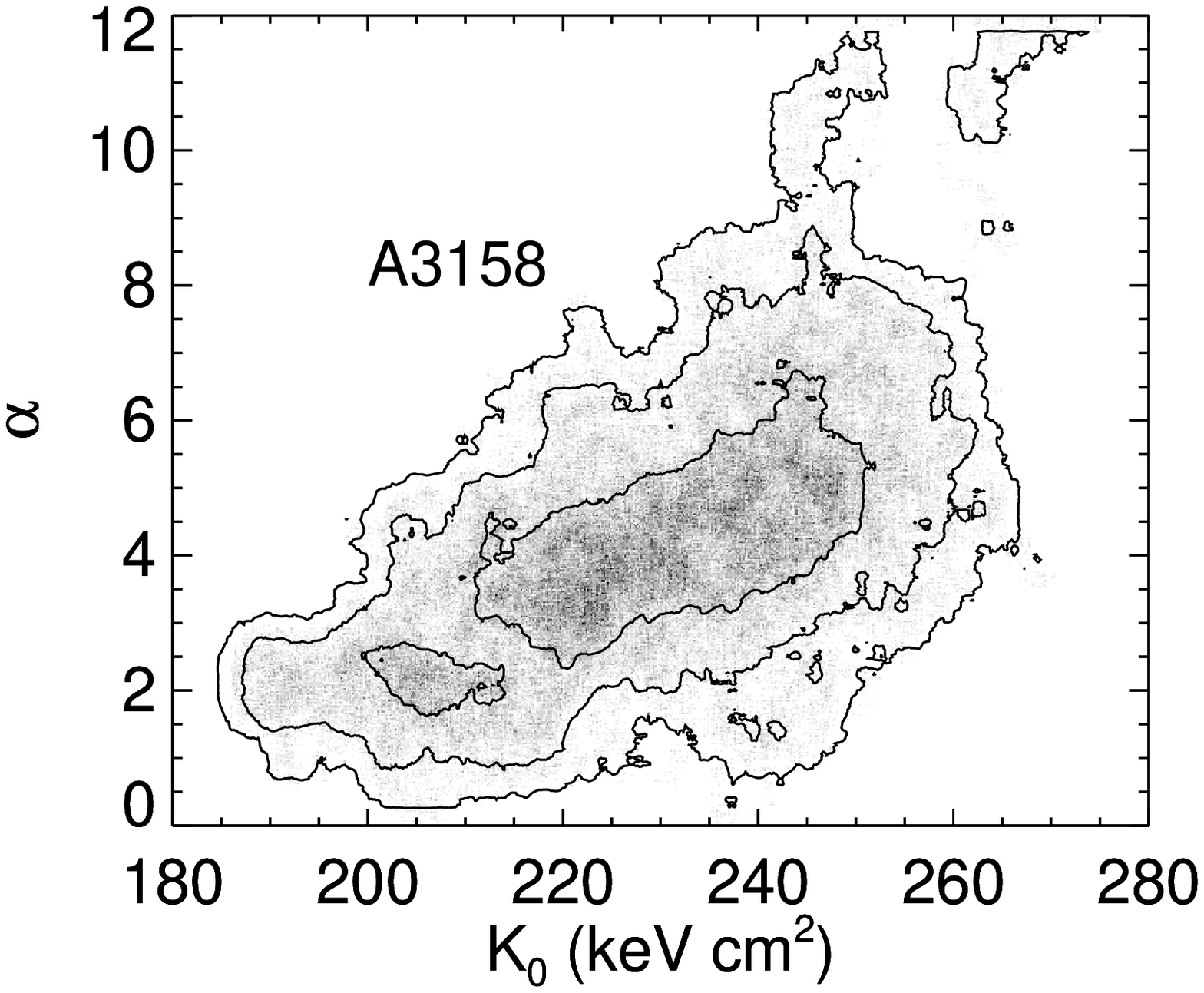}
  \includegraphics[width=0.32\textwidth]{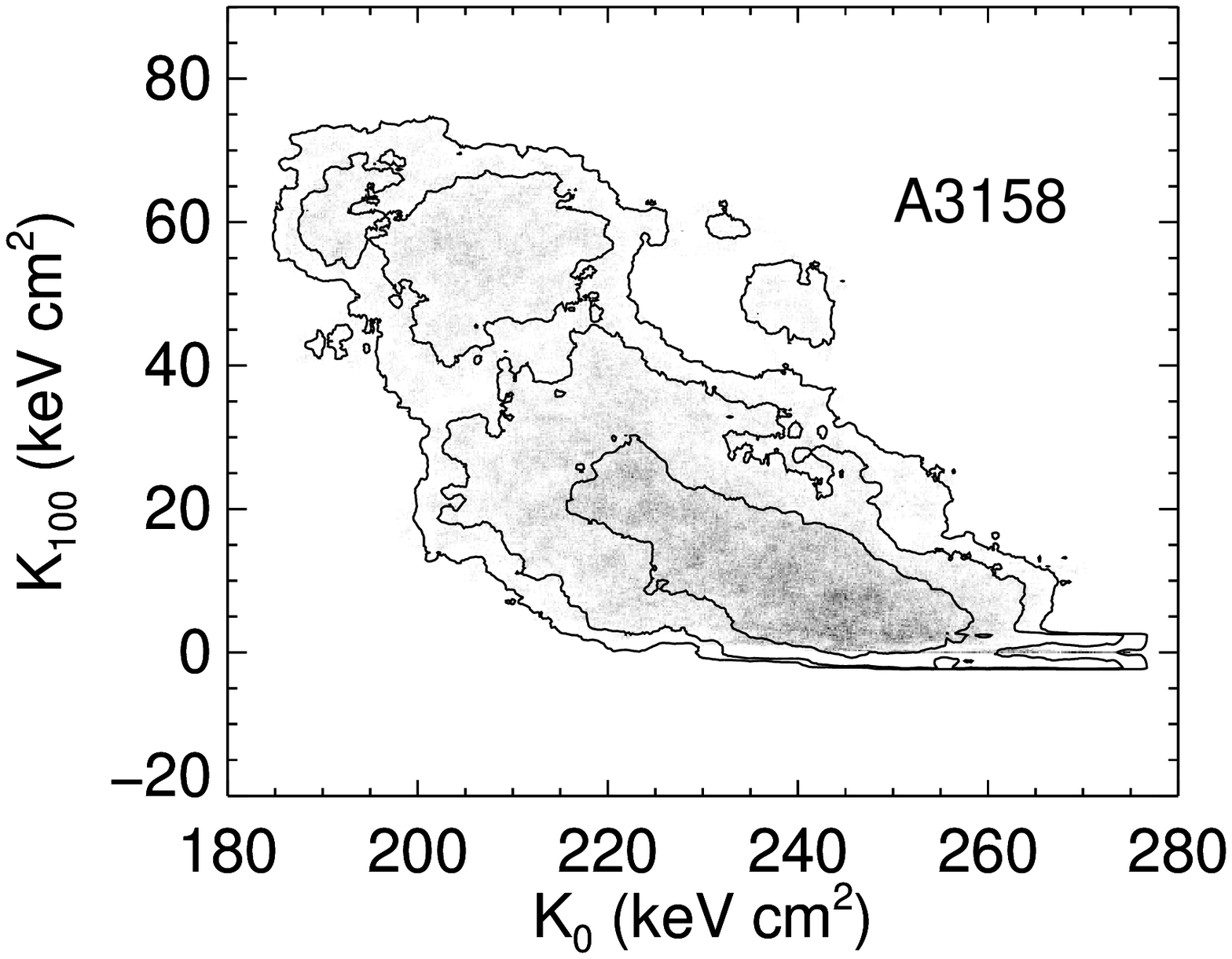}
  \includegraphics[width=0.32\textwidth]{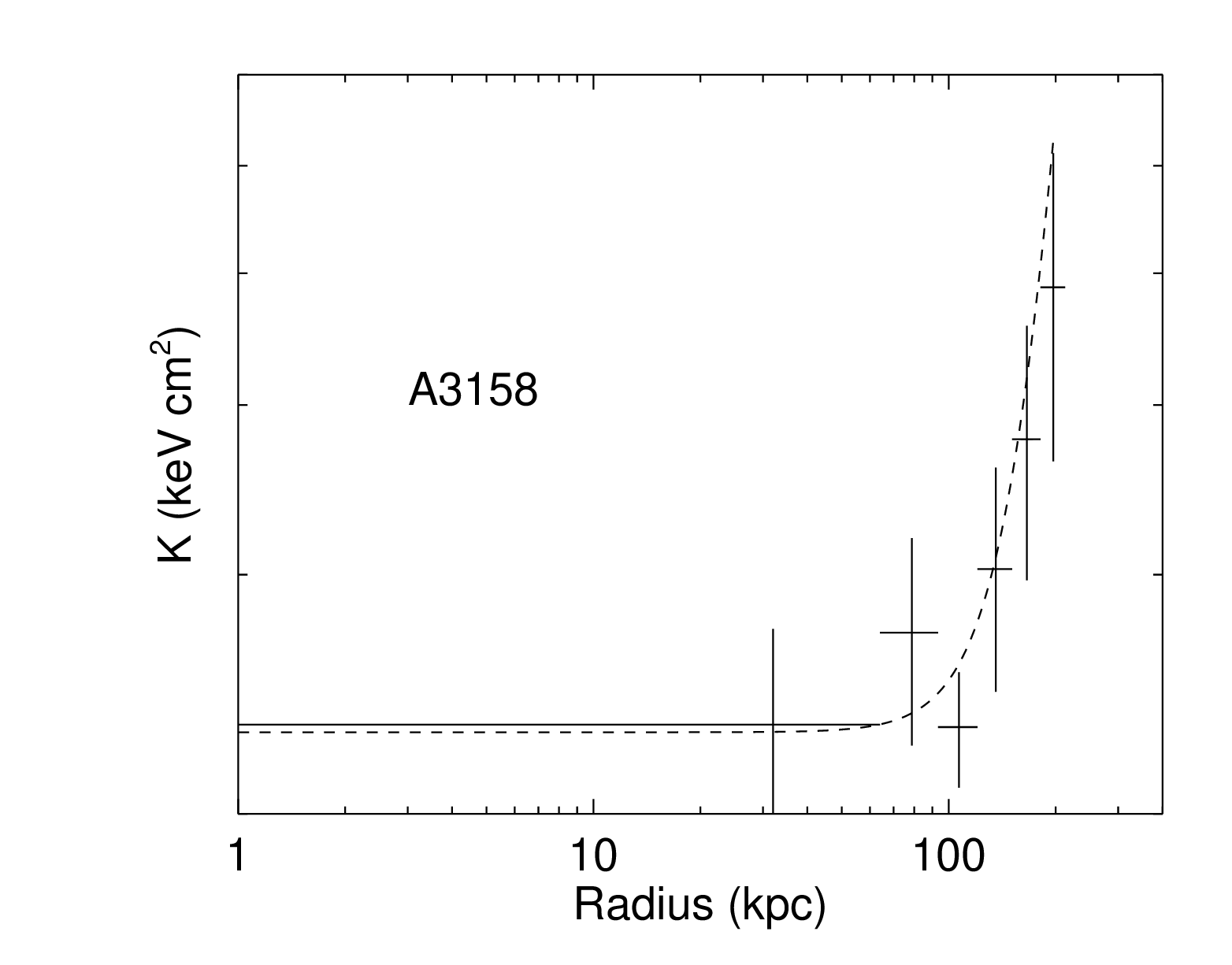}\\   
  \caption{(Contd.)}
  \label{fig:flat_core_entr_fit_samp1}
\end{figure*}

\setcounter{figure}{4}
\begin{figure*}
 \centering
  \includegraphics[width=0.32\textwidth]{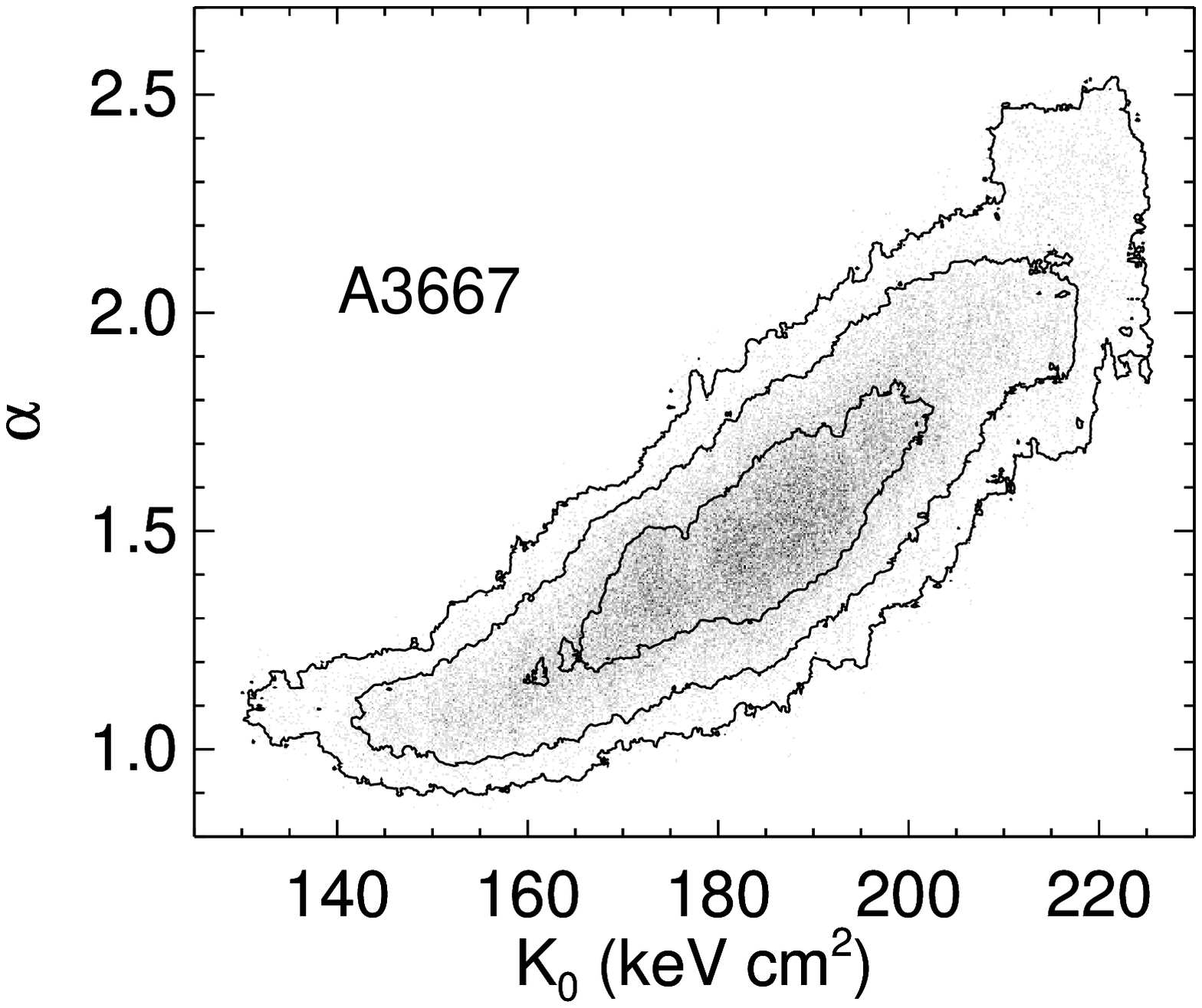}
  \includegraphics[width=0.32\textwidth]{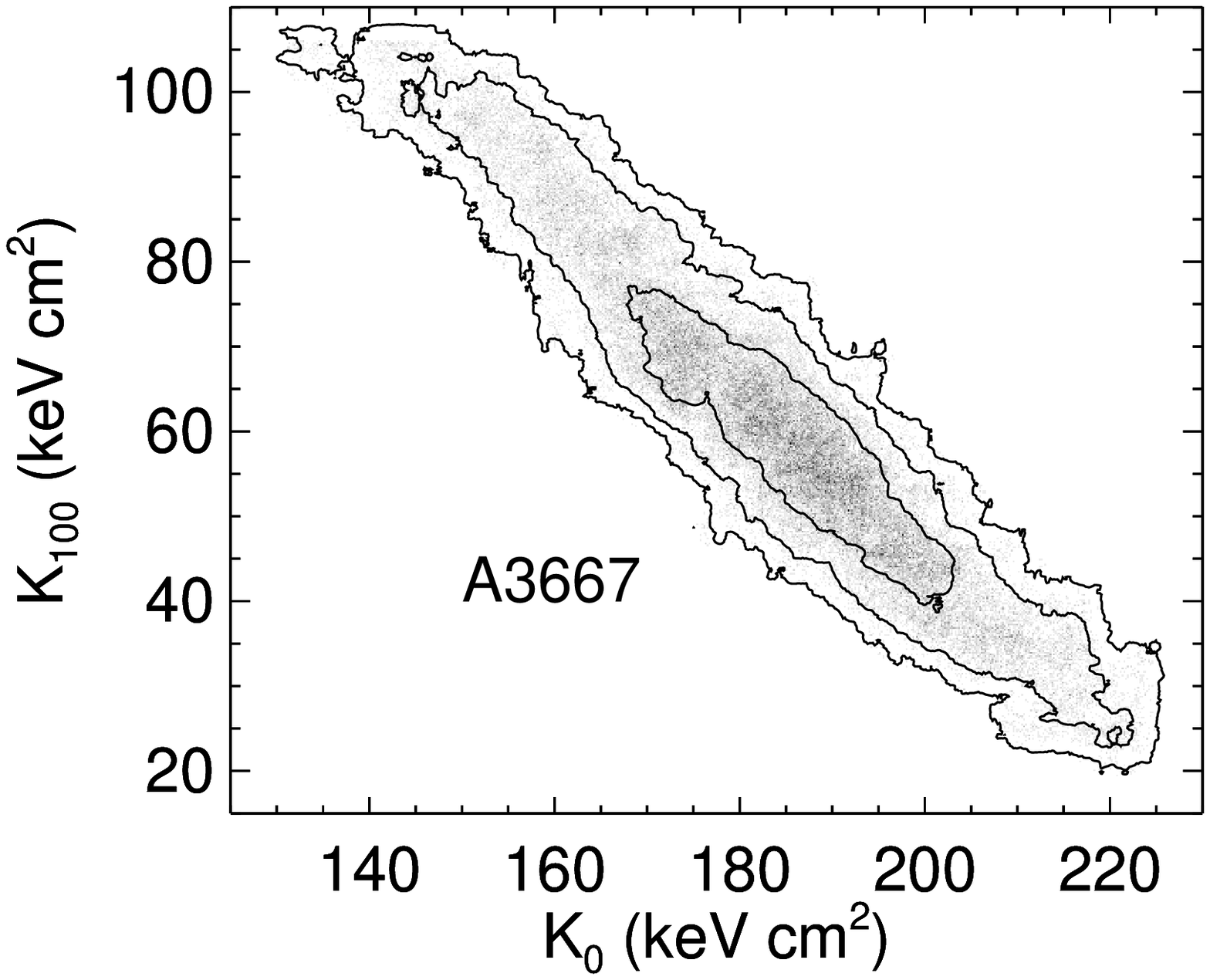}
  \includegraphics[width=0.32\textwidth]{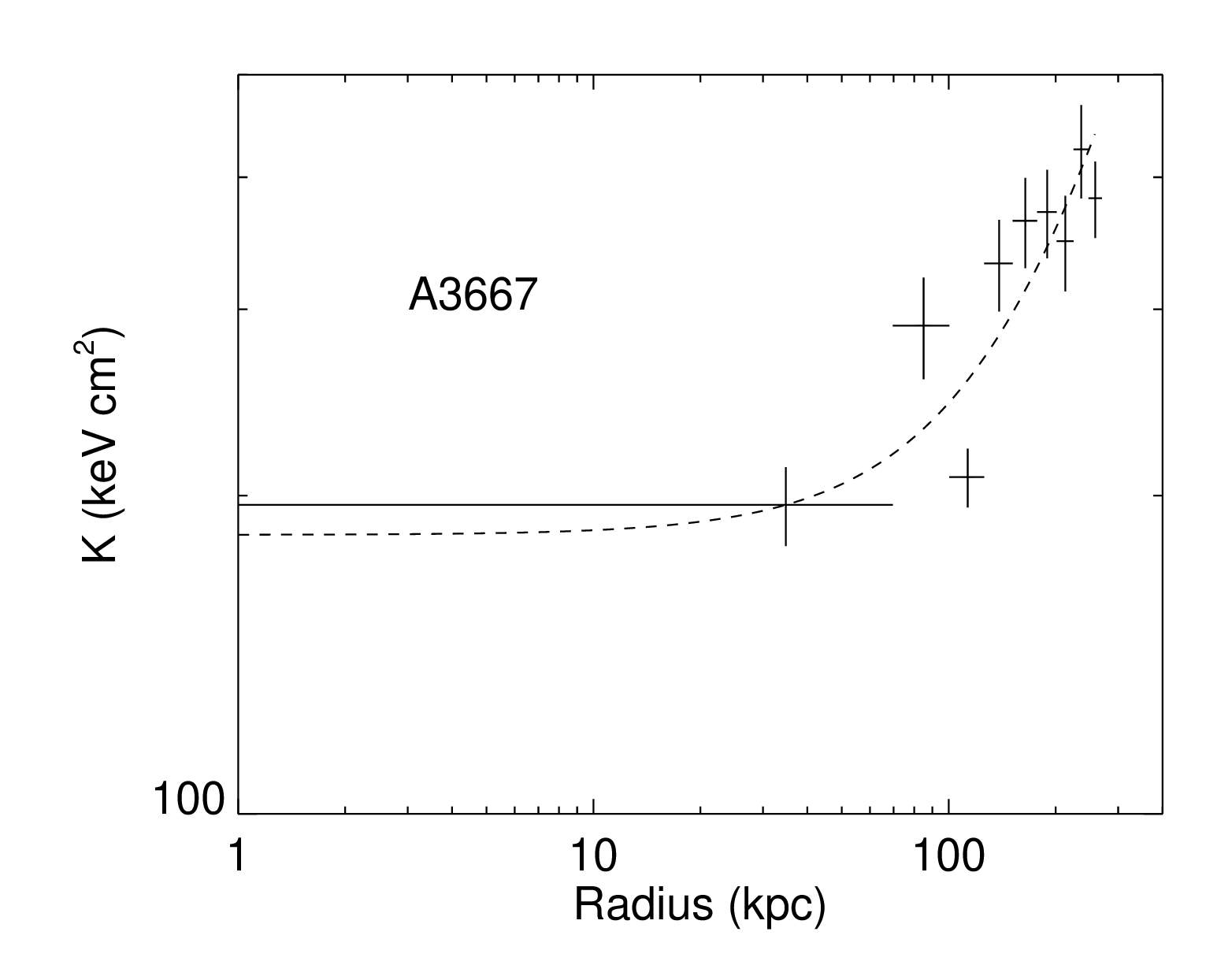}\\  
  \includegraphics[width=0.32\textwidth]{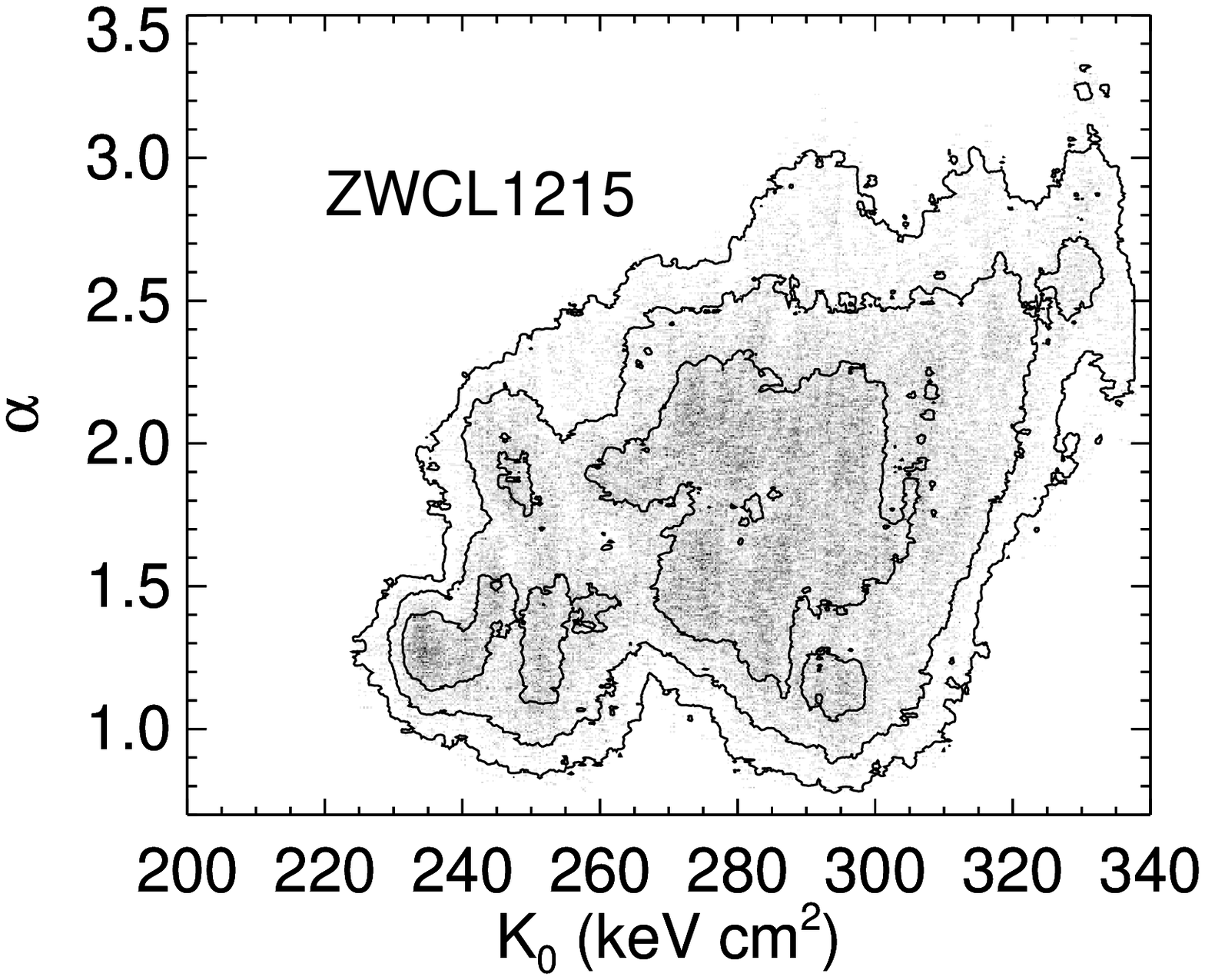}
  \includegraphics[width=0.32\textwidth]{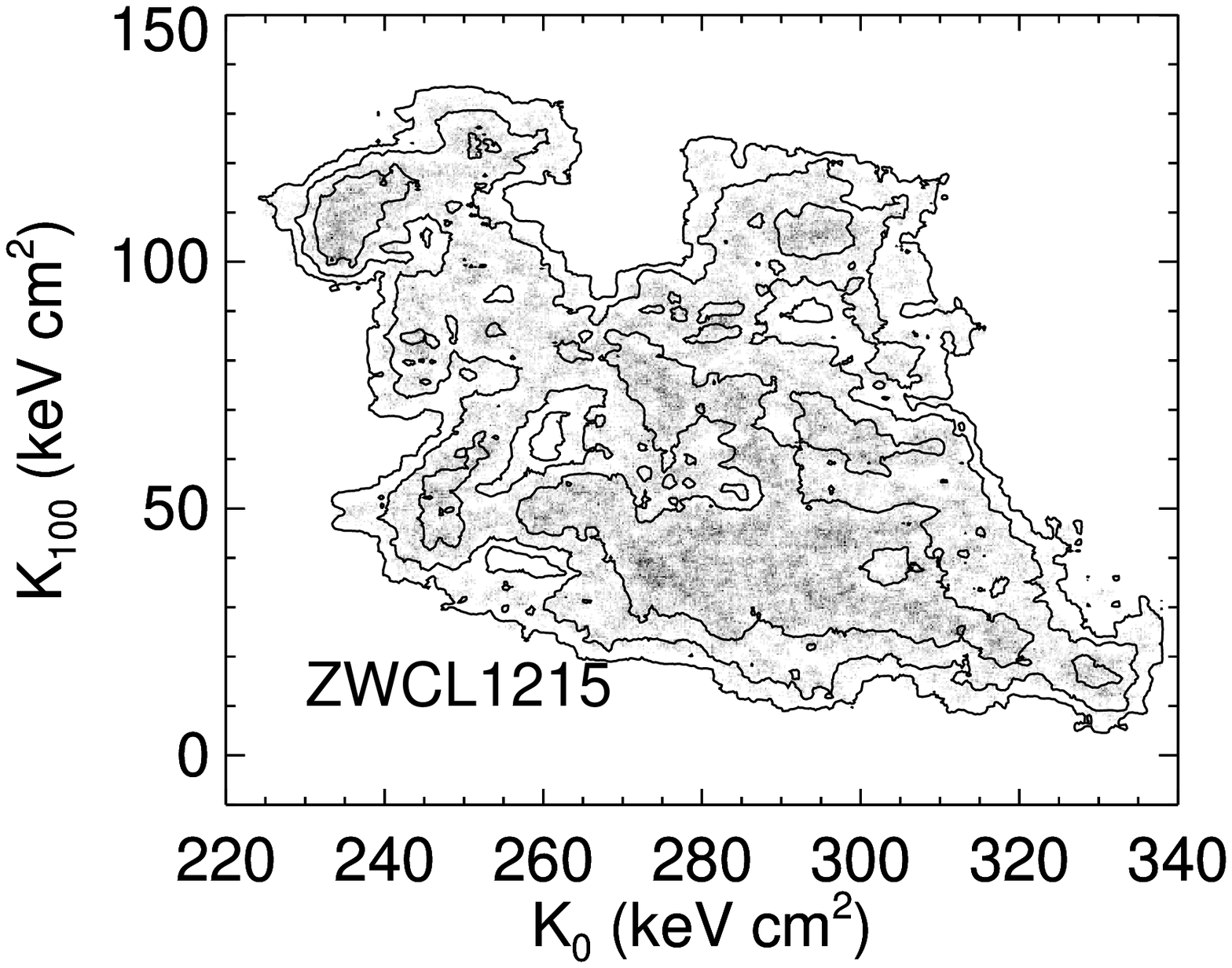}
  \includegraphics[width=0.32\textwidth]{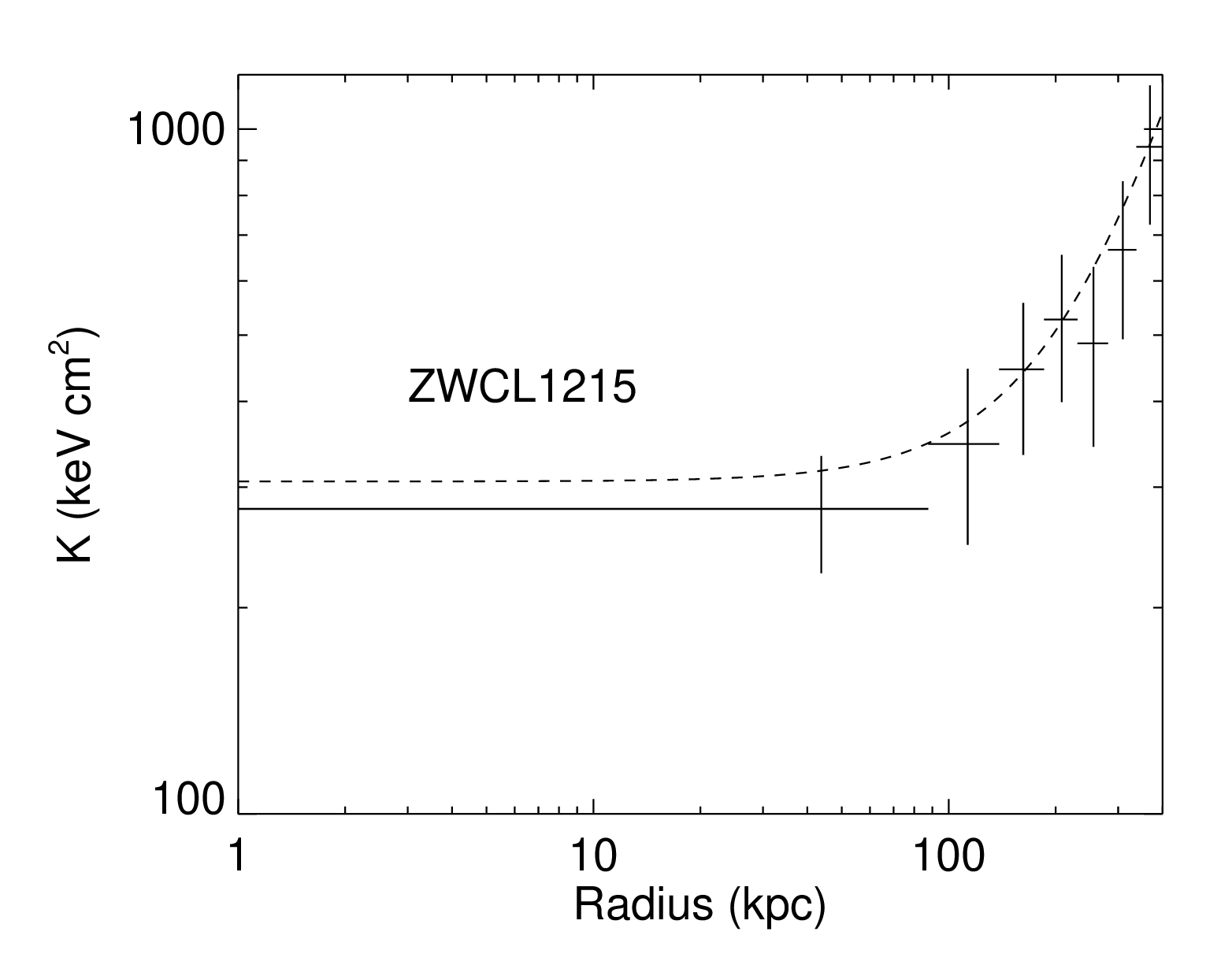}\\
  \includegraphics[width=0.32\textwidth]{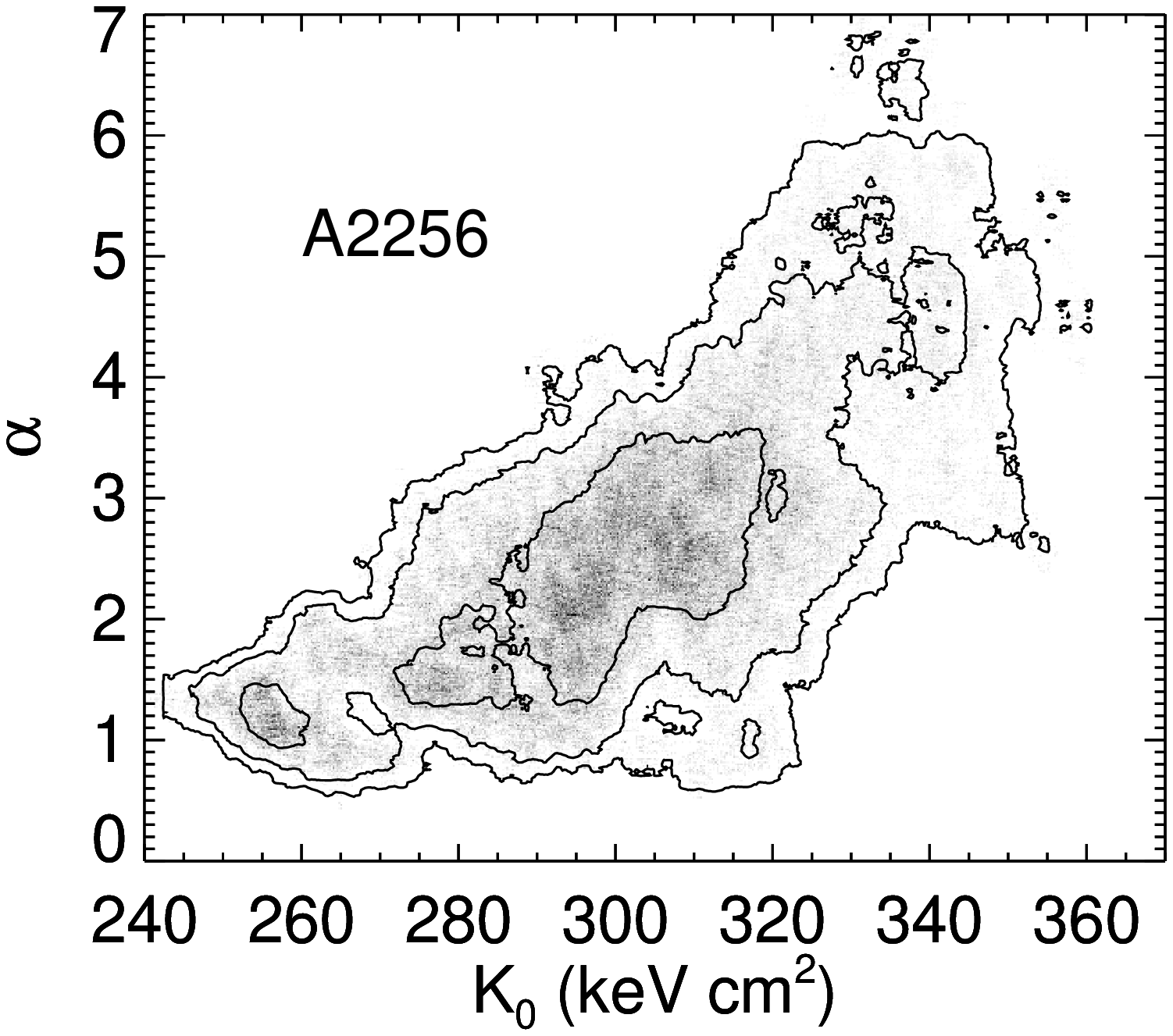}
  \includegraphics[width=0.32\textwidth]{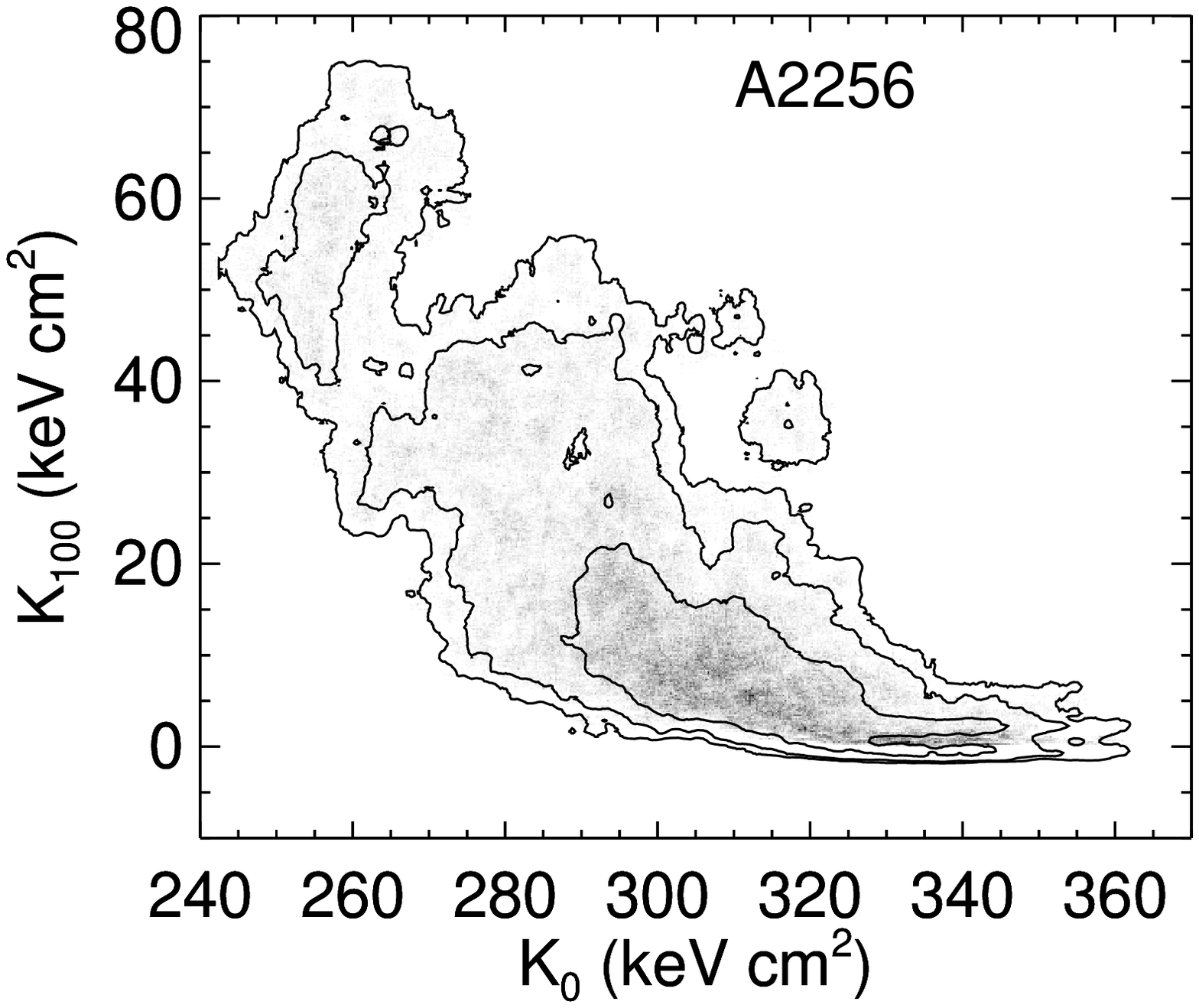}
  \includegraphics[width=0.32\textwidth]{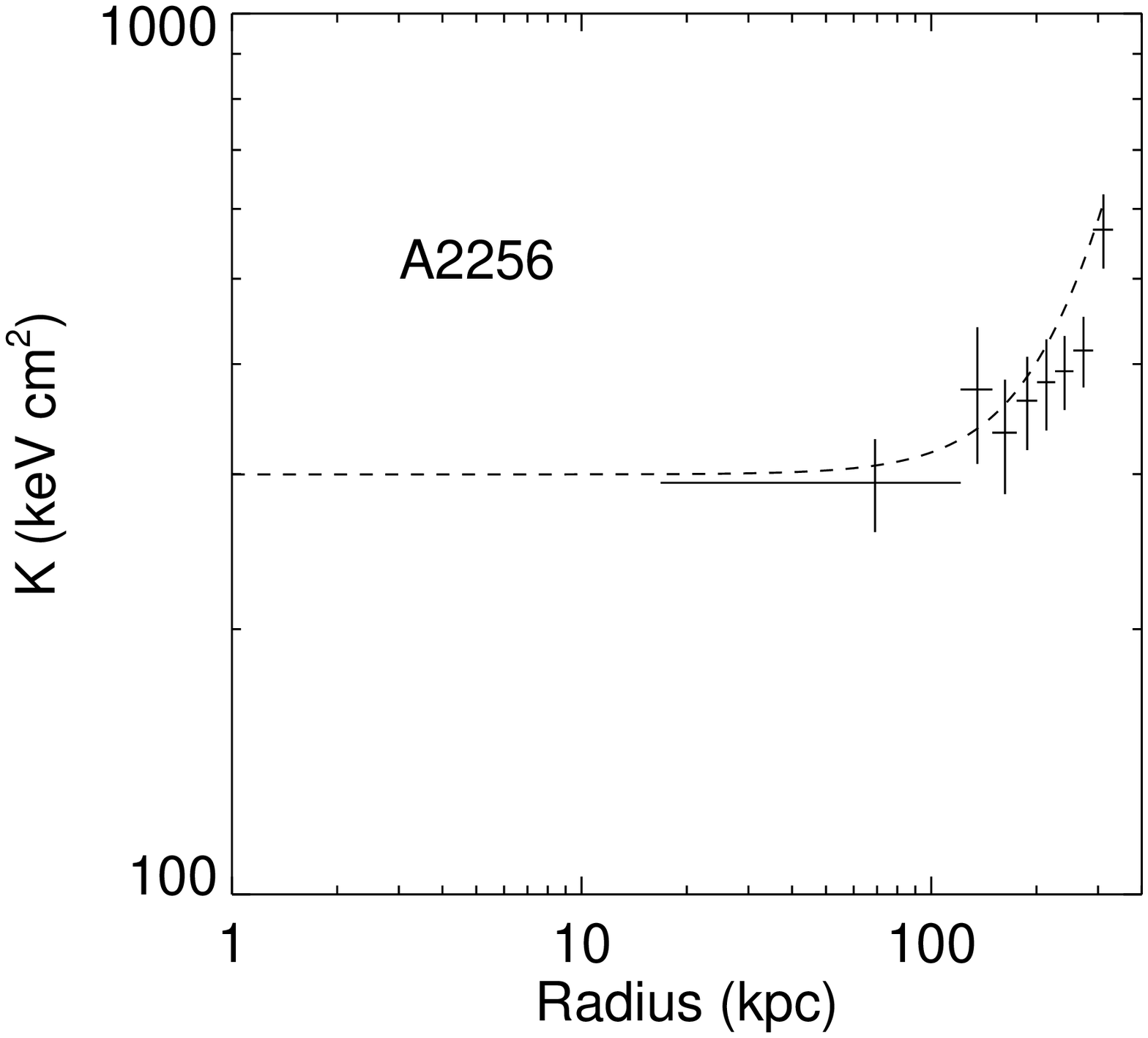}\\
  \caption{(Contd.)}
  \label{fig:flat_core_entr_fit_samp1}
\end{figure*}

\begin{table}
 \caption{Flat-core entropy model (eq. \ref{eq:flat_core_model}) for the FC (flat-core) and PL (power law) samples. 
 Here, DOF = number of annuli $-$ number of fit parameters (3 for a flat-core fit).
}
\label{tab:flat_core_results_smpl1}
\vskip 0.5cm
\centering
{\scriptsize
\begin{tabular}{c c c c c}
\hline
Cluster Name & $K_{0}$ & $K_{100}$ & $\alpha$ & $\chi^{2}_{\rm red}$ (DOF)\\
 & (keV cm$^{2}$) & (keV cm$^{2}$) & & \\
\hline
\hline
 FC sample & & & & \\
\hline
 A133 & 13.4$\pm$0.4 & 226.7$\pm$10.8 & 1.76$\pm$0.05 & 3.85 (5)\\
A1650 & 79.6$\pm$10.0 & 108.7$\pm$15.3 & 1.62$\pm$0.25 & 0.53 (4) \\
A1795 & 28.7$\pm$3.0 & 107.2$\pm$17.3 & 2.03$\pm$0.44 & 1.06 (11)\\
A2142 & 47.8$\pm$11.8 & 150.3$\pm$14.1 & 1.17$\pm$0.17 & 0.57 (6) \\
A2204 & 7.8$\pm$0.6 & 258.7$\pm$15.8 & 1.98$\pm$0.08 & 3.34 (4) \\
A2244 & 50.4$\pm$13.4 & 116.8$\pm$15.9 & 0.96$\pm$0.16 & 1.067 (4) \\
A2597$^\dag$ & 11.2$\pm$0.8 & 100.7$\pm$1.6 & 1.40$\pm$0.04 & 1.17 (8)\\
Hydra-A & 11.5$\pm$1.3 & 116.5$\pm$2.3 & 1.25$\pm$0.06 & 18.53 (6) \\
 A754 & 263.0$\pm$24.0 & 60.3$\pm$25.8 & 1.82$\pm$0.41 & 1.07 (10)\\
A2256 & 299.8$\pm$21.0 & 17.8$\pm$15.7 & 2.52$\pm$1.02 & 8.36 (5) \\
A3158 & 229.7$\pm$17.0 & 21.2$\pm$17.9 & 4.32$\pm$1.95 & 1.69 (3) \\
A3667 & 183.6$\pm$16.2 & 61.1$\pm$17.1 & 1.51$\pm$0.28 & 3.71 (6)\\
ZWCL1215 & 305.7$\pm$68.0 & 54.2$\pm$32.7 & 1.90$\pm$0.57 & 0.95 (5) \\
\hline
PL sample & & & & \\
\hline
  A85 & 3.4$\pm$0.7 & 168.9$\pm$2.2 & 1.00$\pm$0.02 & 1.67 (8)\\
 A478 & 3.5$\pm$0.8 & 124.8$\pm$2.4 & 1.03$\pm$0.03 & 1.71 (13) \\
A2029 & -1.7$\pm$0.6 & 177.2$\pm$1.8 & 0.81$\pm$0.01 & 4.32 (12)\\
A3112 & 2.0$\pm$0.4 & 135.1$\pm$1.5 & 0.95$\pm$0.01 & 12.74 (8)\\
\hline
$^\dag$Test cluster
\end{tabular}}
\end{table}

The entropy profiles of the cluster sample were fitted with the flat-core model (eq.~\ref{eq:flat_core_model}), using the 
entropy covariance matrices (see eq. \ref{eq:fisher}) and the jMCMC method, as described in section \ref{S:cent_entr_prof}. 
The resulting $K_0$, $K_{100}$, $\alpha$ chains obtained for individual clusters plotted in the $K_0$-$\alpha$ and 
$K_0$-$K_{100}$ planes along with their flat-core profile fits, are shown in Figs. \ref{fig:flat_core_entr_fit_samp1} \ref{fig:flat_core_entr_fit_samp2}. 
Almost all clusters of the sample show a positive correlation between $K_0$ and $\alpha$, and an anti-correlation 
between $K_0$ and $K_{100}$. This (anti-)correlation is not sample correlation across clusters but only holds for 
individual clusters. The similarity of this correlation trend across clusters signifies the robustness of our technique 
and data quality, indicating that the parameters of these clusters are well determined in comparison to those where
the confidence regions show large scatter. Even in latter cases the general trends in correlation are present.

For the non-cool-core clusters with large errors in their entropy profiles (viz., A754, A3667, A3158, ZWCL1215 and A2256), 
the $K_0$-$\alpha$ and $K_0$-$K_{100}$ probability distributions are found to be highly irregular and do not show any particular trend. 
The expectation values of $K_0$, $K_{100}$ and $\alpha$ (with the associated variances) and the reduced chi-squared values 
obtained for the individual clusters are given in Table \ref{tab:flat_core_results_smpl1}. For a small subsample of the clusters (power law or PL sample 
comprising of A85, A2029, A478 and A3112; the remaining sample will be called flat-core or FC sample), 
the entropy flattening in the core was not that obvious (Table \ref{tab:f_test_results} shows that a flat-core fit is preferred over a single power-law even 
for the PL sample).  The results of flat core entropy model fitting for the PL subsample are given in Fig. 
\ref{fig:flat_core_entr_fit_samp2} and Table \ref{tab:flat_core_results_smpl1}. The PL sample also shows a positive 
correlation between $K_0$ and $\alpha$, and an anti-correlation between $K_0$ and $K_{100}$. The cluster A2029 is found to show an inverted 
entropy core; i.e., the entropy profile seems to steepen instead of flattening near the centre. The mean value of $K_0$ for this cluster is found 
to be negative and, therefore, is unphysical.

Since a flat-core profile (eq.~\ref{eq:flat_core_model}) provides a better fit to entropy data of most of our clusters (F-test is shown in Table \ref{tab:f_test_results}), 
we have moved the discussion of single and double power law fits to Appendix \ref{S:sgl_powlaw_fit} \& \ref{S:dbl_powlaw_fit}, respectively.

\setcounter{figure}{5}
\begin{figure*}
 \centering  
  \includegraphics[width=0.32\textwidth]{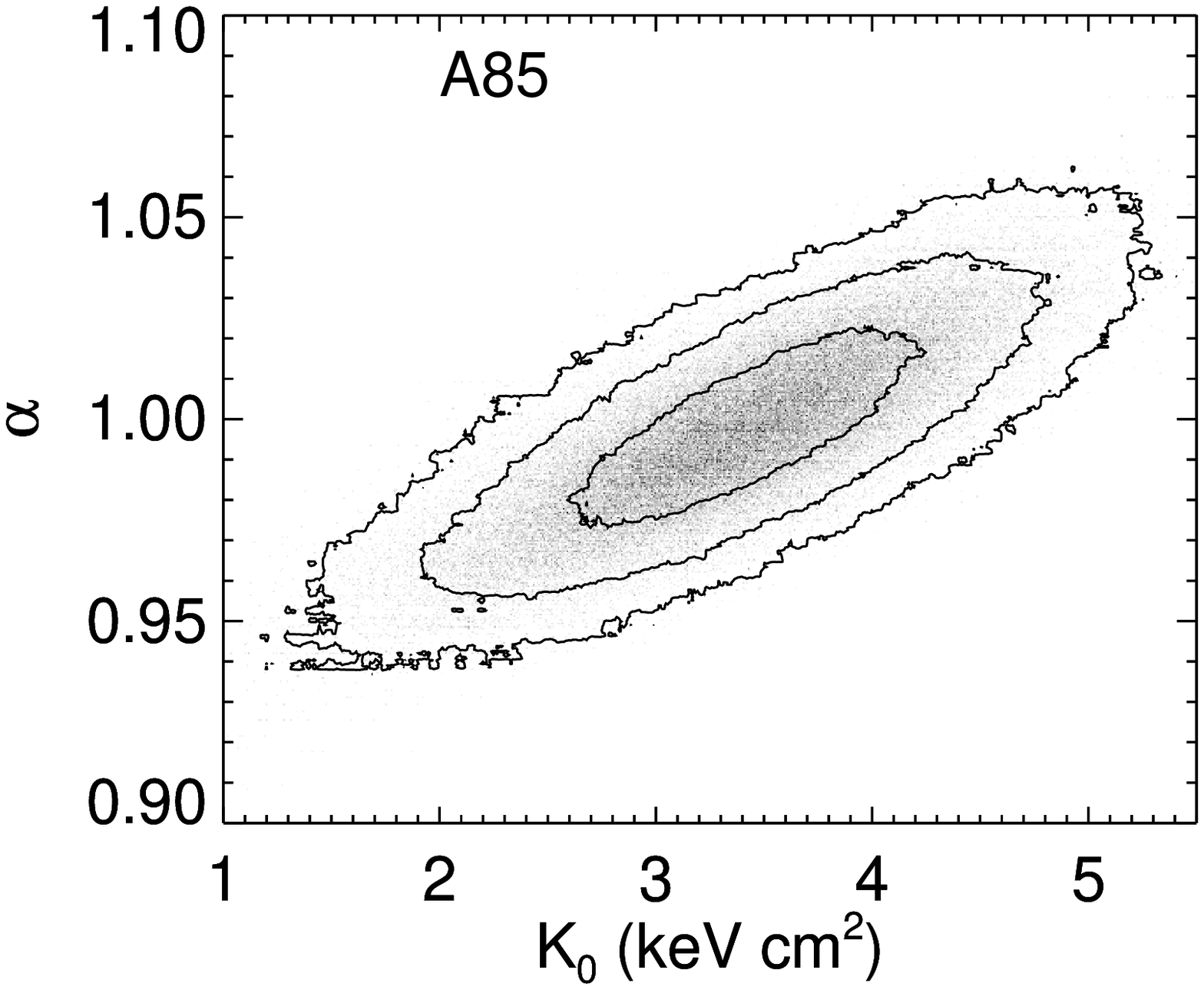}
  \includegraphics[width=0.32\textwidth]{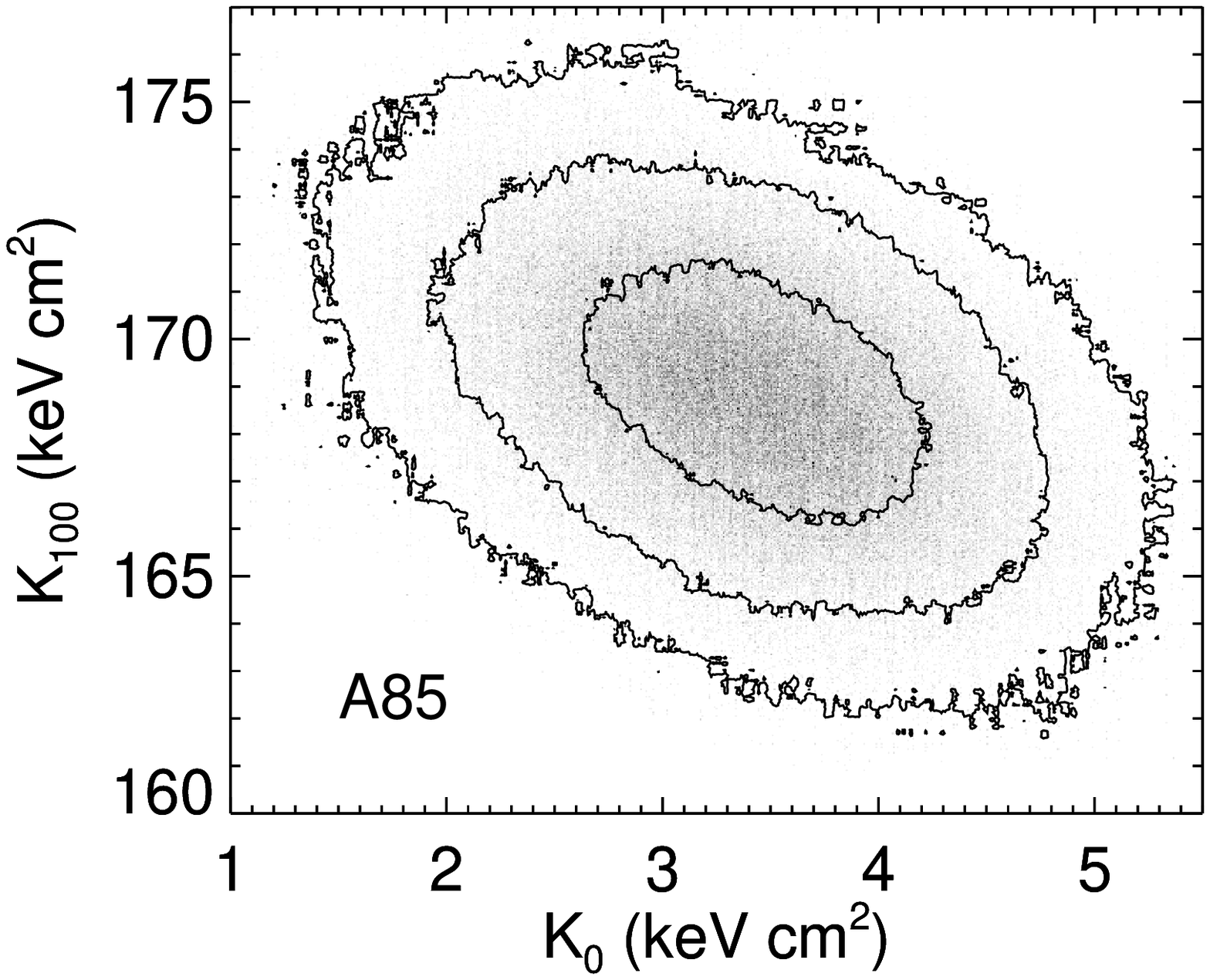}
  \includegraphics[width=0.32\textwidth]{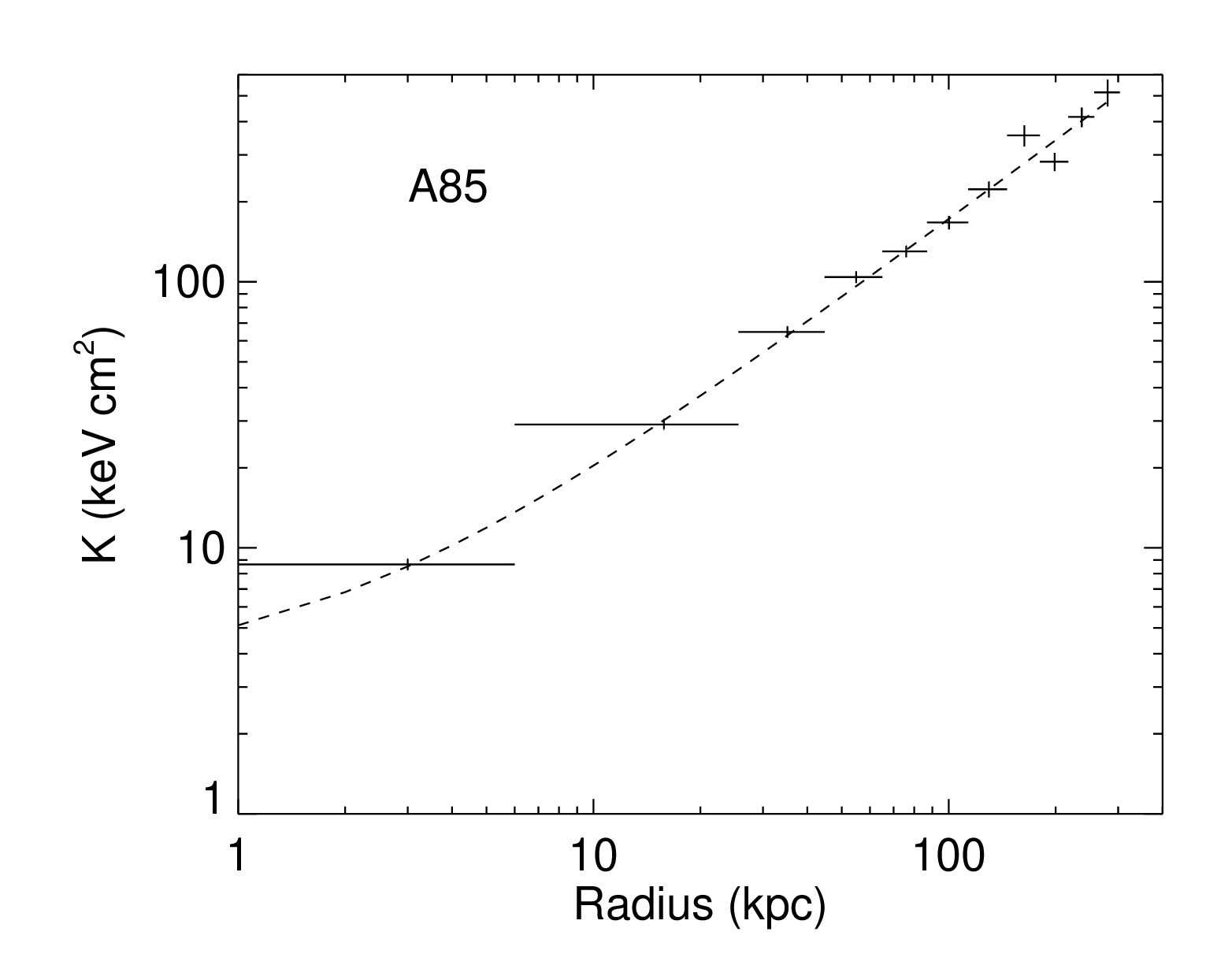}\\
  \includegraphics[width=0.32\textwidth]{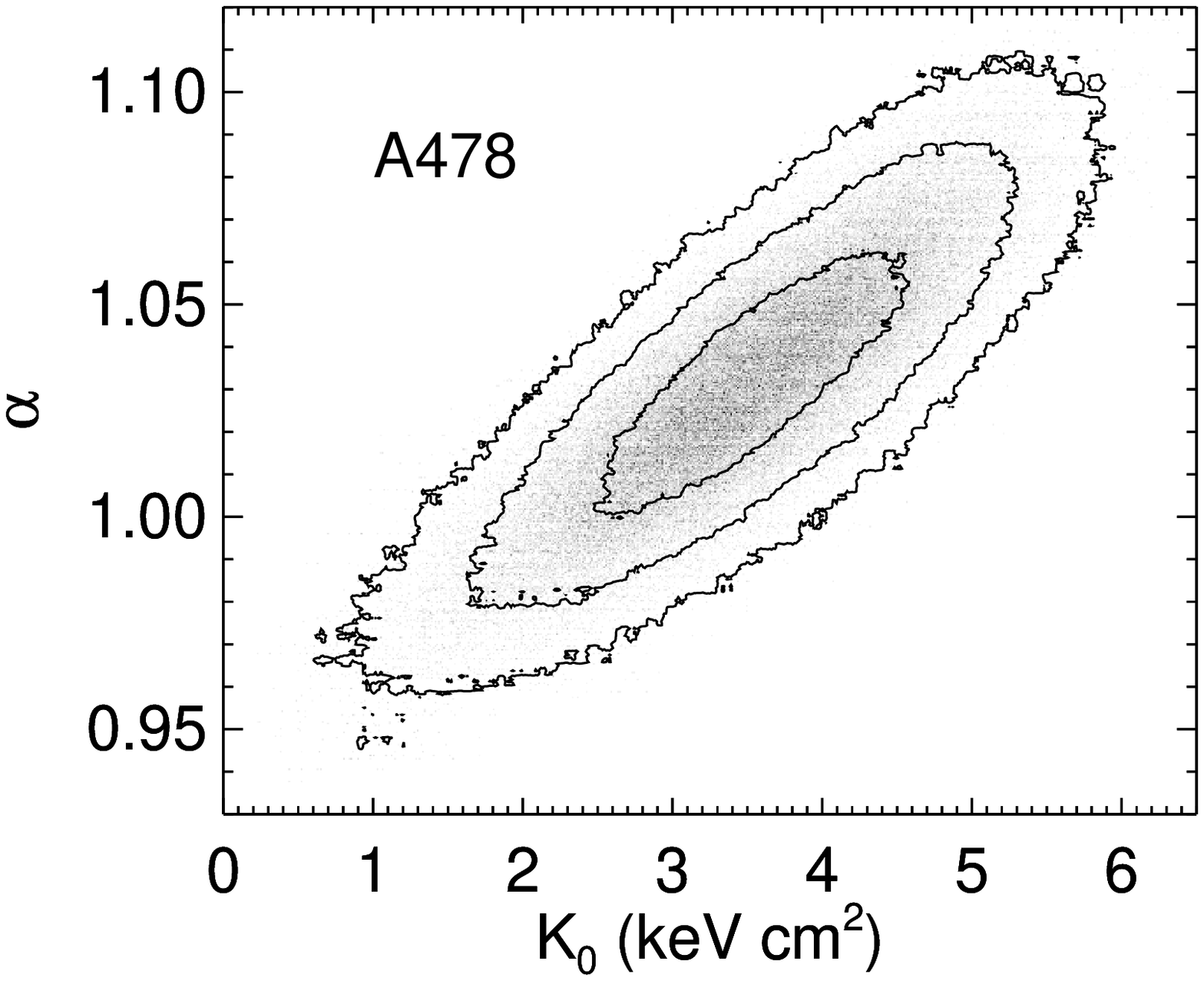}
  \includegraphics[width=0.32\textwidth]{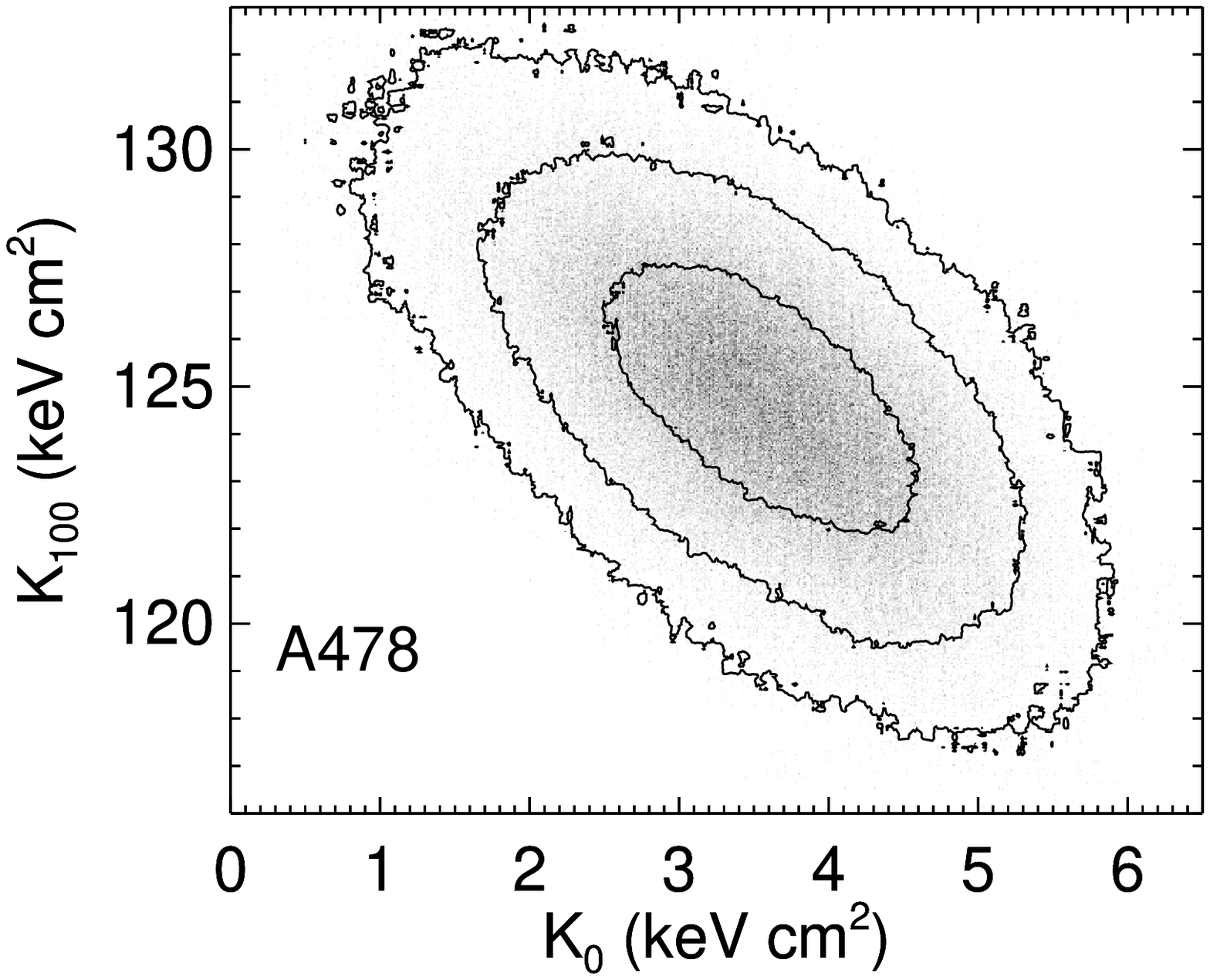}
  \includegraphics[width=0.32\textwidth]{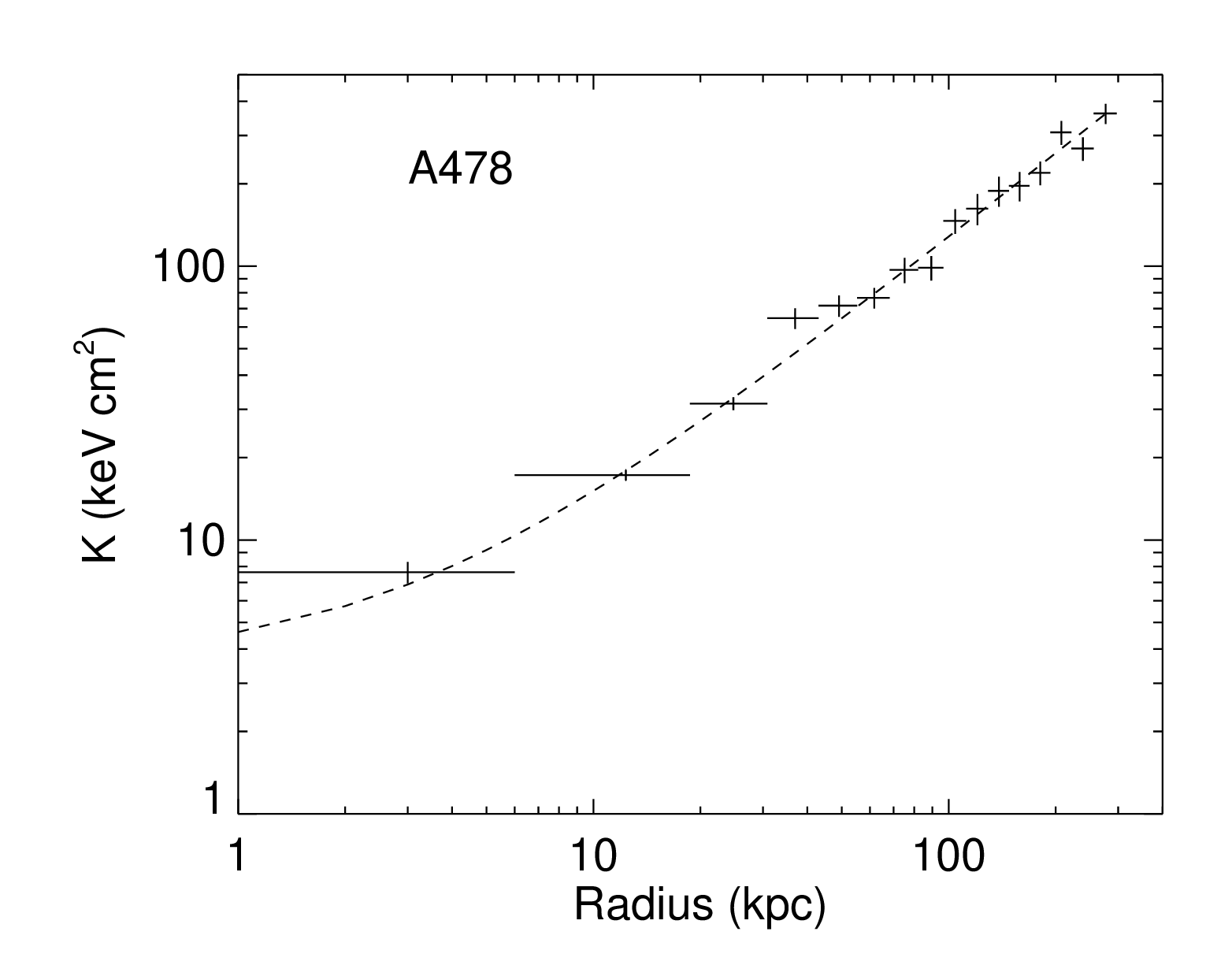}\\
  \includegraphics[width=0.32\textwidth]{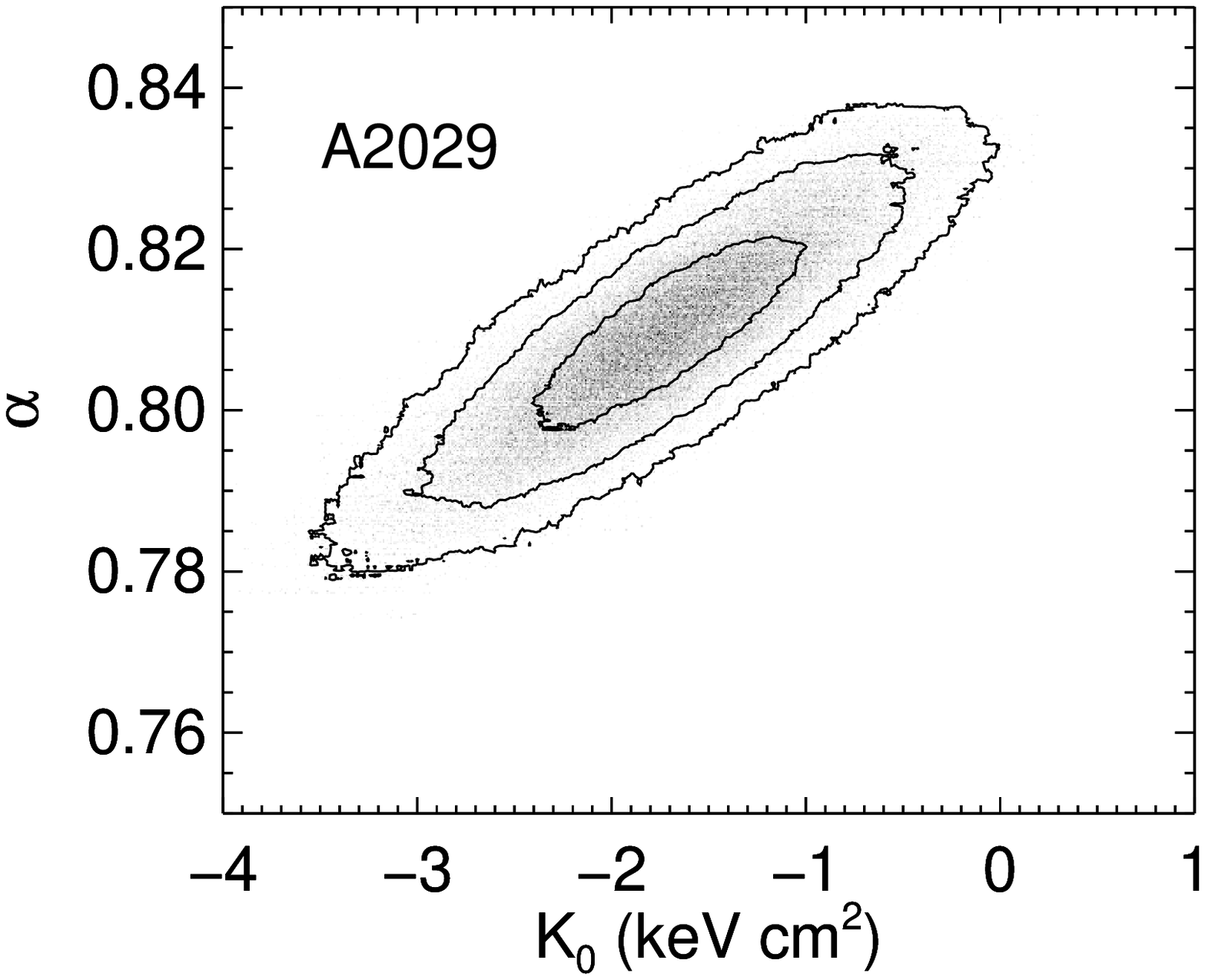}
  \includegraphics[width=0.32\textwidth]{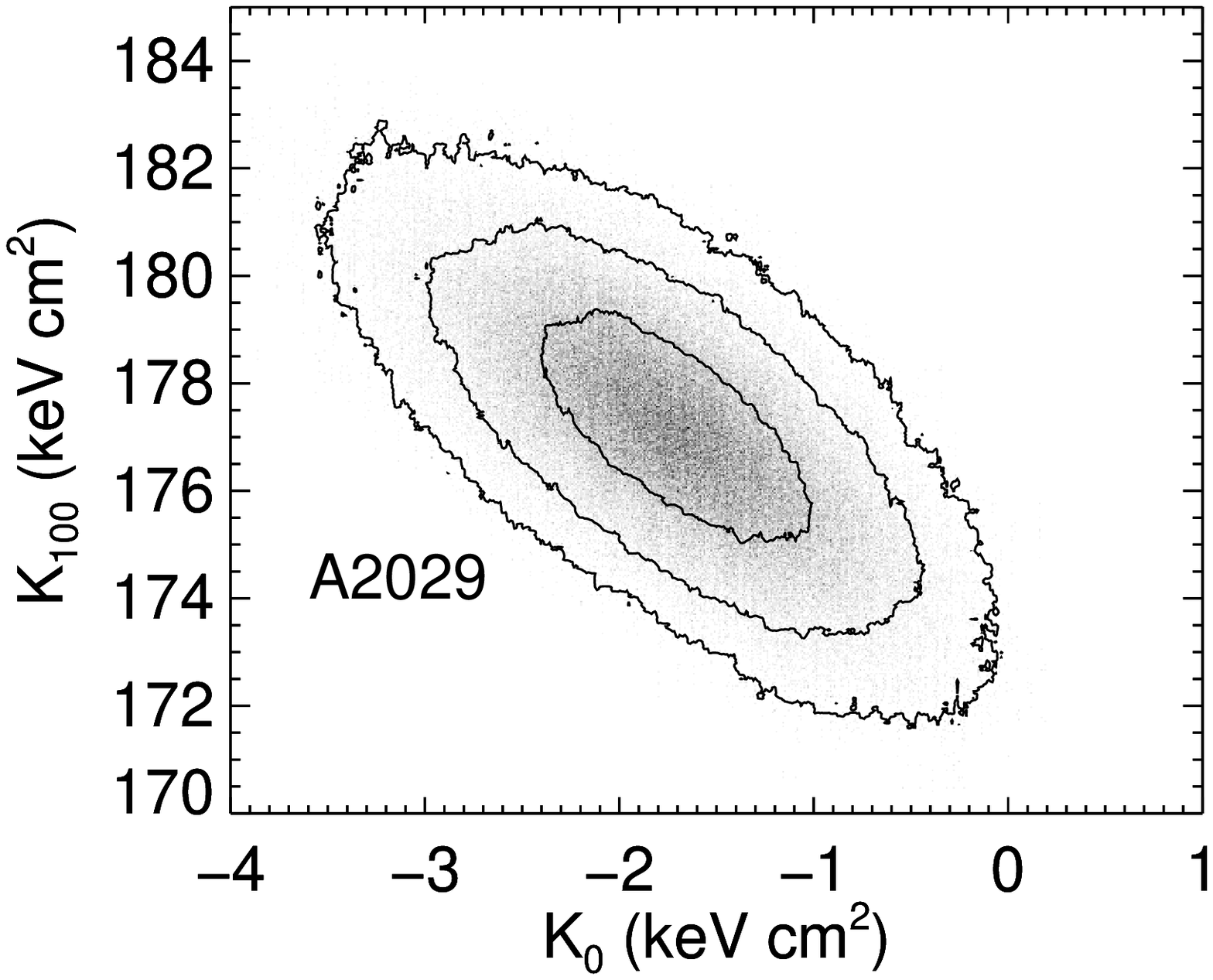}
  \includegraphics[width=0.32\textwidth]{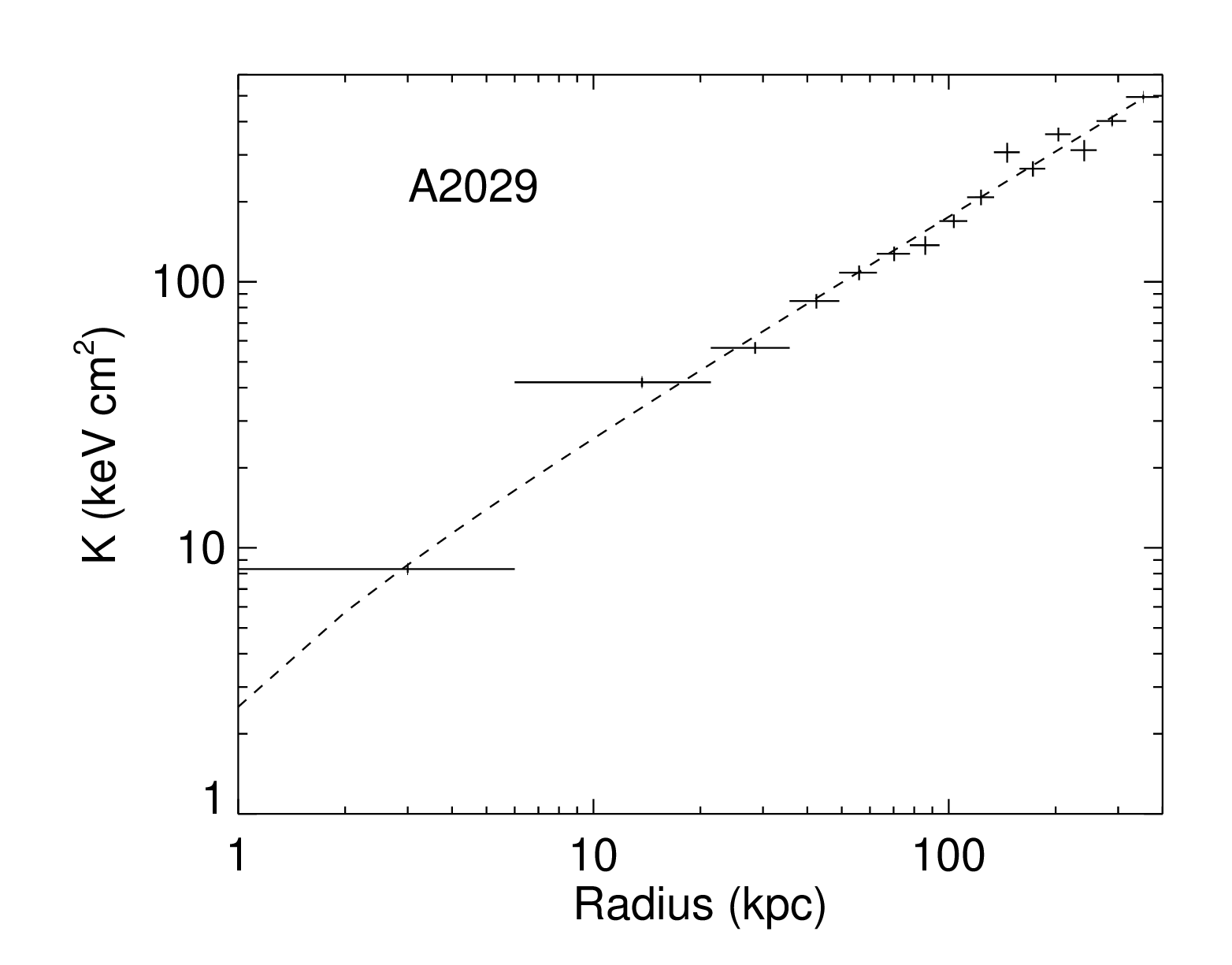}\\  
  \includegraphics[width=0.32\textwidth]{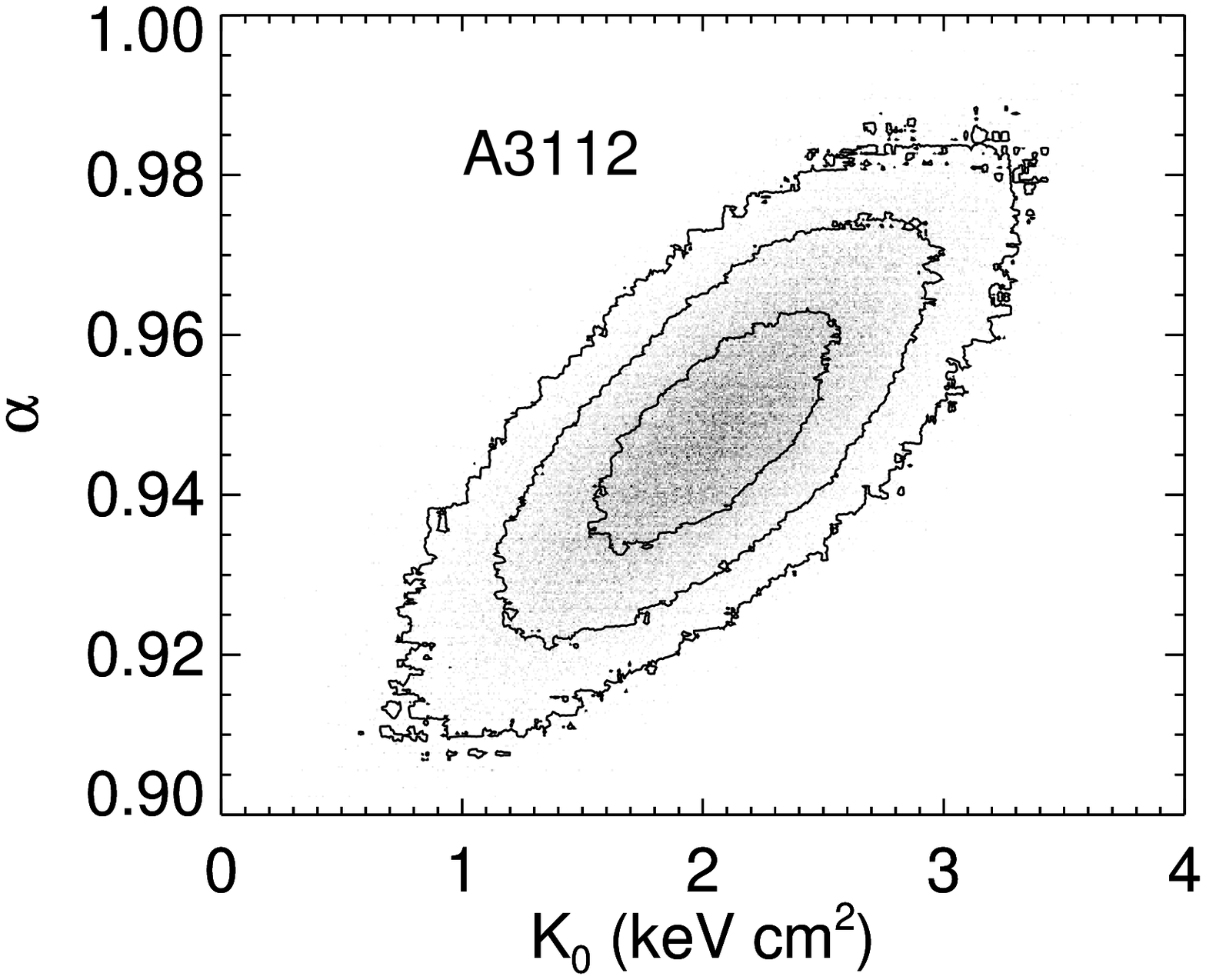}
  \includegraphics[width=0.32\textwidth]{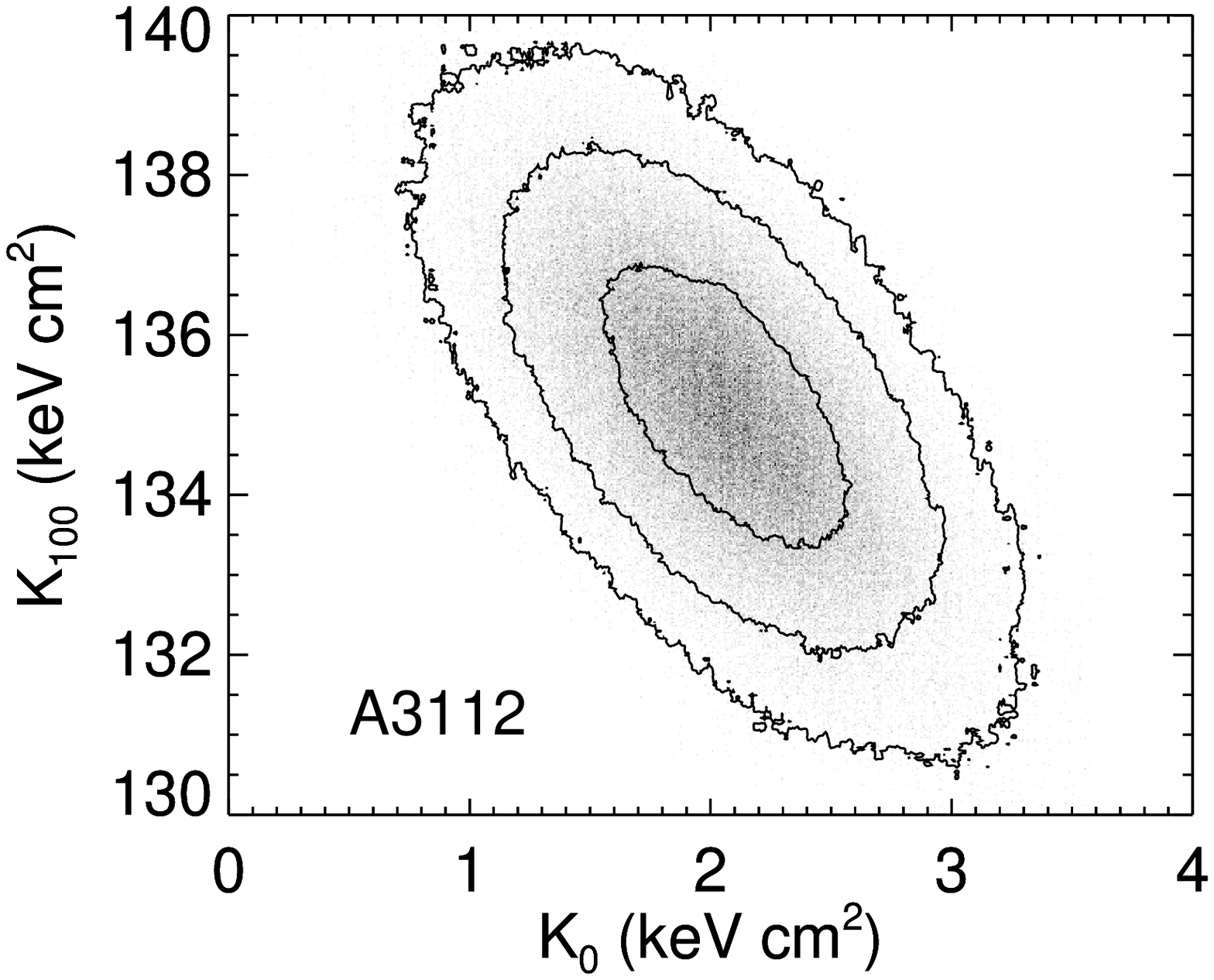}
  \includegraphics[width=0.32\textwidth]{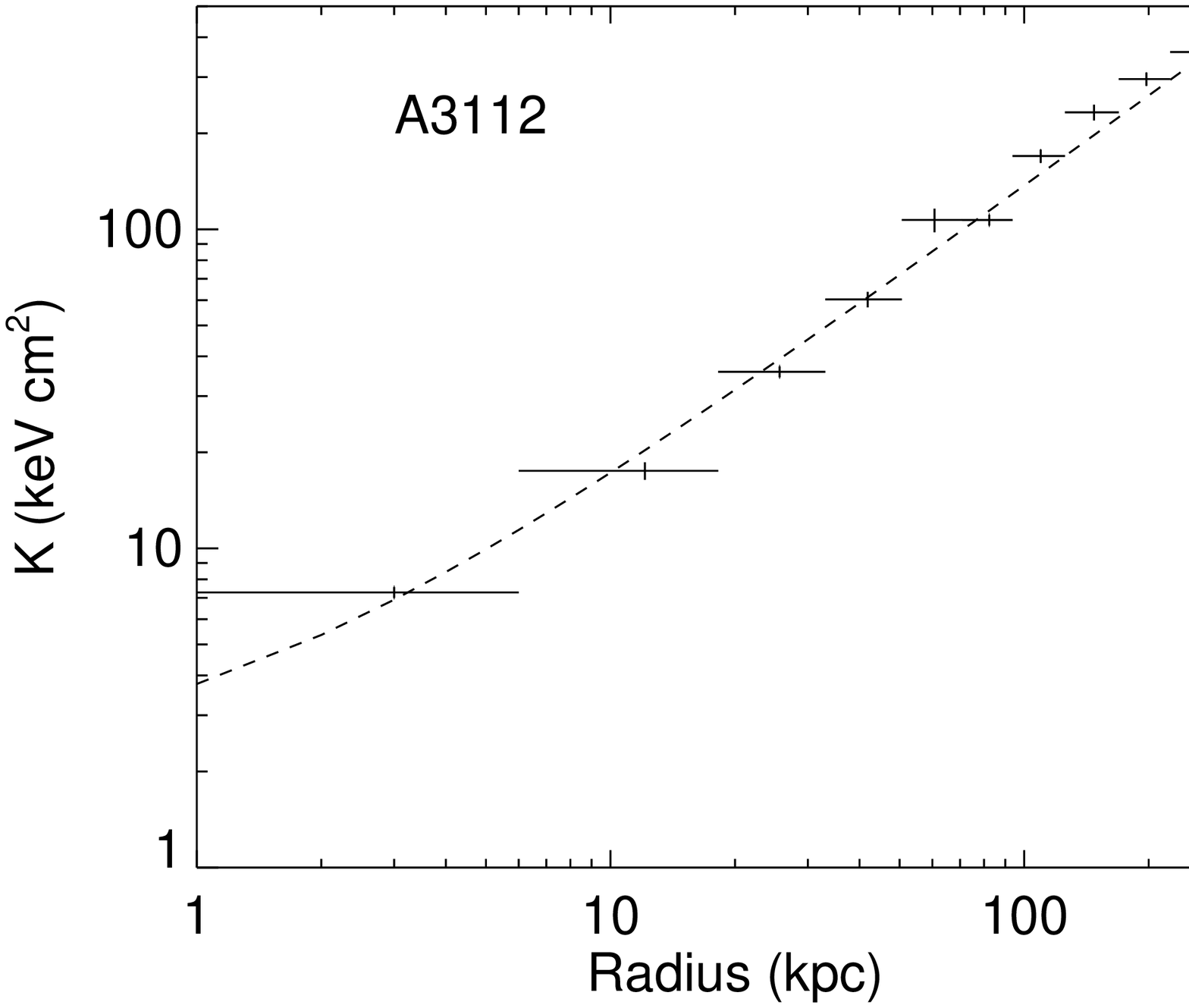}\\  
  \caption{The $K_0$-$\alpha$ and $K_0$-$K_{100}$ marginalized probability distributions obtained from the flat-core entropy model (eq.~\ref{eq:flat_core_model}) fitting for the PL sample. 
   The contours mark the 50\%, 90\% and 99\% inclusion levels based on the density of points starting from the innermost contour outwards.
   Greyscale denotes PDF density.}
  \label{fig:flat_core_entr_fit_samp2}
\end{figure*}

\section{Discussion}
\label{S:discussion}

As mentioned in the introduction, entropy, cooling time and $t_{\rm cool}/t_{\rm ff}$ profiles carry useful information about the
physical processes in cluster cores. These profiles are sensitive to processes such as thermal conduction and 
multiphase condensation. In this section we discuss the implications of our results.

\subsection{Comparison with previous works}
The inverse method of peeling off spherically symmetric shells to obtain deprojected spectra does not take care of covariance of parameters corresponding to different shells. Using jMCMC method for data analysis enables clear visualization of the model parameter space. The 
method, however, has not been much used for analysing X-ray data of clusters. A joint Bayesian analysis of the \textit{Chandra} and \textit{ROSAT} data of the galaxy group NGC4325 was carried out by \cite{russ07} using the MCMC method. The analysis was performed using a \textquoteleft{}forward fitting\textquoteright{} approach in which the gas density and temperature were assumed to have known parametrised functional forms. \citet{piz03} also used parametric forms to deduce density/temperature profiles from the X-ray spectrum of A1795. While using parametrized models for density and temperature has several advantages, such as multiple datasets can be fitted together and the models can be extrapolated to large distances, its main disadvantage is that it introduces inherent biases in the results. Our method can be improved for 
poor quality data by introducing regularization priors which impose smoothness on density and temperature profiles (e.g., \citealt{cro06}). \citet{ame07} applied
MCMC maximum likelihood fitting to mock thermal Sunyaev-Zeldovich (tSZ) and X-ray surface-brightness maps from numerical simulations to deduce 
the de-projected density 
and temperature profiles (without using X-ray spectra). Even here, regularization was found to be essential to mitigate temperature oscillations.

We note that \textit{DSDEPROJ} (an inverse method for spectral deprojection; \citealt{russ08}) may suffer from an intrinsic instability due to the following possibility: when the contribution from the outermost shell is subtracted from the next, any over/under estimate of photon counts---especially at high energies that best constrain the temperature---may cause an under/over estimate of photon counts deduced for the next shell. Since the correction appears with opposite signs in adjacent radial bins, albeit with diminishing volume factors, it may lead to an alternating over/under estimation of photon counts across the radial bins that may show up as oscillating inferred temperature, even if the gas is not multiphase. This effect may be more pronounced for clusters with poor quality data with large Poisson noise. This problem may be avoided if we do not treat the data in annuli separately, but jointly across all radial bins. 
\citet{russ08} have argued that the oscillations mentioned above may also be caused by fitting a single-temperature model to a multiphase gas. It is straightforward to include multi-temperature plasma in our MCMC method, provided we have sufficiently high-quality data.

In a recent paper \cite{san14} have described a new MCMC based code MBPROJ that is in spirit very similar to our method. 
Unlike our method, which works only for those clusters that have high quality data since we use multiple spectral bins to fit the spectrum, by choosing only three spectral bins, MBPROJ can work even for clusters that have only a few thousand counts. 
It does that by assuming hydrostatic equilibrium; however, this assumption can be relaxed in the code to make it similar to 
our approach. To the best of our knowledge MBPROJ has not been used to analyze the entropy profiles of a cluster sample as we have done in this work.  

\subsection{Entropy core versus cusp}

As discussed in section~\ref{S:method_comp}, our method agrees with the method adopted in \citetalias{pan14} for simulated and the test cluster. Although our method 
does not agree with \citetalias{cav09}, the differences are well explained by the fact that not deprojecting temperature leads to an increase in temperature 
inferred at the centre of clusters. Therefore, the method adopted in \citetalias{cav09} is inherently biased to measure higher entropies in  cluster cores, 
while \citetalias{pan14} seems to be unbiased. However, the method of \citetalias{pan14} does not account for the propagation of errors as the layers of a cluster are peeled 
off. 

Given that the results from our analysis for the test cluster and simulated spectra agree with \citetalias{pan14}'s method,  in Fig. \ref{fig:flat_core_prof_full_sampl}, 
in complete 
contrast, we find that most of the clusters analyzed by us are actually 
better described by a cored power law model. Although, this is in overall agreement with  \citetalias{cav09}, our tests show that our core entropy values 
are smaller than theirs for the coolest clusters (see top-left panel of Fig. \ref{fig:comp_voit}).
Note that all the non-cool core clusters (shown in red in Fig. \ref{fig:flat_core_prof_full_sampl}; defined here as clusters with $K_0> 100$ keV cm$^2$) are consistent with flat cores. 
As expected, their central entropies are also higher. The blue curves corresponding to cool cores in Fig. \ref{fig:flat_core_prof_full_sampl}, however, 
while mostly displaying flat cores, do have a sub-sample of about  four clusters that are somewhat consistent with a single power law without any 
core (Fig. \ref{fig:flat_core_entr_fit_samp2}). 
However, their core entropy is so small ($K_0 \simeq 1$ keV cm$^2$) that they could be well described as overall single power laws. 
Therefore, the question of core versus cusp depends somewhat on the scales that we are interested in.

\setcounter{figure}{6}
\begin{figure*}
 \centering
  \includegraphics[width=0.32\textwidth]{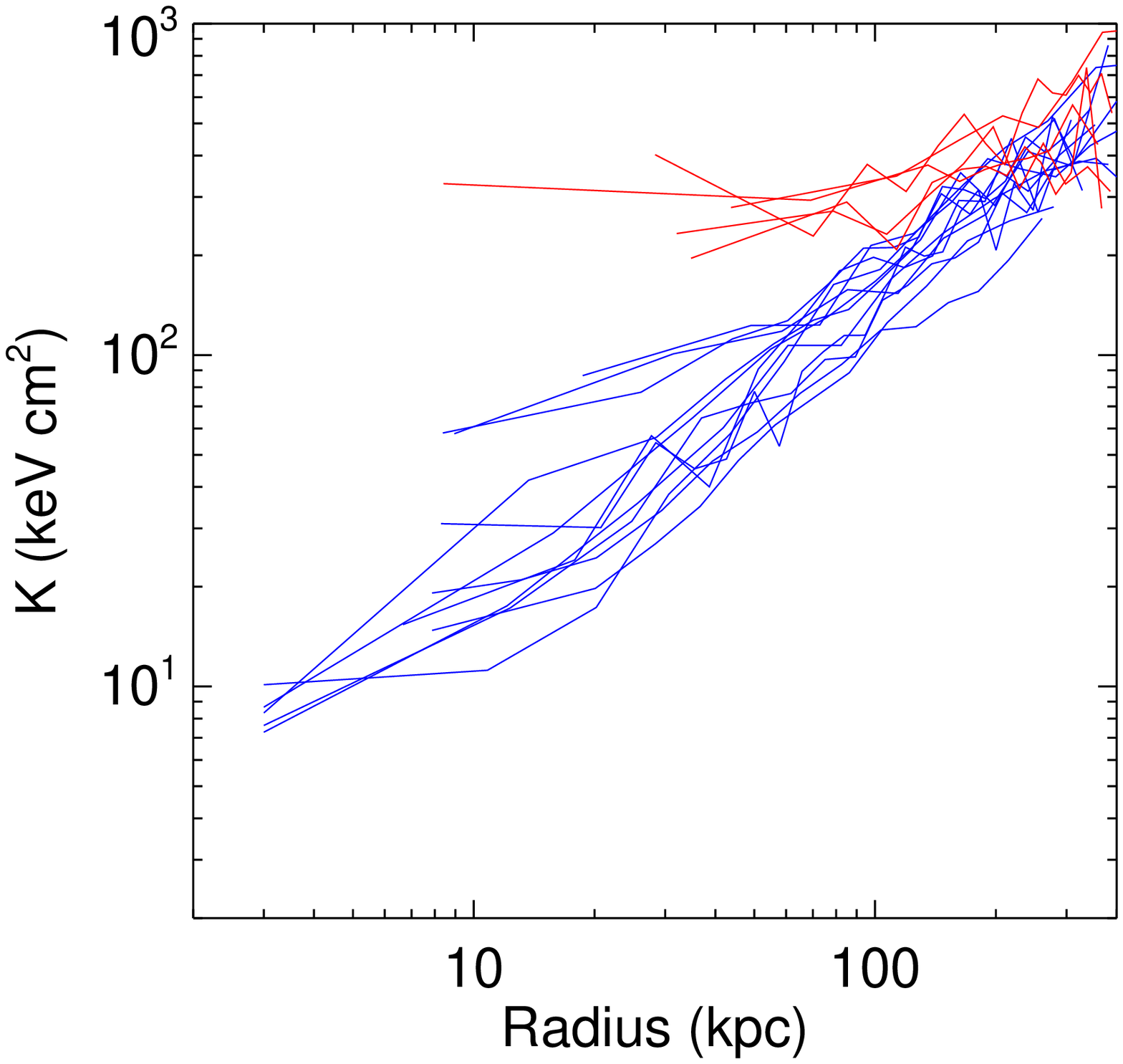}
  \includegraphics[width=0.32\textwidth]{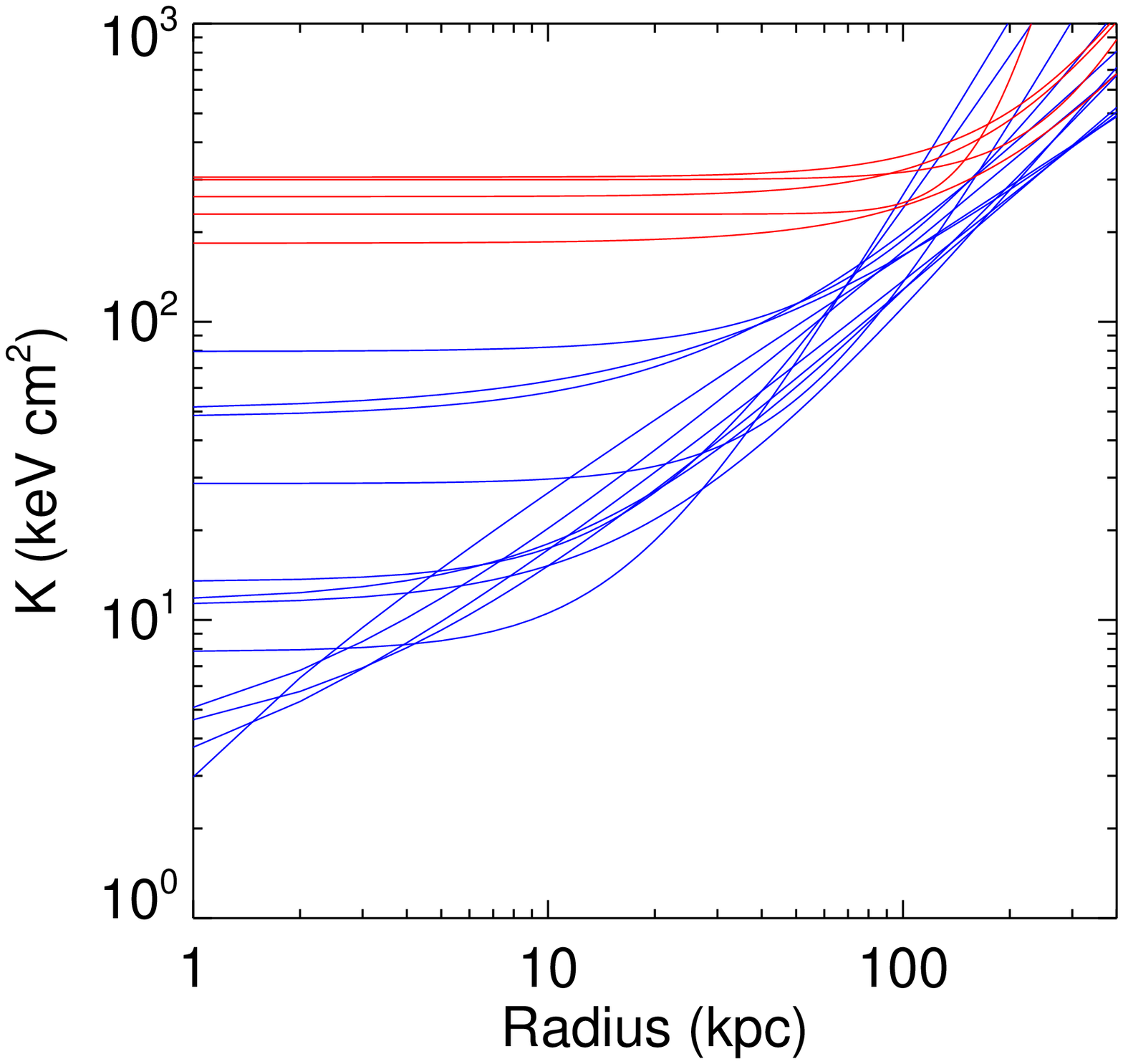}
  \includegraphics[width=0.32\textwidth]{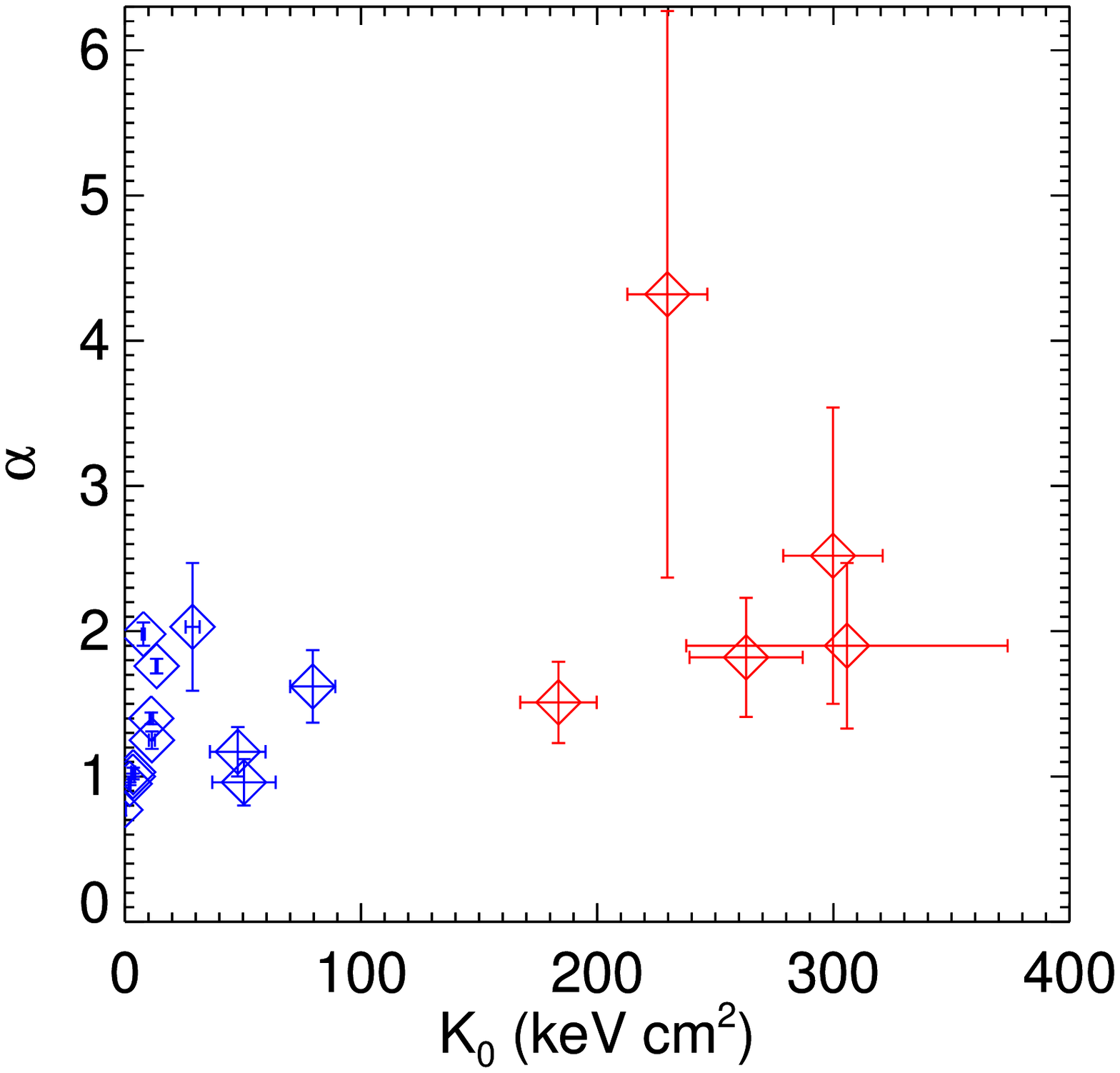}
  \caption{Entropy profiles of the full cluster sample obtained from the jMCMC fits (left) and the their flat-core 
  profile fits (centre). The best-fit K$_{0}$ vs. $\alpha$ values obtained from flat-core fits 
  is shown in the right panel. Cool-core ($K_0 < 100$ keV cm$^2$) and non-cool-core ($K_0 > 100$ keV cm$^2$) 
  clusters are shown using blue and red colors, respectively.}
  \label{fig:flat_core_prof_full_sampl}
\end{figure*}

Another discrepancy between \citetalias{pan14} and \citetalias{cav09} is the presence or absence of bimodality in the central values of entropy in clusters (see also \citealt{pra10}). 
Although our sample 
of clusters is small, the second panel of Fig.~\ref{fig:flat_core_prof_full_sampl} shows an apparent gap between the red and blue curves close to 
the centre of the clusters. This is indicated more clearly (with error bars) in the third panel. Most of the blue points are clustered at small 
$K_0$, with smaller values of $\alpha$ (this is in contrast to \citealt{san09} who find a steeper entropy in cool cores). 
Then we see an apparent gap at about $K_0 \simeq 100 \, {\rm keV cm^2}$. The statistical significance of 
this gap is marginal due to small number of clusters in our sample, but seems to be in general agreement with the results of \citetalias{pan14}.
Much  further work is required to explore the issue of bimodality, and in the future we shall add more clusters  to our sample to 
adequately address it. 

While our results are broadly consistent with \citetalias{cav09}, our method is a significant improvement in terms of how we handle statistics. We are unable to fully
explain our lack of agreement with \citetalias{pan14}, even though our tests show that our results match well on simulated data and a sample cluster, despite the 
difference in the two approaches. Most clusters in our sample have cores and not cusps; especially, the non-cool clusters display only cores. A 
few cool-core clusters are probably flat cored too, but with very small core radius, so our fits are only marginally able to capture the core.

\setcounter{figure}{7}
\begin{figure*}
 \centering
  \includegraphics[width=0.32\textwidth]{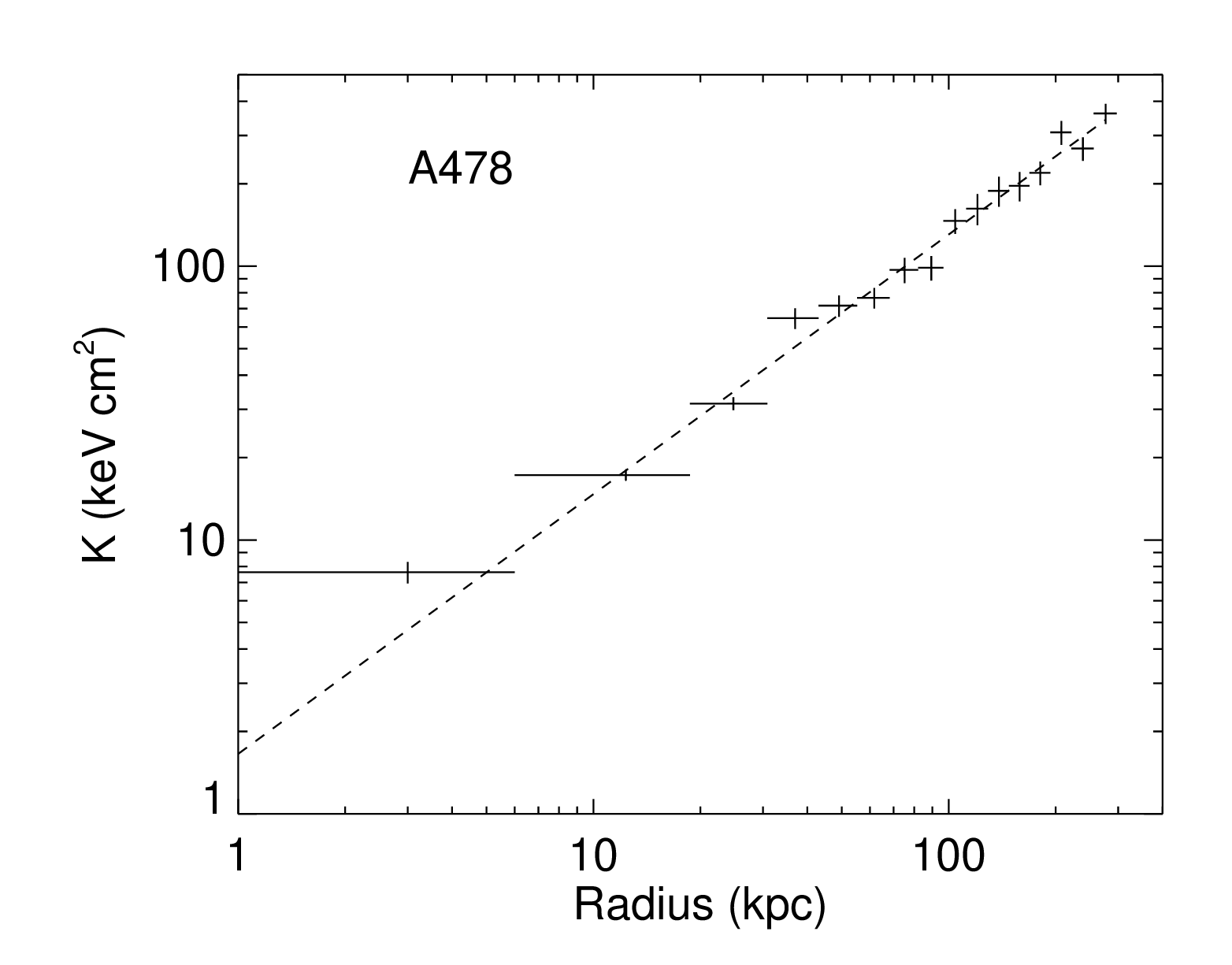}
  \includegraphics[width=0.32\textwidth]{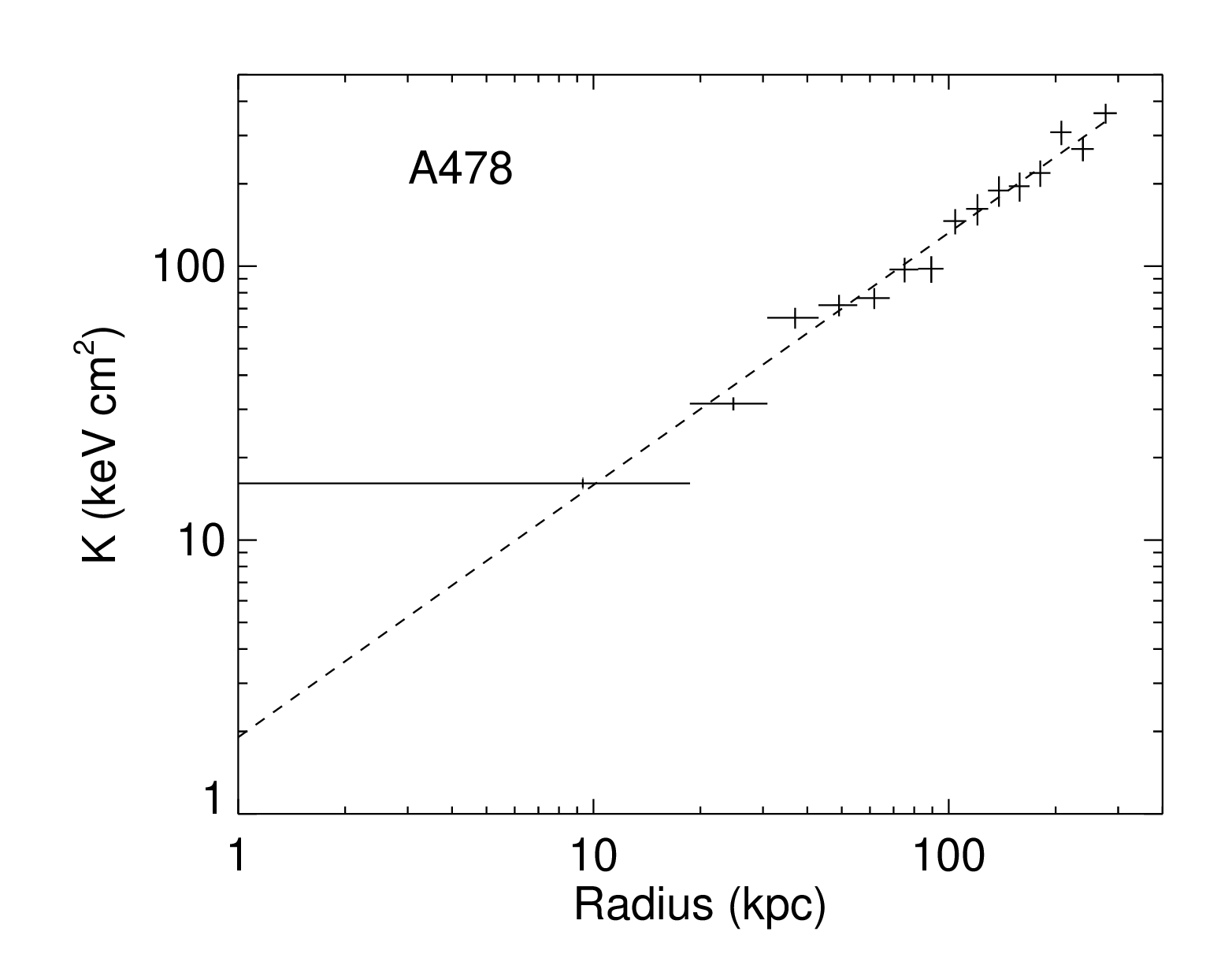}
   \caption{Entropy profiles of the cluster A478 with small (left) and large (right) central radial bins, fitted with single 
   power law models (dotted lines). An excess 
   entropy (flattening) is seen in the innnermost  annulus when a smaller radial bin size is used.}
  \label{fig:large_radial_bin_effect}
\end{figure*}

\citetalias{pan14}'s results are in complete disagreement with \citetalias{cav09}, and the former have argued that the difference is mainly 
due to the use of projected temperature profiles, large central radial bins (which might smooth out central gradients 
leading to flattening of entropy profiles) and assumption of single phase gas in the central regions by the latter. 
The first issue has been addressed in the present work as we have used deprojected temperature profiles in our analysis. In contrast to 
\citetalias{pan14}, in our analysis  
we found that using larger central radial bins for some of our clusters actually favours power law instead of flat-core 
profiles; an example of this is shown in Fig. \ref{fig:large_radial_bin_effect}.

There are some differences between our and \citetalias{pan14}'s approach when it comes to fitting entropy profiles in the core. We, like \citetalias{cav09}, fit the parametric 
forms (power law, double power law, flat core) for {\it individual} clusters but \citetalias{pan14} use a {\it common power law for all the clusters and groups in 
their sample} (see their Figs. 2 \& 3). From the left panel of Fig.~\ref{fig:flat_core_prof_full_sampl} we can see that a single power law indeed provides 
a good {\it common} fit for all the lowest entropy profiles. However, as we show in detail, individual cluster entropy profiles almost always flatten towards 
the center. Another, and potentially more significant, difference is that our sample has clusters with temperature greater than 3 keV (see Table 1 in 
\citealt{san10}) but \citetalias{pan14}'s sample has several groups with temperature $\leq 1.2$ keV (see their Fig. 3). Their groups sample shows a good agreement
with a single power law in the core compared to the full sample shown in their Fig. 2. The lowest entropy clusters (which are well fitted by 
a power law) in \citetalias{pan14} may be dominated by small scale coronae associated with the massive central galaxy (e.g., see \citealt{sun09}; see also Fig. 2 in \citealt{sha12b}) rather than 
the typical large cool cores.

\subsection{Cooling time \& free-fall time}

\setcounter{figure}{8}
\begin{figure*}
 \centering
  \includegraphics[width=0.32\textwidth]{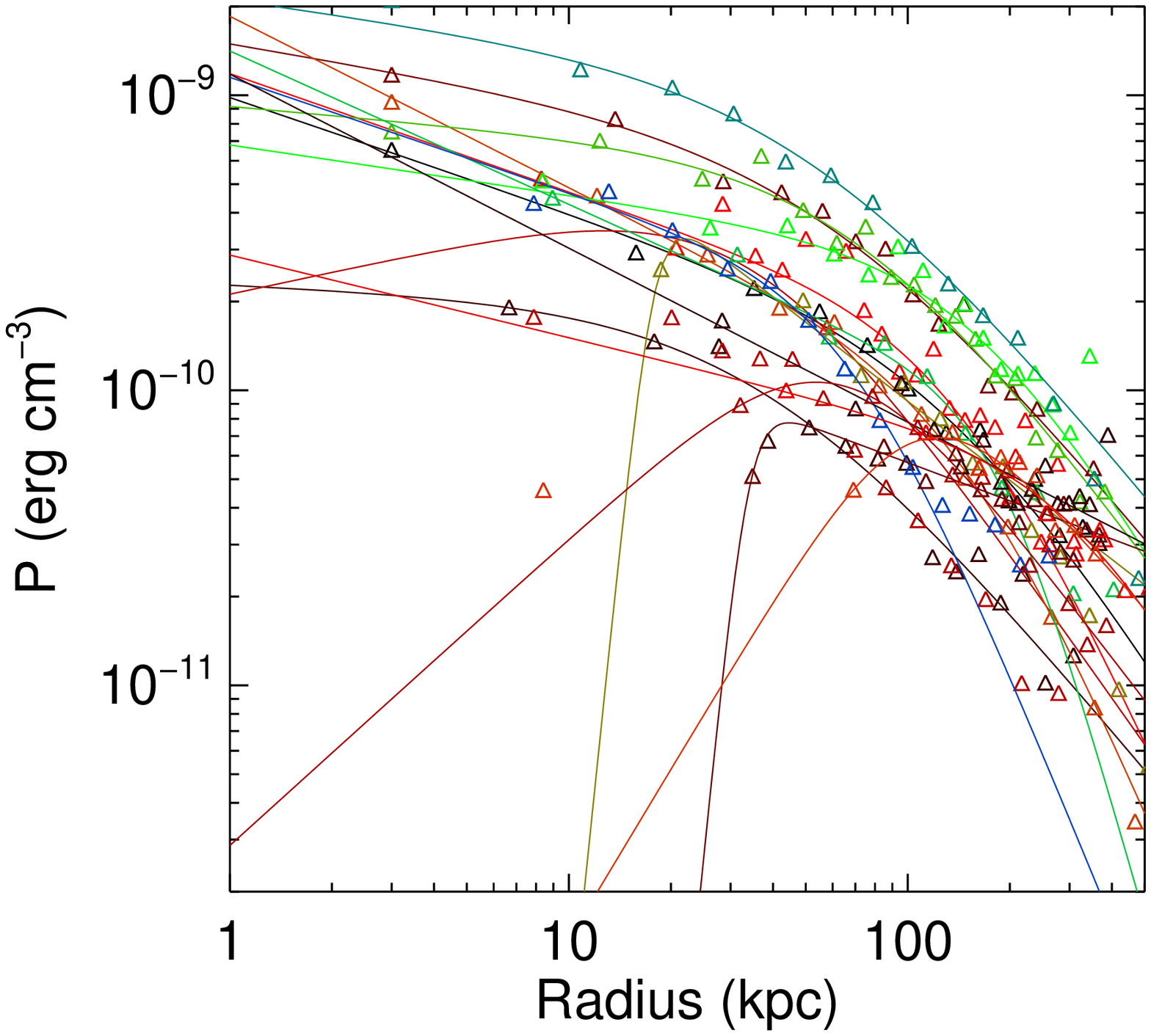}
  \includegraphics[width=0.32\textwidth]{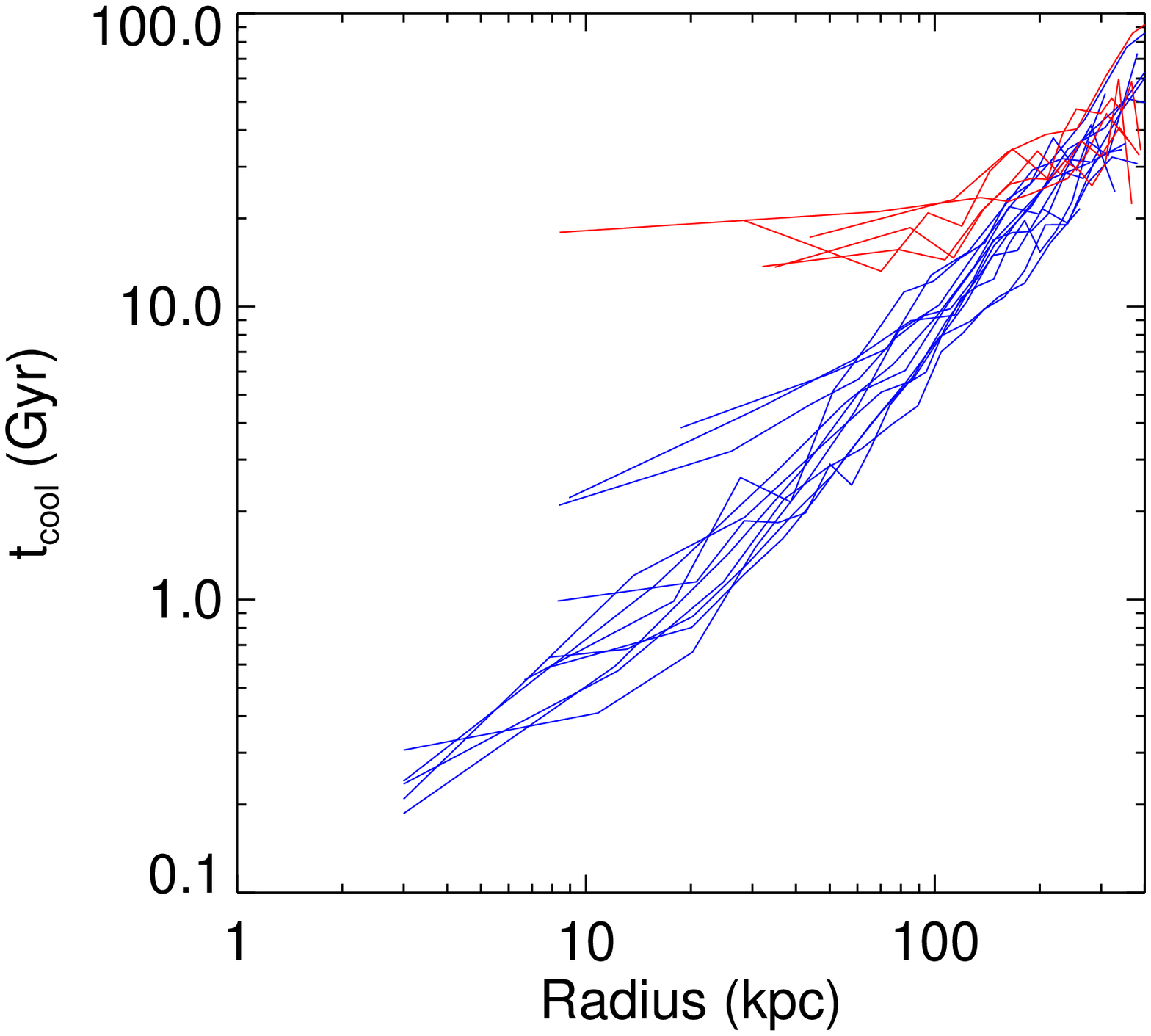}
  \includegraphics[width=0.32\textwidth]{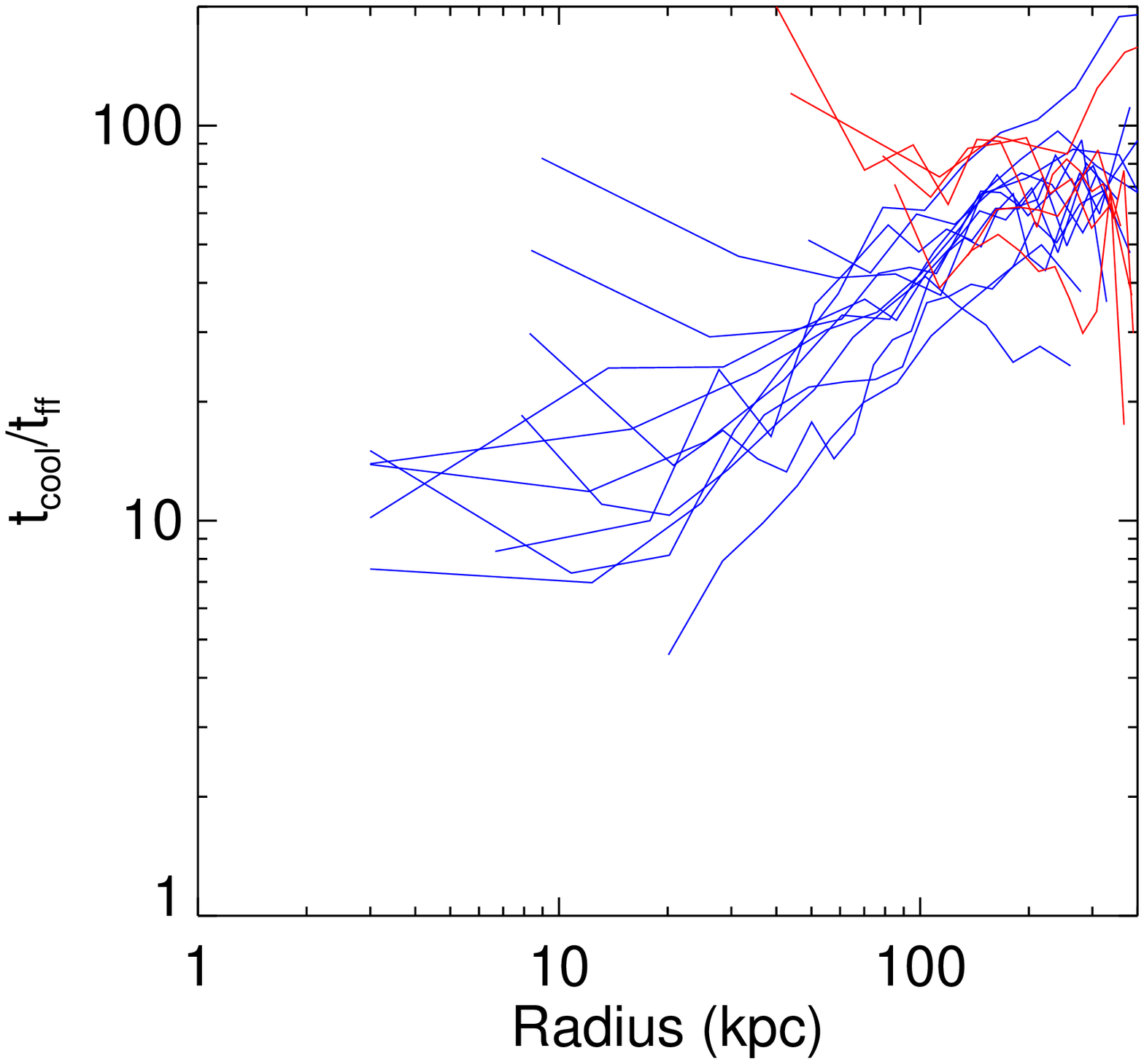}
  \caption{Left: The pressure profiles of the full cluster sample obtained from the jMCMC analysis (triangles) with the best-fit models (eq. \ref{eq:pfit}) 
  shown using lines. 
  Different colors are used for different clusters. Middle and Right: The $t_{\rm cool}$ and $t_{\rm cool}/t_{\rm ff}$ profiles of the full cluster sample. Again, 
  cool-core and non-cool-core clusters are shown using blue and red colors.}
  \label{fig:press_prof_full_sampl}
\end{figure*}

The cooling time and the gravitational free-fall time are important parameters that govern the properties of cluster cores (e.g., \citealt{mcc12,sha12a}).
\citet{mit09} argue that the central cooling times better characterize the properties of cluster cores. 
\citet{raf08} find that the cooling time ($t_{\rm cool}$)
at 12 kpc (rather than the central cooling time or cooling time at the location of the central dominant galaxy) better delineates the star-forming
versus non-star-forming clusters. This radius is close to the minimum of $t_{\rm cool}/t_{\rm ff}$ for most cool-core clusters. Since this radius is 
better resolved, and the cooling time at the very center is ambiguous, min($t_{\rm cool}/t_{\rm ff}$) is a good parameter to describe the cool cores. 
We therefore obtain the cooling time and 
$t_{\rm cool}/t_{\rm ff}$ profiles from our best-fit  shell density and temperature values. While the cooling time is just a function of density, temperature and 
elemental abundance (the quantities obtained from X-ray observations) and is easy to obtain (middle panel of Fig. \ref{fig:press_prof_full_sampl}), the free-fall time requires 
the knowledge of gravitational acceleration profile. One way to
obtain gravitational acceleration is by assuming hydrostatic equilibrium (e.g., see \citealt{ras06}). While this is a good assumption for relaxed, cool-core 
clusters, non-cool clusters are generally merging clusters in which this assumption fails.

We obtain pressure profiles by combining density and temperature data. Since these pressure values are not always decreasing with radius 
(which is impossible for hydrostatic equilibrium), we fit an empirical form to obtain smooth pressure profiles
\begin{equation}
\label{eq:pfit}
P(x)=\frac{P_0 }{ (x/a)^{\alpha_1} + (x/a)^{\alpha_2} },
\end{equation}
which is simpler compared to the `universal' pressure profile suggested by \citet{arn10}. The left panel of Fig. \ref{fig:press_prof_full_sampl} shows the pressure data and the best-fit
pressure profiles for our cluster sample. Pressure (both data points and fits) decreases towards the centre for some non-cool clusters because of bulk flows 
and large turbulent velocities. Gravitational acceleration is obtained by differentiating pressure fits as 
$$
g_{\rm HSE} \equiv -\frac{1}{\rho} \frac{dP}{dr},
$$
where $\rho$ is the gas mass density. Since pressure fits can have $dp/dr>0$, $g_{\rm HSE}$ can be negative, resulting in an imaginary 
$t_{\rm ff} \equiv (2 r /g)^{1/2}$; we ignore such meaningless points in our analysis.
Again, the red color corresponds to non-cool core clusters, while blue is for cool-core clusters. 
Blue curves extend more towards smaller radii than the red ones because high/low central entropy/density means that non-cool clusters are not very 
bright at the centre and hence are difficult to resolve. 

We show $t_{\rm cool}/t_{\rm ff}$ profiles for our cluster sample in the right panel of Fig. \ref{fig:press_prof_full_sampl}. From various recent studies
(\citealt{wag14,ban14,tre15,voi15}),
it appears that this ratio plays a key role in determining the amount of cold gas and AGN feedback in cluster cores. Cold gas condensation and signatures of AGN 
feedback (in form of radio emission and X-ray cavities) are expected if this ratio falls below a critical value close to 10 (\citealt{mee15,cho15}). Moreover, 
clusters are not expected to fall too much below this threshold as extreme feedback due to cold gas condensation is triggered in that case which pushes
the gas out and maintains min($t_{\rm cool}/t_{\rm ff}$) close to the critical value. Indeed, in all our cases $t_{\rm cool}/t_{\rm ff} \gtrsim 4$.

\setcounter{figure}{9}
\begin{figure*}
 \centering
  \includegraphics[width=0.32\textwidth]{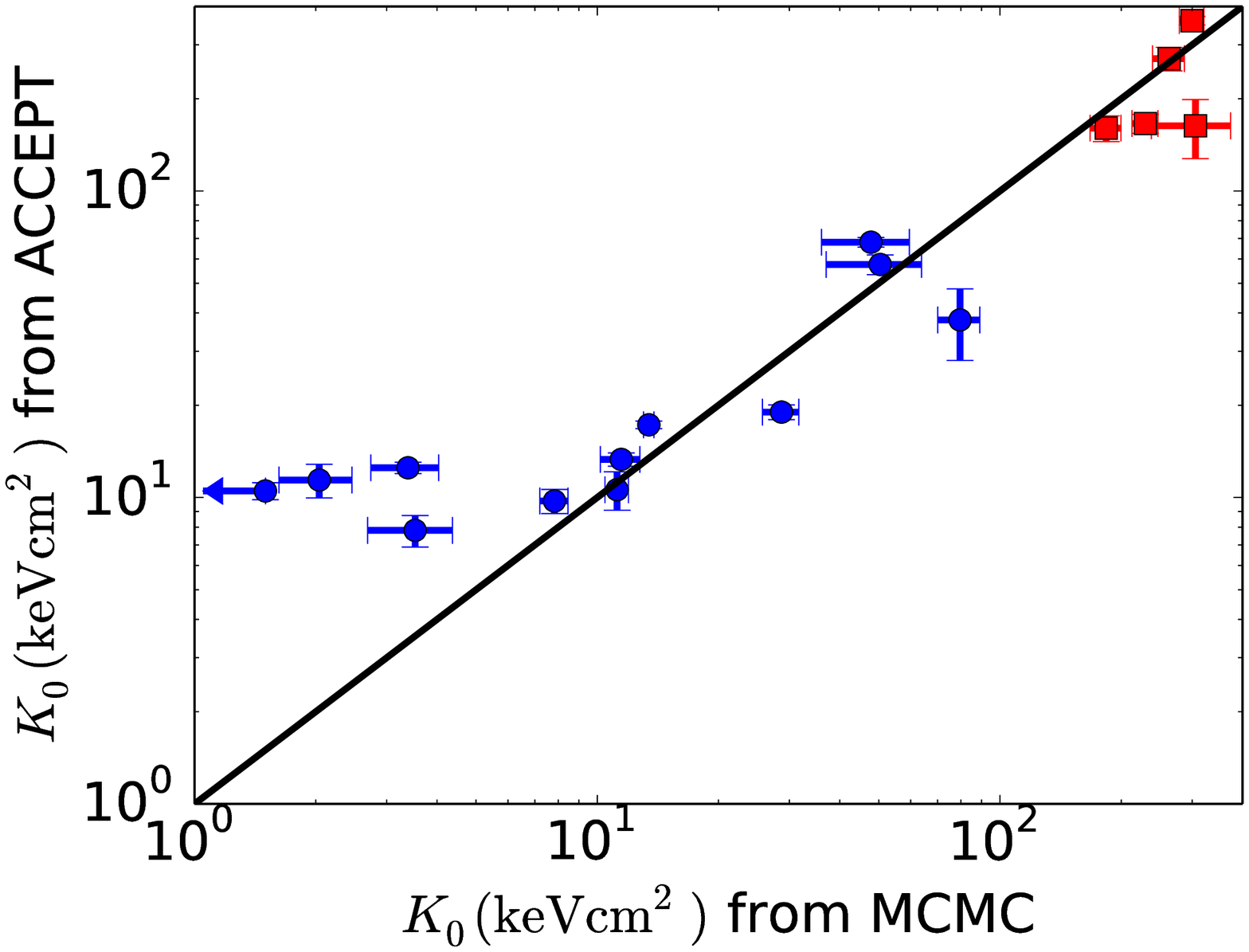}
  \includegraphics[width=0.32\textwidth]{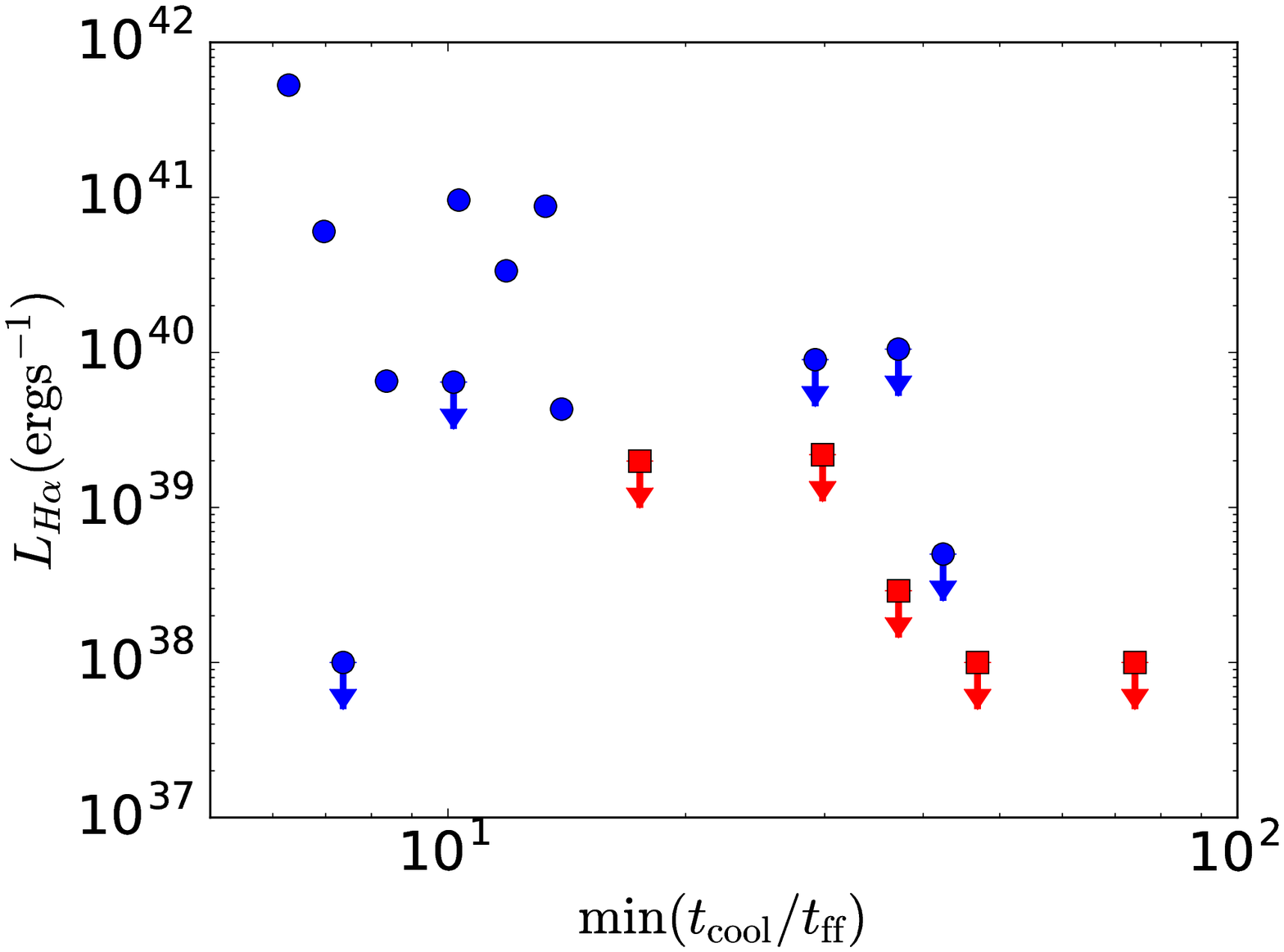}
  \includegraphics[width=0.32\textwidth]{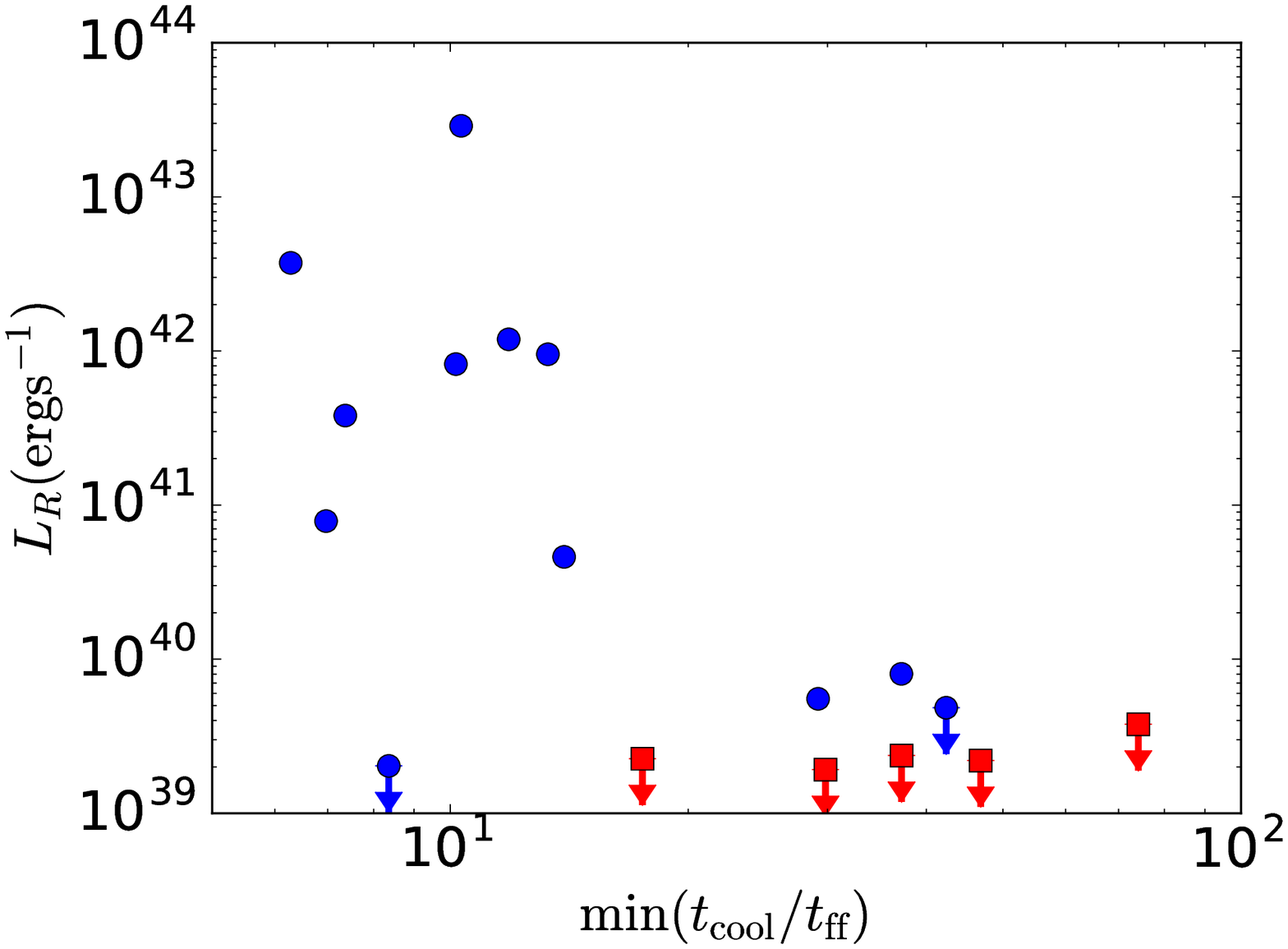}
  \includegraphics[width=0.32\textwidth]{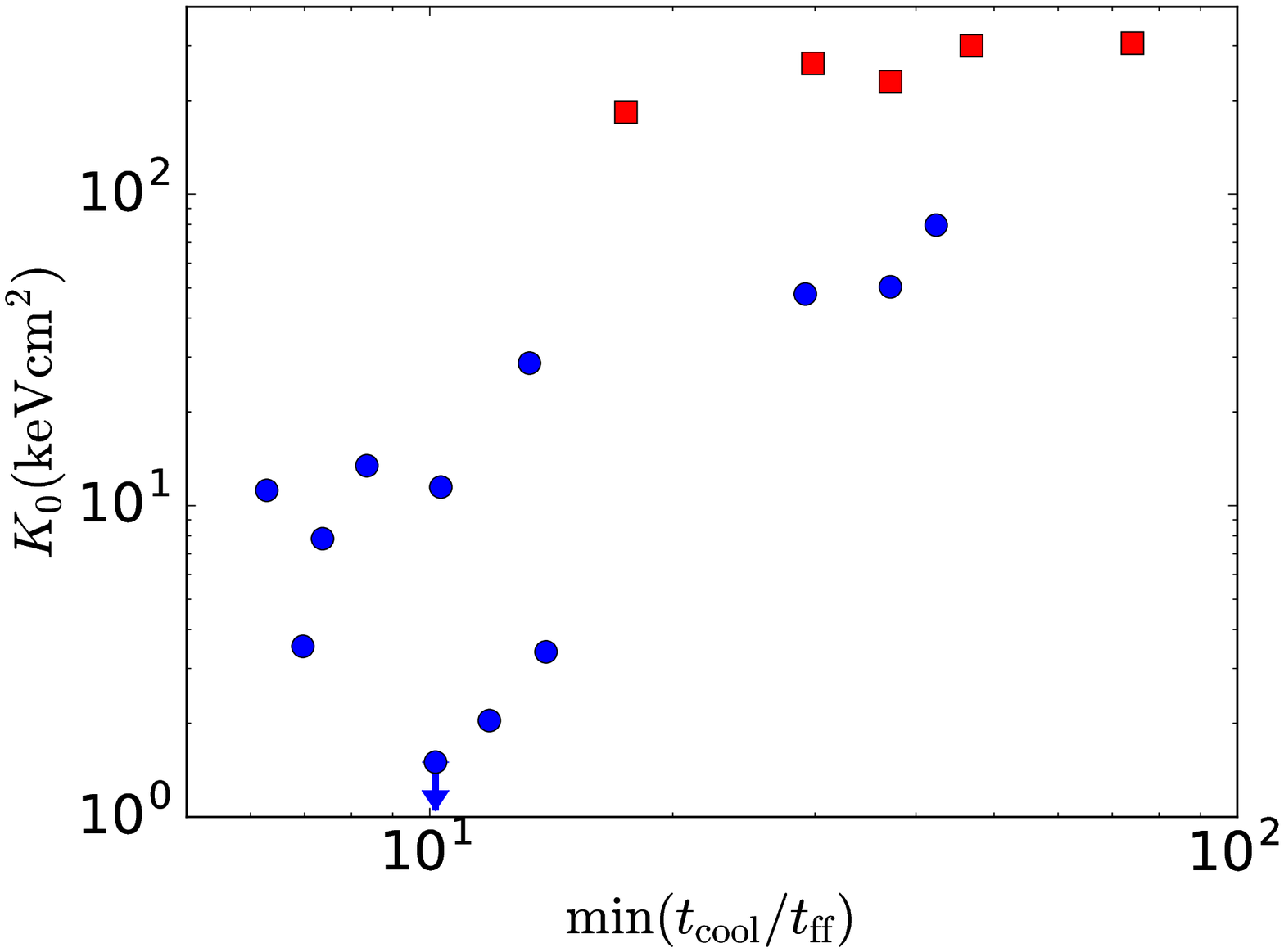}
    \includegraphics[width=0.32\textwidth]{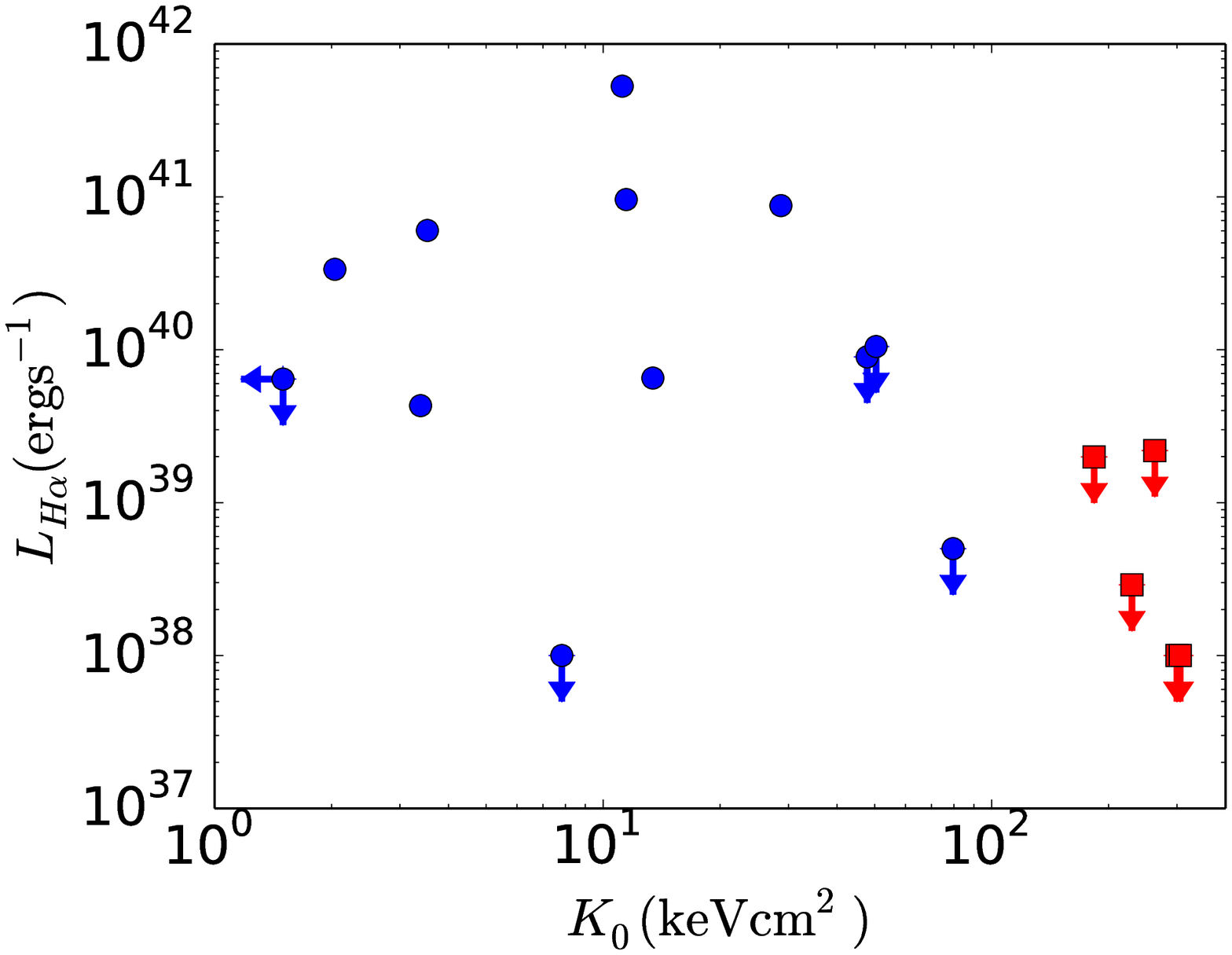}
  \includegraphics[width=0.32\textwidth]{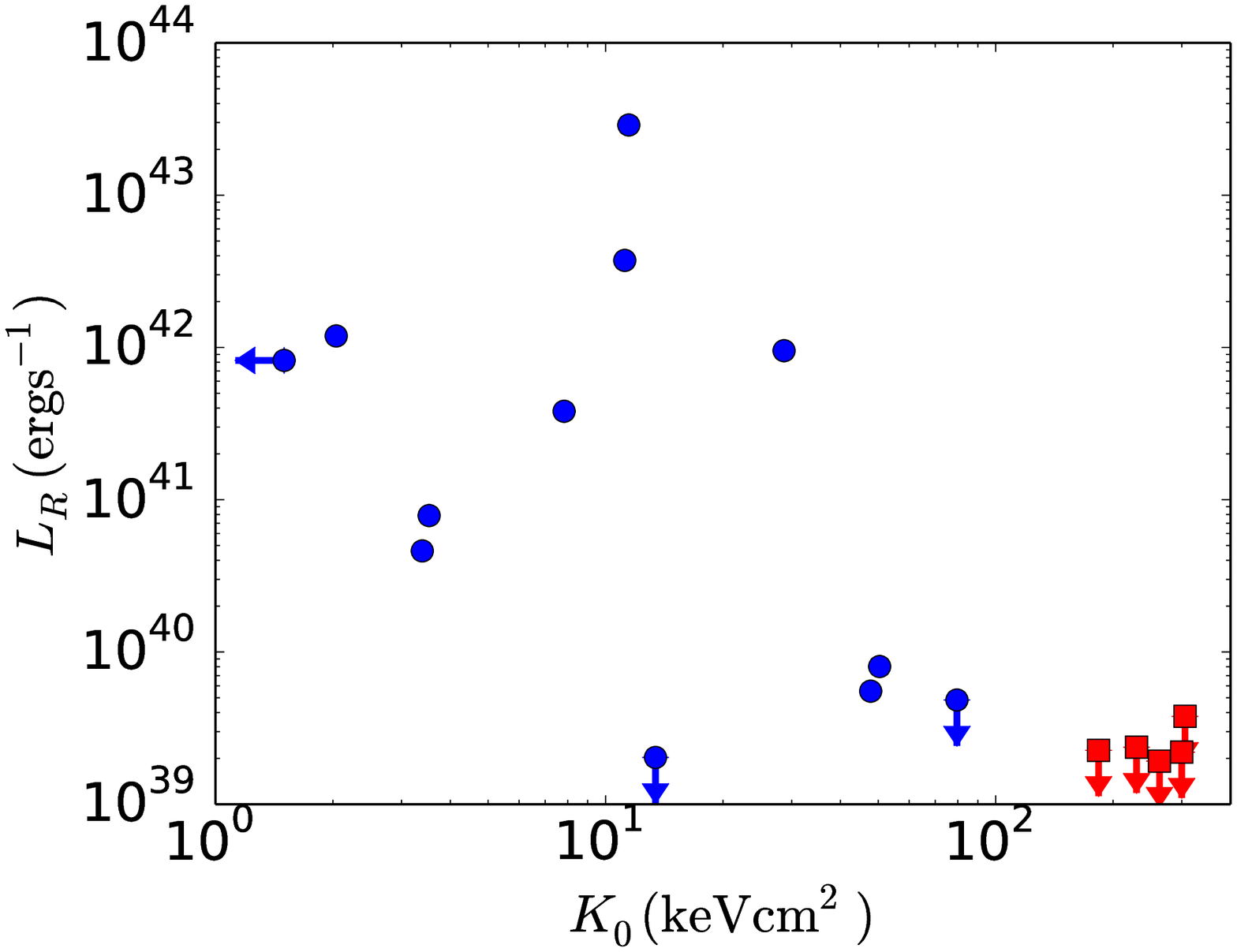}
  \caption{Correlation between various important quantities derived from our analysis ($K_0$ and min$(t_{\rm cool}/t_{\rm ff})$ are derived from jMCMC; see middle panel of Fig. \ref{fig:flat_core_prof_full_sampl} and the right panel of Fig. \ref{fig:press_prof_full_sampl}) and from the ACCEPT sample (where radio and $H\alpha$ luminosities are listed). Although our sample is a small subset of their, our trends are generally similar to \citet{voi15a}. Top-left panel shows that the core entropy values for our coolest cores are systematically smaller than those in the ACCEPT sample (solid black line is the 1:1 line). We have a smaller fraction of non-cool cores compared to \citetalias{cav09}. }
  \label{fig:comp_voit}
\end{figure*}

The top-left panel of Fig. \ref{fig:comp_voit} shows the core entropy derived from \citetalias{cav09} (ACCEPT sample) as a function of the same quantity derived from our jMCMC
analysis. For core entropy ($K_0$) larger than 10 keV cm$^2$ the two methods agree with each other. However, for the lowest core entropies obtained from 
our jMCMC analysis the ACCEPT core entropies seem to flatten out (i.e., are significantly larger). We can speculate on the saturation of core entropy in \citetalias{cav09}'s 
analysis. First, of course, \citetalias{cav09} uses projected temperatures which overestimates the entropy in the core. However, this is not large enough to explain
the observed discrepancy in the coolest cores. \citetalias{cav09} fit the flat-core entropy profile (eq. \ref{eq:flat_core_model}) over {\it all} their radial shells but {\it we integrate
the PDF over the outermost radial shells} and then obtain the marginalized densities and temperatures for the inner shells (extending up to $\lesssim 300$ 
kpc). Only after this, do we fit a flat-core entropy profile over these shells. The entropy values in the outermost shells for the discrepant coolest cores show 
a steepening of the entropy profile in the outermost shells. Thus, having to fit larger entropy value for the outermost shells may lead to an overestimate of 
the core entropy ($K_0$) in \citetalias{cav09}. More work is needed to fully understand this discrepancy.

Different panels of Fig. \ref{fig:comp_voit} show correlation of X-ray properties derived from our analysis and observations from other wavebands (obtained from 
ACCEPT\footnote{\url{www.pa.msu.edu/astro/MC2/accept/}} tables; \citealt{cav09}), namely, $H\alpha$ and radio. We also compare our results 
with those of \citet{voi15a}. Like them, we find that the {\it amount} of $H\alpha$ luminosity is anti-correlated with min($t_{\rm cool}/t_{\rm ff}$), albeit 
with some scatter. A similar trend is seen for radio luminosity but the spread is larger. Anti-correlation is not as strong with the core entropy ($K_0$).
The large scatter is consistent with the cyclic behaviour observed in cool core simulations (compare with Fig. 14 of \citealt{pra15}; note that these are 
data from a single simulation and do not span as large a range in $K_0$ and min[$t_{\rm cool}/t_{\rm ff}$] as seen in our Fig. \ref{fig:comp_voit}). 
One aspect in which we differ from \citet{voi15a} is that, unlike them, we do not see $H\alpha$ emitting systems with min($t_{\rm cool}/t_{\rm ff} \gtrsim
10$); they have several $H\alpha$ emitting clusters with largest min$(t_{\rm cool}/t_{\rm ff} ) \gtrsim 20$. This discrepancy is closely related to the saturation
of core entropy for coolest clusters in \citetalias{cav09} (and hence under/over-estimation of core density/cooling time), discussed in the previous paragraph.
 Other factor that can systematically bias their $t_{\rm cool} \propto T/\Lambda(T)$ high is the fact that they use {\it projected} temperatures which are larger 
than the actual temperatures (which our jMCMC method faithfully reproduces) in spherical shells (see left panels of Figs. 
\ref{fig:100ks_sim_different_method} \& \ref{fig:test_clus_different_method}). Another difference is the computation of $t_{\rm ff}$ which 
depends on how well the pressure fit approximates the pressure data (see left panel of Fig. \ref{fig:press_prof_full_sampl}). \citet{voi15} introduce a
singular isothermal potential with $\sigma=250$ km s$^{-1}$ at all their cluster centers; this can lead to a shorter $t_{\rm ff}$ and a higher 
$t_{\rm cool}/t_{\rm ff}$. One more minor difference from \citet{voi15} 
is that we have two clusters with 
min$(t_{\rm cool}/t_{\rm ff})<10$ with no $H\alpha$ detection. Sorting out these disagreements is left for future. 

\section{Conclusions}
\label{S:conclusions}
Following are the key conclusions of our paper:
\begin{itemize}
\item jMCMC method is a statistically accurate method to recover the density and temperature profiles in cluster cores, which accounts for covariance
between density and temperature of different shells.
\item Most cluster cores favour a flat entropy core and not a power law (neither a single or a double power law), on the scales of 10 kpc. 
A single power law is clearly favoured in only one of the clusters in our sample. One must specify a scale when talking about flattening of the entropy profile 
as some clusters and groups show a decrease in entropy at scales $\lesssim$ few kpc, corresponding to small coronae associated with BCGs.
\item The entropy and cooling time distributions may indicate bimodality but more clusters are required to reach definitive conclusions.
\item The minimum value of $t_{\rm cool}/t_{\rm ff}$ appear to govern the presence of $H\alpha$ and radio emission in cluster cores. The {\it amount} of 
cold gas is anti-correlated with min($t_{\rm cool}/t_{\rm ff}$), but with a large scatter.
\item The core entropy for coolest cores is systematically overestimated by \citetalias{cav09} as compared to our jMCMC method, resulting in a lower density, a longer
cooling time, and a larger min$(t_{\rm cool}/t_{\rm ff})$. The discrepancies in the observational reconstruction of thermodynamic profiles need to be sorted 
out before a detailed comparison with theoretical models can be made.

\end{itemize}

\label{lastpage}

\bigskip
\noindent {\bf Acknowledgements:}  This work is partly supported by the DST-India grant no. Sr/S2/HEP-048/2012 (which also supports KL) and an India-Israel
joint research grant (6-10/2014[IC]). We thank the anonymous referee for suggestions which substantially improved our paper.

\bigskip

\appendix

\section{Optimizing jMCMC}
\label{S:optmz_jmcmc}
  We performed various tests on A2597 to optimize the jMCMC algorithm. Based on the results of these tests we made certain choices that apply to the 
rest of the cluster sample analyzed in the paper. For most of our analyses the elemental abundance, $Z$, in all the annuli spectra was assumed to be 
equal to the average elemental abundance of the cluster obtained from the analysis of the cluster's full spectrum in XSPEC. To check for the effect of 
a variable $Z$, we added $Z_i$ as an additional parameter for the $i^{\rm th}$ shell along with $n_i$ and $T_i$. To keep the size of model spectra 
(${\bf M}_{iJ}$) manageable we reduced the number of $n$ and $T$ grid points to half for this test (i.e., from 500 and 150 to 250 and 75; see section 
\ref{S:MCMC}). The $Z$ parameter was added as a third-axis to this box with 5 linearly distributed values between 0.2 to 1.0 times the solar elemental abundance value. 

Having too many annuli in our analysis makes it computationally slow, and can lead to large errors in the parameters. On the other hand, having few annuli can 
lead to the inner regions of the cluster not being resolved properly. Therefore, we tested for the effect 
of varying the number of annuli used for the analysis. Minimum $\sim$17000 counts in each annulus led to 13 annuli in A2597. The minimum counts were increased 
to $\sim$30000 and decreased to $\sim$12000 to have a total of 6 and 18 annuli, respectively. 
We carried out the jMCMC analysis of A2597 with 6, 13 and 18 annuli. We also investigated the effect of changing the length of MCMC chains. The results from all 
these tests are discussed below.\\

\begin{itemize}

\setcounter{figure}{10}
\begin{figure*}
 \includegraphics[width=0.33\linewidth]{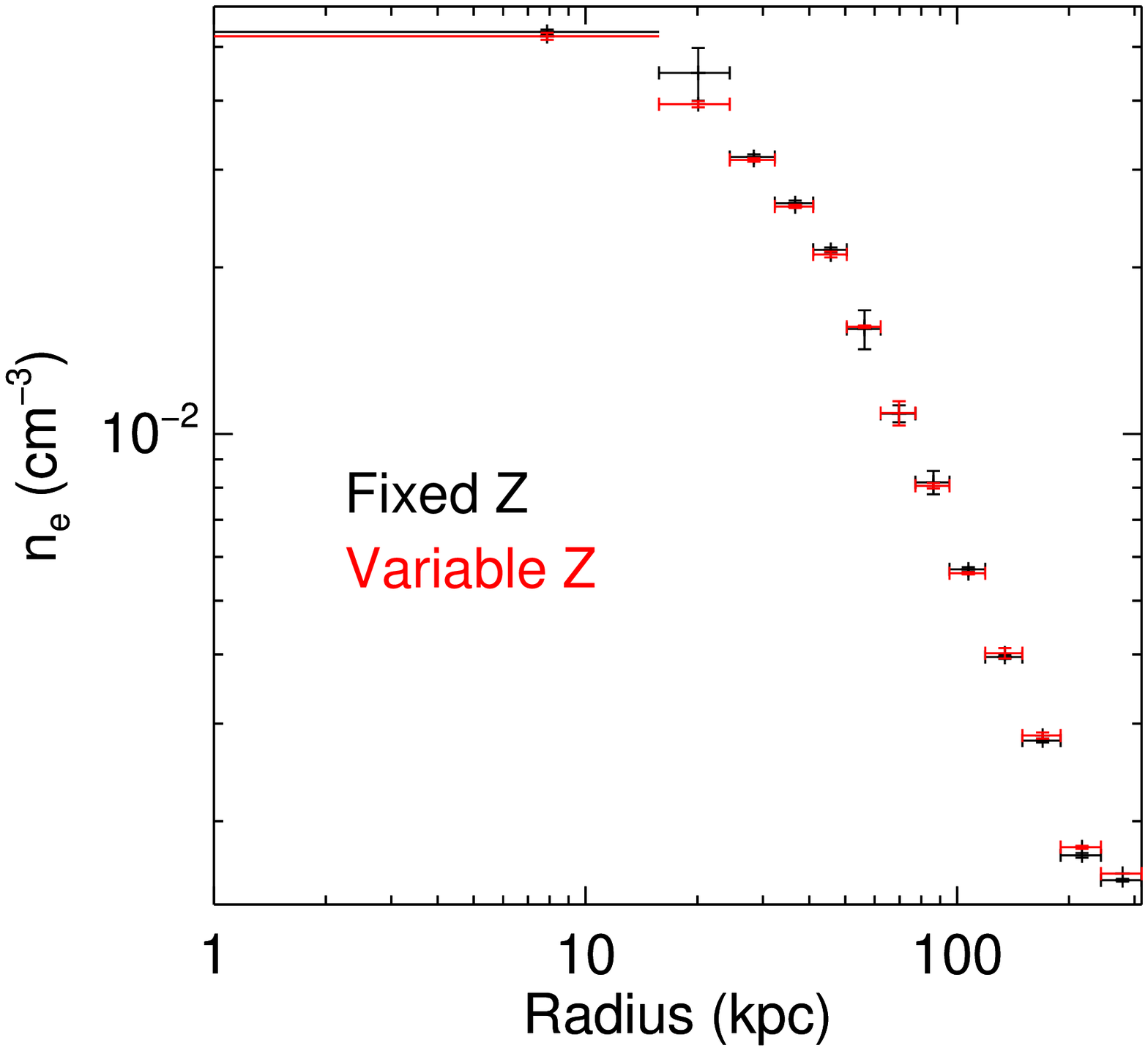}
 \includegraphics[width=0.33\linewidth]{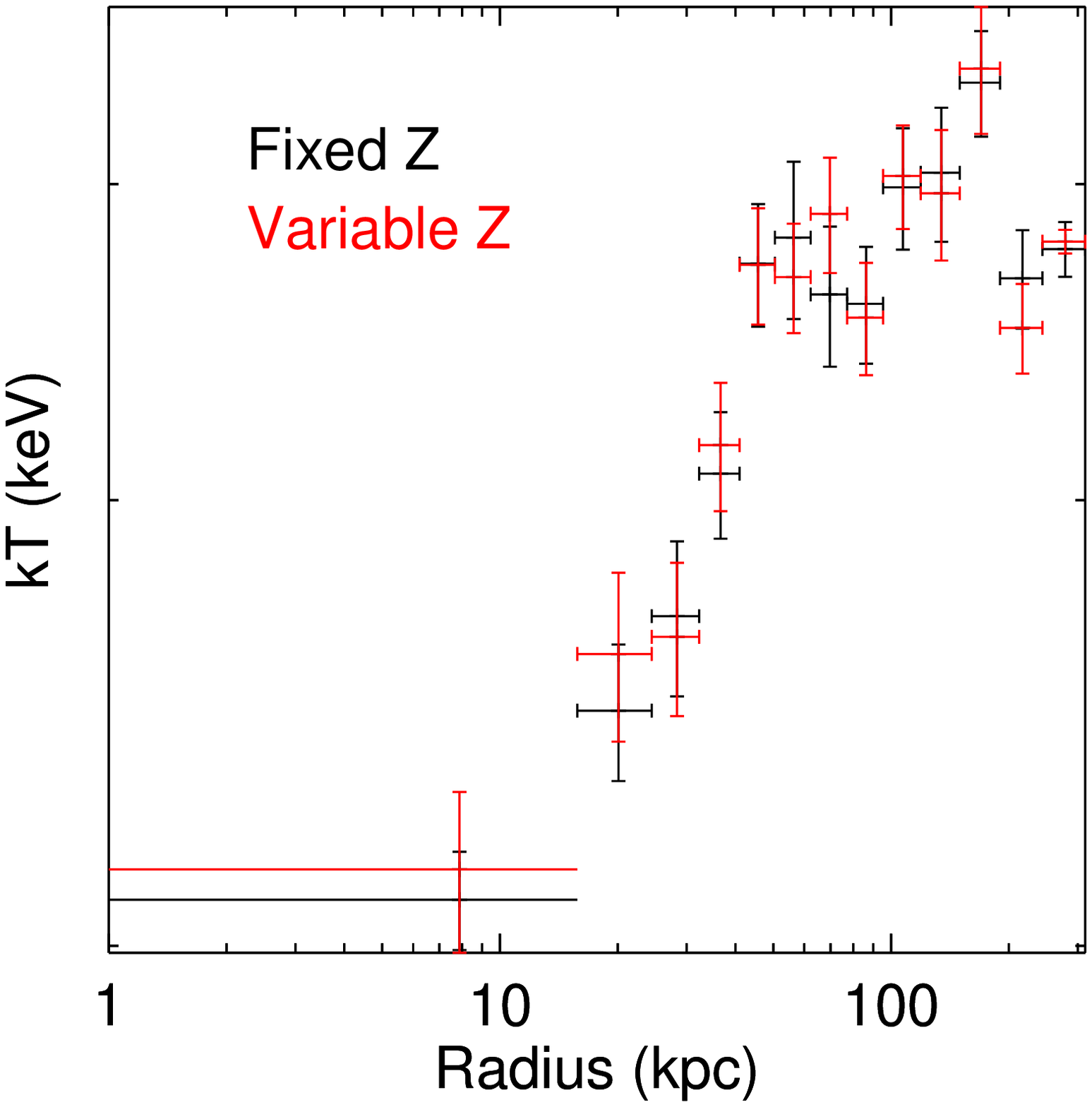}
 \includegraphics[width=0.33\linewidth]{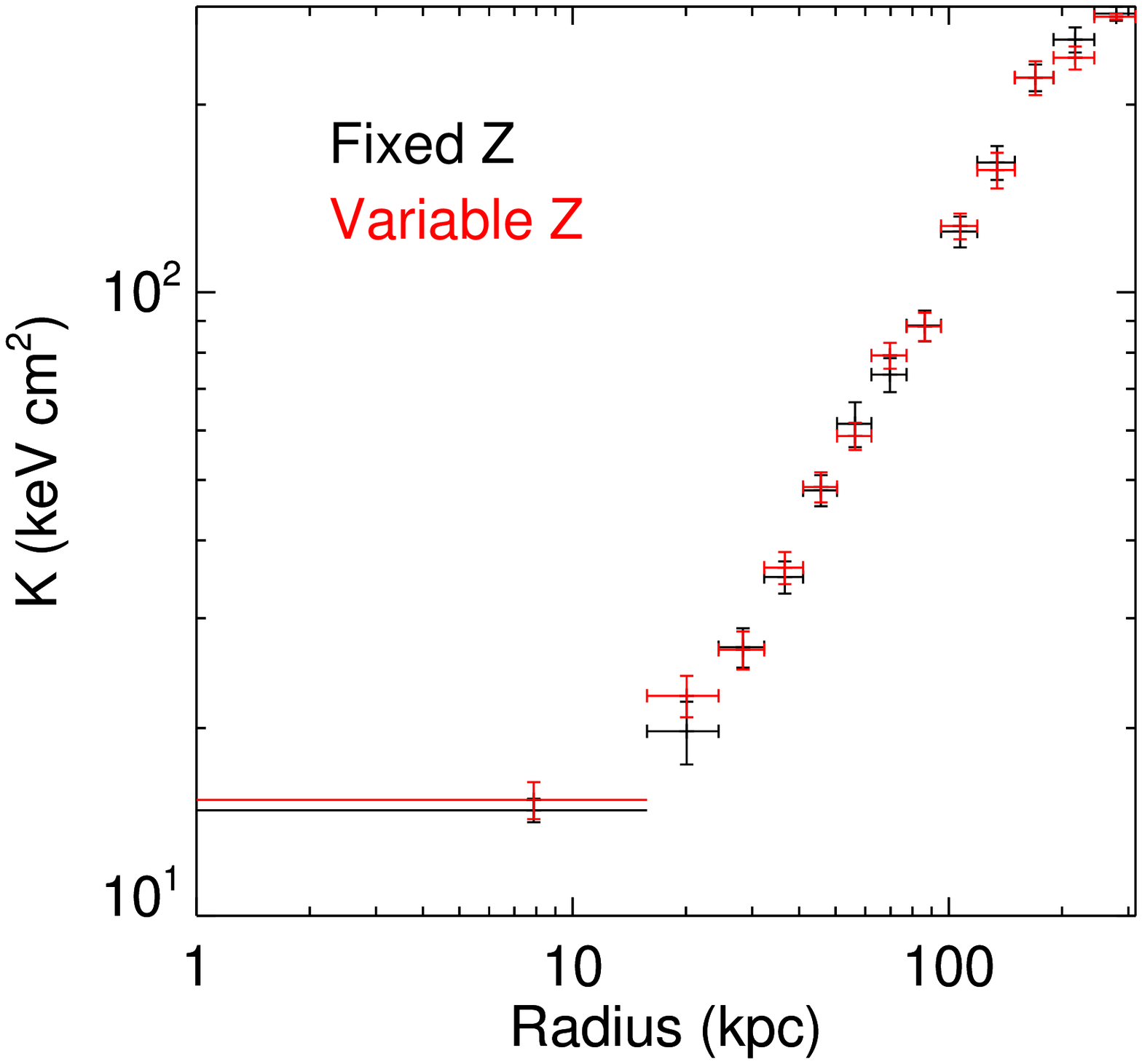}
\caption{Electron density, temperature and entropy profiles obtained from the jMCMC analysis of the Chandra observations of the 
test cluster A2597. Profiles obtained with fixed and variable elemental abundance are shown using black and red colors, 
respectively.}
\label{fig:test_clus_13shell_res}
\end{figure*}

\item {{\it Fixed vs. variable elemental abundance}:} 
The electron number density, temperature and entropy profiles obtained from the jMCMC analysis of the test cluster A2597 are shown in 
Fig.~\ref{fig:test_clus_13shell_res}. The profiles obtained with a variable elemental abundance are also shown. The figure shows 
that a free elemental abundance does not affect the density and entropy profiles significantly. 
As we are mostly interested in the entropy profiles of the clusters, 
we have {\it fixed the elemental abundance} for the analysis of the remaining clusters at their respective average values obtained 
from spectrum integrated over all shells. This is essentially done for computational reasons.

\setcounter{figure}{11}
\begin{figure}
 \includegraphics[width=0.98\linewidth]{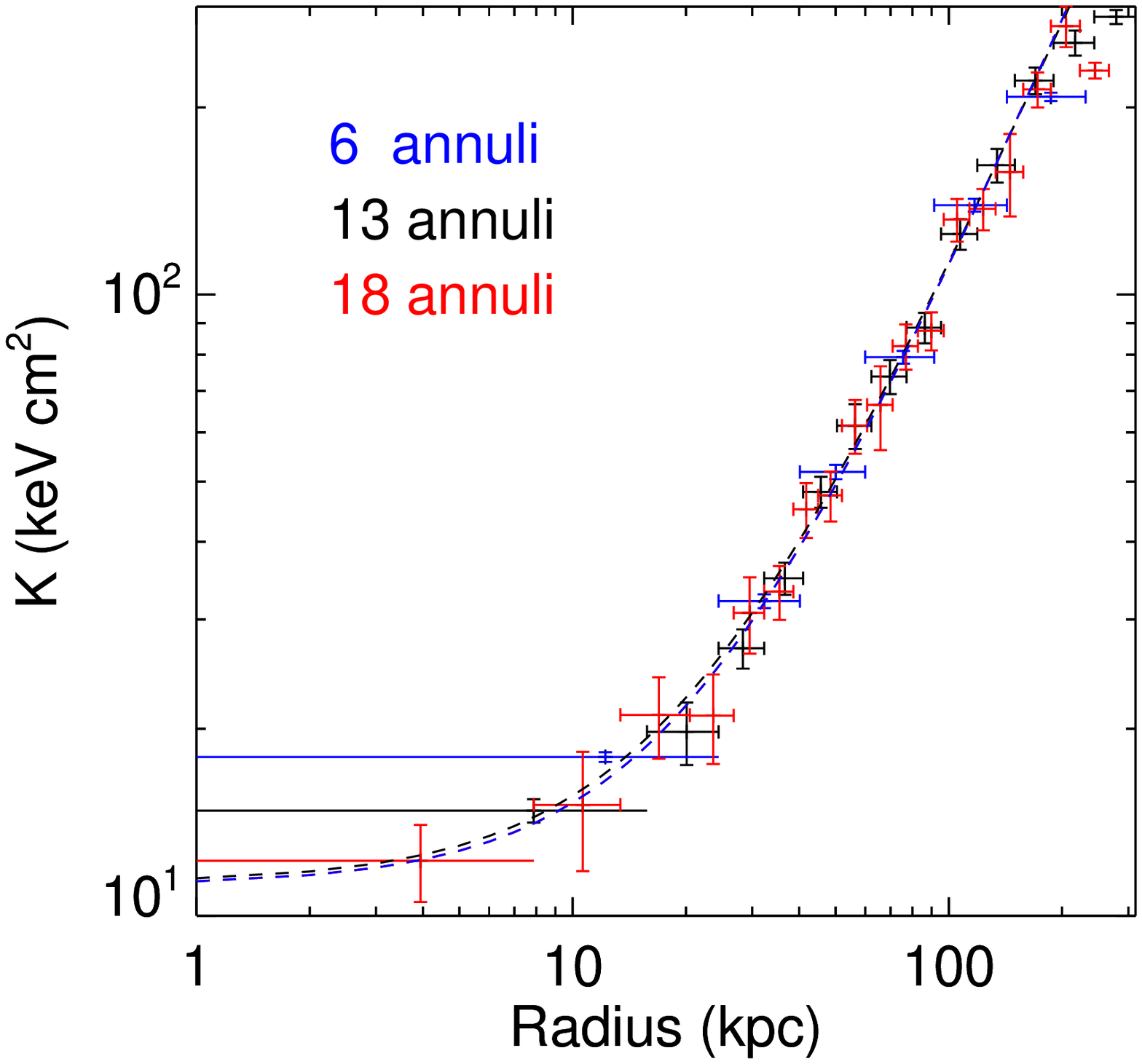}
\caption{Entropy profiles obtained from the jMCMC analysis of the 
test cluster using 6 (blue), 13 (black) and 18 (red) annuli. The flat-core model fits are also shown using dashed lines.}
\label{fig:test_clus_different_annuli_res}
\end{figure}

\item {{\it The number of annuli}:} The test cluster was analyzed with different number of annuli and the resulting entropy profiles 
are shown in Fig. \ref{fig:test_clus_different_annuli_res}. It can be seen that varying the number of annuli does not change the 
general shape of the entropy profile, especially at resolved radii. However, note that with just 6 annuli one can not constrain the 
profile shape very well. For example, the flattening of the entropy profile towards the centre seen with 13 and 18 annuli, does not seem 
to be very prominent in the profile with just 6 annuli. On the other hand, having 18 annuli leads to large errors in the inner entropy values 
and therefore can lead to a poor fit. In general, for the rest of the sample, we try to optimise in terms of best resolution of the cluster core profiles.

\setcounter{figure}{12}
\begin{figure*}
 \includegraphics[width=0.33\linewidth]{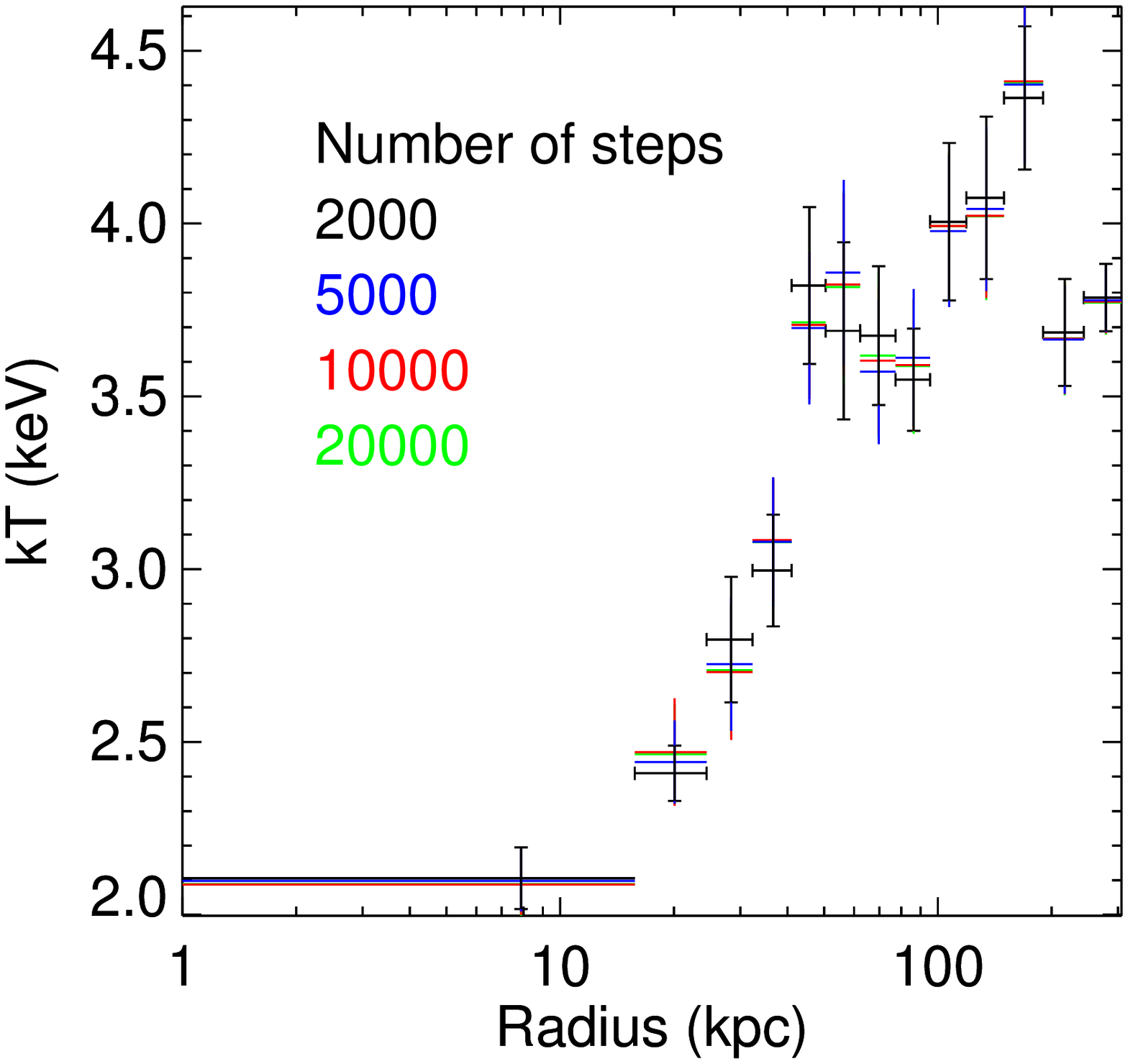}
 \includegraphics[width=0.33\linewidth]{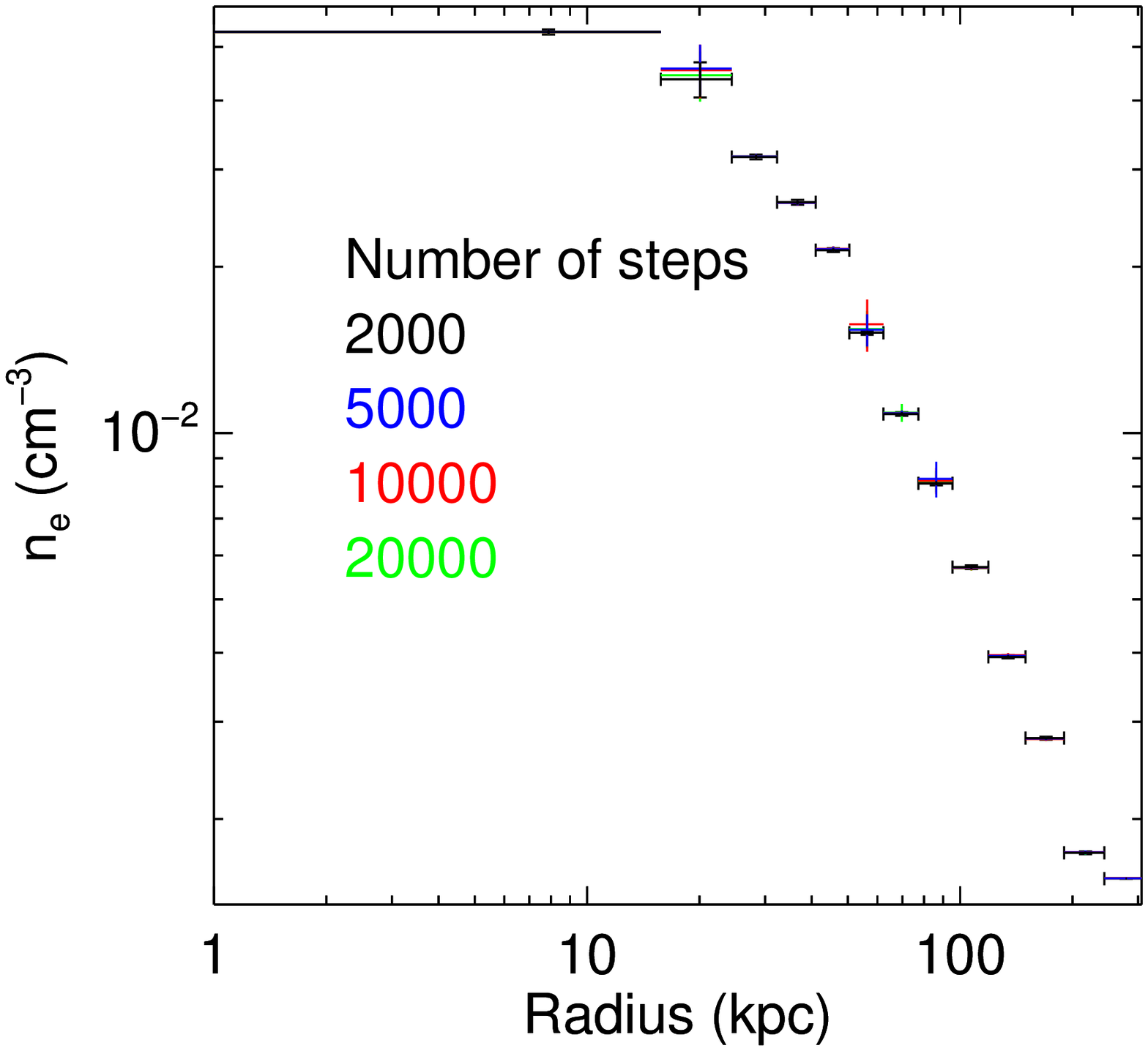}
 \includegraphics[width=0.33\linewidth]{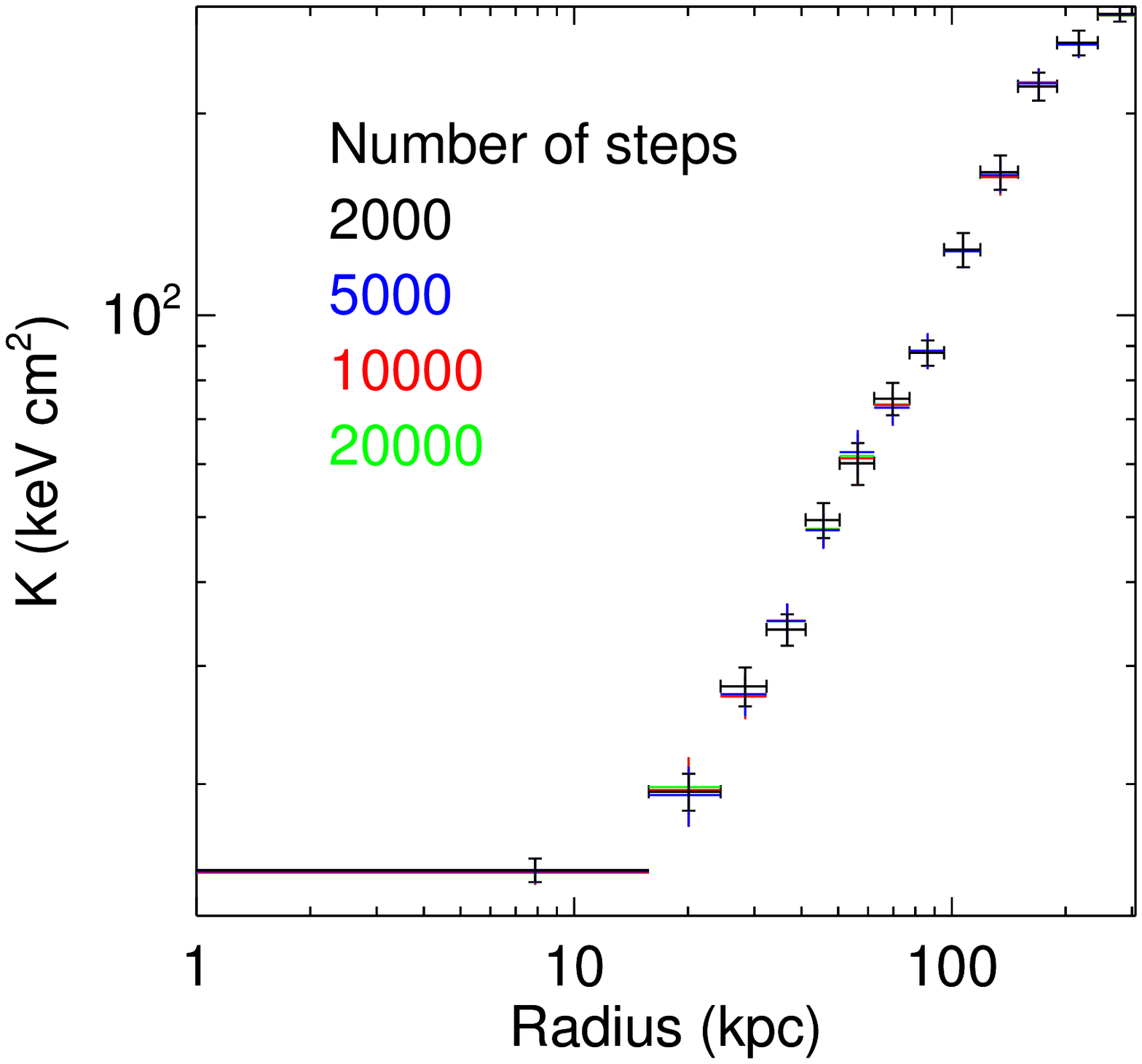}
\caption{Temperature, electron density and entropy profiles obtained from the jMCMC analysis of the test 
cluster using 2000 (black), 5000 (blue), 10000 (red) and 20000 (green) steps along each 
parameter direction in the MCMC random walk.}
\label{fig:test_clus_different_cl_res}
\end{figure*}

\item {{\it Number of steps}:} We tried different numbers of steps (2000, 5000, 10000 and 20000) in the MCMC random walk along {\it each} parameter direction 
for the test cluster, and the temperature, electron density and entropy profiles obtained are shown in Fig. \ref{fig:test_clus_different_cl_res}. The profiles 
are found to be mostly consistent with each other. However, it can be seen that using 2000 steps gives slightly larger error bars, especially in the 
temperature profile, and the mean values also seem to be slightly shifted from the rest of the profiles. Profiles for larger number of steps are found to be almost 
identical. Although for the test cluster 5000 steps seems to be an optimum value at which the results seem to be converging, considering that 
many of our clusters have poor count statistics, we chose 100000 steps as a safe value for our analysis.

\end{itemize}

\section{Single power law model}
\label{S:sgl_powlaw_fit}

\begin{table}
 \caption{Single power law entropy model fitting (eq. \ref{eq:spl_model}) for PL sample.}
\label{tab:sgl_powlaw_results}
\vskip 0.5cm
\centering
{\scriptsize
\begin{tabular}{c c c c}
\hline
Cluster Name & $K_{1}$ & $\alpha_1$ & $\chi^{2}_{\rm red}$ (DOF)\\
 & (keV cm$^{2}$) & & \\
\hline
\hline
  A85 & 173.3$\pm$2.1 & 0.92$\pm$0.01 & 4.09 (9)\\
 A478 & 130.5$\pm$1.9 & 0.95$\pm$0.01 & 2.65 (14)\\
A2029 & 173.3$\pm$1.2 & 0.84$\pm$0.01 & 4.75 (13)\\
A3112 & 144.6$\pm$0.9 & 1.44$\pm$0.01 & 18.24 (9)\\
\hline
\end{tabular}}
\end{table}

Since the entropy profiles of PL sample  showed almost linear shapes in the log-log plot, single power law model 
(eq. \ref{eq:spl_model}) fits were tested for these clusters.  The resulting $K_1$-$\alpha1$ probability distributions, along with the 
single power law profile fits are shown in Fig. \ref{fig:sgl_pl_entr_fit_samp2}. All the clusters of PL sample
show a weak positive correlation between $K_1$ and $\alpha_1$, except for A2029, which does not show any correlation 
between them. The expectation values of 
$K_1$ and $\alpha1$ (with the associated variances) and the reduced chi-squared values 
obtained from the fits are given in Table \ref{tab:sgl_powlaw_results}. The flat-core and single power law model 
fits for PL sample were compared using F-test and the results are given in Table \ref{tab:f_test_results}. 
Flat-core fits seem to be preferred compared to single power law fits for all the clusters of PL sample (this is of course true for the FC sample). 
However,  since for the cluster A2029, flat-core model fitting leads to negative $K_0$, single power law 
fit may be a preferred.

\setcounter{figure}{13}
\begin{figure*}
 \centering  
  \includegraphics[width=0.32\textwidth]{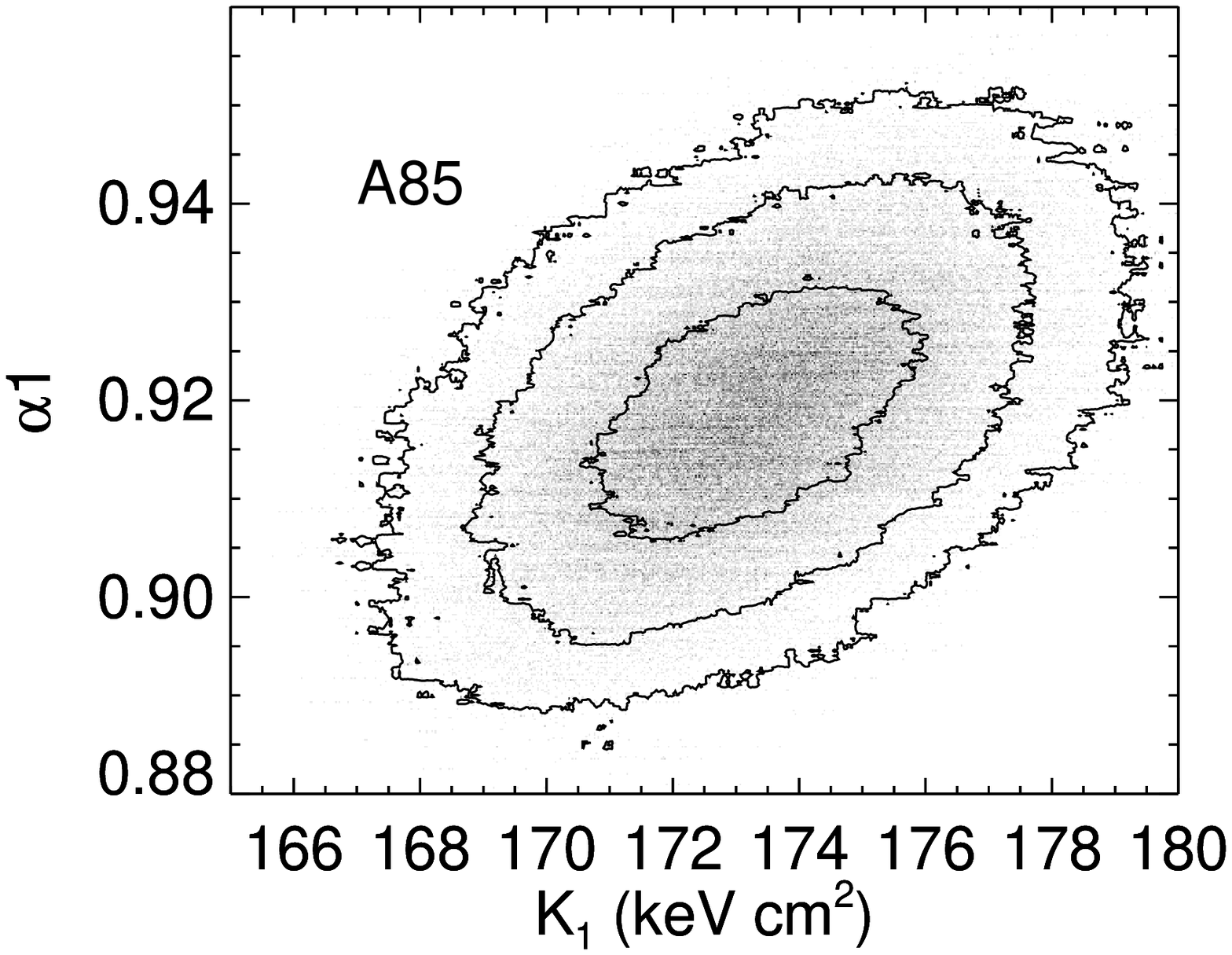}
  \includegraphics[width=0.32\textwidth]{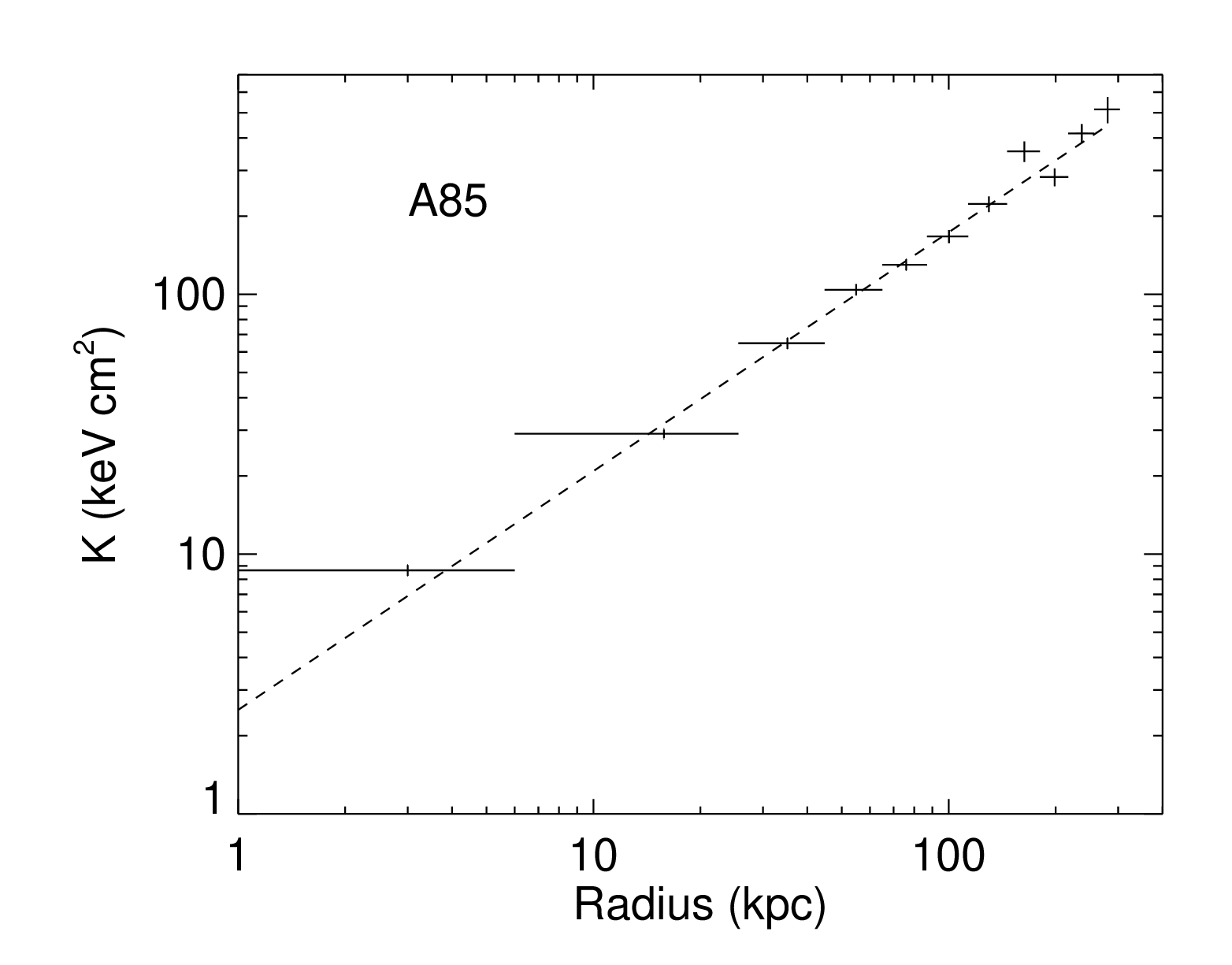}\\
  \includegraphics[width=0.32\textwidth]{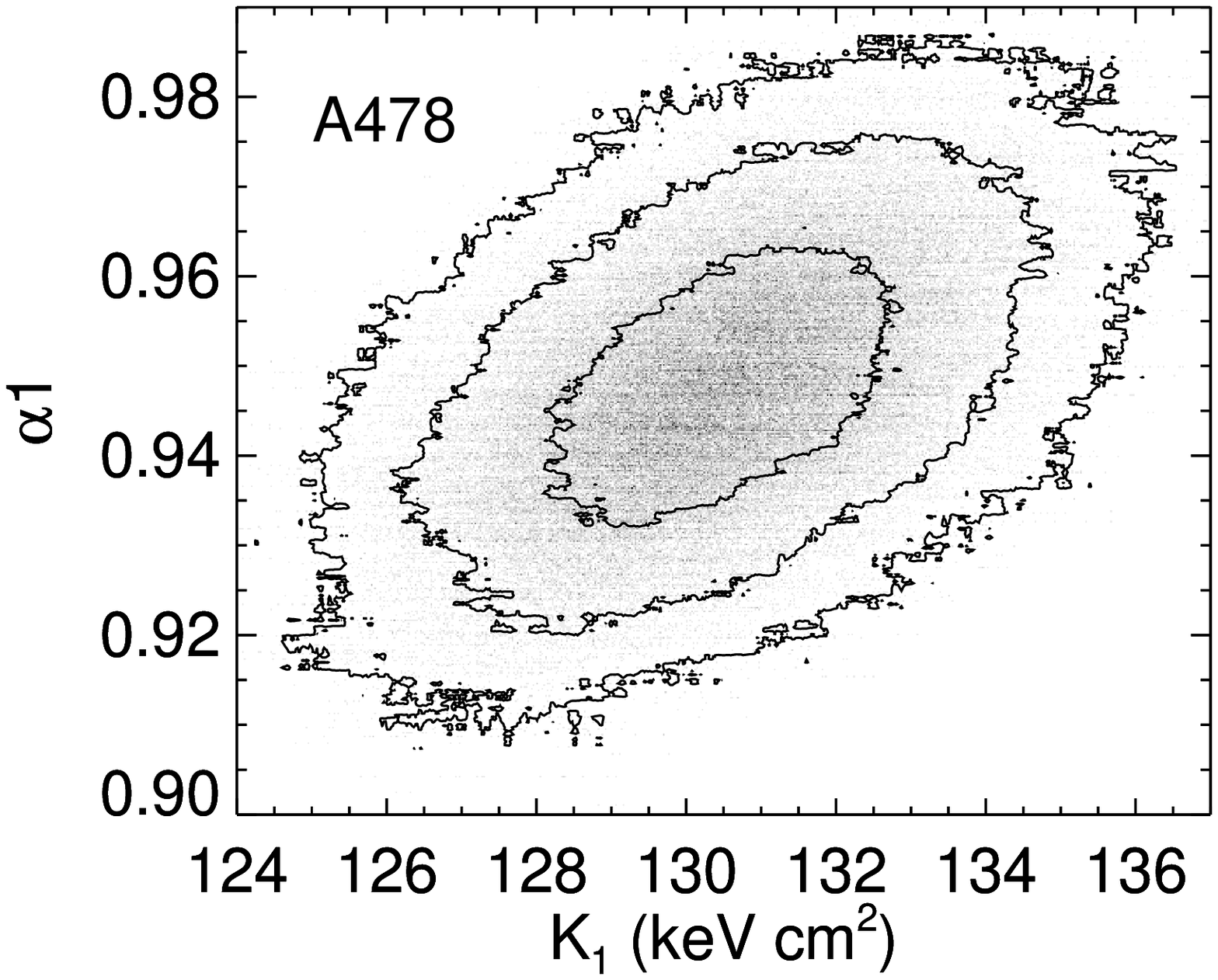}
  \includegraphics[width=0.32\textwidth]{figures/A478_entropy_sgl_pow_law_fit_using_mcmc.ps}\\
  \includegraphics[width=0.32\textwidth]{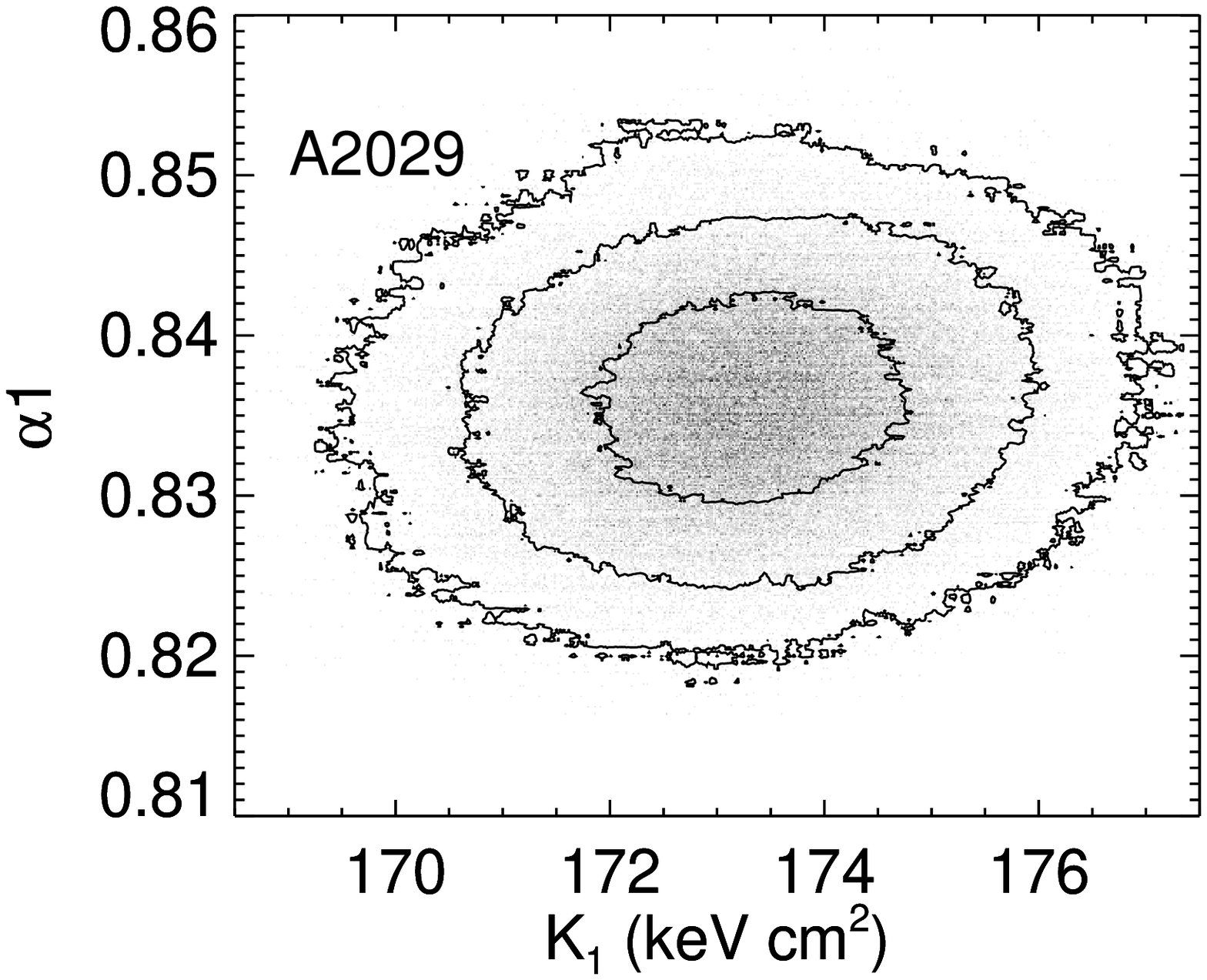}
  \includegraphics[width=0.32\textwidth]{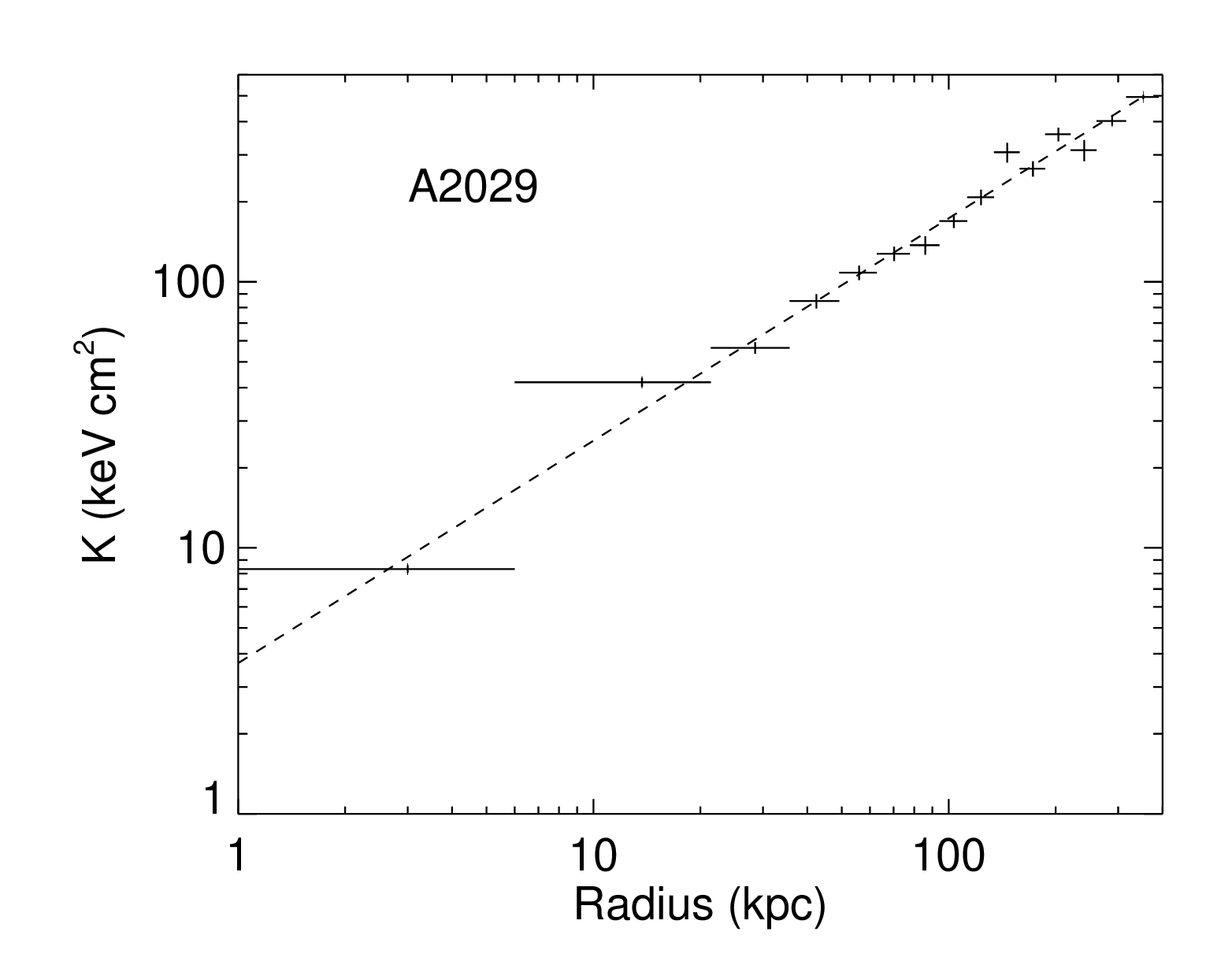}\\   
  \includegraphics[width=0.32\textwidth]{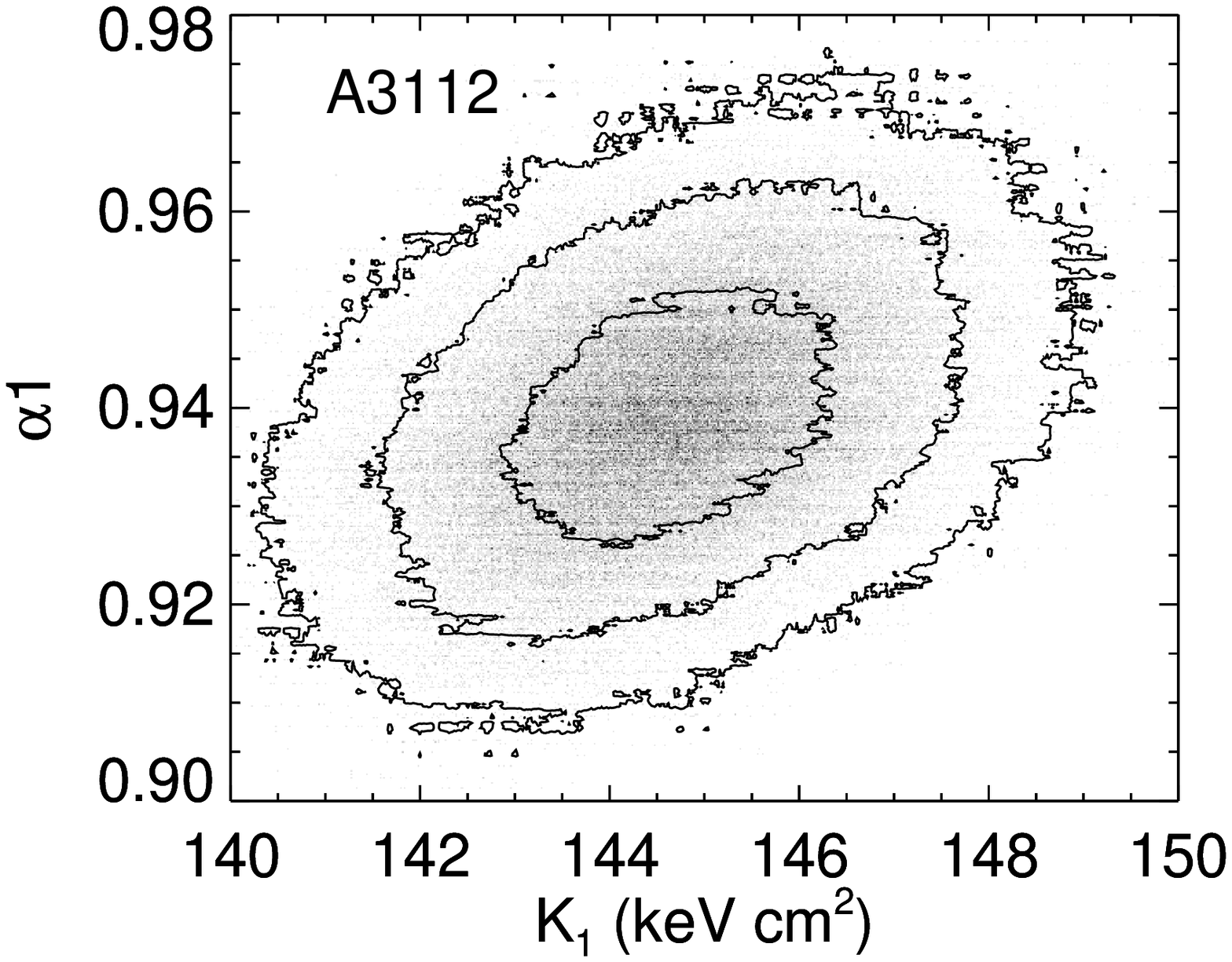}
  \includegraphics[width=0.32\textwidth]{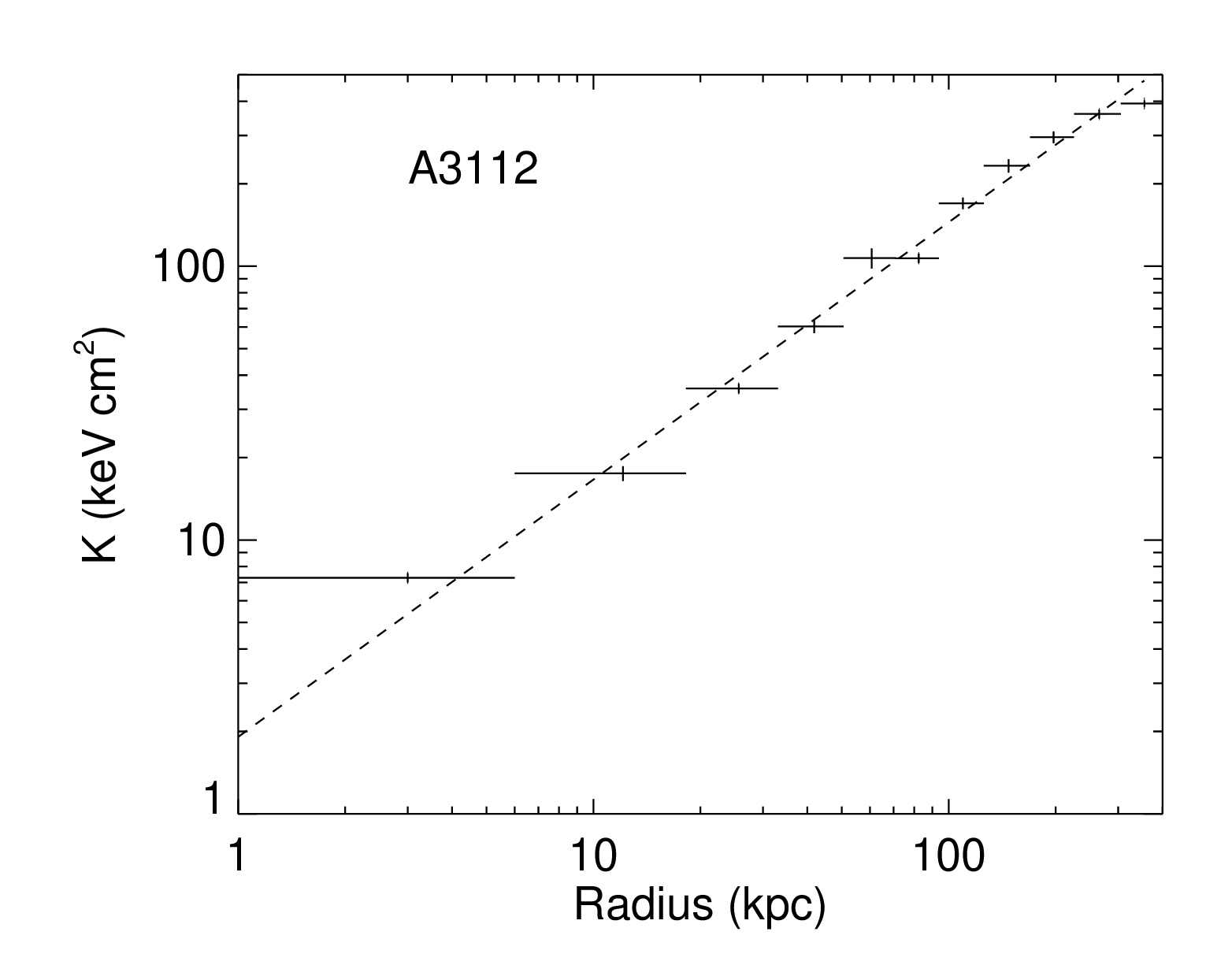}\\ 
  \caption{The $K_1$-$\alpha_1$ probability distributions obtained from the single power law entropy model fitting for PL sample. 
   The contours mark the 50\%, 90\% and 99\% inclusion levels based on the density of points starting from the innermost contour outwards. 
   Greyscale denotes the PDF density.}
  \label{fig:sgl_pl_entr_fit_samp2}
\end{figure*}

\section{Double power law model}
\label{S:dbl_powlaw_fit}

\begin{table}
\setlength{\tabcolsep}{1pt}
 \caption{Double power law entropy model fitting (eq. \ref{eq:dbl_model}).}
\label{tab:dbl_powlaw_results}
\vskip 0.5cm
\centering
{\scriptsize
\begin{tabular}{c c c c c c}
\hline
Cluster & $K_{1}$ & $\alpha_1$ & $K_2$ & $\alpha_2$ & $\chi^{2}_{\rm red}$ (DOF)\\
Name& (keV cm$^{2}$) & & (keV cm$^{2}$) & & \\
\hline
\hline
  A85&41.5$\pm$29.7&0.49$\pm$0.20&129.2$\pm$30.5&1.12$\pm$0.10&6.37 (7)\\
 A133&17.7$\pm$3.1&0.09$\pm$0.06&227.7$\pm$10.3&1.86$\pm$0.09&5.21 (4)\\
 A478&11.9$\pm$8.0&0.29$\pm$0.18&116.0$\pm$8.7&1.07$\pm$0.05&2.40 (12)\\
A1650&130.6$\pm$28.4&0.26$\pm$0.13&58.7$\pm$27.6&2.34$\pm$0.81&1.35 (3)\\
A1795&53.4$\pm$17.2&0.26$\pm$0.15&80.6$\pm$23.9&2.26$\pm$0.57&0.59 (10)\\
A2029& 0.4$\pm$0.3&0.33$\pm$0.72&169.9$\pm$12.4&0.87$\pm$0.13&8.89 (11)\\
A2142&120.9$\pm$37.8&0.39$\pm$0.16&75.2$\pm$39.2&1.94$\pm$0.94&1.44 (5)\\
A2204&10.3$\pm$2.5&0.09$\pm$0.08&261.3$\pm$14.1&2.03$\pm$0.09&5.24 (3)\\
A2244s&139.7$\pm$18.9&0.38$\pm$0.10&25.6$\pm$18.5&2.32$\pm$1.08&2.06 (3)\\
A2597&16.1$\pm$4.2&0.12$\pm$0.08&95.3$\pm$4.8&1.46$\pm$0.07&1.65 (7)\\
A3112&5.2$\pm$4.8&0.18$\pm$0.15&131.9$\pm$5.1&0.96$\pm$0.02&18.24 (7)\\
Hydra-A&16.2$\pm$4.6&0.11$\pm$0.09&102.8$\pm$4.8&1.30$\pm$0.07&3.94 (5)\\
 A754&293.2$\pm$29.9&0.23$\pm$0.15&39.0$\pm$22.4&1.78$\pm$0.52&1.31 (9)\\
A2256&298.2$\pm$22.4&0.15$\pm$0.11&12.0$\pm$12.8&2.47$\pm$1.06& 4.63 (4)\\ 
A3158&261.6$\pm$33.8&0.22$\pm$0.17&0.3$\pm$25.6&3.23$\pm$2.00&5.36 (2)\\
A3667&211.9$\pm$18.6&0.15$\pm$0.10&35.6$\pm$16.6&1.83$\pm$0.47&4.97 (5)\\
ZWCL1215&305.9$\pm$38.0&0.32$\pm$0.18&39.1$\pm$25.1&1.68$\pm$0.61& 0.41 (4)\\
\hline
\end{tabular}}
\end{table}

One of the main results of \citetalias{pan14} has been that the entropy profiles of all clusters in their sample were found to be best-fitted 
using single/double power law models. Therefore, we have also tried double power law (eq. \ref{eq:dbl_model}) 
fits for our full cluster sample\footnote{The flat-core model can also be interpreted as a special case of the double power law 
model with $K_1 = K_0$, $\alpha_1 = 0$,  $K_2 = K_{100}$ and $\alpha_2 = \alpha $.}. To avoid any degeneracy between the two power laws, 
we have forced the conditions $\alpha_1 < \alpha_2$ and $\alpha_1, \alpha_2 > 0$ , so that the first power law fits the central region 
(where almost all the clusters 
show flattening) and the second power law fits the outer parts of the clusters. The resulting $K_1$-$\alpha_1$ and $K_2$-$\alpha_2$  
probability distributions, along with the double power law profile fits are shown in Fig. \ref{fig:dbl_pl_entr_fit}. 
Nearly all the clusters of the sample show a positive correlation between $K_1$ and $\alpha_1$, and a negative correlation between $K_2$ 
and $\alpha_2$. The cluster, A2029, which showed an inverted core in the flat-core fit, shows a poor fit with the double power law 
model; the $K_1$-$\alpha_1$ probability distribution for this cluster seems to be highly irregular although the $K_2$-$\alpha_2$ plot looks 
symmetric with no visible correlation between the two quantities. The expectation values of 
$K_1$, $\alpha_1$, $K_2$ and $\alpha2$ (with errors estimated using variances) and the reduced chi-squared values 
obtained from the fits are given in Table \ref{tab:dbl_powlaw_results}. A comparison of the flat-core and double power law model fits 
for the full sample based on F-test is given in Table \ref{tab:f_test_results}. 
For all the clusters of the sample, except for four, flat-core model is found to provide significantly better fit than the double power law model.

\setcounter{figure}{14}
\begin{figure*}
 \centering
  \includegraphics[width=0.32\textwidth]{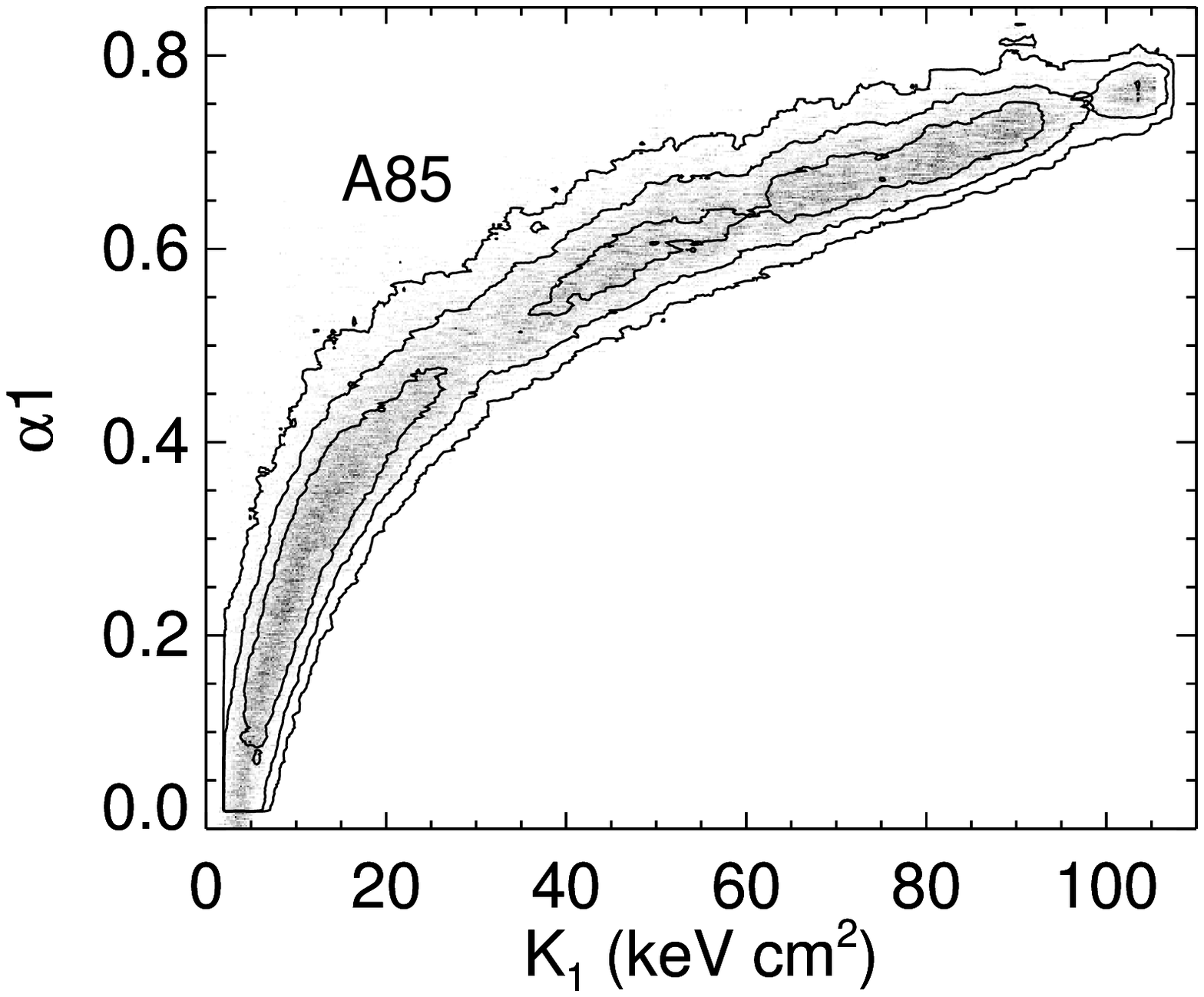}
  \includegraphics[width=0.32\textwidth]{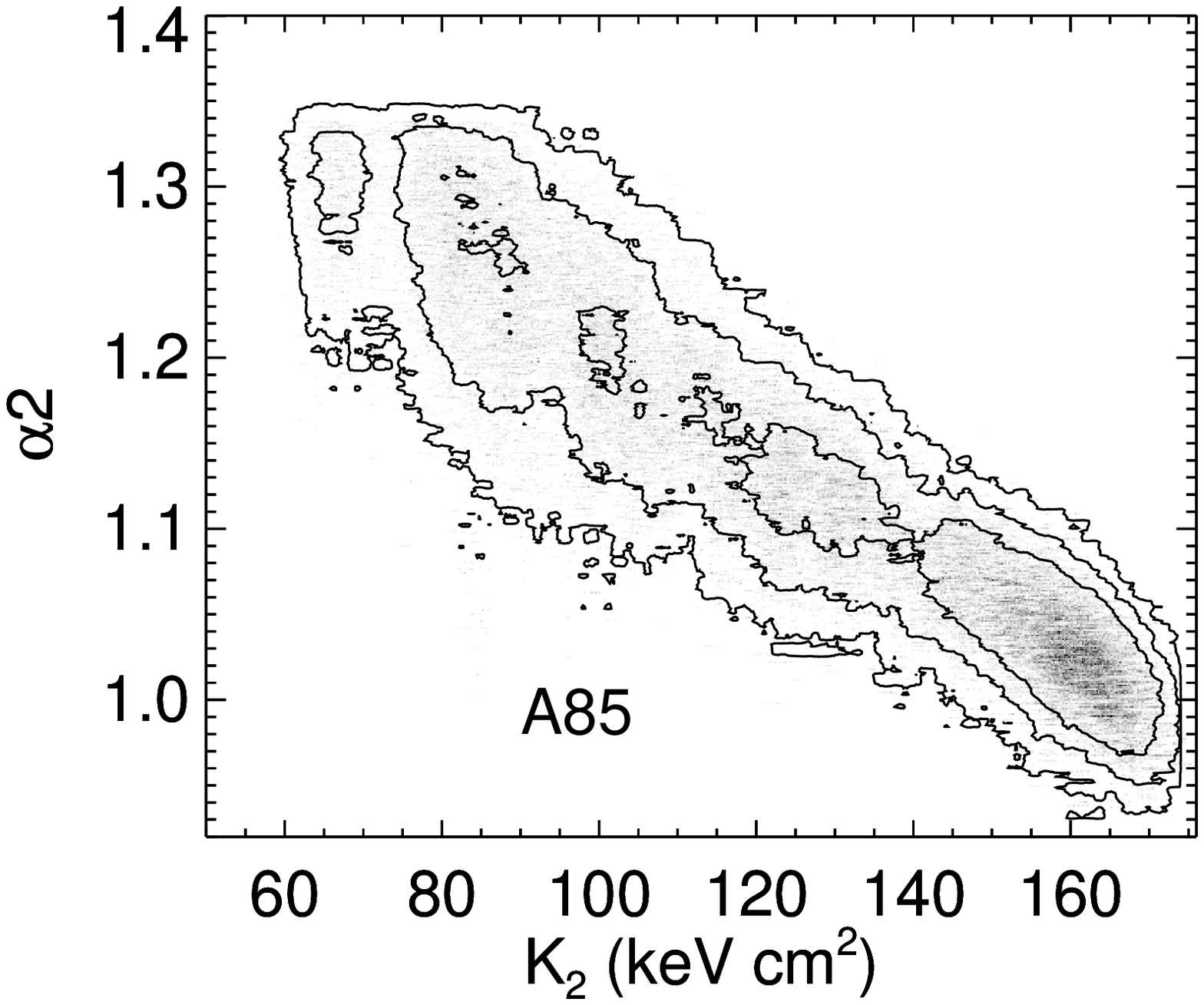}
  \includegraphics[width=0.32\textwidth]{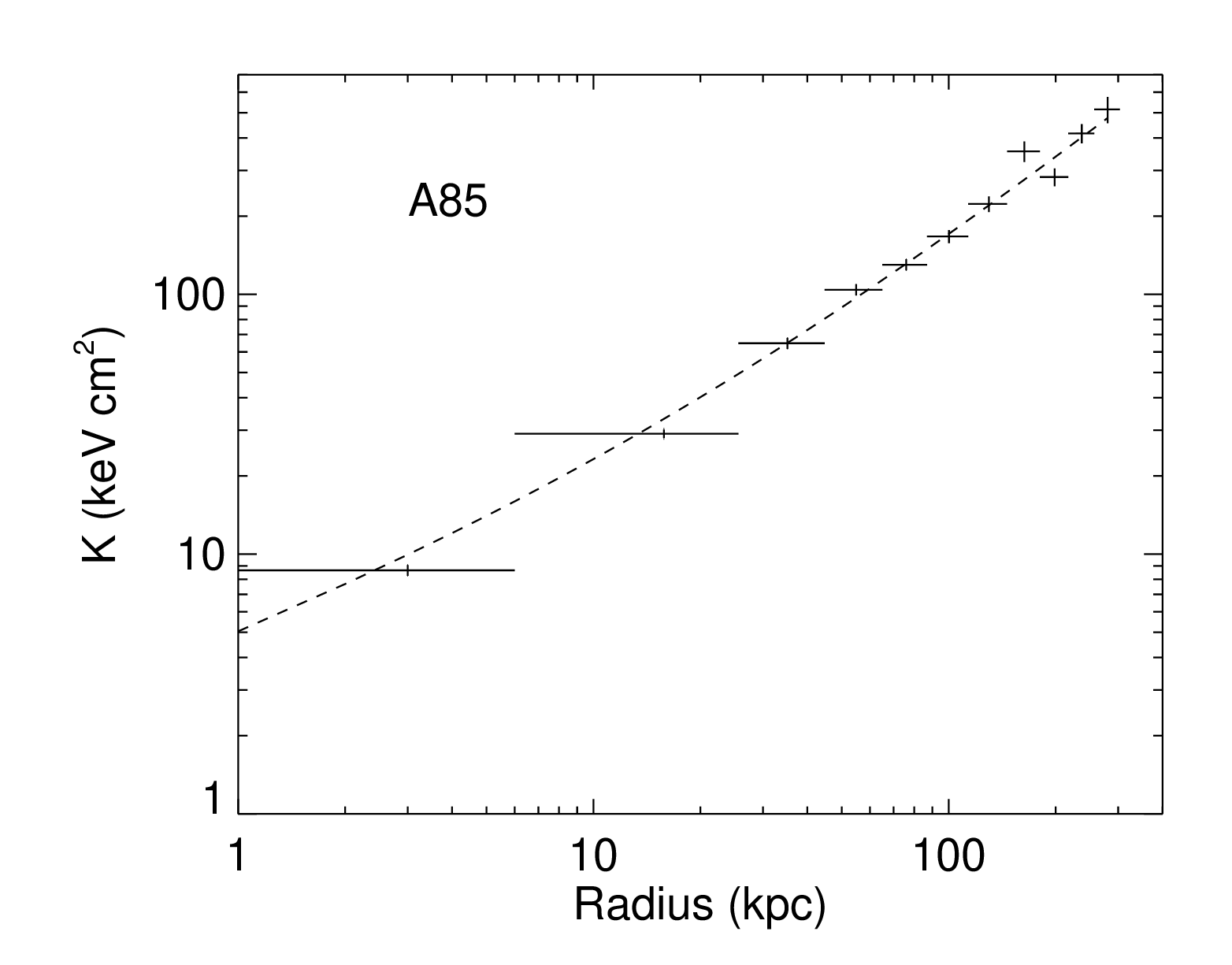}\\
  \includegraphics[width=0.32\textwidth]{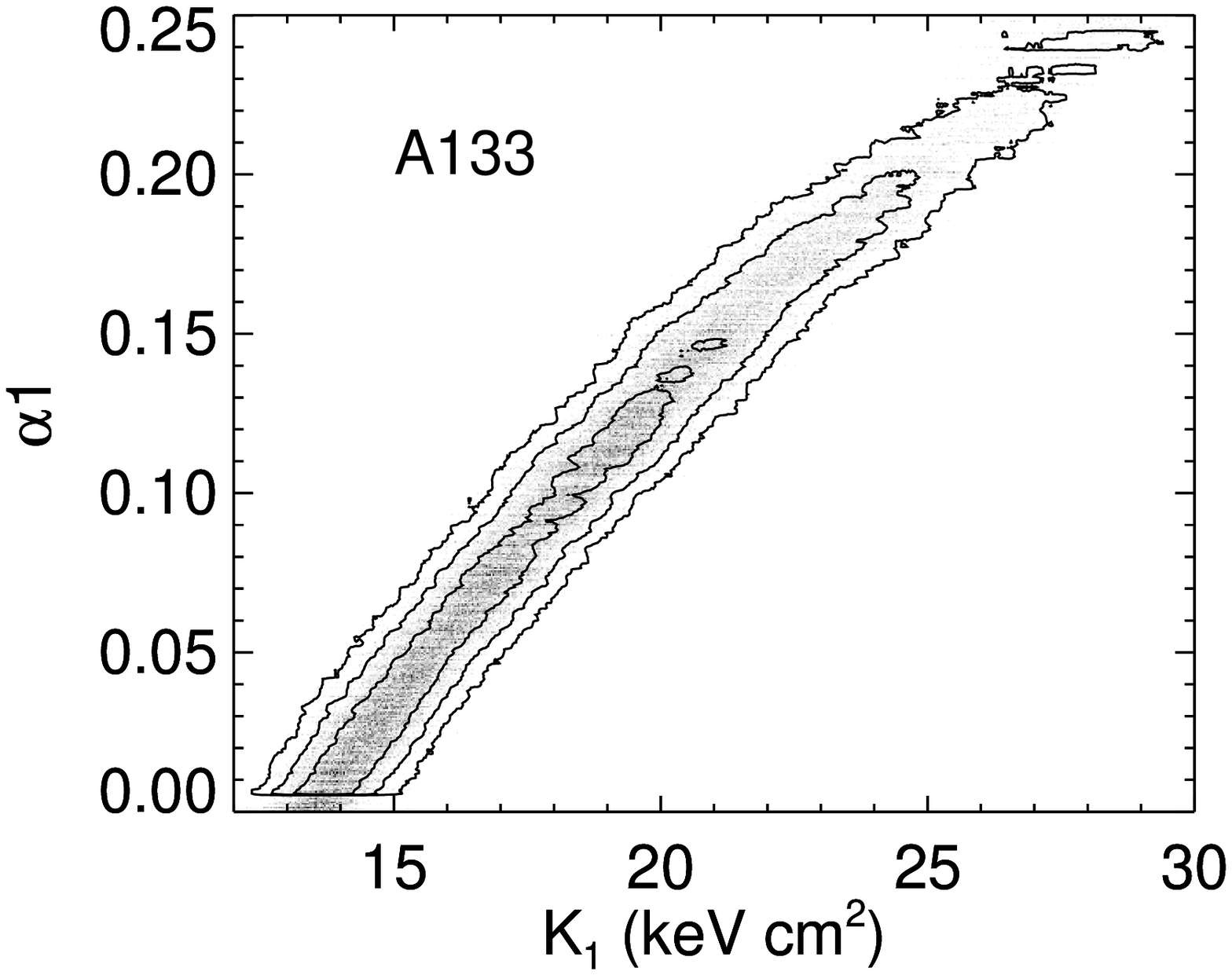}
  \includegraphics[width=0.32\textwidth]{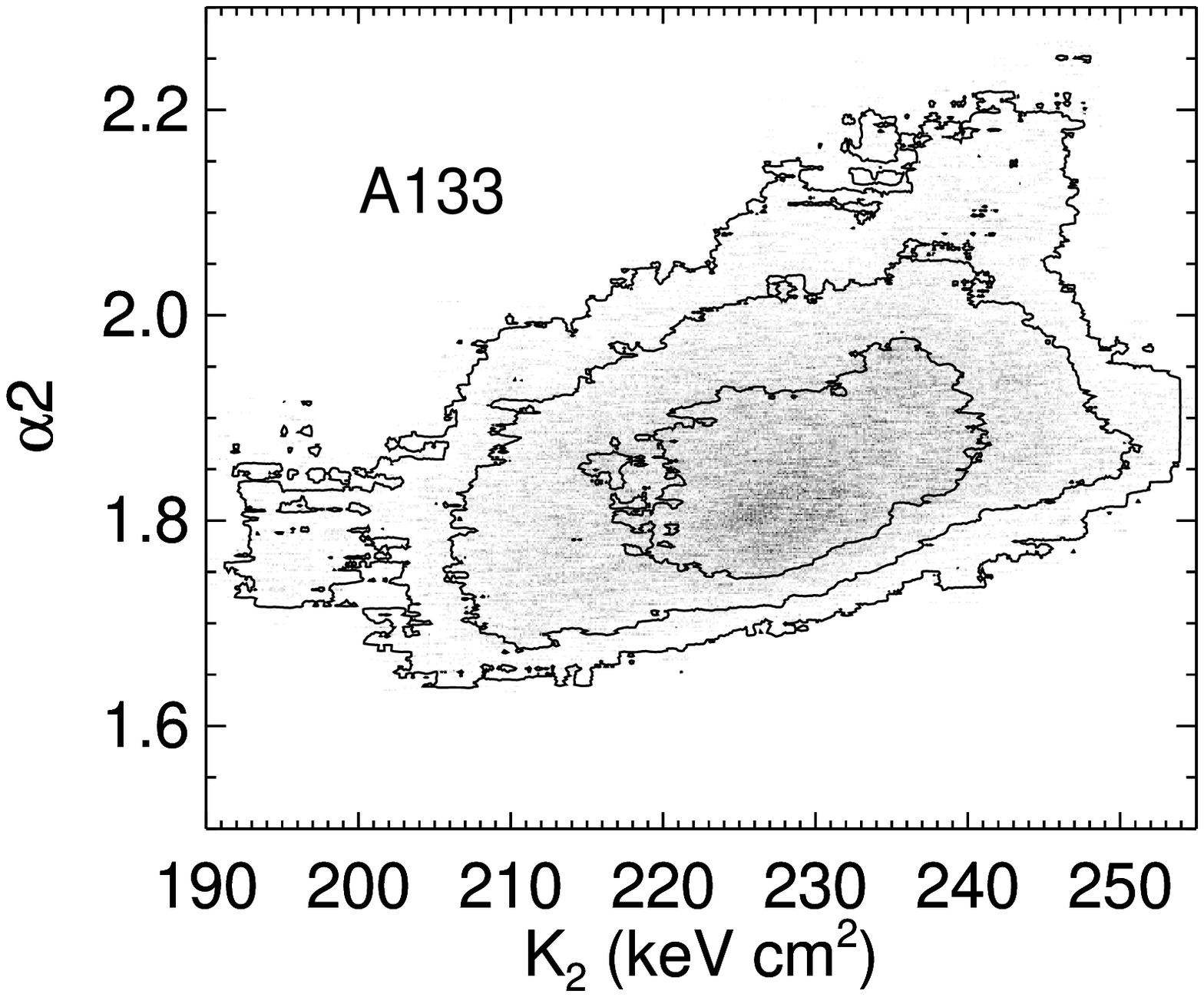}
  \includegraphics[width=0.32\textwidth]{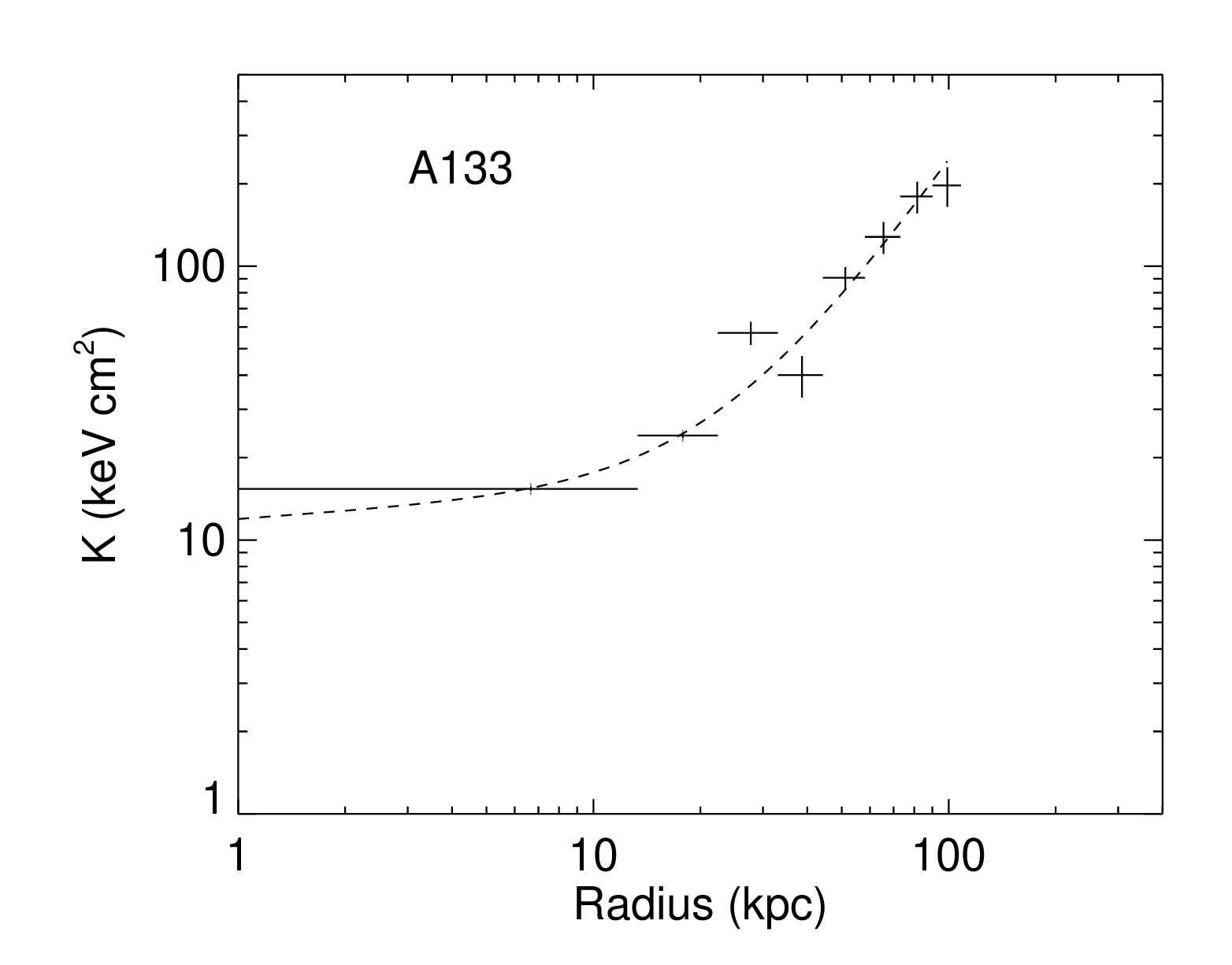}\\
  \includegraphics[width=0.32\textwidth]{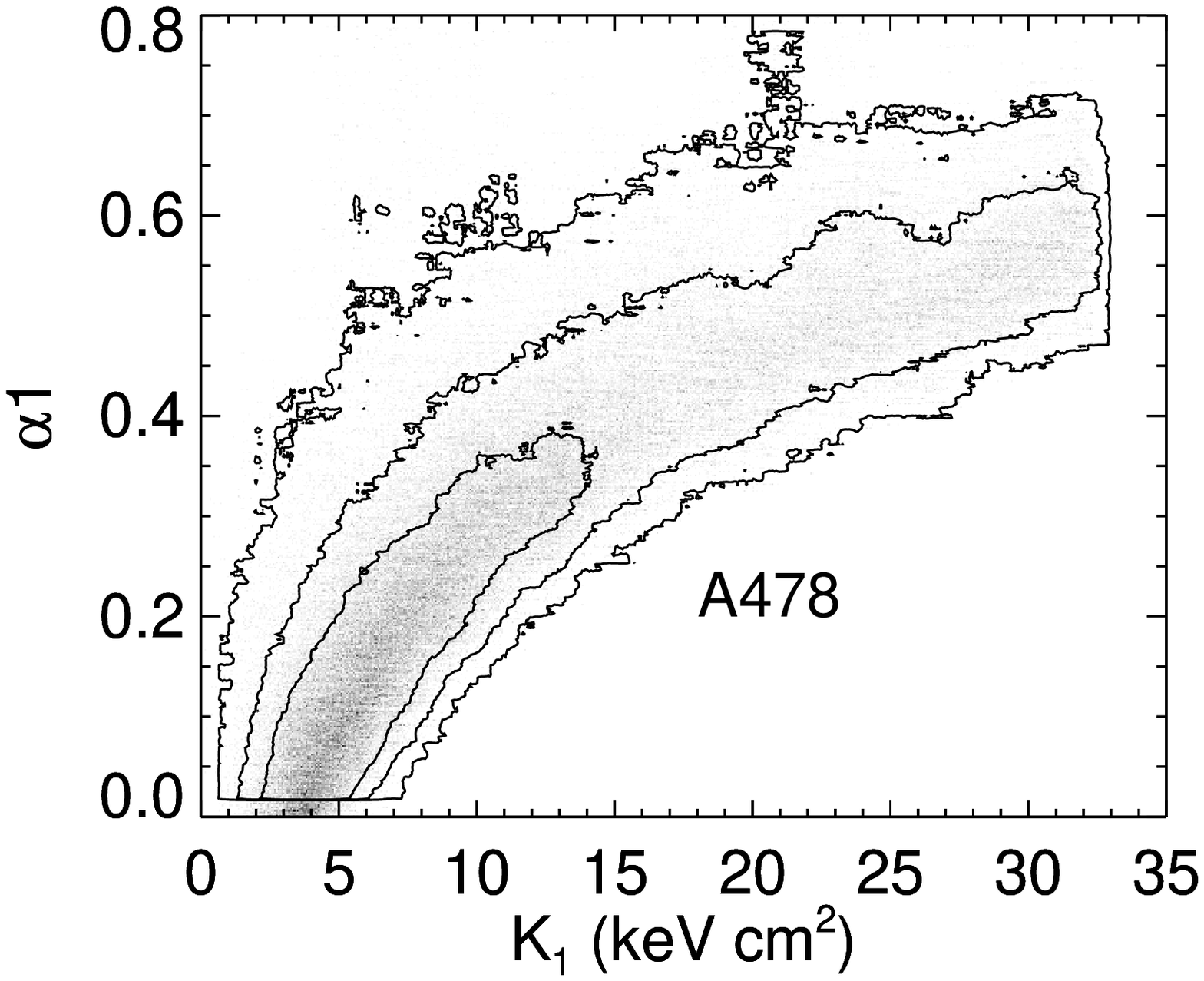}
  \includegraphics[width=0.32\textwidth]{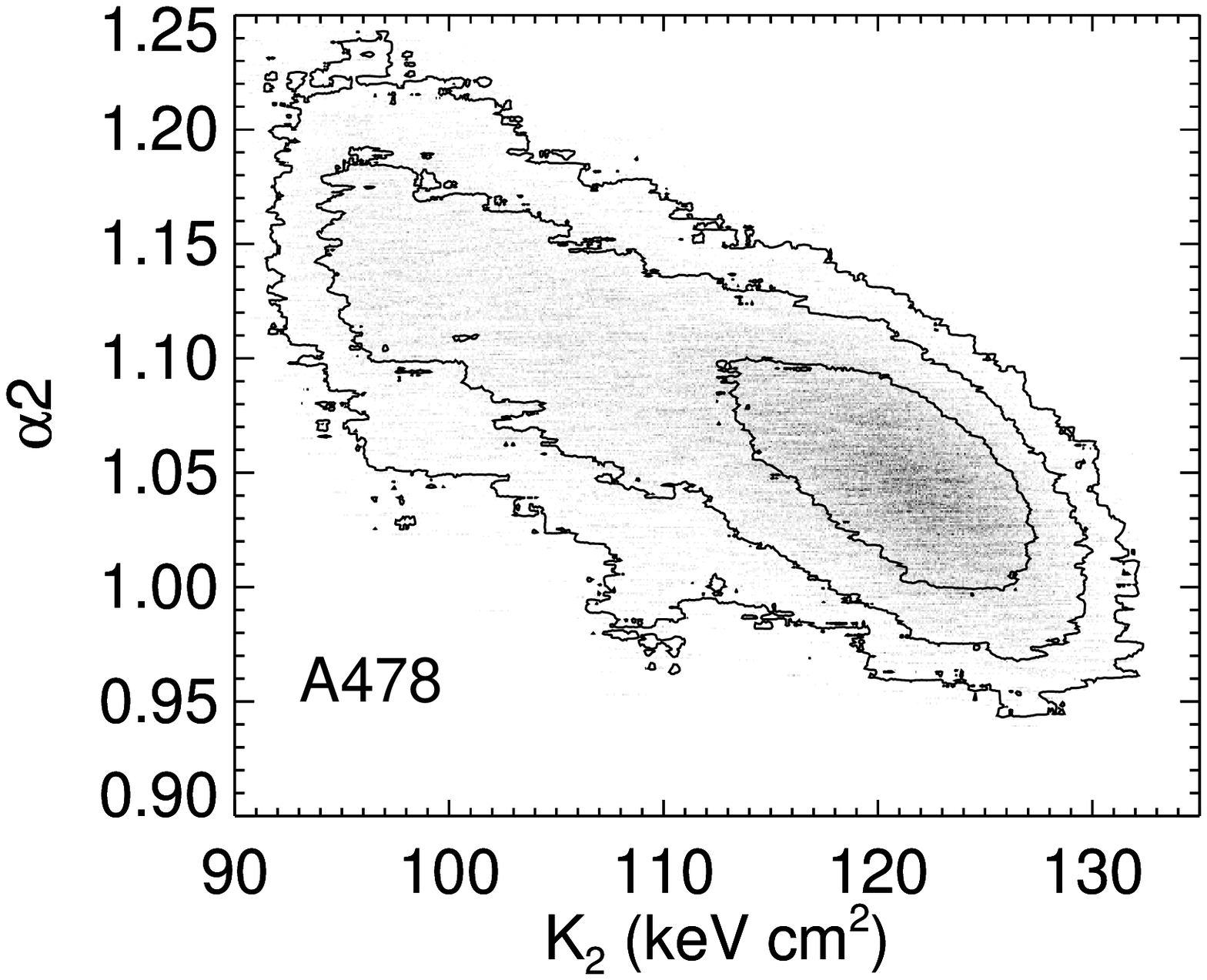}
  \includegraphics[width=0.32\textwidth]{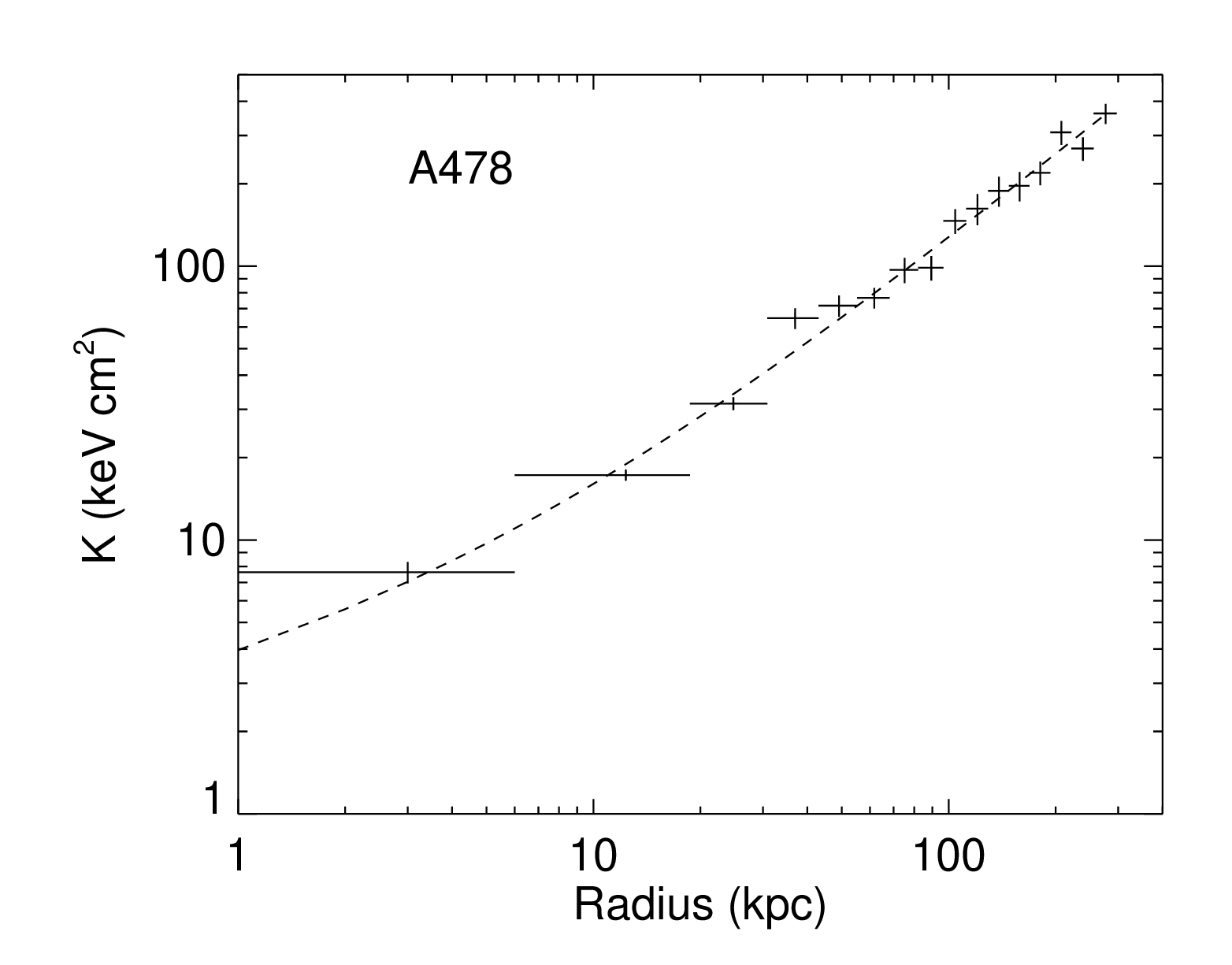}\\
  \includegraphics[width=0.32\textwidth]{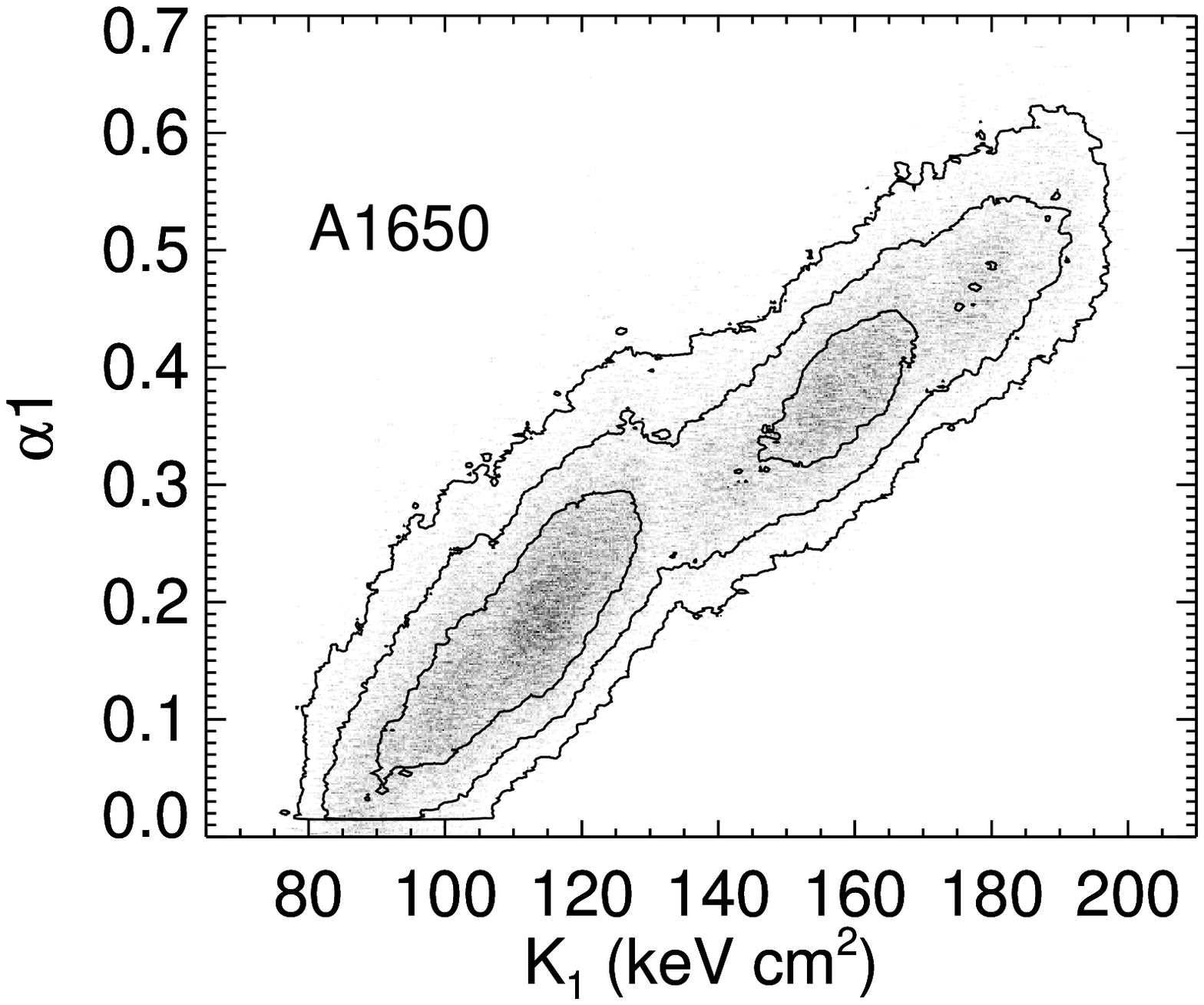}
  \includegraphics[width=0.32\textwidth]{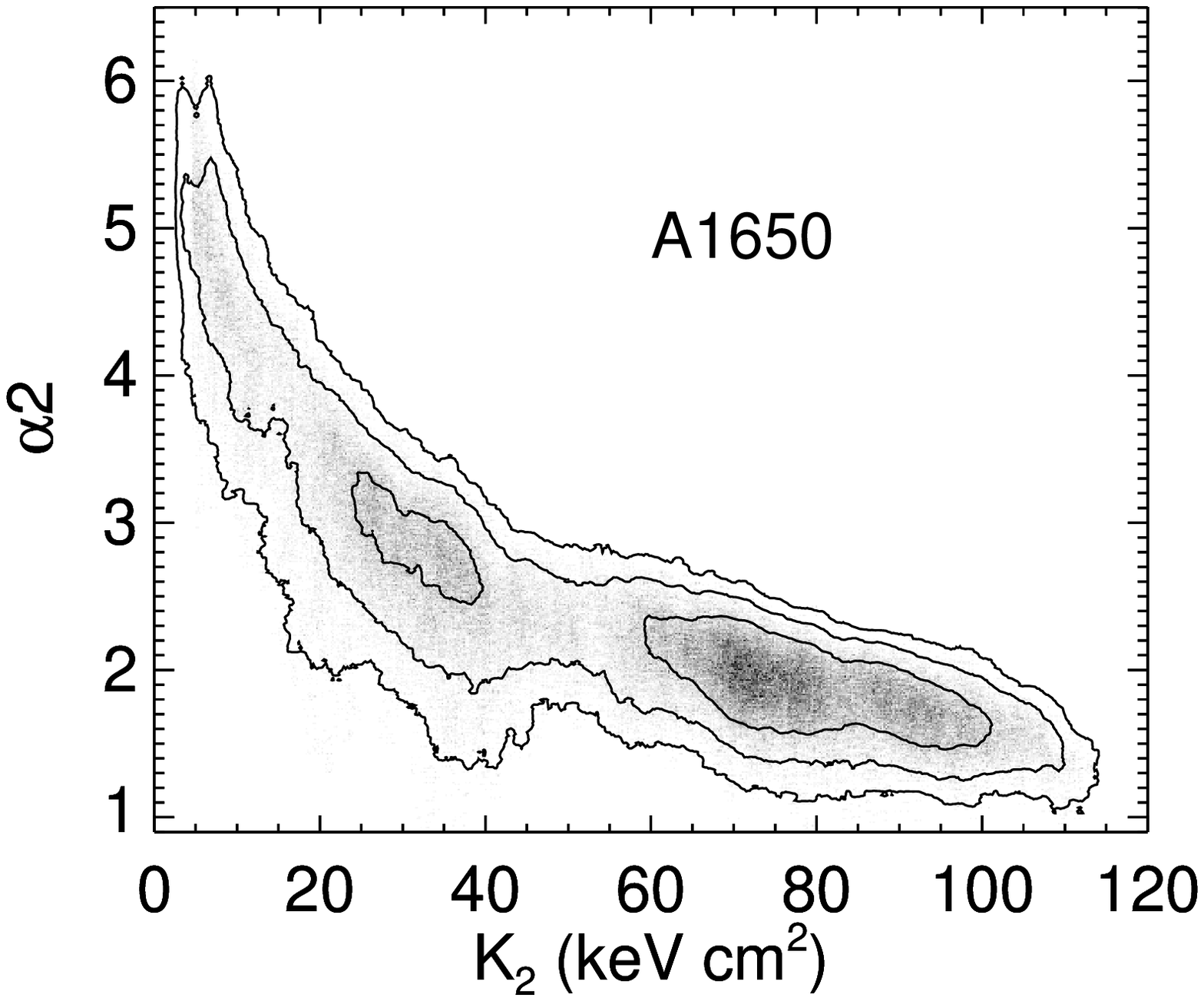}
  \includegraphics[width=0.32\textwidth]{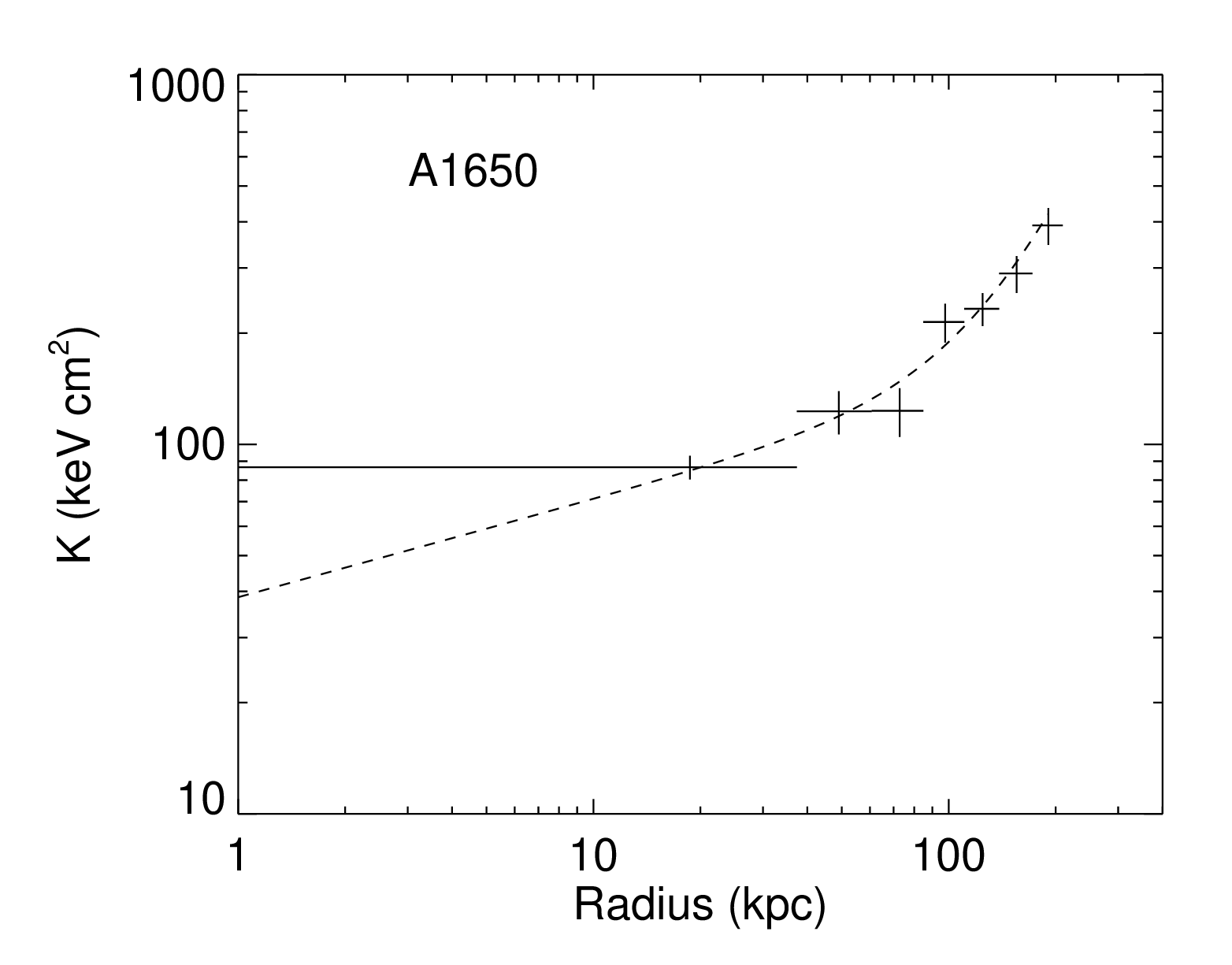}\\
  \includegraphics[width=0.32\textwidth]{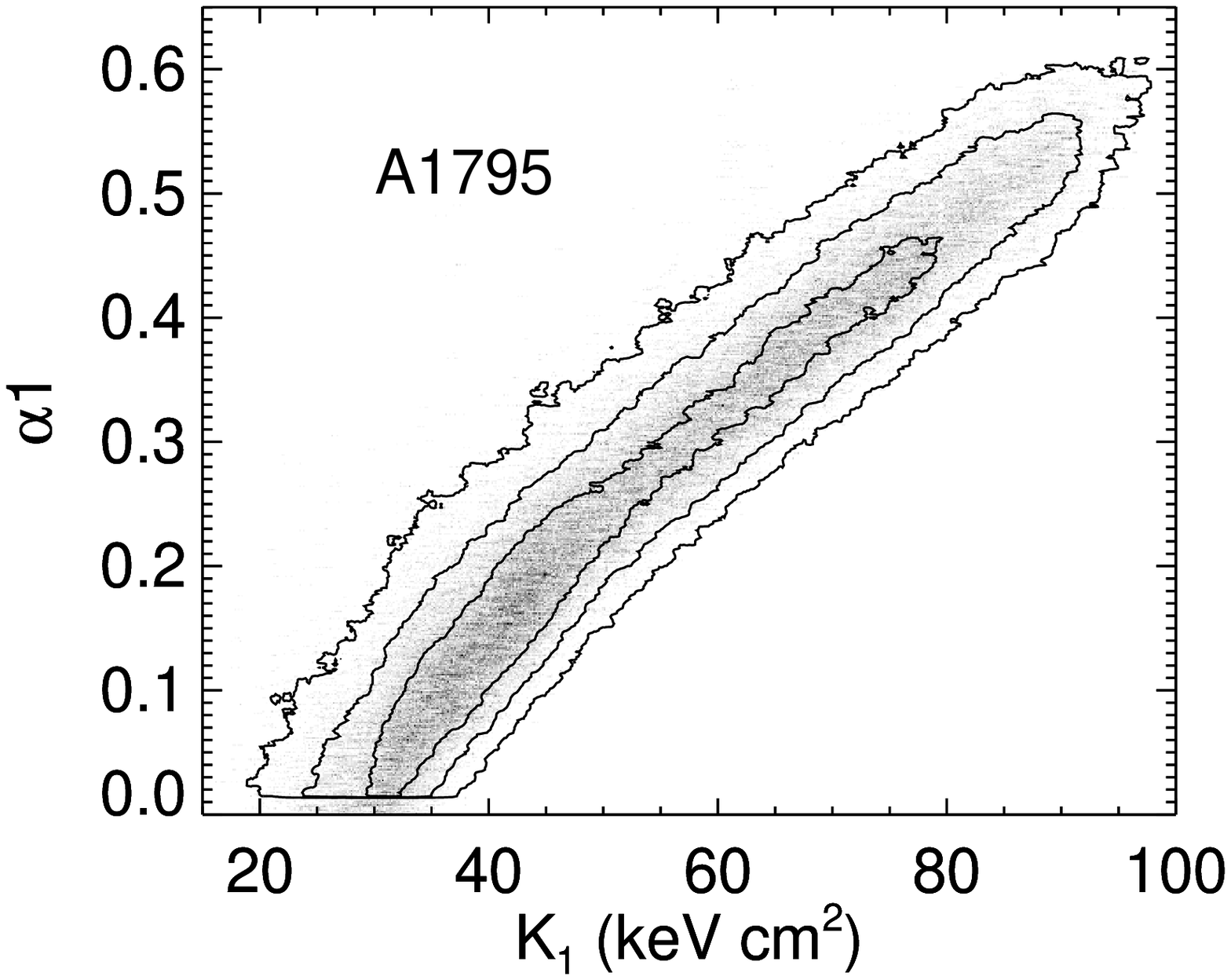}
  \includegraphics[width=0.32\textwidth]{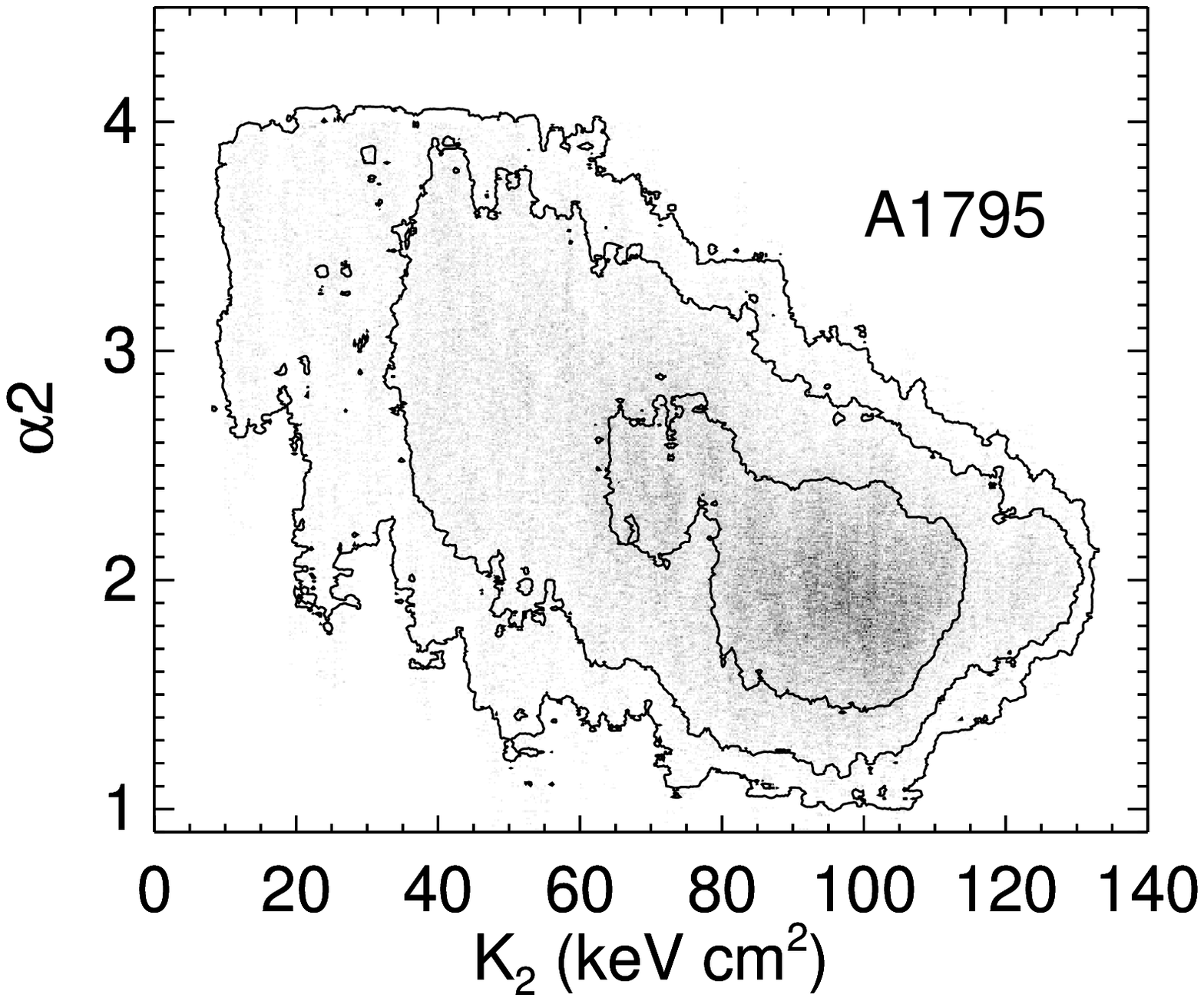}
  \includegraphics[width=0.32\textwidth]{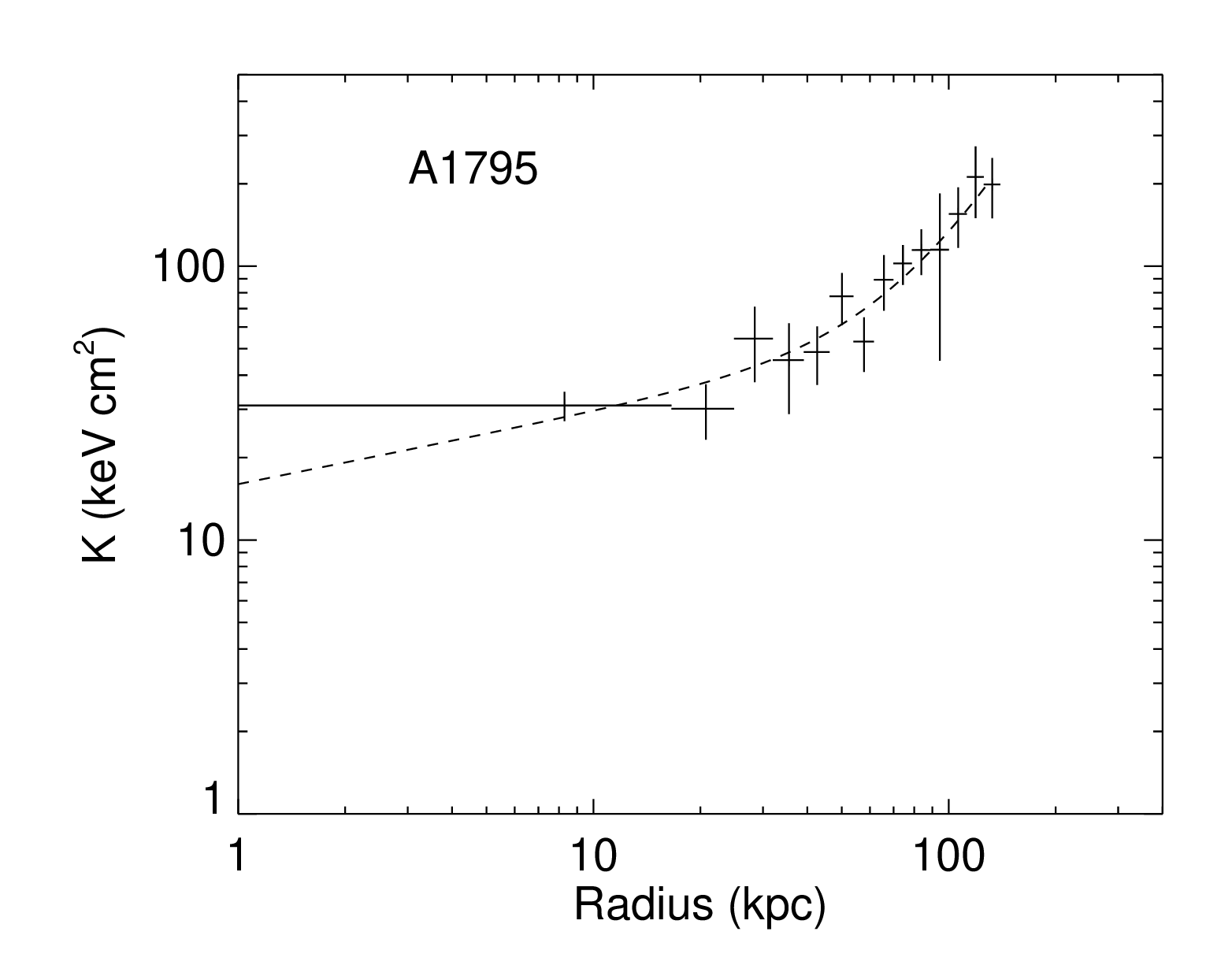}\\
  \caption{The $K_1$-$\alpha_1$ and $K_2$-$\alpha_2$ marginalized probability distributions obtained from the double power law entropy model fitting for the full cluster sample. The contours mark the 50\%, 90\% and 99\% inclusion levels based on the density of points starting from the innermost contour outwards. Greyscale denotes the 
   PDF density.}
  \label{fig:dbl_pl_entr_fit}
\end{figure*}  

\setcounter{figure}{14}
\begin{figure*}
 \centering
  \includegraphics[width=0.32\textwidth]{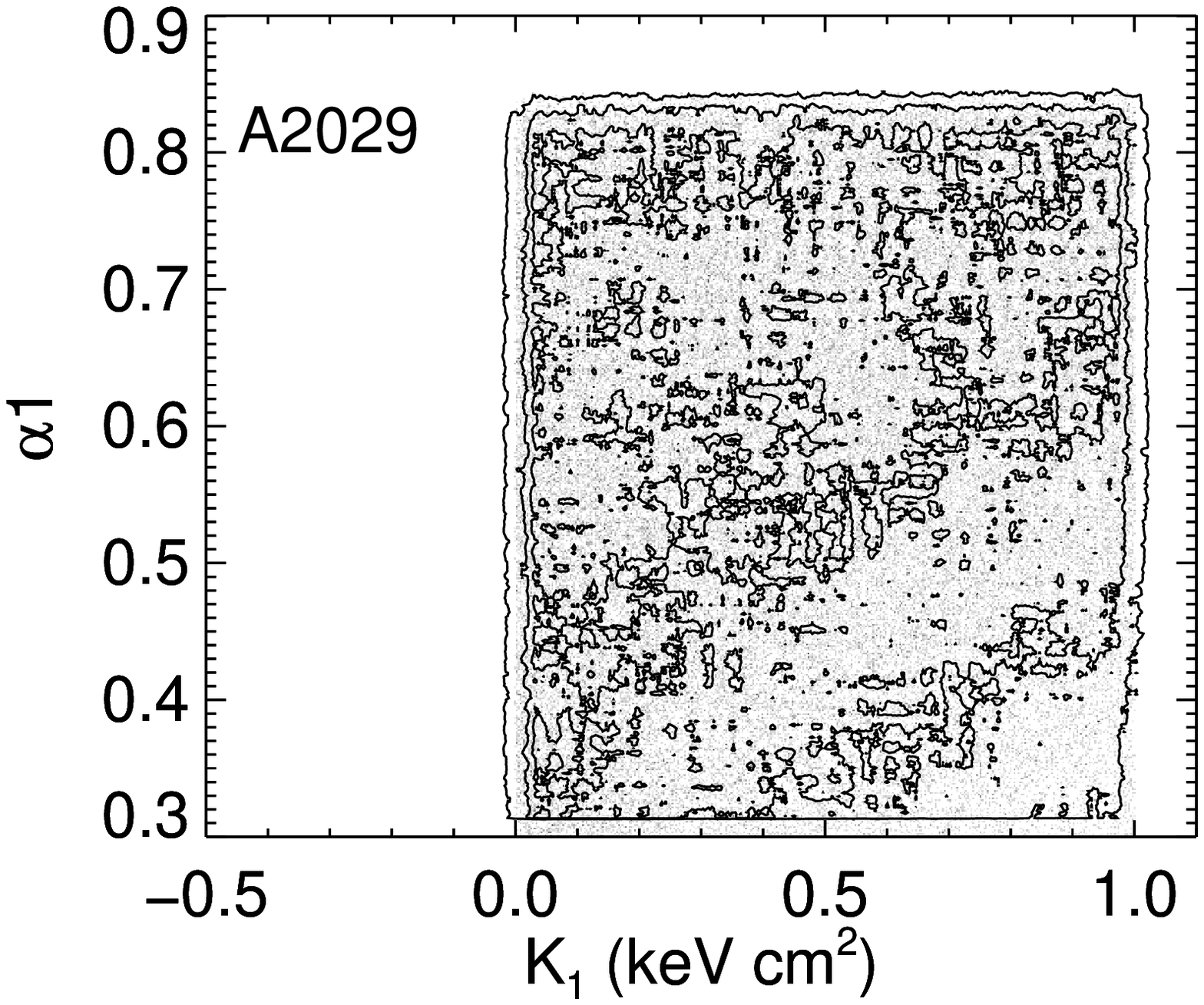}
  \includegraphics[width=0.32\textwidth]{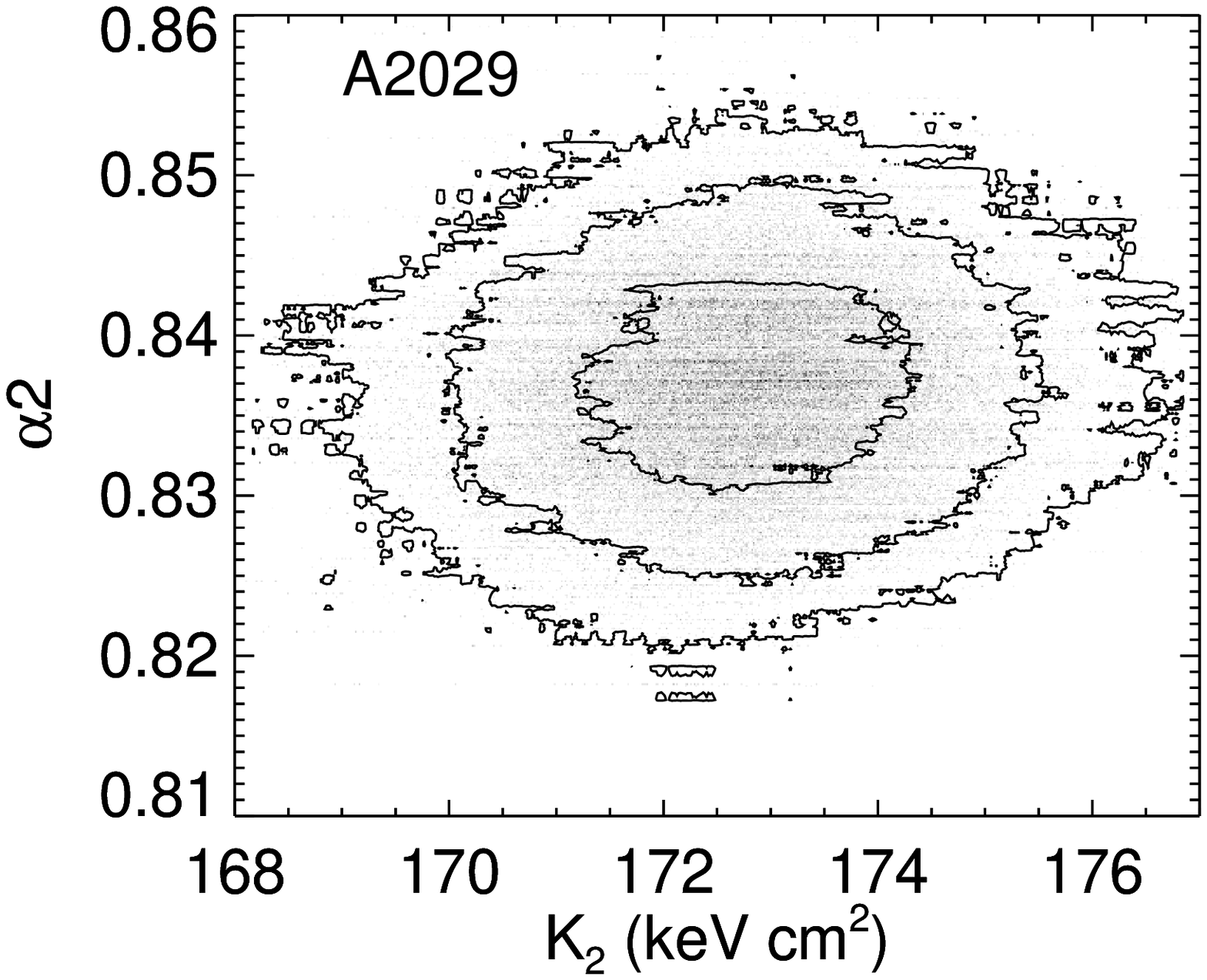}
  \includegraphics[width=0.32\textwidth]{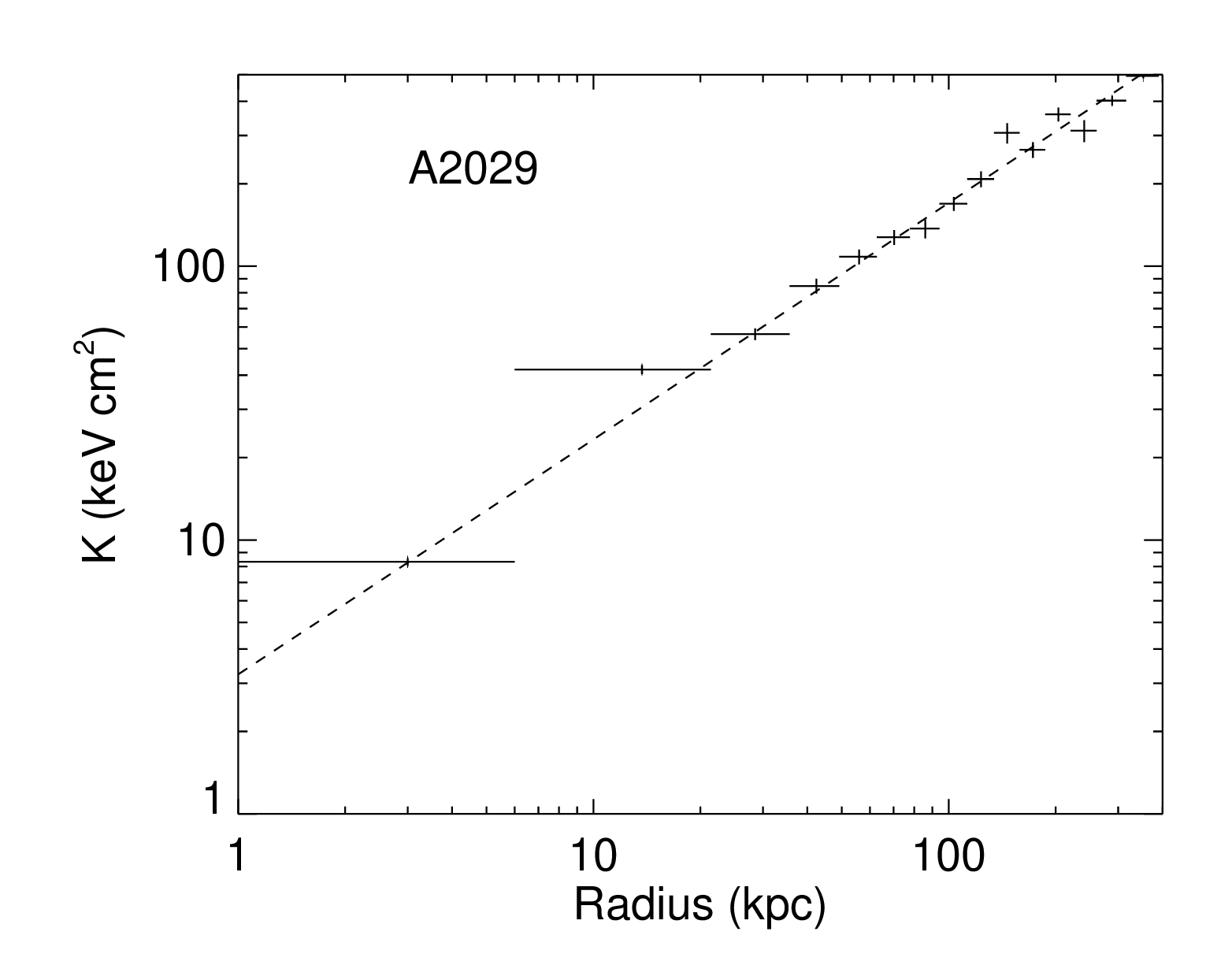}\\
  \includegraphics[width=0.32\textwidth]{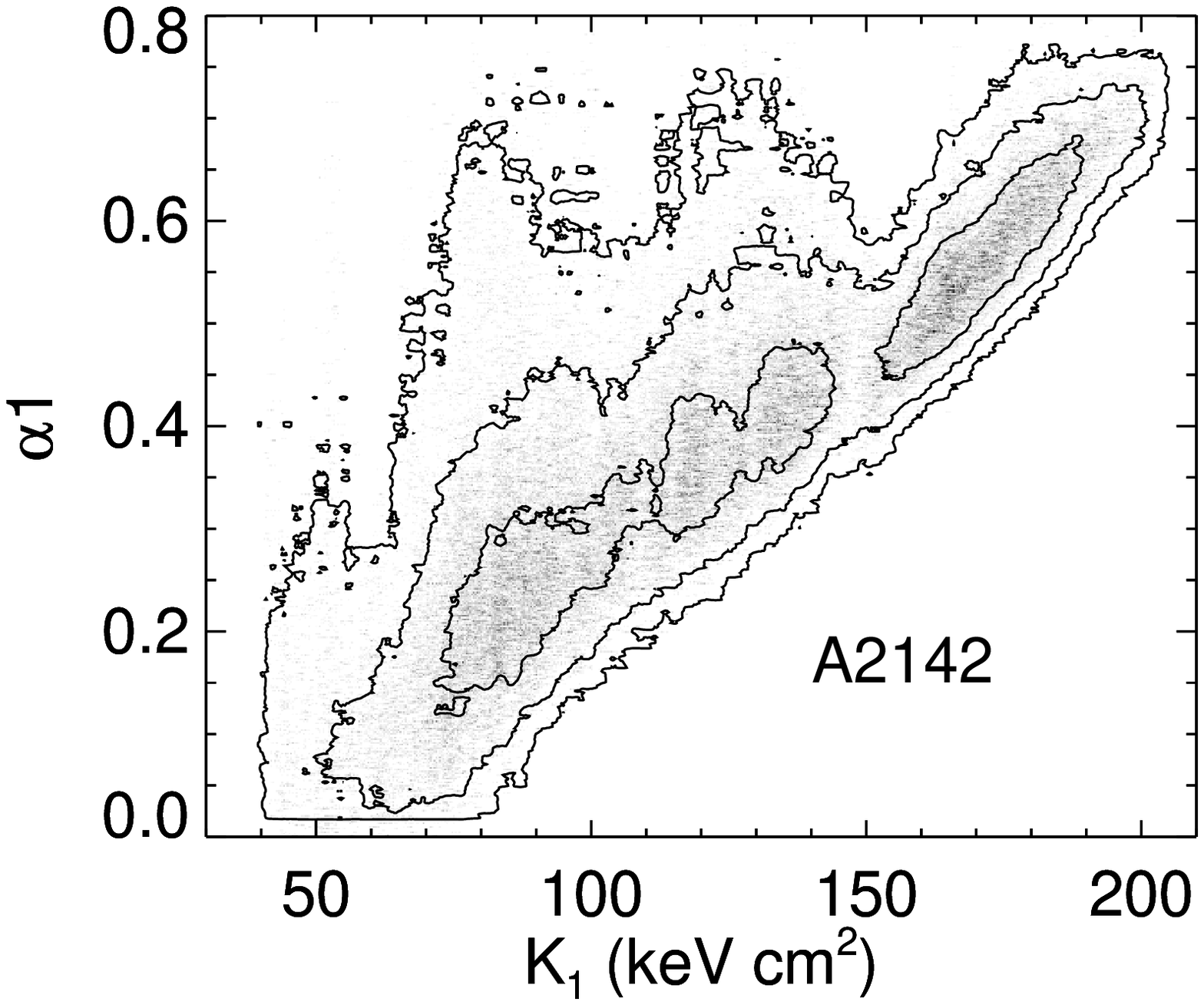}
  \includegraphics[width=0.32\textwidth]{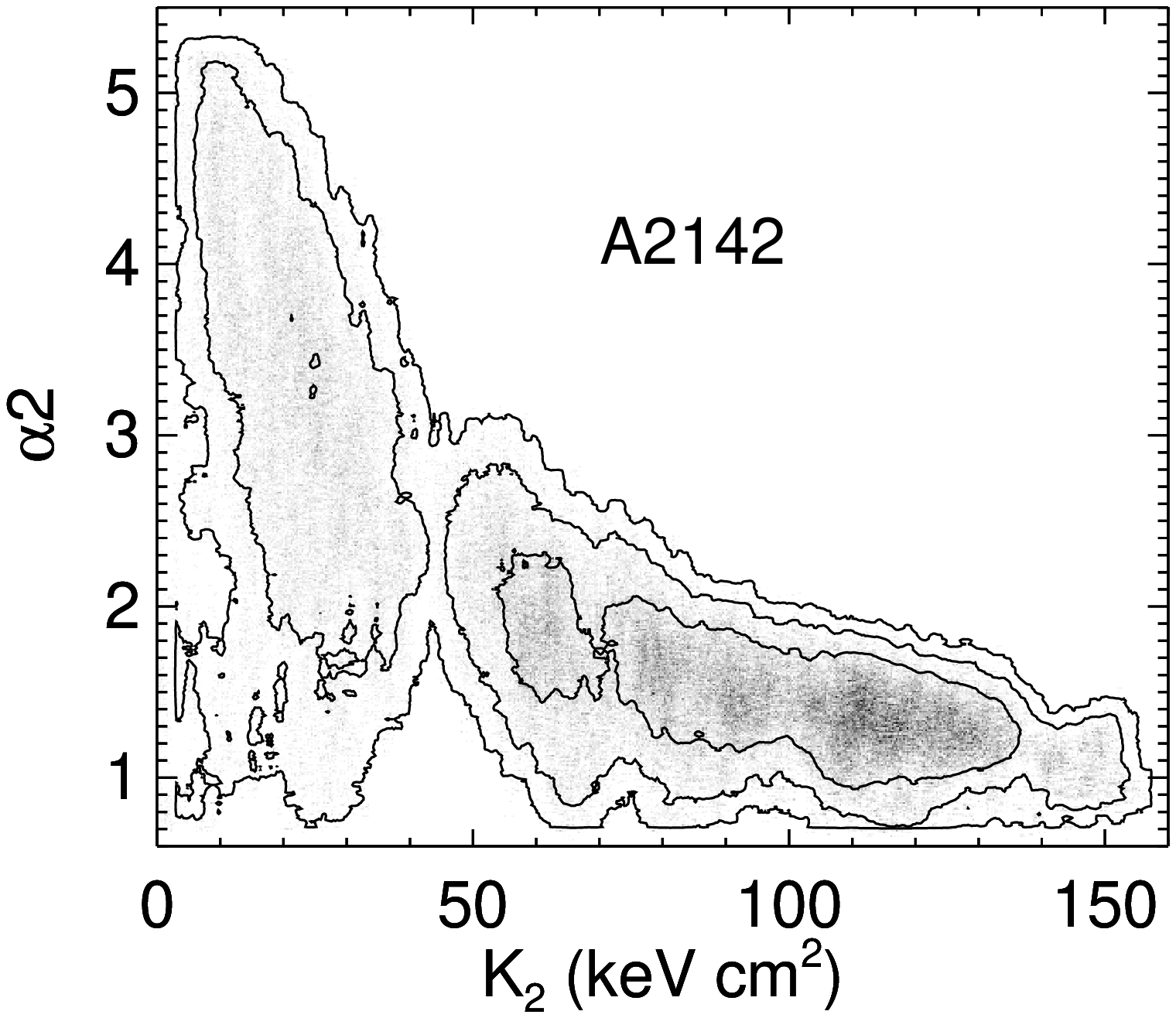}
  \includegraphics[width=0.32\textwidth]{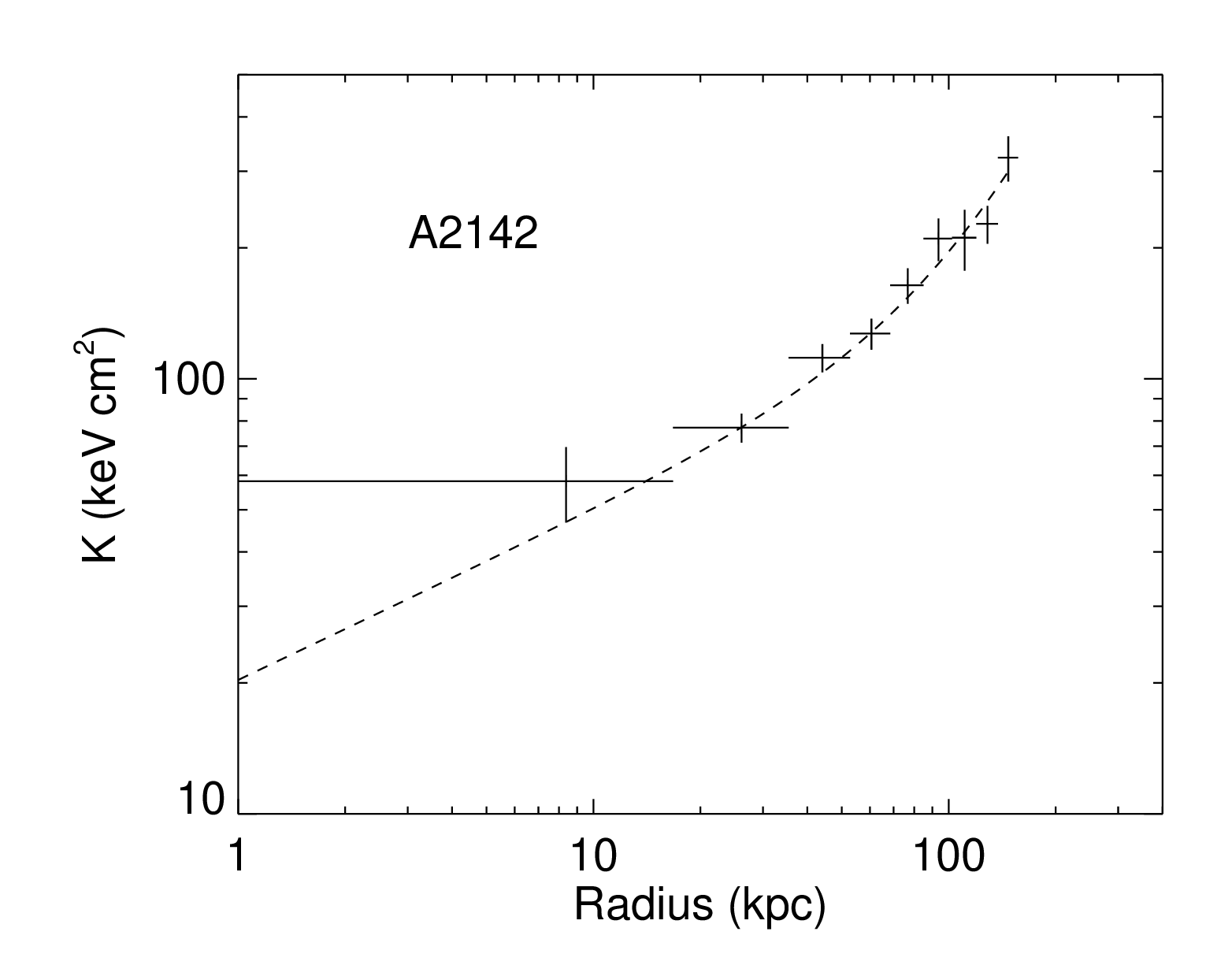}\\
  \includegraphics[width=0.32\textwidth]{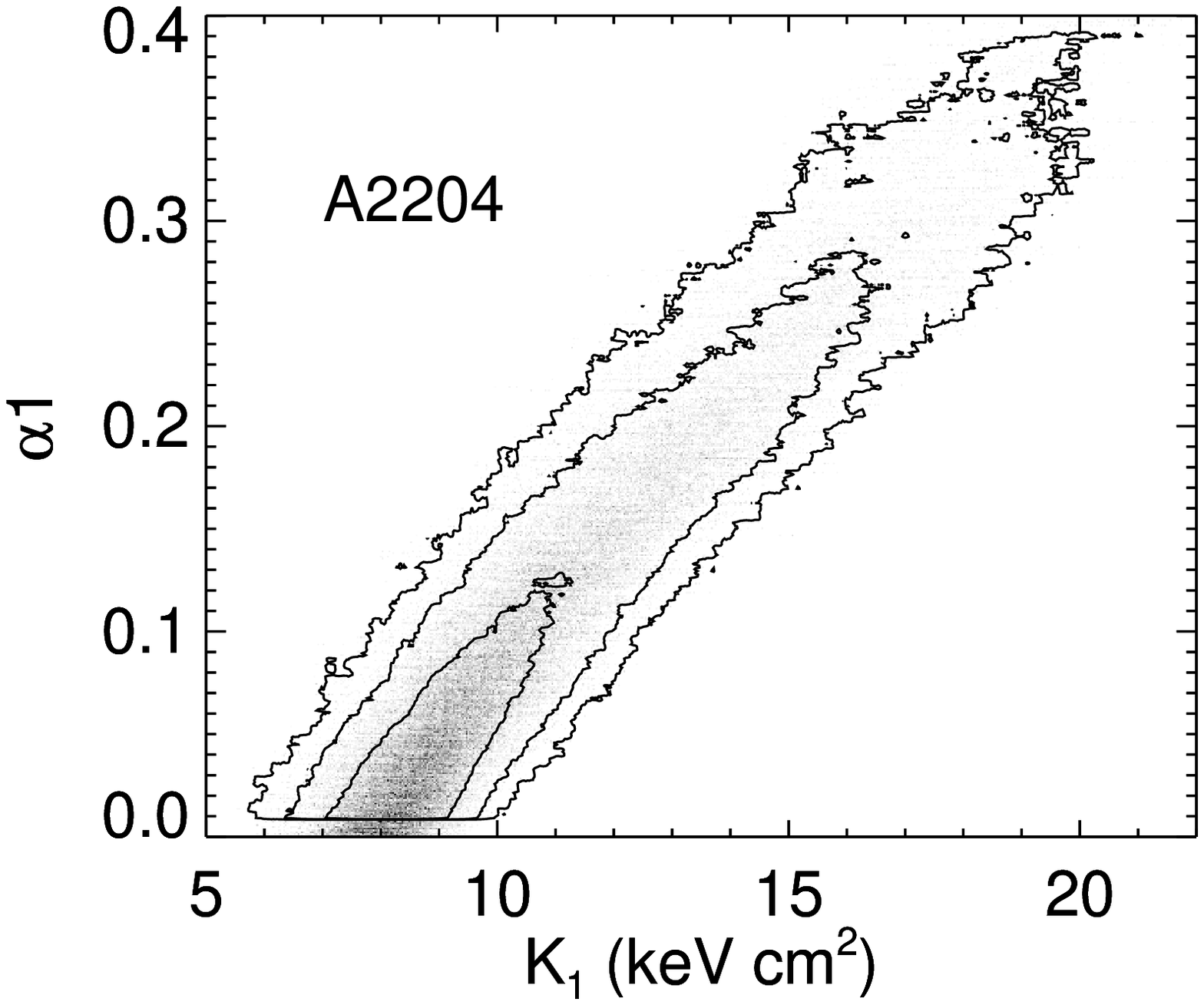}
  \includegraphics[width=0.32\textwidth]{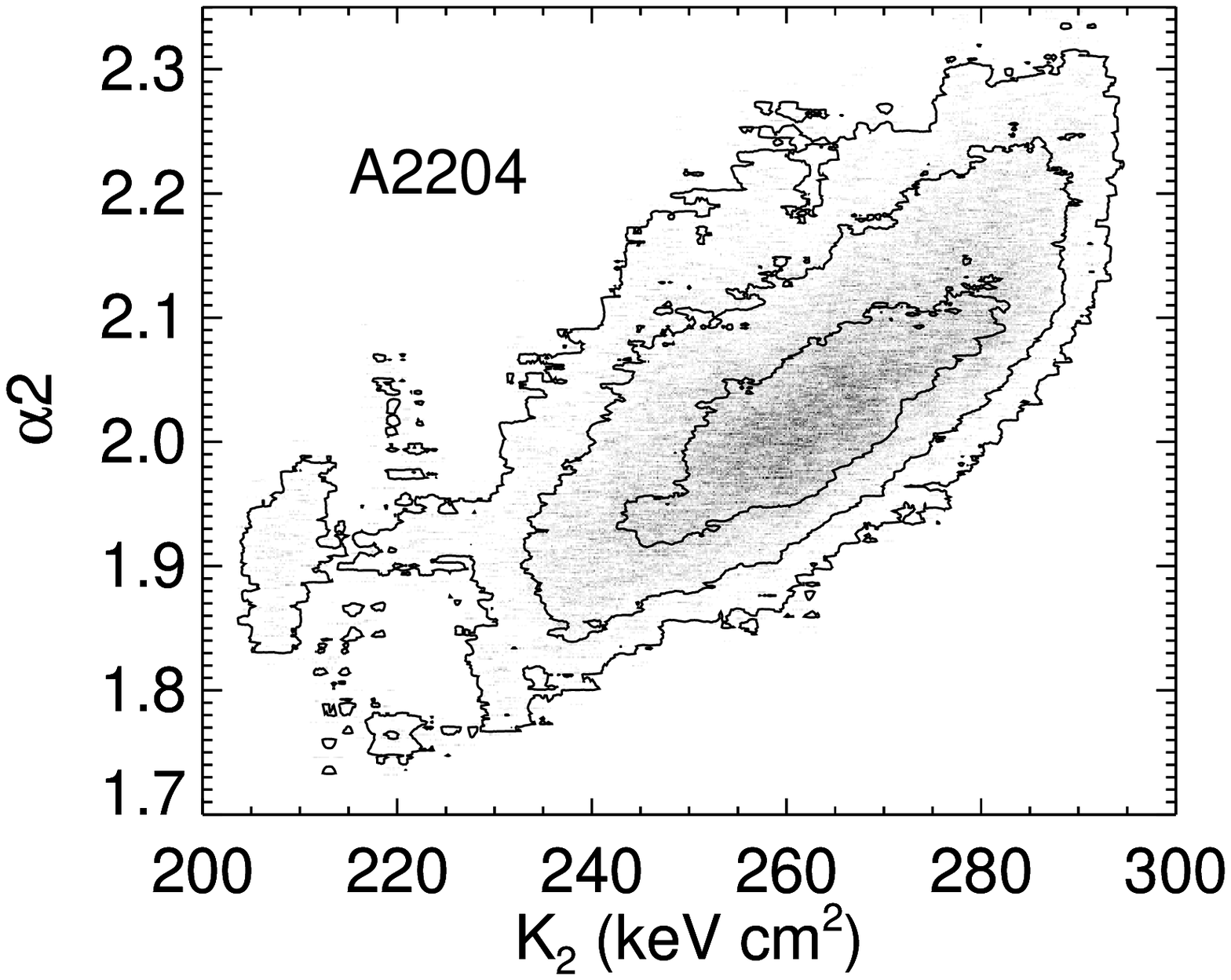}
  \includegraphics[width=0.32\textwidth]{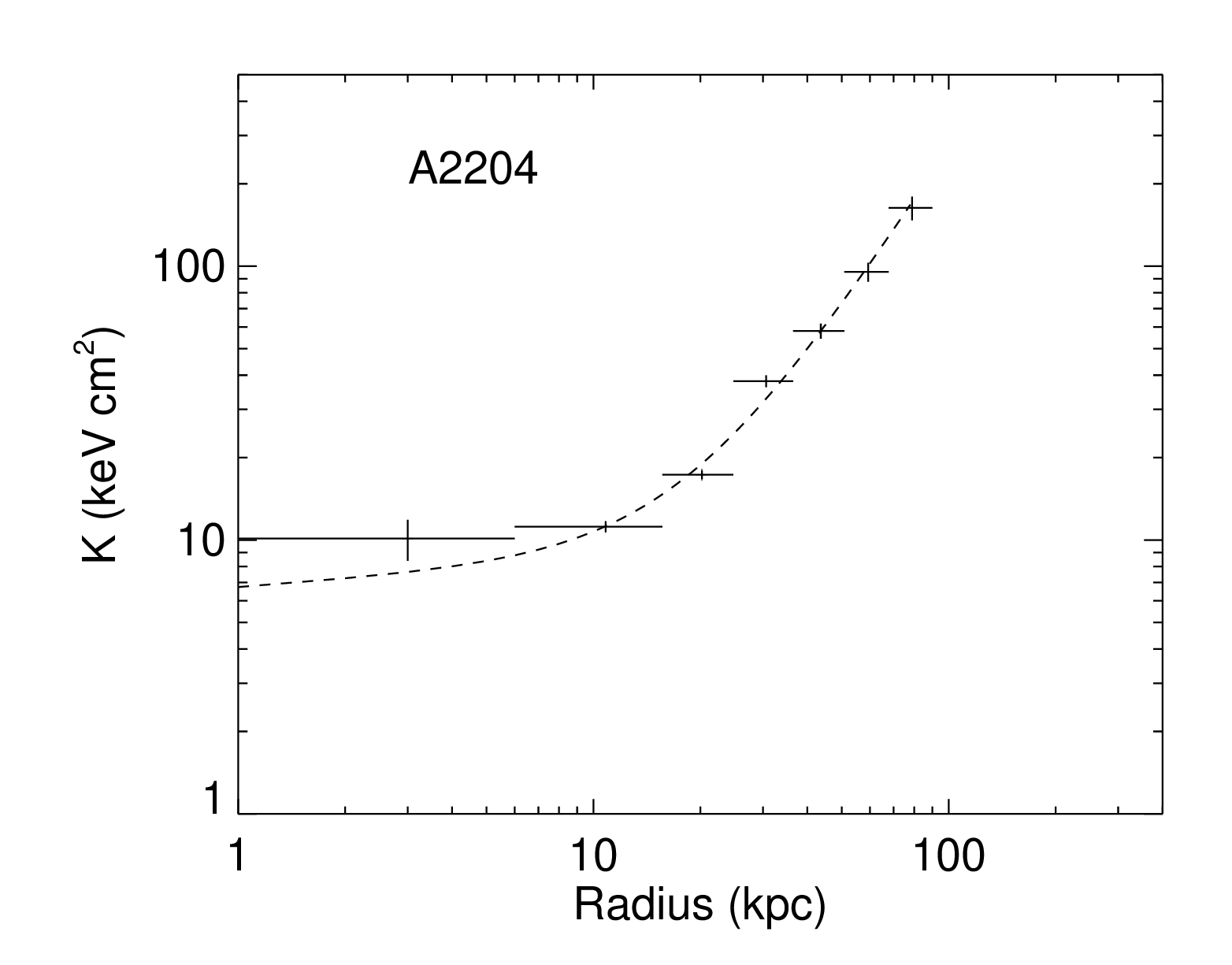}\\
  \includegraphics[width=0.32\textwidth]{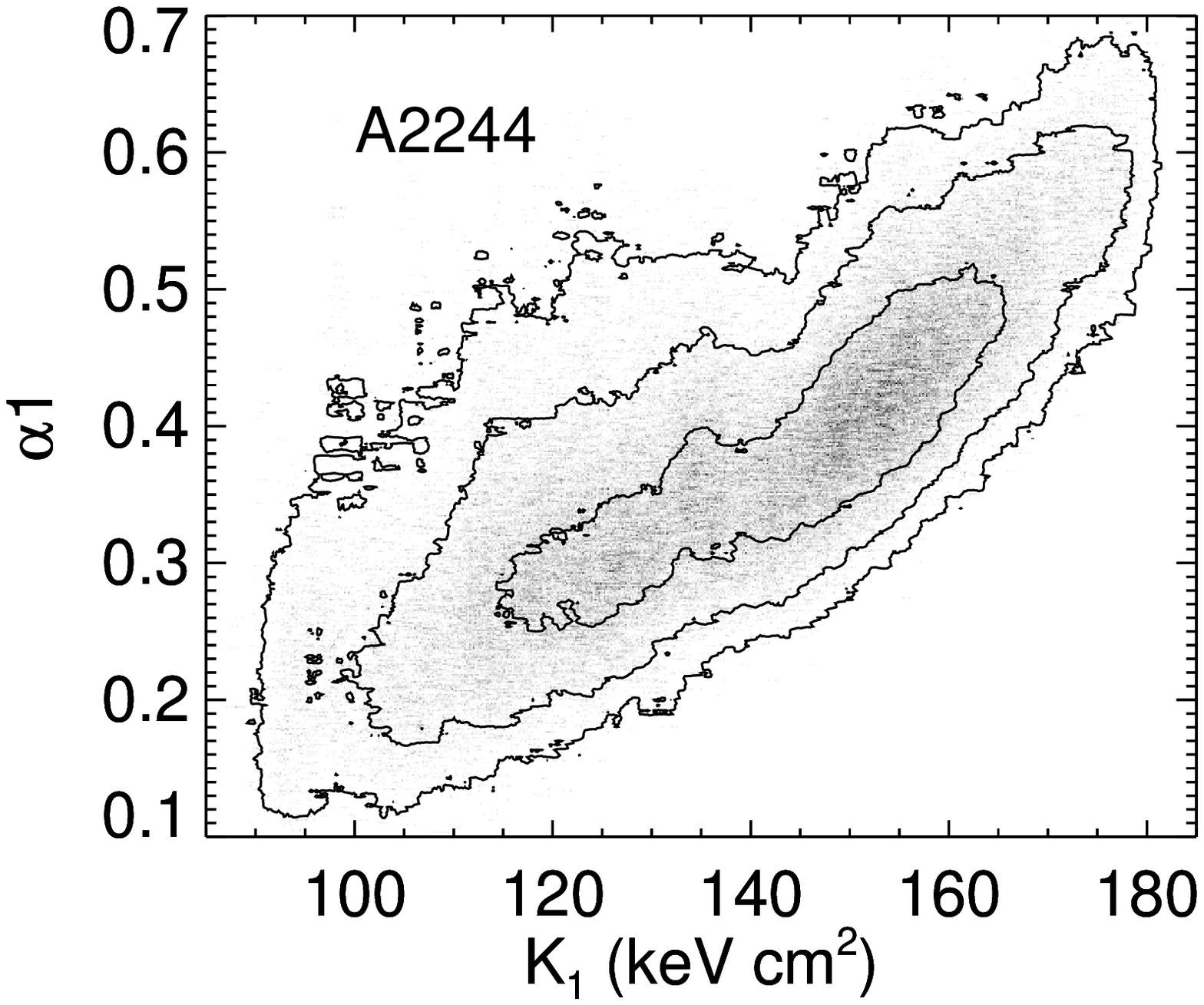}
  \includegraphics[width=0.32\textwidth]{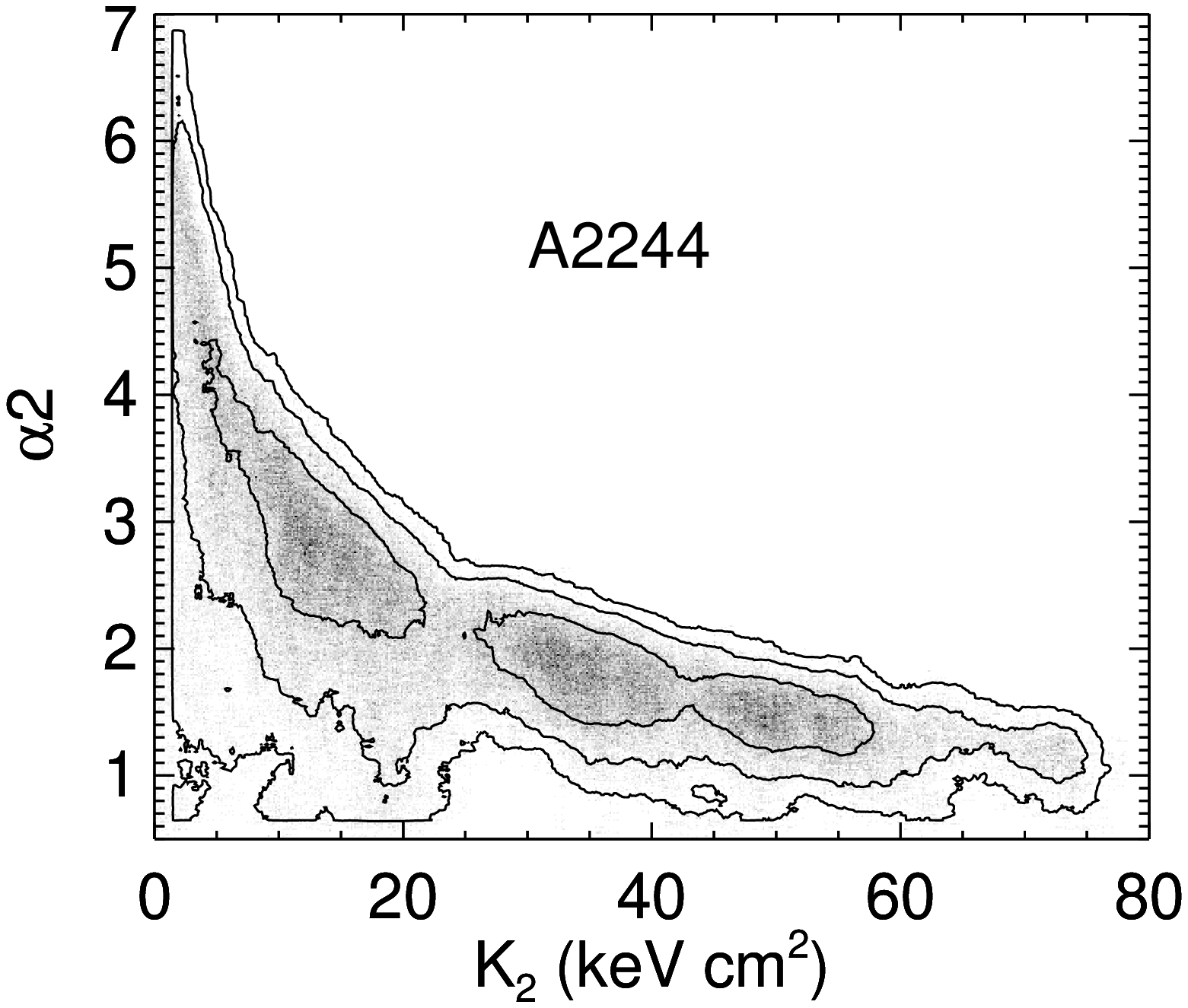}
  \includegraphics[width=0.32\textwidth]{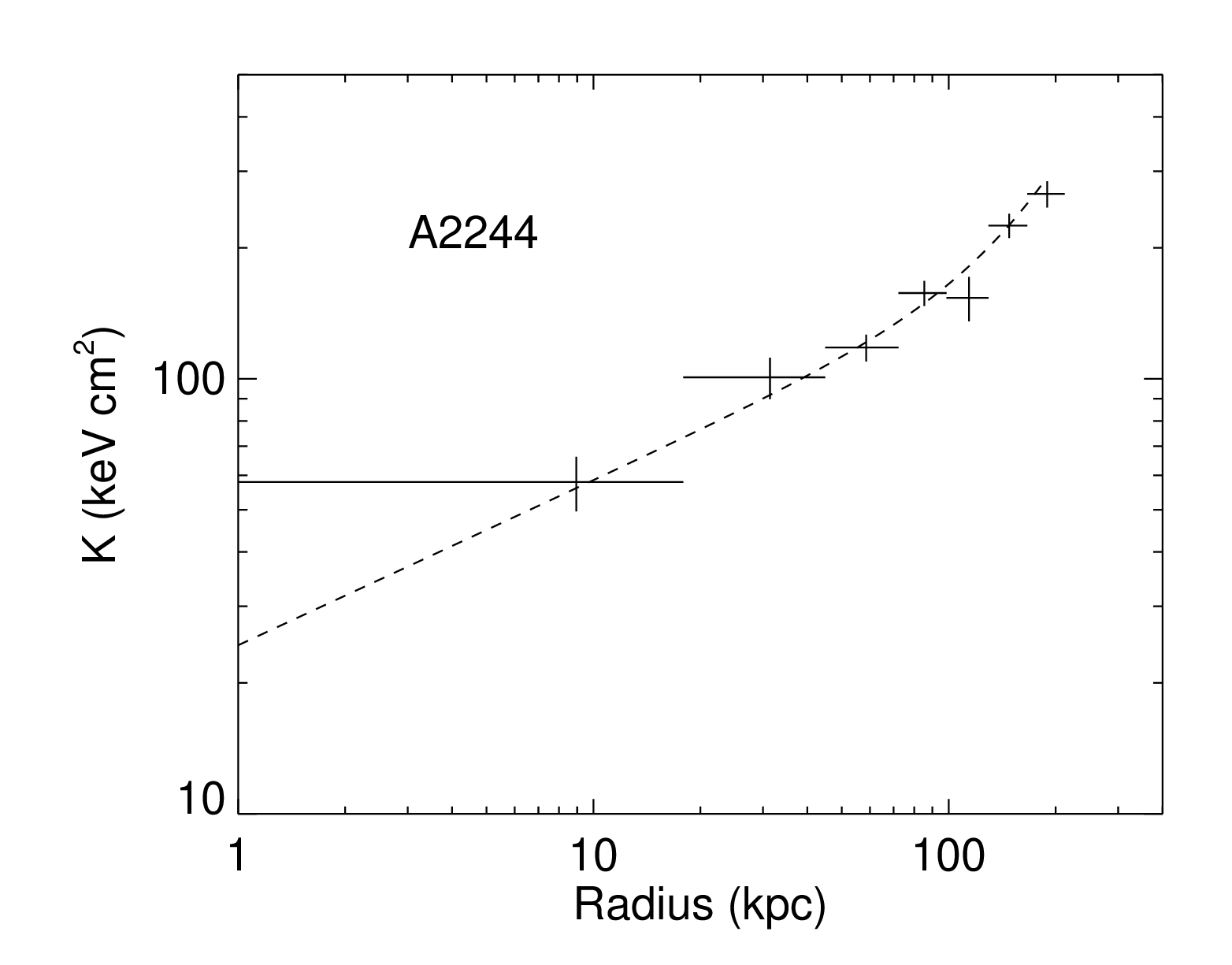}\\
  \includegraphics[width=0.32\textwidth]{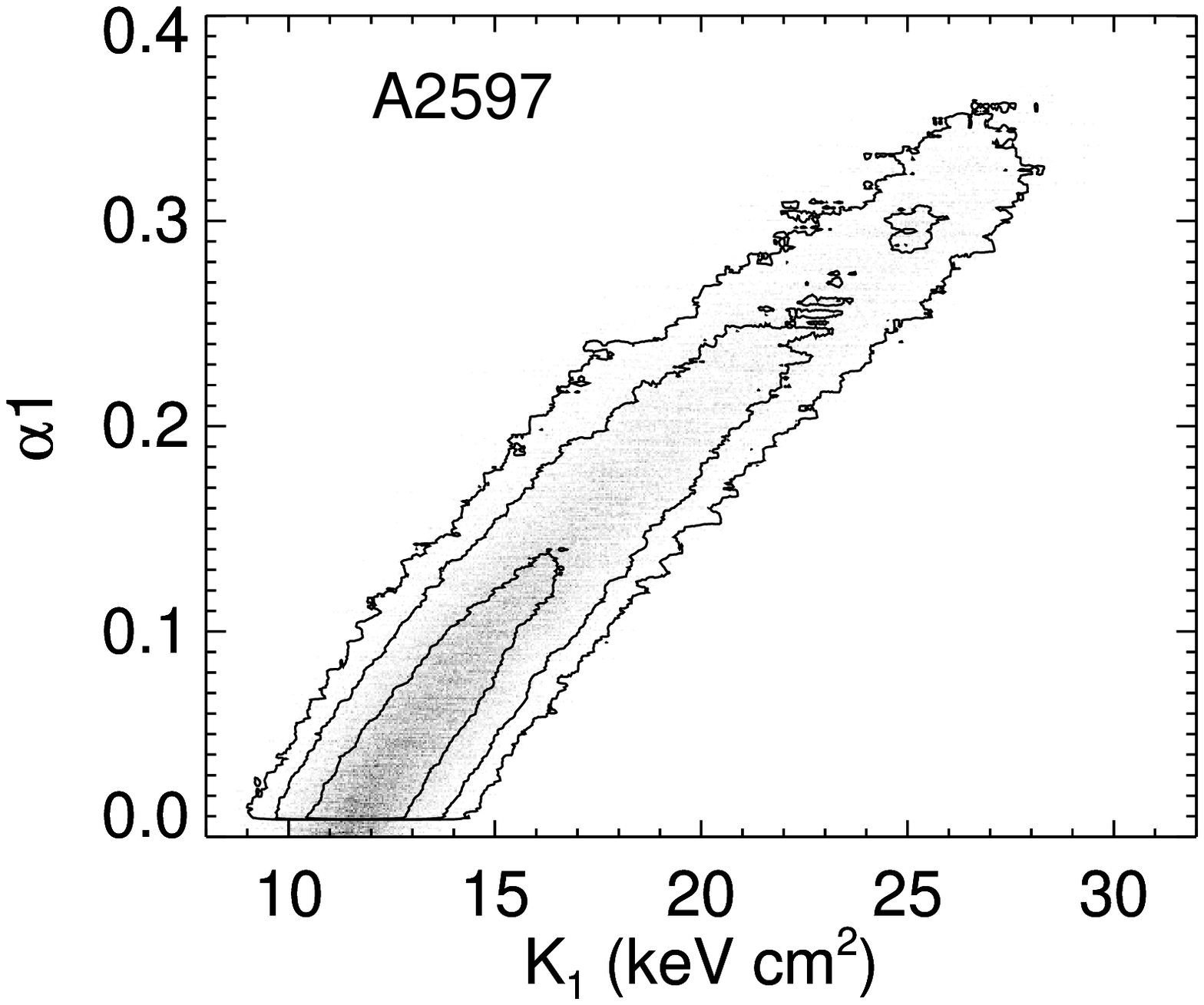}
  \includegraphics[width=0.32\textwidth]{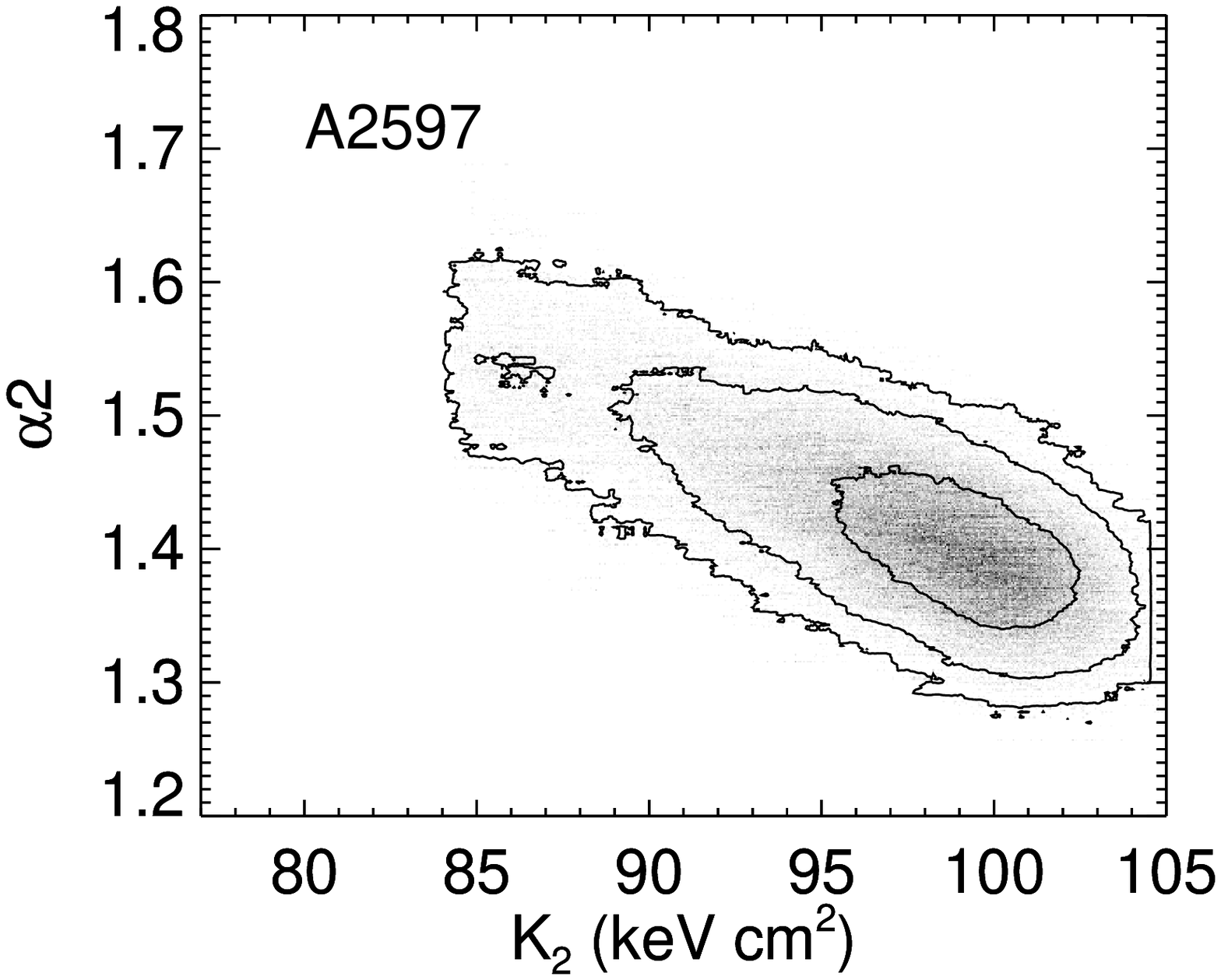}
  \includegraphics[width=0.32\textwidth]{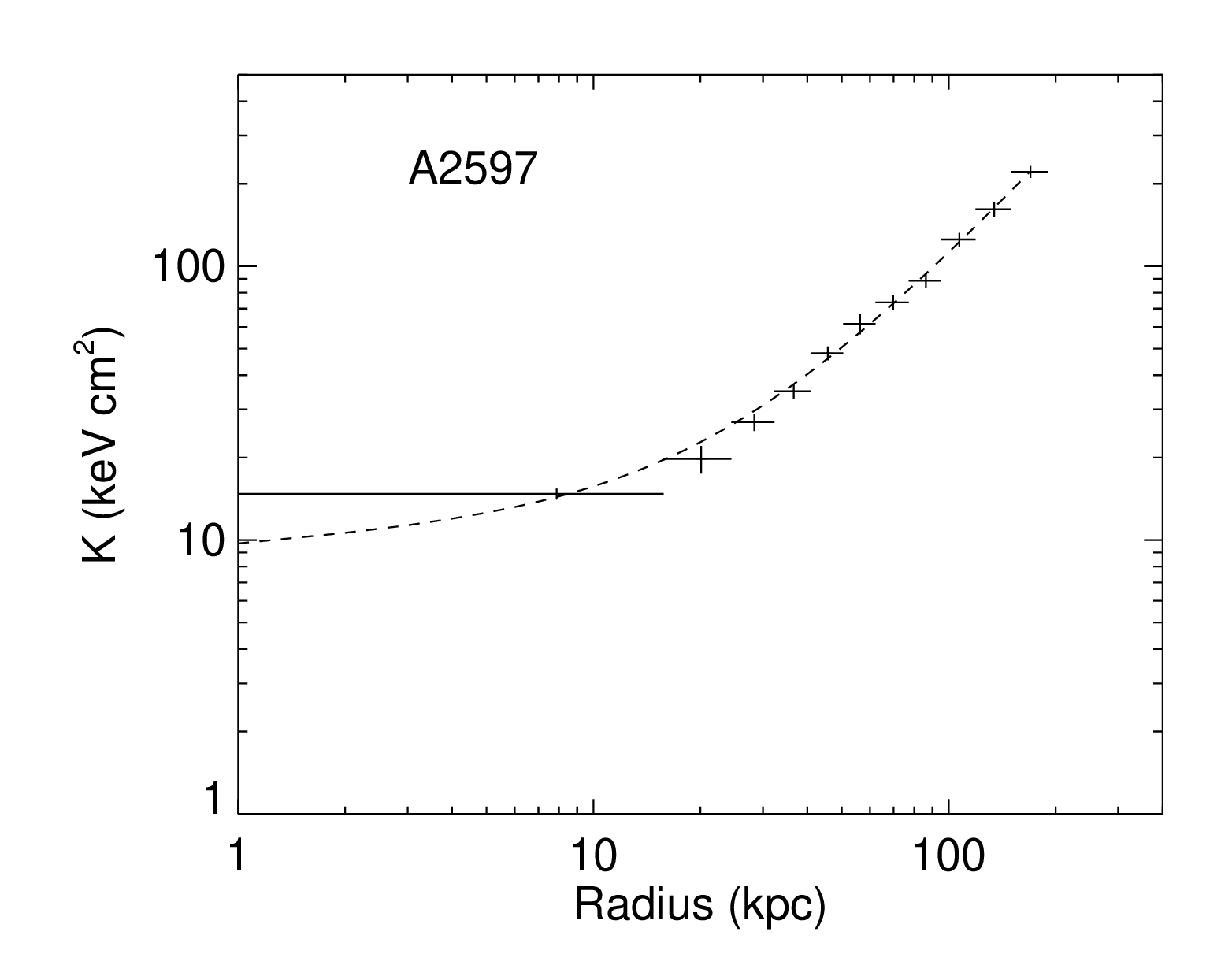}\\
  \caption{(Contd.)}
  \label{fig:dbl_pl_entr_fit}
\end{figure*}

\setcounter{figure}{14}
\begin{figure*}
 \centering
  \includegraphics[width=0.32\textwidth]{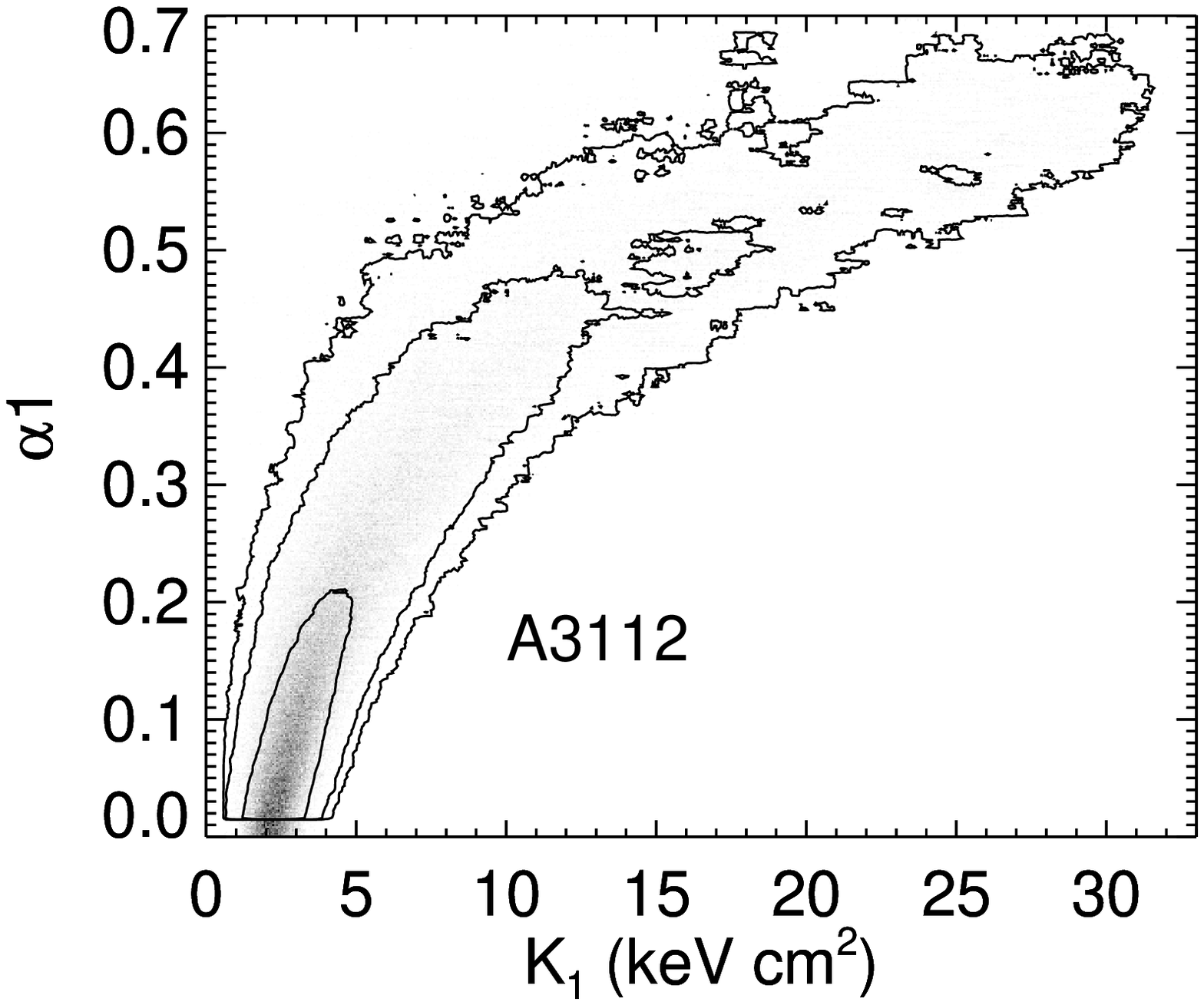}
  \includegraphics[width=0.32\textwidth]{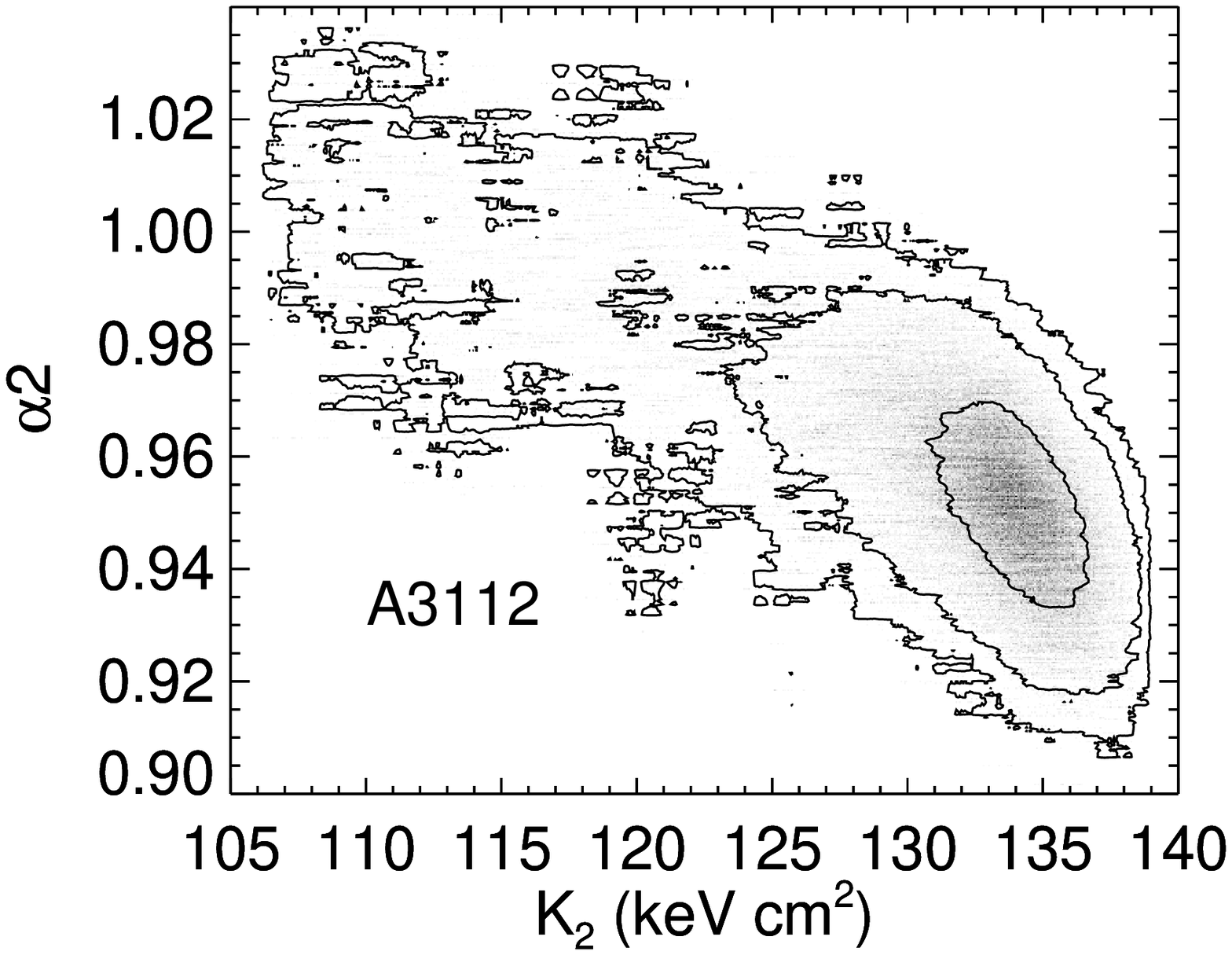}
  \includegraphics[width=0.32\textwidth]{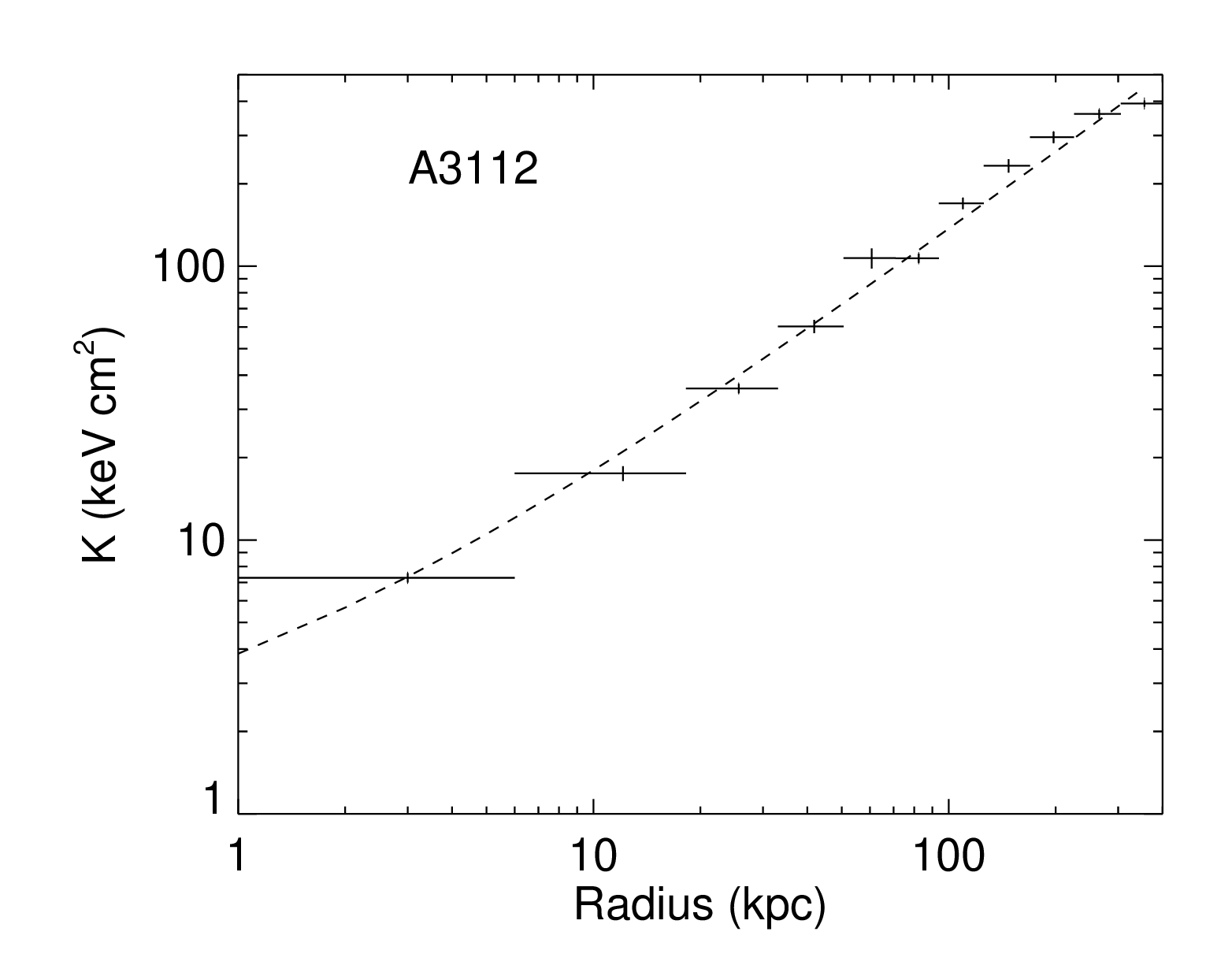}\\
  \includegraphics[width=0.32\textwidth]{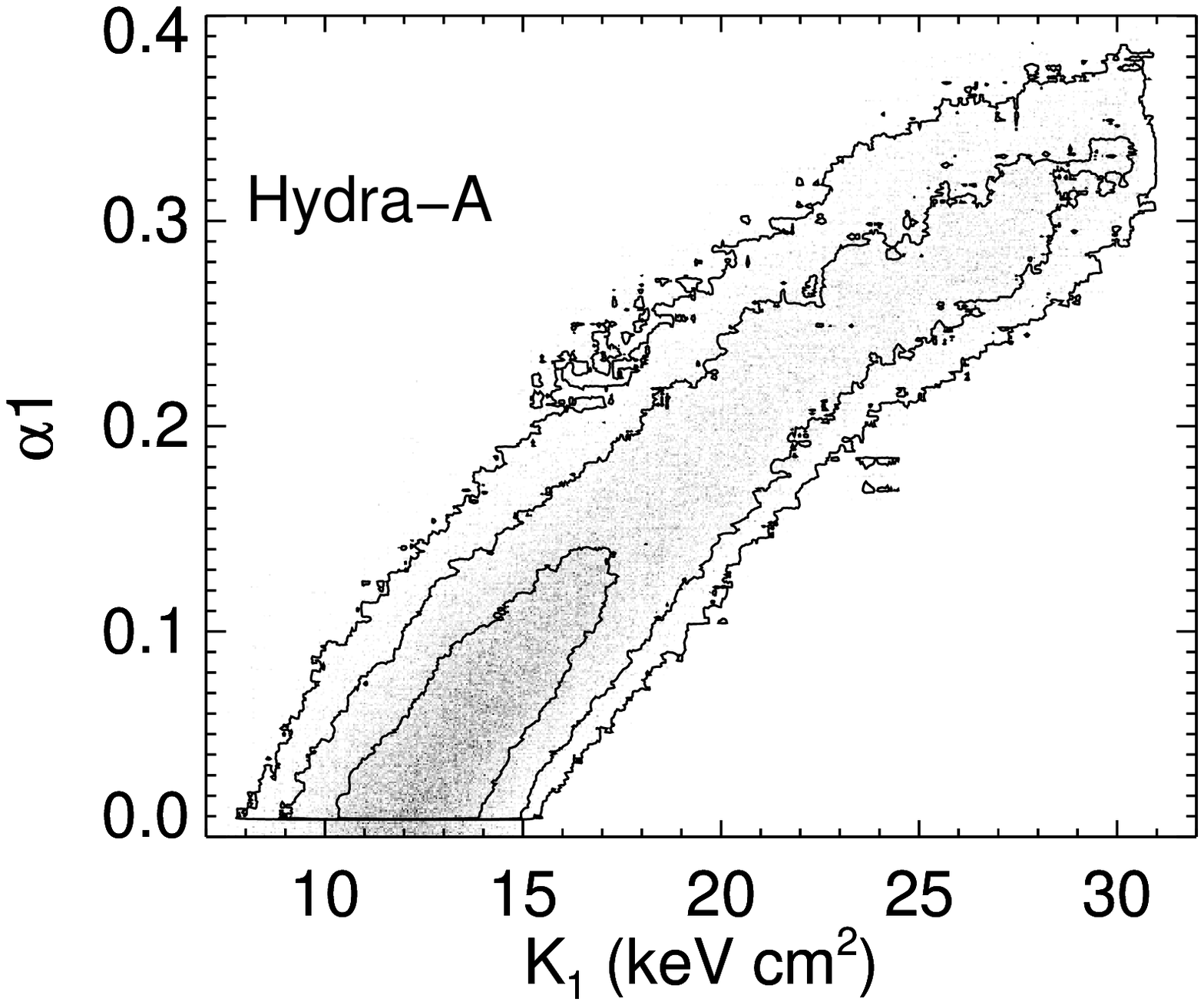}
  \includegraphics[width=0.32\textwidth]{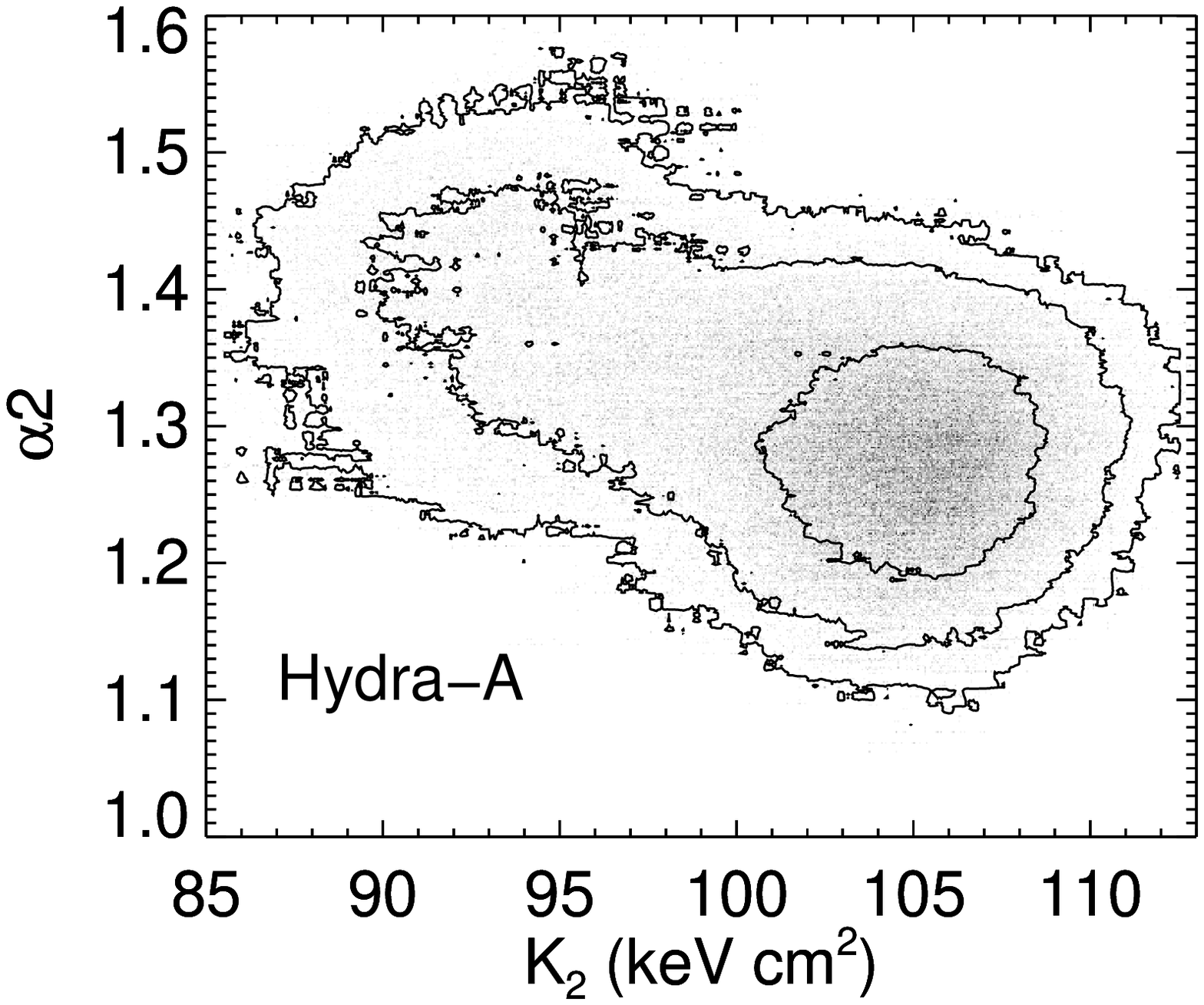}
  \includegraphics[width=0.32\textwidth]{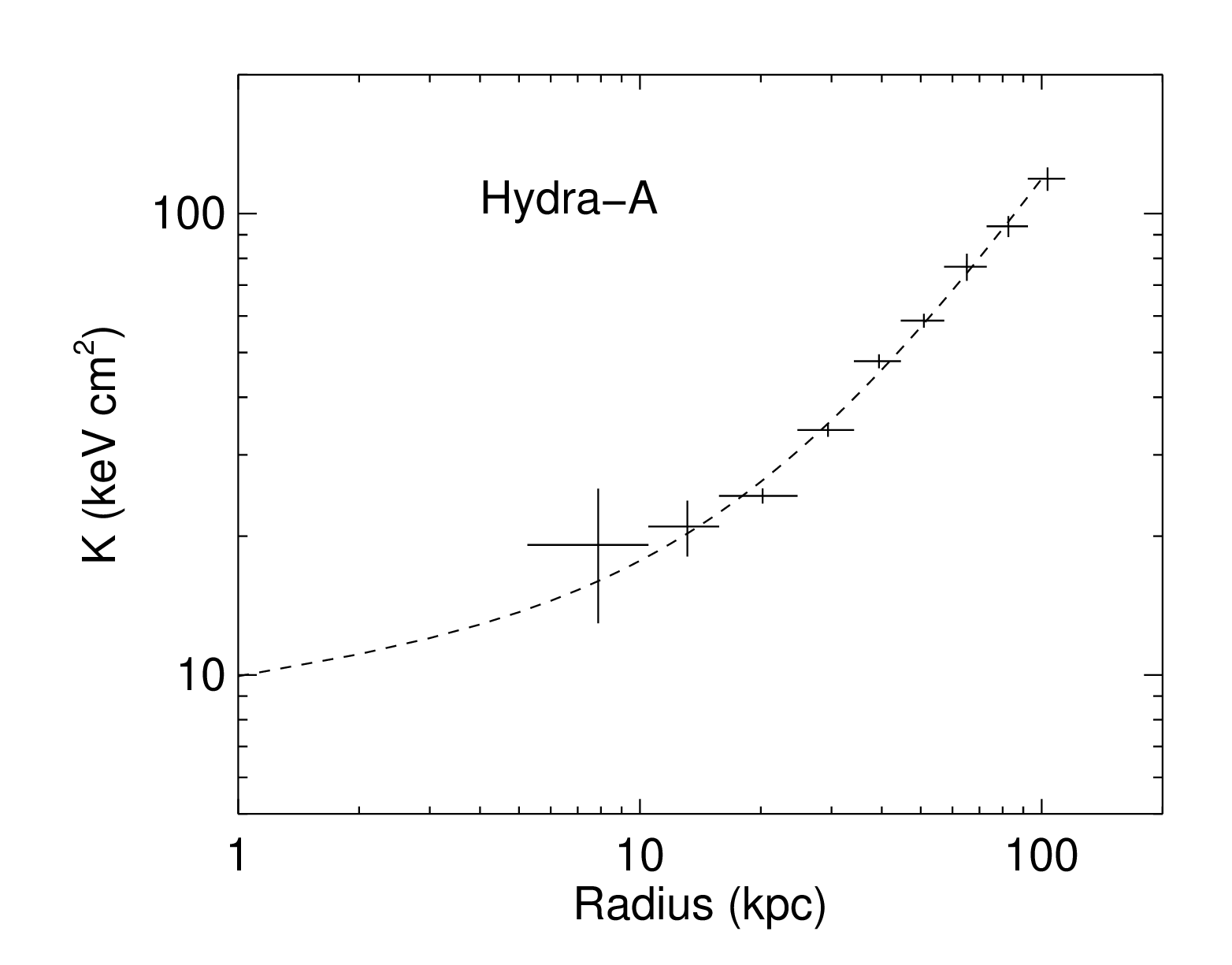}\\
  \includegraphics[width=0.32\textwidth]{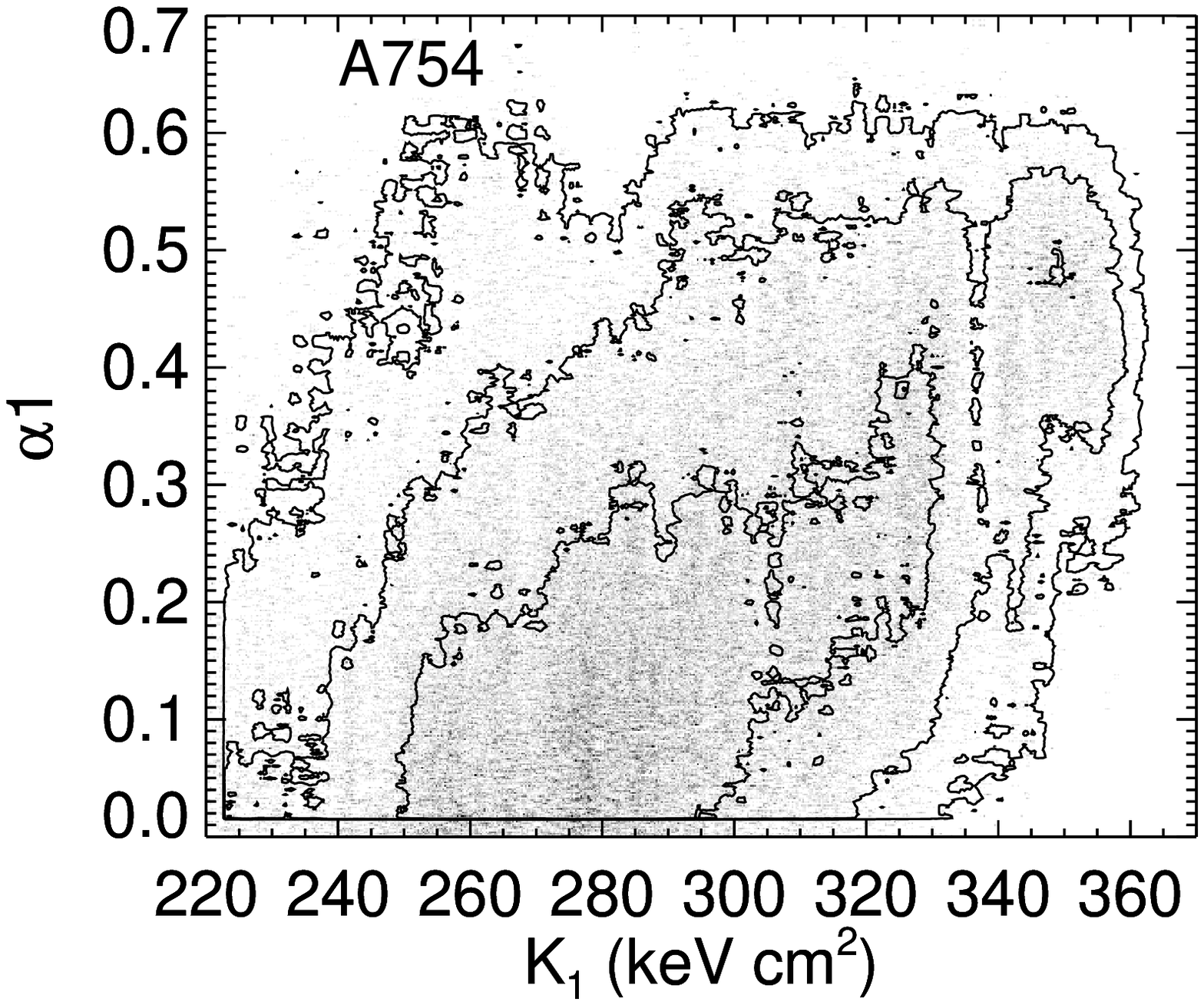}
  \includegraphics[width=0.32\textwidth]{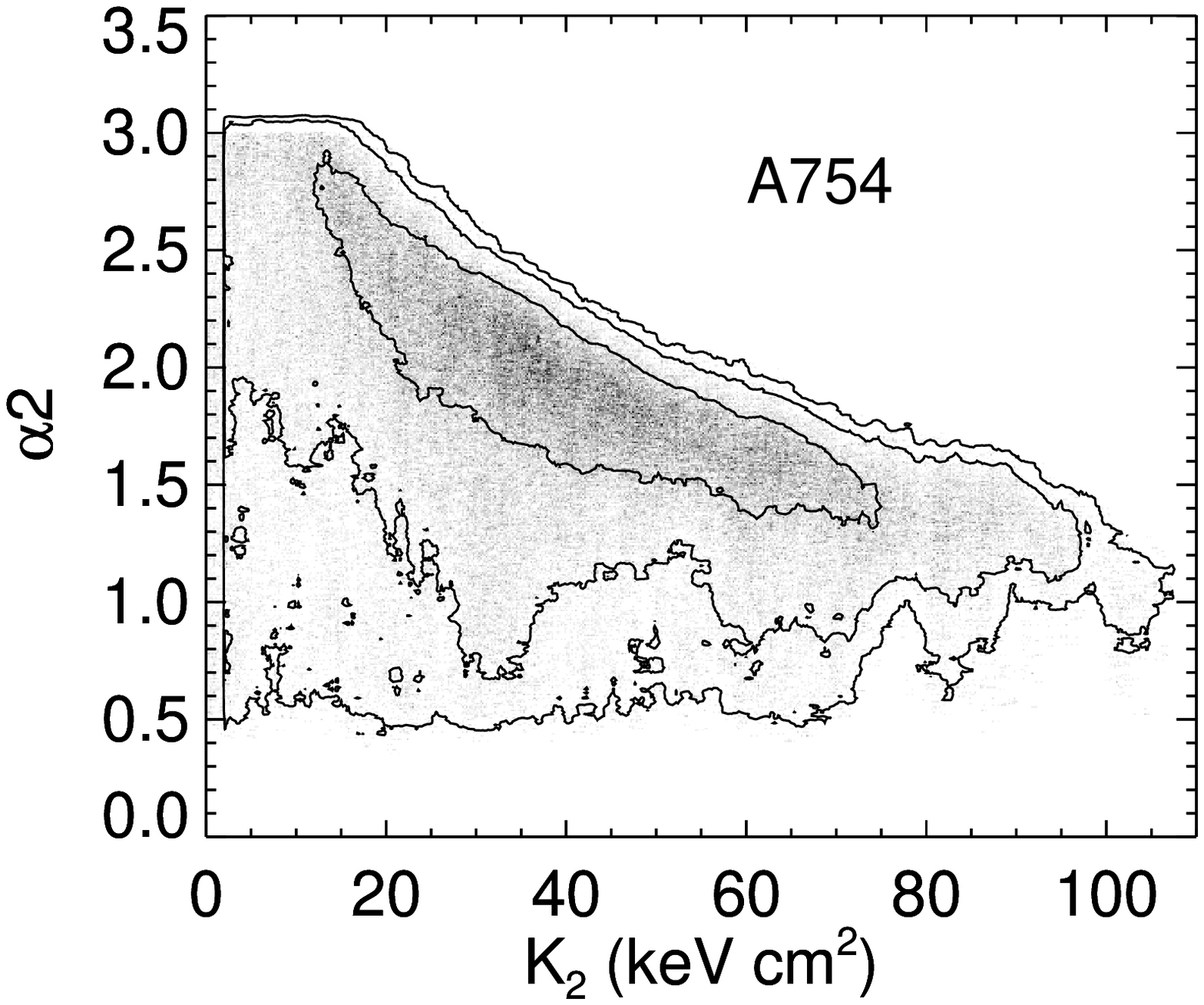}
  \includegraphics[width=0.32\textwidth]{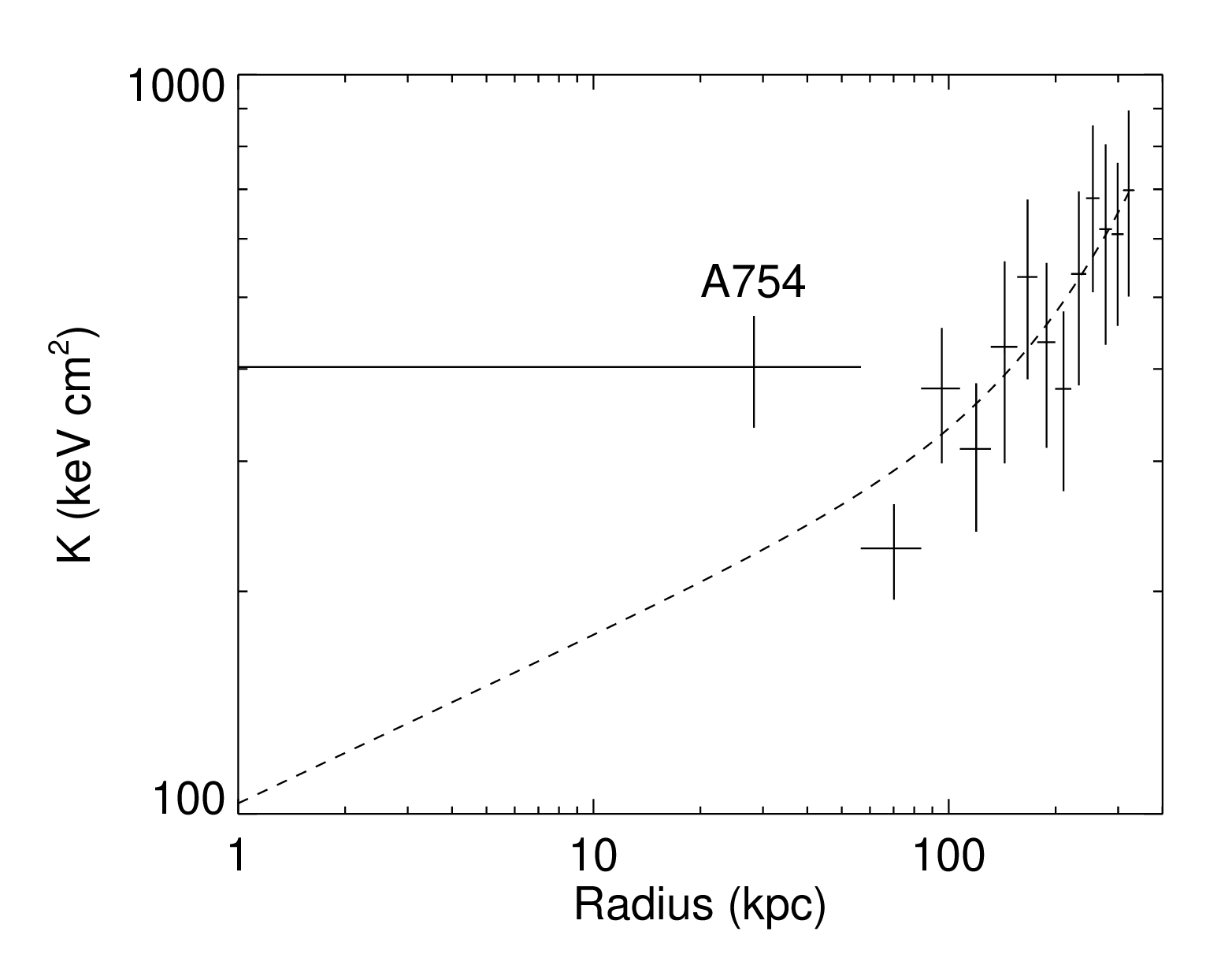}\\
  \includegraphics[width=0.32\textwidth]{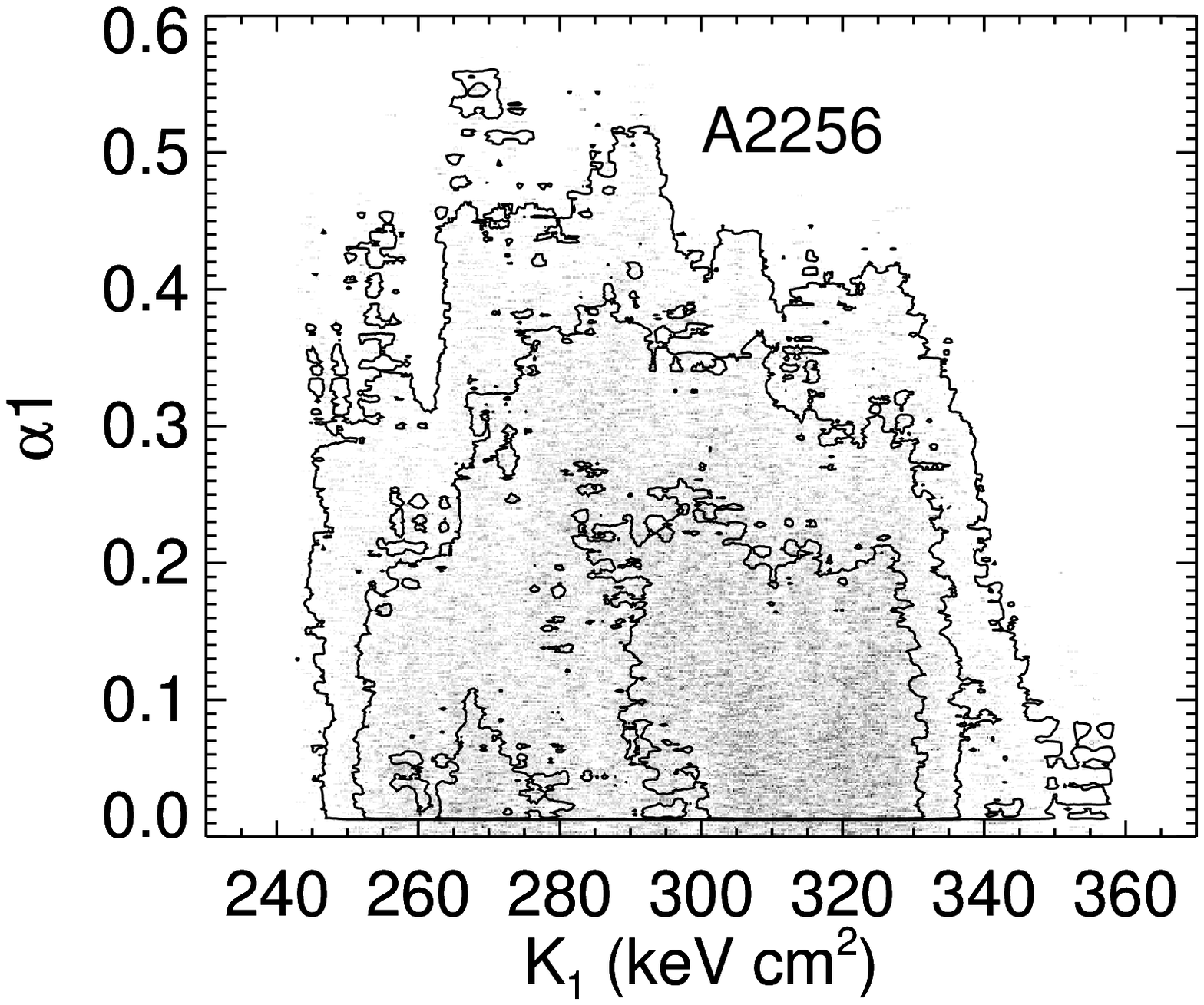}
  \includegraphics[width=0.32\textwidth]{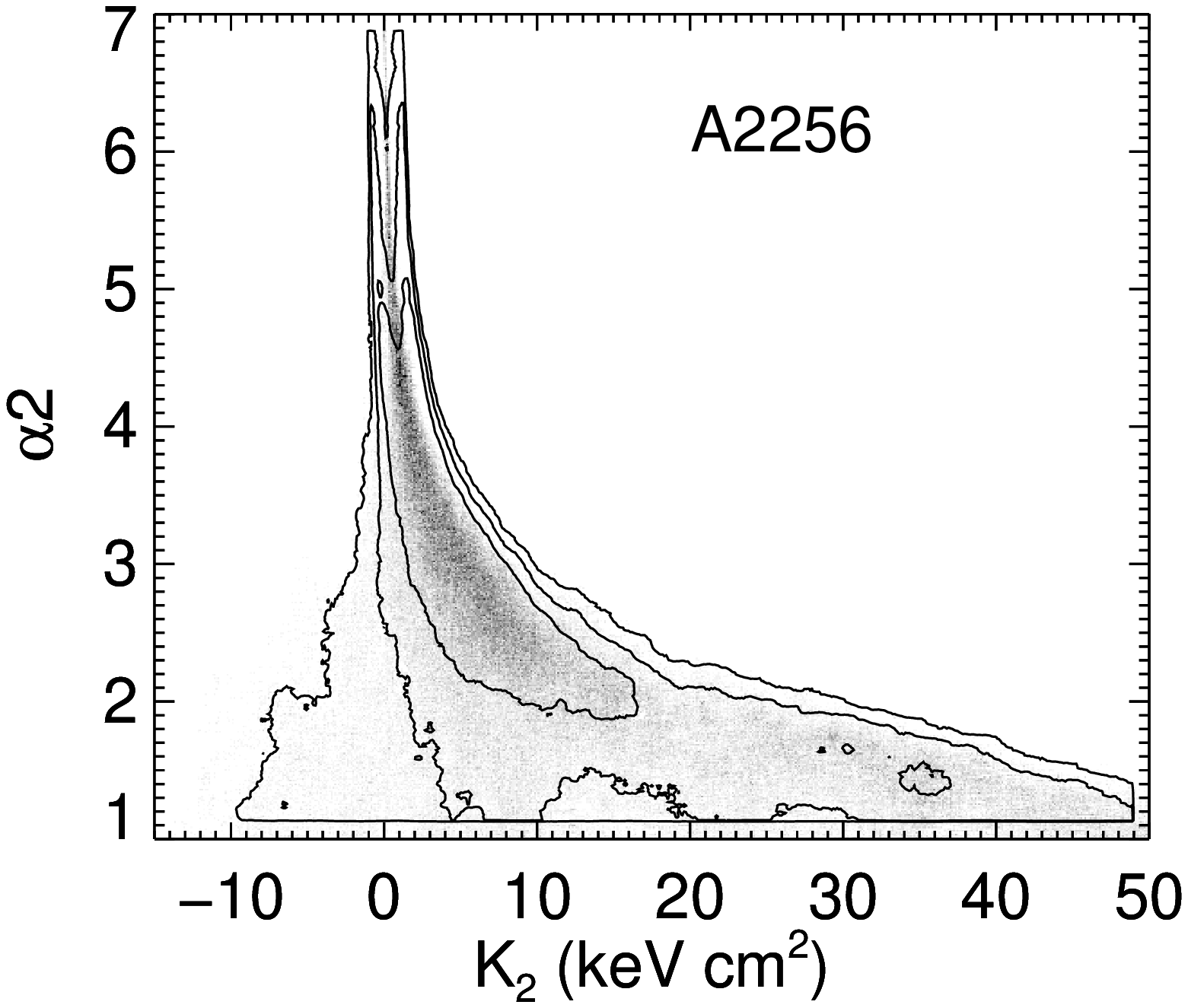}
  \includegraphics[width=0.32\textwidth]{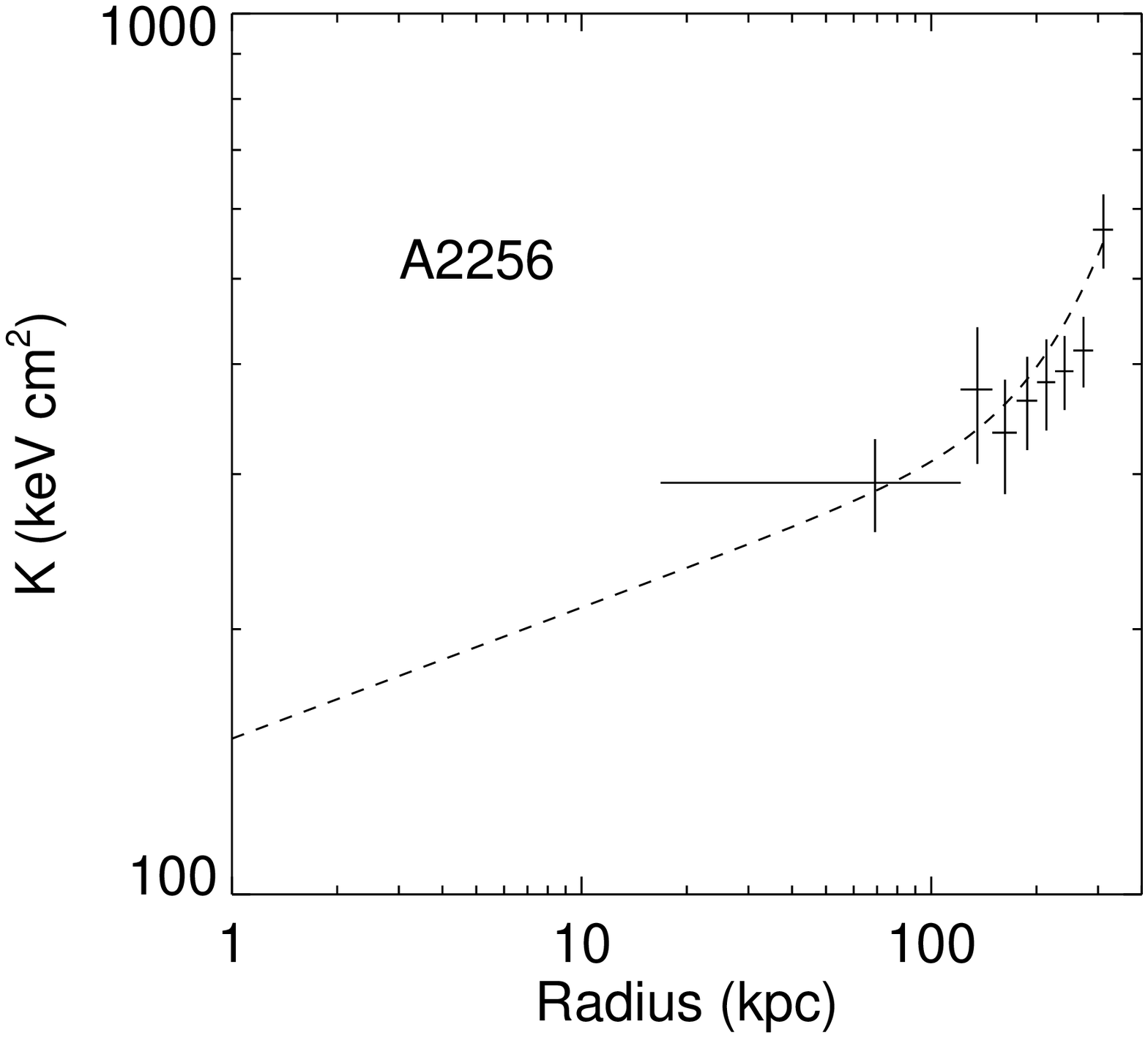}\\
  \includegraphics[width=0.32\textwidth]{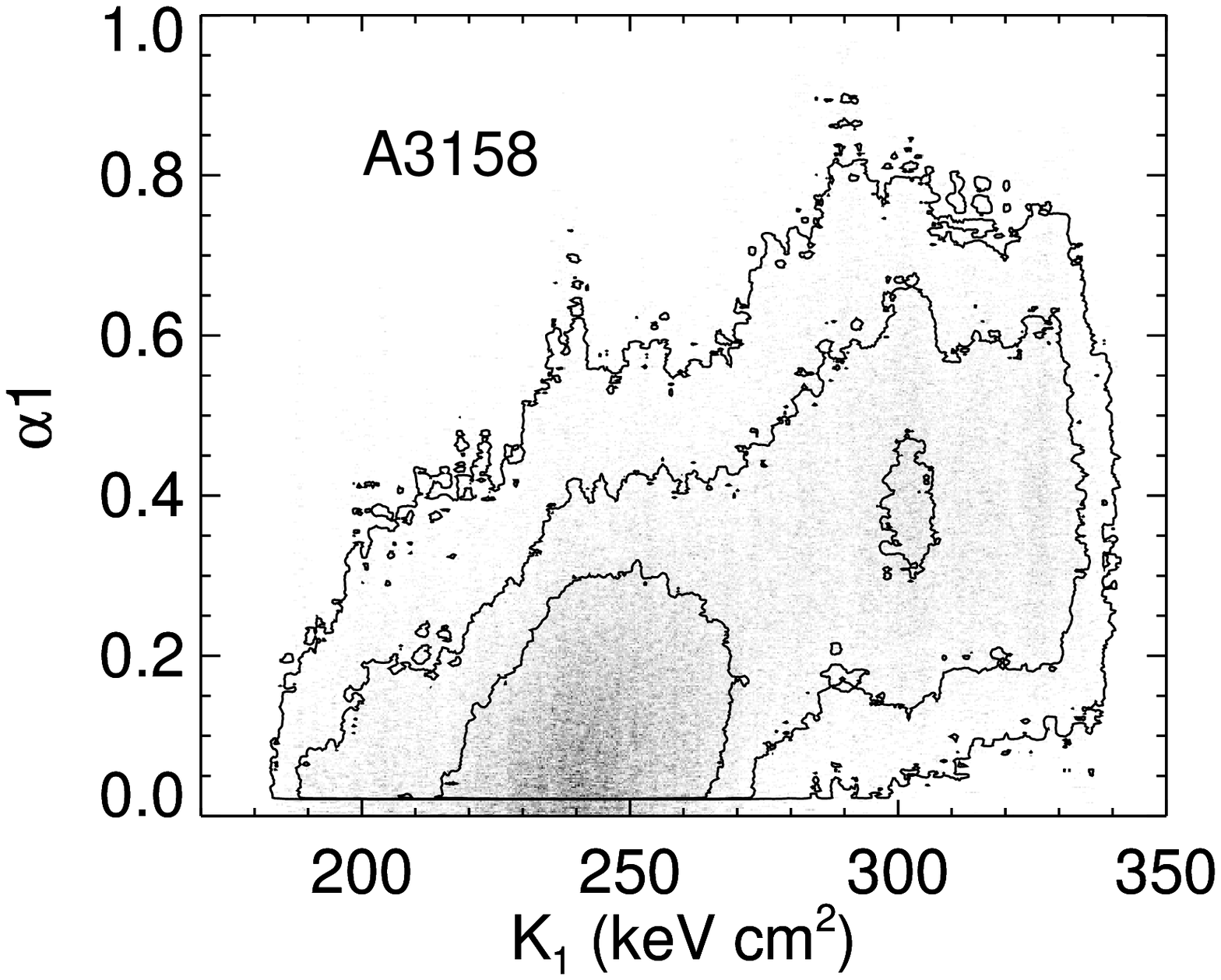}
  \includegraphics[width=0.32\textwidth]{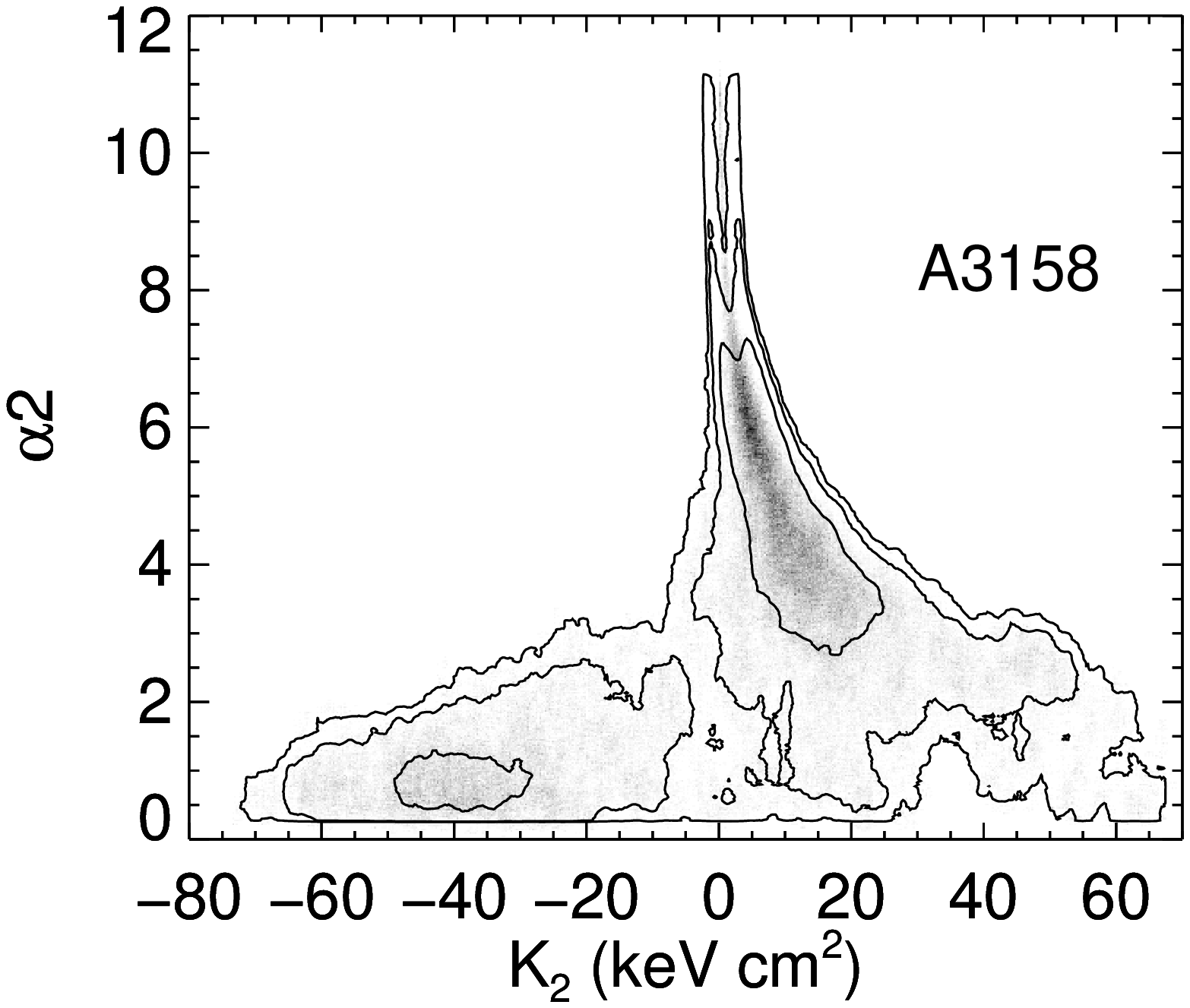}
  \includegraphics[width=0.32\textwidth]{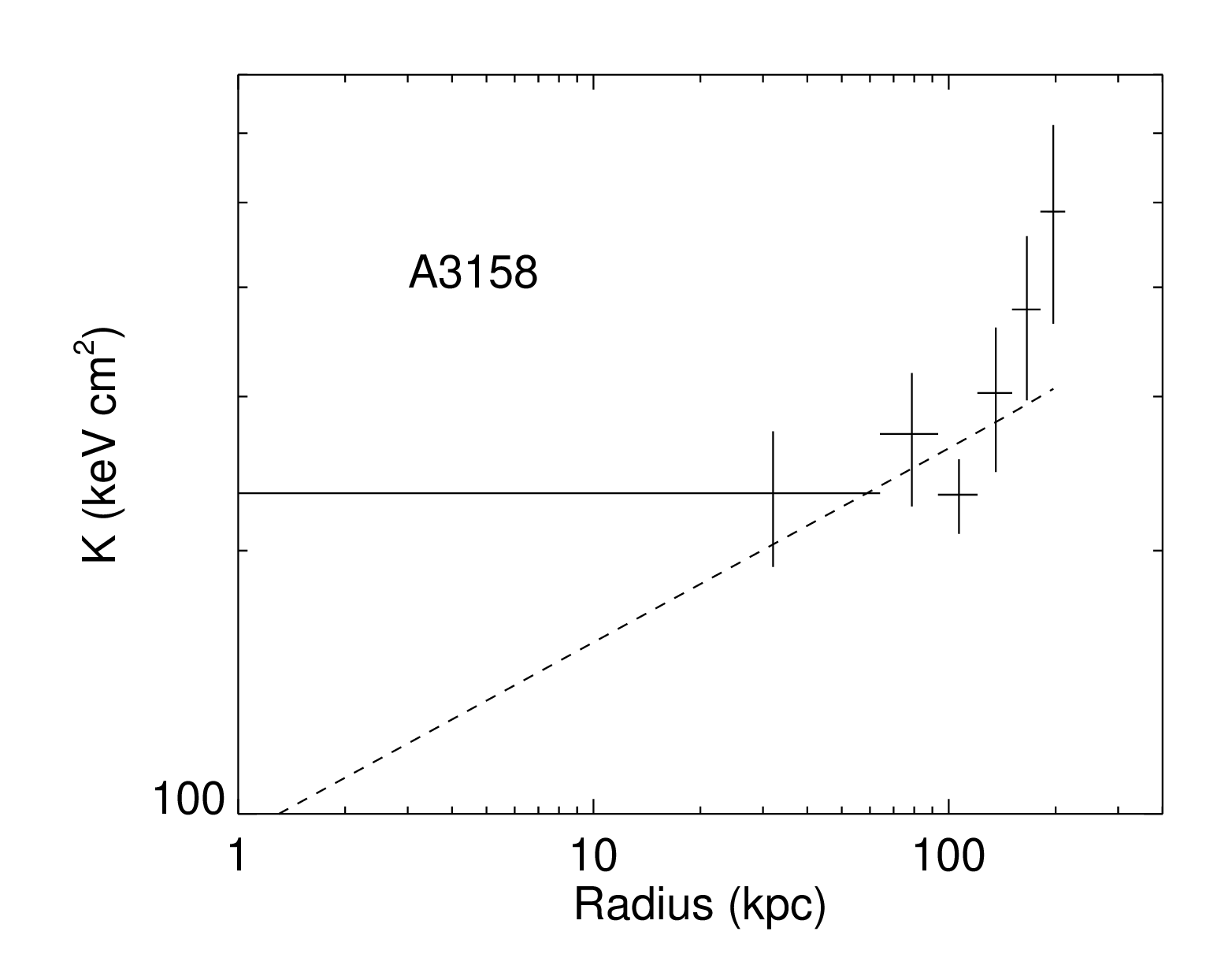}\\
  \caption{(Contd.)}
  \label{fig:dbl_pl_entr_fit}
\end{figure*}

\setcounter{figure}{14}
\begin{figure*}
 \centering
  \includegraphics[width=0.32\textwidth]{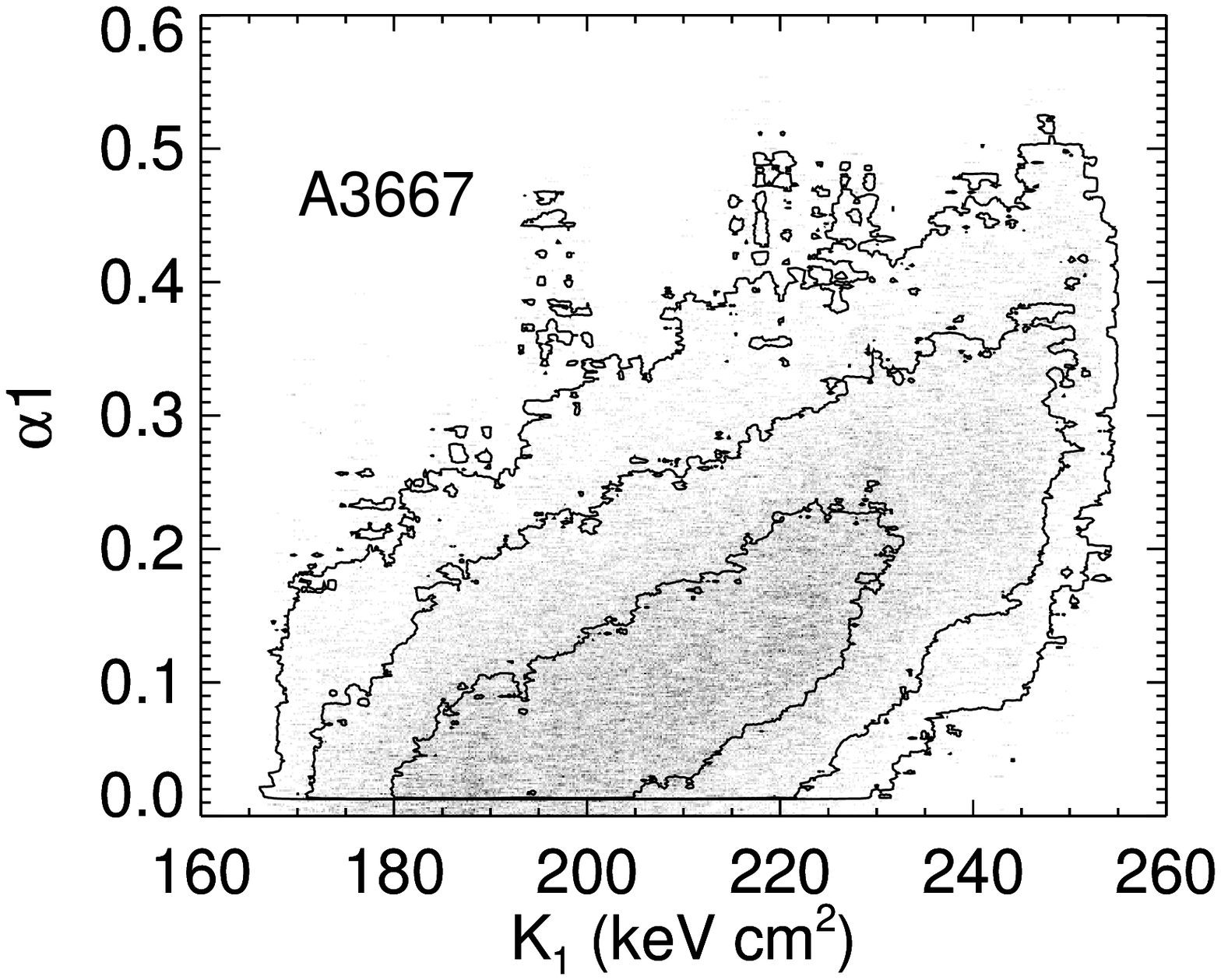}
  \includegraphics[width=0.32\textwidth]{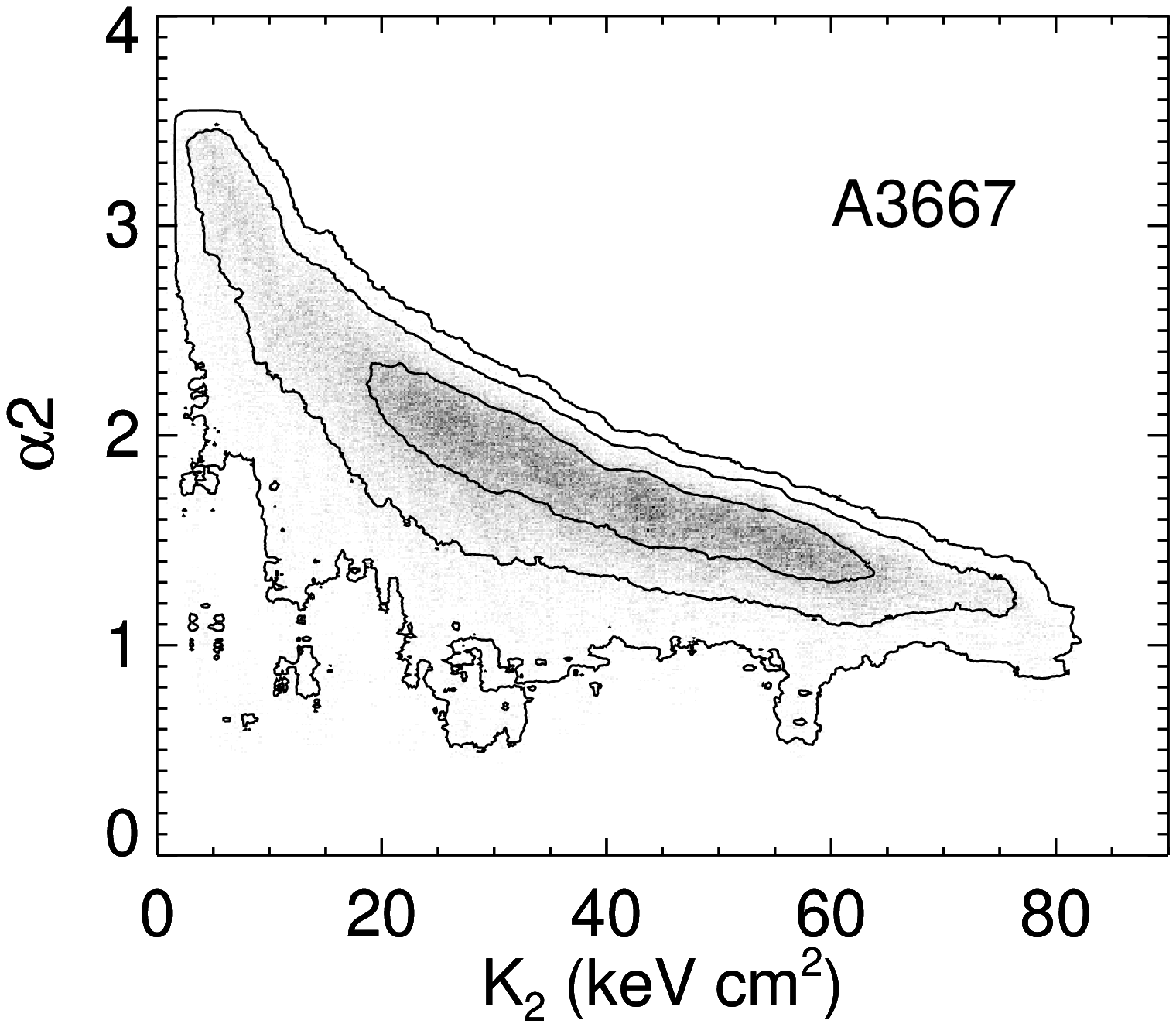}
  \includegraphics[width=0.32\textwidth]{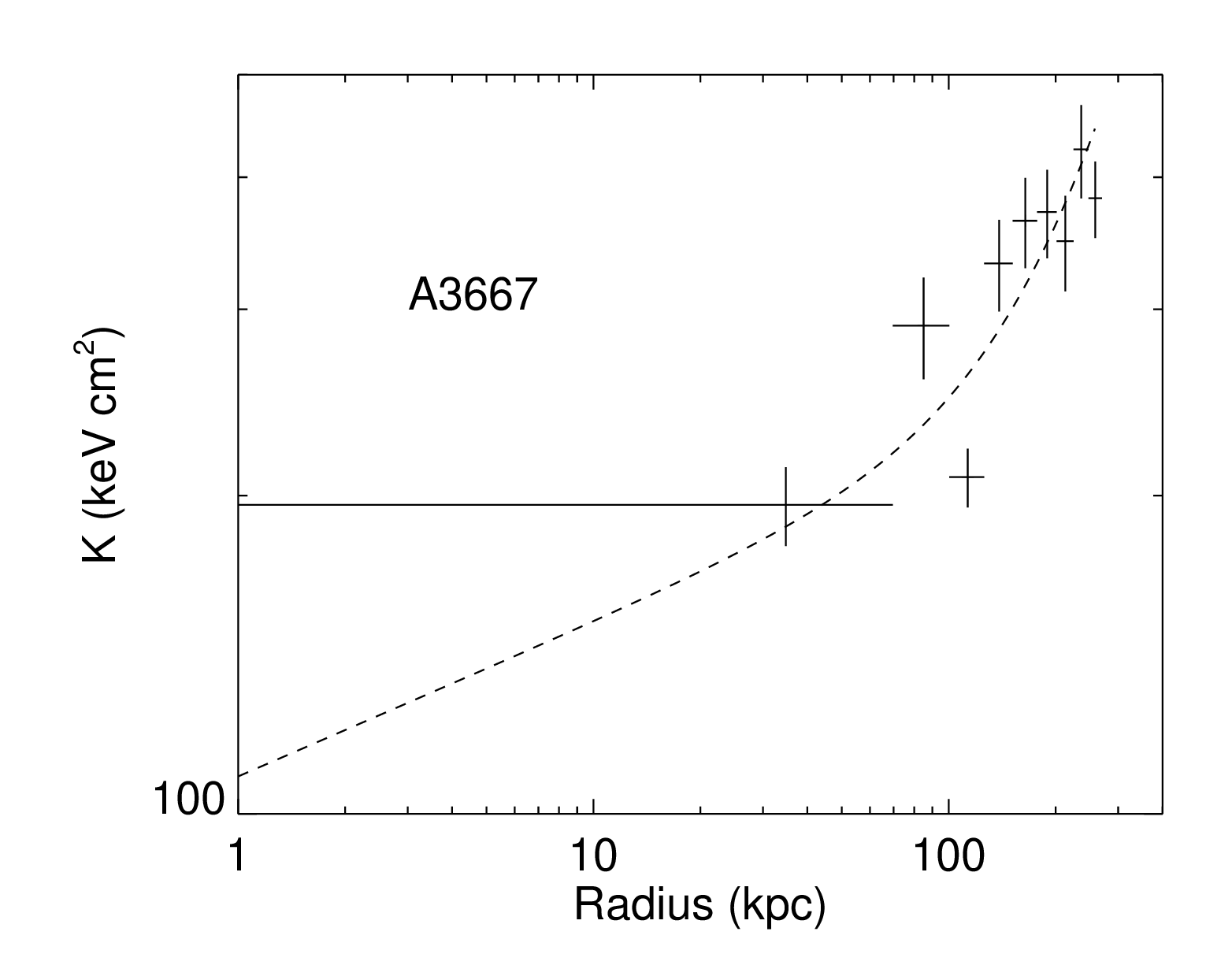}\\
  \includegraphics[width=0.32\textwidth]{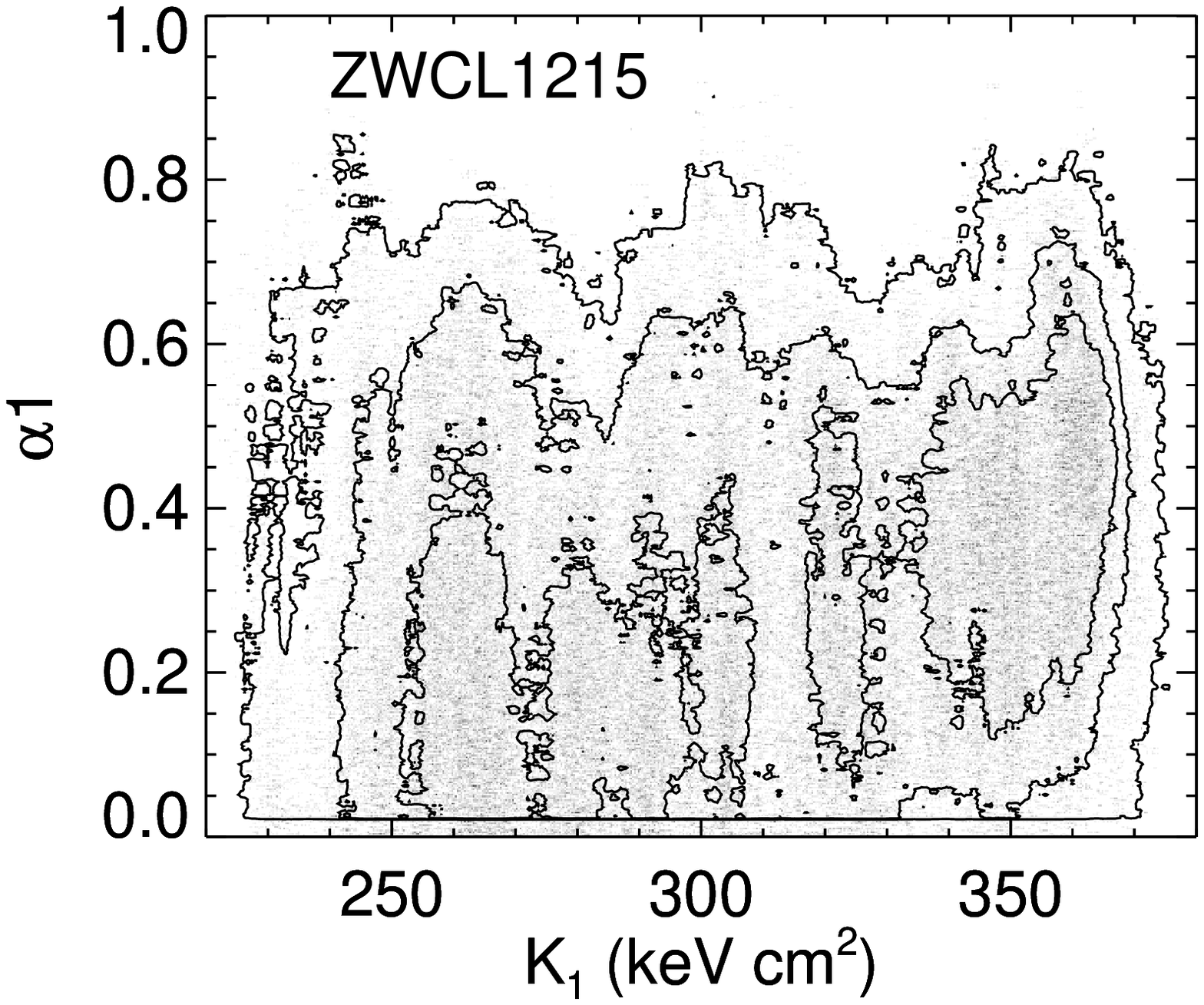}
  \includegraphics[width=0.32\textwidth]{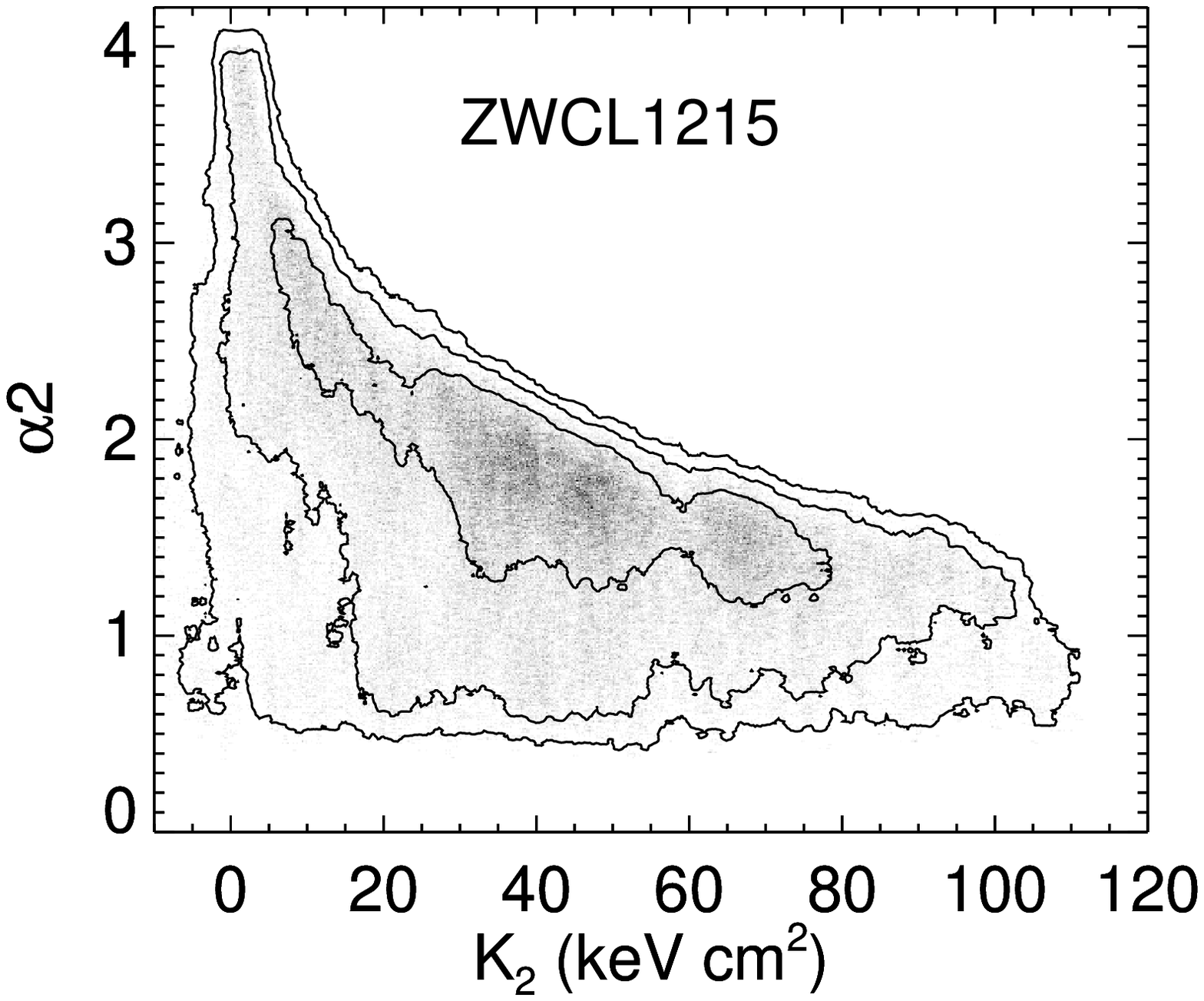}
  \includegraphics[width=0.32\textwidth]{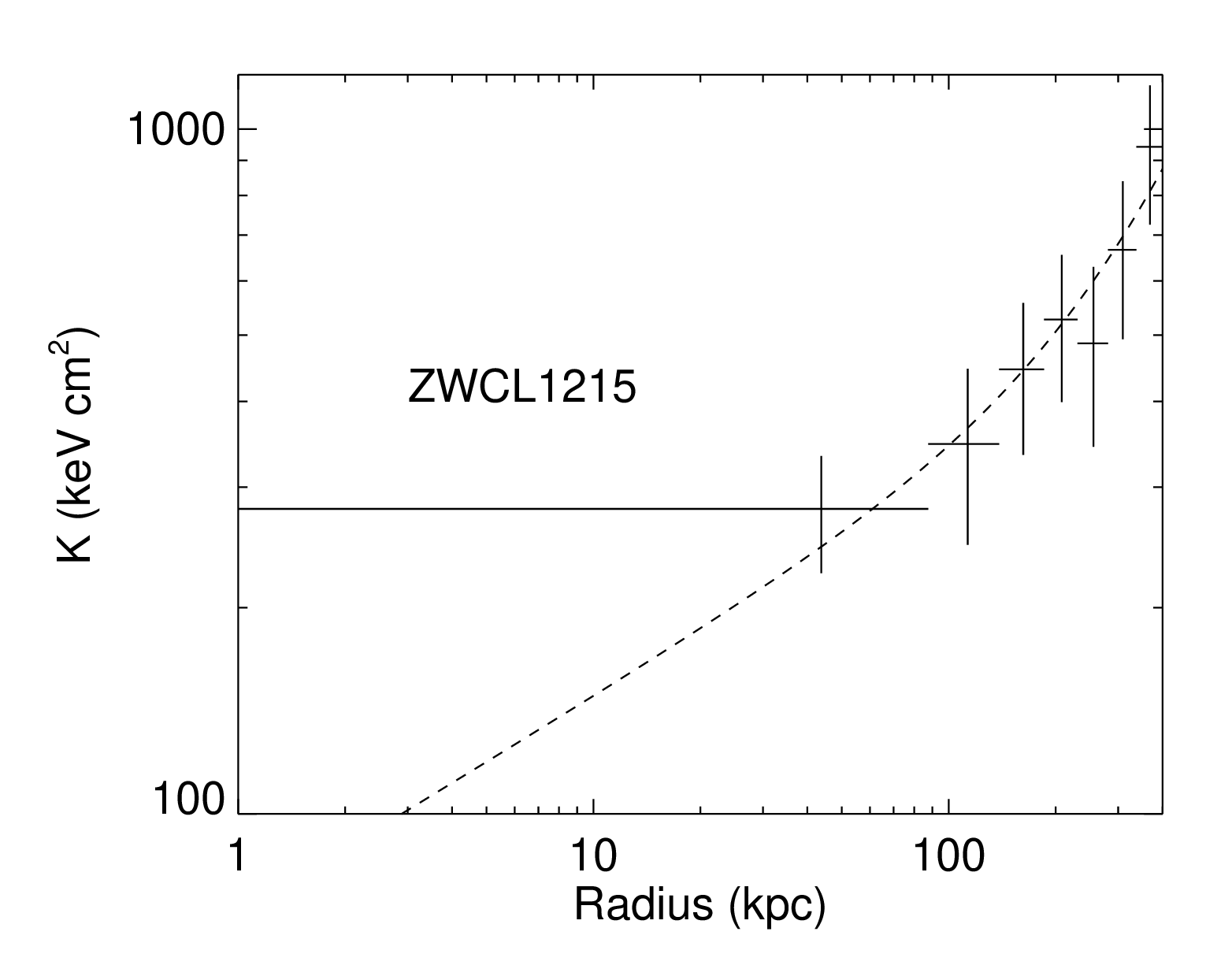}\\
  \caption{(Contd.)}
  \label{fig:dbl_pl_entr_fit}
\end{figure*}

\begin{table}
 \caption{Upper Panel: Comparison of results from single-power law (SPL) and flat-core (FC) models based on F-test for the PL sample.
 Lower Panel: Comparison of results from flat-core (FC) and double power law (DPL) models based on F-test for the complete sample.
 F-value is calculated as F$=((\chi^2_1-\chi^2_2)/(DOF_1-DOF_2))/(\chi^2_2/DOF_2)$. The last column gives the probability 
 that model 2 is an improvement over model 1.}
\label{tab:f_test_results}
\vskip 0.5cm
\centering
{\scriptsize
\begin{tabular}{c c c c c c c}
\hline
Cluster & $\chi^2_1$ & DOF$_1$ & $\chi^2_2$ & DOF$_2$ & F-value & Prob. (\%) \\
Name & & & & & & \\
\hline
\hline
 & (SPL) & & (FC) & & & \\
\hline
  A85 & 36.81  &  9 &  13.36 &  8 & 14.04 & 99.44 \\
 A478 & 37.10  &  14 &  22.23 & 13 & 8.70 & 98.87 \\
A2029  & 61.75  & 13 &  51.84 & 12 & 2.29 & 84.43 \\
A3112 & 164.16 &  9 & 101.92 &  8 & 4.89 & 94.20 \\
\hline
 & (FC) & & (DPL) & & & \\
\hline
  A85 & 13.36  &  8 &  44.59 &  7 & -4.90 & 0.00 \\
 A133 & 19.25  &  5 &  20.84 &  4 & -0.31 & 0.00 \\
 A478 & 22.23  & 13 &  28.80 & 12 & -2.73 & 0.00 \\
A1650 &  2.12  &  4 &   4.05 &  3 & -1.43 & 0.00 \\
A1795 & 11.66  & 11 &   5.90 & 10 &  9.76 & 98.92 \\
A2029 & 51.84  & 12 &  97.79 & 11 & -5.17 & 0.00 \\
A2142 &  3.42  &  6 &   7.20 &  5 & -2.63 & 0.00 \\
A2204 & 13.36  &  4 &  15.72 &  3 & -0.45 & 0.00 \\
A2244 &  4.28  &  4 &   6.18 &  3 & -0.92 & 0.00 \\
A2597 &  9.36  &  8 &  11.55 &  7 & -1.32 & 0.00 \\
A3112 & 101.92 &  8 & 127.68 &  7 & -1.41 & 0.00 \\
Hydra-A & 111.18 & 6 & 19.75 &  5 & 23.15 & 99.52 \\
 A754 & 10.70  & 10 &  11.79 &  9 & -0.83 & 0.00 \\
A2256 & 41.80  &  5 &  18.52 &  4 &  5.03 & 91.16 \\
A3158 &  5.07  &  3 &  10.72 &  2 & -1.05 & 0.00 \\ 
A3667 & 22.26 &   6 &  24.85 &  5 & -0.52 & 0.00 \\
ZWCL1215 & 4.75 & 5 &   1.64 &  4 &  7.59 & 94.88 \\
\hline
\end{tabular}}
\end{table}


\begin{thebibliography}{}

\bibitem[\protect\citeauthoryear{Allen et al.} {2011}]{all11}
Allen S. W., Everard A. E., \& Mantz A. B. 2011, ARAA, 49, 409

\bibitem[\protect\citeauthoryear{Ameglio et al.} {2007}]{ame07}
Ameglio S., Borgani S., Pierpaoli E., \& Dolag K. 2007, \mnras, 382, 397

\bibitem[\protect\citeauthoryear{Anders 
\& Grevesse}{1989}]{and89} Anders E., Grevesse N., 1989, GeCoA, 53, 197 

\bibitem[\protect\citeauthoryear{Anders 
\& Ebihara}{1982}]{and82} Anders E., Ebihara M., 1982, GeCoA, 46, 2363

\bibitem[\protect\citeauthoryear{Anderson \& Bregman} {2010}]{and10}
Anderson M. \& Bregman J. N. 2010, ApJ, 714, 320

\bibitem[\protect\citeauthoryear{Arnaud}{1996}]{arn96} Arnaud 
K.~A., 1996, ASPC, 101, 17

\bibitem[\protect\citeauthoryear{Banerjee \& Sharma}{2014}]{ban14} 
Banerjee N., Sharma P., 2014, MNRAS, 443, 687

\bibitem[\protect\citeauthoryear{Arnaud}{2010}]{arn10} Arnaud 
M.,  Pratt G.~W., Piffaretti R., B\"ohringer H., Croston J.~H., \& Pointecouteau E., 2010, A\&A, 517, A92

\bibitem[\protect\citeauthoryear{Cavagnolo et al.} {2009}]{cav09}
Cavagnolo K. W., Donahue M., Voit G. M., \& Sun M. 2009, ApJS, 182, 12 

\bibitem[\protect\citeauthoryear{Choudhury \& Sharma}{2015}]{cho15} 
Choudhury P.~P., Sharma P. 2016, \mnras, 457, 2554

\bibitem[\protect\citeauthoryear{Colless et al.} {2001}]{col01}
Collies M., Dalton G., Maddox S. et al. 2001, MNRAS, 328, 1039

\bibitem[\protect\citeauthoryear{Croston et al.} {2006}]{cro06}
Croston J.~H., Arnaud M.~A., Pointecouteau E., Pratt G.~W., 2006, A\&A, 459, 1007

\bibitem[\protect\citeauthoryear{Croston et al.} {2008}]{cro08}
Croston J. H., Pratt G. W., B\"ohringer H. et al. 2008, A\&A, 487, 431

\bibitem[\protect\citeauthoryear{Davis et al.} {1985}]{dav85}
Davis M., Efstathiou G., Frenk C. S., \& White S. D. M. 1985, ApJ, 292, 371

\bibitem[\protect\citeauthoryear{Donahue et 
al.}{2006}]{don06} Donahue M., Horner D.~J., Cavagnolo K.~W., 
Voit G.~M., 2006, ApJ, 643, 730 

\bibitem[\protect\citeauthoryear{Eke et al.} {1996}]{eke96}
Eke V. R., Cole S., \& Frenk C. S. 1996, MNRAS, 282, 263

\bibitem[\protect\citeauthoryear{Gaspari et al.} {2013}]{gas13}
Gaspari M., Ruszkowski M., \& Oh S.~P., 2013, MNRAS, 432, 3401 

\bibitem[\protect\citeauthoryear{Hastings}{1970}]{hast70}
Hastings W. K., 1970, Biometrika, 57, 97 

\bibitem[\protect\citeauthoryear{Hudson et al.} {2010}]{hud10}
Hudson D. S., Mittal R., Reiprich T. H. et al. 2010, A\&A, 513, A37

\bibitem[\protect\citeauthoryear{Kaiser et al.} {1991}]{kai91}
Kaiser N., 1991, ApJ, 383, 104

\bibitem[\protect\citeauthoryear{Kriss, Cioffi, 
\& Canizares}{1983}]{kri83} Kriss G.~A., Cioffi D.~F., Canizares C.~R., 1983, ApJ, 272, 439 

\bibitem[\protect\citeauthoryear{McCourt et al.} {2012}]{mcc12}
McCourt M., Sharma P., Quataert E., Parrish I. J. 2012, MNRAS, 419, 3319

\bibitem[\protect\citeauthoryear{McNamara \& Nulsen} {2007}]{mcn07}
McNamara B. R., Nulsen P. E. J. 2007, ARAA, 45, 117

\bibitem[\protect\citeauthoryear{Meece et al.}{2015}]{mee15} 
Meece G.~R., O'Shea B. W., \& Voit, G. M., 2015, ApJ, 808, 43

\bibitem[\protect\citeauthoryear{Metropolis et al.}{1953}]{met53} Metropolis, N., 
Rosenbluth, A.~W., Rosenbluth, M.~N., Teller, A.~H., 
\& Teller, E.\ 1953, The Journal of Chemical Physics, 21, 1087 

\bibitem[\protect\citeauthoryear{Morrison 
\& McCammon}{1983}]{mor83} Morrison R., McCammon D., 1983, ApJ, 270, 119 

\bibitem[\protect\citeauthoryear{Mittal et al.}{2009}]{mit09}
Mittal R., Hudson D. S., Reiprich T. H., \& Clarke T. 2009, A\&A, 501, 835

\bibitem[\protect\citeauthoryear{Navarro et al.}{1996}]{nav96}
Navarro J.~F., Frenk C.~S., \& White S.~D.~M., 1996, 462, 563

\bibitem[\protect\citeauthoryear{Panagoulia et al.}{2014}]{pan14}
Panagoulia E. K., Fabian A. C., \& Sanders J. S. 2014, MNRAS, 438, 2341

\bibitem[\protect\citeauthoryear{Pizzolato et al.} {2003}]{piz03}
Pizzolato F., Molendi S., Ghizzardi S., De Grandi S. 2003, \apj, 592, 62

\bibitem[\protect\citeauthoryear{Pizzolato \& Soker} {2005}]{piz05}
Pizzolato F. \& Soker N., 2005, ApJ, 632, 821

\bibitem[\protect\citeauthoryear{Prasad et al.} {2015}]{pra15}
Prasad D., Sharma P., \& Babul A. 2015, ApJ, 811, 108

\bibitem[\protect\citeauthoryear{Pratt et al.} {2010}]{pra10}
Pratt G. W., Arnaud M., Piffaretti R. et al. 2010, A\&A, 511, A85

\bibitem[\protect\citeauthoryear{Press et al.}{2003}]{press03} William H. Press, Saul A. Teukolsky, William T. Vetterling, and Brian P. Flannery. 2007. Numerical Recipes 3rd Edition: The Art of Scientific Computing (3 ed.). Cambridge University Press, New York, NY, USA

\bibitem[\protect\citeauthoryear{Putman et al.}{2012}]{put12}
Putman M. E., Peek J. E. G., \& Joung, M. R. 2012, ARAA, 2012, 50, 491

\bibitem[\protect\citeauthoryear{Rafferty et al.} {2008}]{raf08}
Rafferty D.~A., McNamara B.~R., Nulsen P.~J.~R., 2008, ApJ, 687, 899

\bibitem[\protect\citeauthoryear{Rasia et al.} {2006}]{ras06}
Rasia E., Ettori S., Moscardini L., Mazzotta P., Brogan S., Dolag K., Tormen G., Cheng L.~M., \& Diaferio A., 2006, MNRAS, 369, 2013

\bibitem[\protect\citeauthoryear{Rossetti \& Molendi} {2010}]{ros10}
Rossetti M. \& Molendi S. 2010, A\&A, 520, A83

\bibitem[\protect\citeauthoryear{Russell, Ponman, 
\& Sanderson}{2007}]{russ07} Russell P.~A., Ponman T.~J., Sanderson A.~J.~R., 2007, MNRAS, 378, 1217

\bibitem[\protect\citeauthoryear{Russell, Sanders, 
\& Fabian}{2008}]{russ08} Russell H.~R., Sanders J.~S., Fabian A.~C., 2008, MNRAS, 390, 1207 

\bibitem[\protect\citeauthoryear{Sanders et 
al.}{2014}]{san14} Sanders J.~S., Fabian A.~C., 
Hlavacek-Larrondo J., Russell H.~R., Taylor G.~B., Hofmann F., Tremblay G., 
Walker S.~A., 2014, MNRAS, 444, 1497 

\bibitem[\protect\citeauthoryear{Sanderson et 
al.}{2014}]{san09} Sanderson J.~R., O'Sullivan E., \& Ponman T. J., 2009, MNRAS, 395, 764 

\bibitem[\protect\citeauthoryear{Santos et al.} {2008}]{san08}
Santos J. S., Rosati P., Tozzi P., et al. 2008, A\&A, 483, 35

\bibitem[\protect\citeauthoryear{Santos et 
al.}{2010}]{san10} Santos J.~S., Tozzi P., Rosati P., B{\"o}hringer H., 2010, A\&A, 521, A64 

\bibitem[\protect\citeauthoryear{Sharma et al.} {2012a}]{sha12a}
Sharma P., McCourt M., Quataert E., Parrish I. J. 2012,  MNRAS, 420, 3174

\bibitem[\protect\citeauthoryear{Sharma et al.} {2012b}]{sha12b}
Sharma P., McCourt M., Parrish I. J., \& Quataert, E. 2012,  MNRAS, 427, 1219

\bibitem[\protect\citeauthoryear{Smith et al.}{2001}]{smi01} 
Smith R.~K., Brickhouse N.~S., Liedahl D.~A., Raymond J.~C., 2001, ApJ, 
556, L91

\bibitem[\protect\citeauthoryear{Sun} {2009}]{sun09}
Sun M., 20009, ApJ, 704, 1586

\bibitem[\protect\citeauthoryear{Tozzi \& Norman} {2001}]{toz01}
Tozzi P. \& Norman C.  2001, ApJ, 546, 63

\bibitem[\protect\citeauthoryear{Tremblay et al.} {2015}]{tre15}
Tremblay G., O'Dea C.~P., Baum S.,  et al., 2015, MNRAS, 451, 3768

\bibitem[\protect\citeauthoryear{Vikhlinin et al.}{2005}]{vik05} 
Vikhlinin A., Markevitch M., Murray S.~S., Jones C., Forman W., Van Speybroeck L., 2005, ApJ, 628, 655 

\bibitem[\protect\citeauthoryear{Voit et al.}{2003}]{voi03} 
Voit G.~M., Balogh M.~L., Bower R. G., Lacey C. G., Bryan G.~L., 2005, MNRAS, 364, 909

\bibitem[\protect\citeauthoryear{Voit \& Donahue} {2015}]{voi15a}
Voit G. M., Donahue M., 2015, ApJL, 799, L1

\bibitem[\protect\citeauthoryear{Voit et al.} {2015}]{voi15}
Voit G. M., Donahue M., Bryan G. L., \& McDonald M., 2015, Nature, 519, 203

\bibitem[\protect\citeauthoryear{Wagh et al.} {2014}]{wag14}
Wagh B., Sharma P., \& McCourt M., 2014, MNRAS, 439, 2822
  
\bibitem[\protect\citeauthoryear{White \& Rees}{1978}]{whi78}
White S. D. M. \& Rees M. J. 1978, 183, 341

\end{thebibliography}
\end{document}